\documentclass[11pt]{article}
\usepackage{amsfonts,amsmath,amssymb,stmaryrd,amstext,latexsym,theorem,bbm,tabls,multirow,hhline,amscd}
\usepackage{graphics}
\usepackage{rotating}
\usepackage{pstricks}

\makeatletter
\@addtoreset{equation}{section}

\newtheorem{lemma}{Lemma}[section]
\newtheorem{define}[lemma]{Definition}
\newenvironment{definition}{\begin{define}\rm}{\end{define}}
\newtheorem{note}[lemma]{Notation}
\newenvironment{notation}{\begin{note}\rm}{\end{note}}
\newtheorem{theo}{Theorem}[section]
\newenvironment{theorem}{\begin{theo}\rm}{\end{theo}}
\newtheorem{conj}{Conjecture}[section]
\newenvironment{conjecture}{\begin{conj}\rm}{\end{conj}}
\newtheorem{rem}{Remark}[section]

\makeatother
\newcommand\ds{\displaystyle}
\def\Id{{\rm 1\kern-.28em I}}
\def\D{{\rm /\kern-.65em {\cal D}}}
\def\a{{\alpha}}
\def\b{{\beta}}
\def\c{{\gamma}}
\def\d{{\delta}}
\def\La{{\Lambda}}
\def\g{{\mathfrak{g}}}
\def\h{{\mathfrak{h}}}
\def\z{{\mathfrak{z}}}
\def\k{{\mathfrak{k}}}
\def\p{{\mathfrak{p}}}
\def\n{{\mathfrak{n}}}
\def\u{{\mathfrak{u}}}
\def\an{{\mathfrak{a}}}
\def\dn{{\mathfrak{d}}}
\def\e{{\mathfrak{e}}}

\def\sl{{\mathfrak{sl}}}
\def\so{{\mathfrak{so}}}
\def\su{{\mathfrak{su}}}
\def\R{{\mathbbm{R}}}
\def\Z{{\mathbbm{Z}}}
\def\C{{\mathbbm{C}}}
\def\N{{\mathbbm{N}}}
\def\O{{\mathbbm{O}}}
\def\Q{{\mathbbm{Q}}}
\def\Ad{{\mathrm{Ad}}}
\def\ad{{\mathrm{ad}}}
\def\Span{{\mathrm{Span}}}
\def\Tr{{\mathrm{Tr}}}
\def\eten{{\mathfrak{e}_{10}}}
\def\ginv{{\mathfrak{g}_{\text{inv}}}}

\begin{document}

\begin{titlepage}
 
$\mbox{ }$
\vspace{-1cm}
\begin{flushright}
\begin{tabular}{l}
KUNS-2025\\
NEIP-06-04\\
hep-th/0607136\\
July 2006
\end{tabular}
\end{flushright}

~~\\
~~\\
~~\\

\vspace*{0cm}
    \begin{Large}
       \vspace{1.5cm}
       \begin{center}
        {\bf Hidden Borcherds symmetries in $\Z_n$ orbifolds of M-theory \\[10pt]
        and magnetized D-branes in type 0' orientifolds}
       \end{center}
    \end{Large}

  \vspace{0.8cm}

\begin{center}
          Maxime B{\sc agnoud}$^{\dag}$\footnote{
e-mail address : bagnoud@gauge.scphys.kyoto-u.ac.jp} and
          Luca C{\sc arlevaro}$^{\ddag}$\footnote
           {
e-mail address : Luca Carlevaro@unine.ch} \\

\vspace*{0.8cm}
          $^{\dag}${\it Department of Physics, Kyoto University,
Kyoto 606-8502, Japan}\\
\vspace*{0.5cm}
       $^{\ddag}$ {\it Institut de Physique, Universit\'e de Neuch\^atel,
   CH-2000 Neuch\^atel, Switzerland}

\end{center}

\vfill

\begin{abstract}
\noindent
We study $T^{11-D-q}\times T^q/\Z_n$ orbifold compactifications of eleven-dimensional 
supergravity and M-theory using a purely algebraic method. Given the description of 
maximal supergravities reduced on square tori as non-linear coset $\sigma$-models, 
we exploit the mapping between scalar fields of the reduced theory and directions
in the tangent space over the coset to construct the orbifold action as a non-Cartan 
preserving finite order inner automorphism of the complexified U-duality algebra. 
Focusing on the exceptional serie of Cremmer-Julia groups, we compute the residual 
U-duality symmetry after orbifold projection and determine the reality properties 
of their corresponding Lie algebras. We carry out this analysis as far as the hyperbolic 
$\eten$ algebra, conjectured to be a symmetry of M-theory. In this case the residual 
subalgebras are shown to be described by a special class of Borcherds and Kac-Moody 
algebras, modded out by their centres and derivations. Furthermore, we construct an 
alternative description of the orbifold action in terms of equivalence classes of shift
vectors, and, in $D=1$, we show that a root of $\eten$ can always be chosen as the class 
representative. Then, in the framework of the $E_{10|10}/K(E_{10|10})$ effective 
$\sigma$-model approach to M-theory near a spacelike singularity, we identify these roots 
with brane configurations stabilizing the corresponding orbifolds. In the particular case 
of $\Z_2$ orbifolds of M-theory descending to type 0' orientifolds, we argue that these 
roots can be interpreted as pairs of magnetized D9- and D9'-branes, carrying the 
lower-dimensional brane charges required for tadpole cancellation. More generally, we 
provide a classification of all such roots generating $\Z_n$ product orbifolds for 
$n\leqslant 6$, and hint at their possible interpretation.
\end{abstract}

\vfill
\end{titlepage}
\vfil\eject

\tableofcontents

\section{Introduction}

Hidden symmetries in toroidally compactified maximal supergravity theories
have been known for a long time, since the foundating works of Cremmer and
Julia~\cite{CremJulN8,Cremmer1,Cremmer2,Julia79,Julia82b}. In particular,
the bosonic part of $11D$ supergravity compactified on $T^{11-D}$ for
$10\geqslant D\geqslant 3$ was shown to possess a continuous exceptional 
$E_{11-D|11-D}(\R)$ global symmetry, provided all RR- and NS-NS fields are dualized to
scalars whenever possible \cite{CremJulLuPope1,CremJulLuPope2}.
This serie of exceptional groups also appear as symmetries of the action of type IIA
supergravity compactified on $T^{10-D}$, and the BPS states of the
compactified theory turn out to arrange into multiplets of the Weyl
group of $E_{11-D|11-D}(\R)$~\cite{LuPope1}. From the type IIA point of view, this
continuous symmetry does not preserve the weak coupling
regime in $g_s$, and is thus expected to be broken by quantum effects.
Nevertheless, the authors of~\cite{HullTownsend} have advocated that a discrete version
thereof, namely $E_{11-D|11-D}(\Z)$ remains as an exact quantum
symmetry of both $11D$ supergravity and type II theories, and thus might
provide a guideline for a better understanding of M-theory.

This exceptional serie of arithmetic groups can alternatively be obtained as
the closure of the perturbative T-duality symmetry of IIA theory
compactified on $T^{10-D}$, namely $SO(10-D,10-D,\Z)$, with the
discrete modular group of the $(11-D)$-torus of M-theory, namely 
$SL(11-D,\Z)$. In $D=9$, the latter turns into the expected S-duality
symmetry of the dual IIB\ string theory. In this perspective, the global 
$E_{11-D|11-D}(\Z)$ can be regarded as a unifying group encoding both a target-space 
symmetry, which relates apparently different string backgrounds endowed with isometries, 
and a rigid symmetry of the maximally symmetric space of compactification which
naturally contains a non-perturbative symmetry of type IIB string theory
(which is, by the way, also shared by the heterotic string compactified to
four dimensions). Furthermore, this so-called U-duality symmetry has been conjectured 
to extend to the moduli space of M-theory compactified on $T^{11-D}$ for 
$10\geqslant D\geqslant 3$. In particular, it was shown in~\cite{OberPiol2} how to
retrieve exact $R^{4}$ and $R^4 H^{4g-4}$ corrections as well as topological couplings 
\cite{Ob1}, from M-theory $E_{11-D|11-D}(\Z)$-invariant mass formulae.

In $D=2,1$, the dualization procedure mentioned above is not enough to derive 
the full U-duality symmetry, which has  been conjectured, already some time 
ago \cite{Julia81,Julia82a,Julia86} to be described by the Kac-Moody affine $\e_{9|10}$ 
and hyperbolic $\e_{10|10}$ split forms, that are characterized by an infinite 
number of positive (\textit{real}) and negative/null (\textit{imaginary}) norm roots. 
In a more recent perspective, $\e_{10|10}$ and the split form of $\e_{11}$ have been 
more generally put forward as symmetries of the uncompactified $11D$, type II and 
type I supergravity theories, and possibly  as a fundamental symmetry of M-theory 
itself, containing the whole chain of Cremmer-Julia split algebras, and hinting at 
the possibility that M-theory might prove intrinsically algebraic in nature.

Along this line,~\cite{DamHenNic1} have proposed a tantalizing correspondence between 
classical $11D$ supergravity operators at a given point close to a spacelike singularity 
and the coordinates of a one-parameter 
sigma-model based on the coset $\e_{10|10}/\k(\e_{10|10})$ and describing the dynamics 
of a hyperbolic cosmological billiard. In particular, a class of real roots of $\e_{10}$ 
have been identified, using a BKL expansion~\cite{BKL1,BKL2}, as multiple gradients of 
$11D$ supergravity fields reproducing the truncated equations of motion of the theory.
More recently~\cite{DamNic1}, imaginary roots were shown to correspond to 8th order 
$R^{m}(DF)^{n}$ type M-theory corrections to the classical $11D$ supergravity action.

From this perspective, studying the behaviour of the infinite dimensional U-duality 
symmetry $\e_{10|10}$ on singular backgrounds is a promising direction. 
This paper is the first in a serie of two papers aiming at obtaining
an algebraic characterization of M-theory compactified on various orbifolds. 
More precisely, we are interested in studying the residual
U-duality symmetry that survives the orbifold projection and the way 
the shift vectors defining the orbifold action are related to 
the extended objects necessary for the theory to reduce to a well-defined string theory.
In the second paper, we will show that the way the background metric and matter fields 
are affected by the orbifold singularities is encoded in the algebraic data of 
a particular non-split real form of the U-duality algebra.

In~\cite{Gan2}, an algebraic analysis of a certain class of orbifolds 
of M-theory has already been carried out in a compact version of the setup 
of~\cite{DamHenNic1}. Their work is based on a previous investigation of 
the relation between the moduli space of M-theory in
the neighbourhood of a spacelike singularity and its possible hyperbolic
billiard description~\cite{MotlBanks2}. For their analysis, these authors 
took advantage of a previous work~\cite{Gan1} which helped establishing a 
dictionary between null roots of $\eten$ and certain Minkowskian branes 
and other objects of M-theory on $T^{10}$. Let us briefly recall this correspondence.

In~\cite{Ob1,Ob2,OberPiol1}, a systematic description of the
relation between a subset of the positive roots of $E_{11-D|11-D}$ and BPS
objects in type II string theories and M-theory has been given. In particular, 
they were shown to contribute to instanton corrections to the low-energy effective theory.
In $D=1$, this suggests a correspondence between certain positive real roots of $\eten$ 
and extended objects of M-theory totally wrapped in the compact space (such as Euclidean 
Kaluza-Klein particles, Euclidean M2 and M5-branes, and Kaluza-Klein monopoles). In the
hyperbolic billiard approach to the moduli space of M-theory near a cosmological
singularity, these real roots appear in exponential terms in the low-energy effective Hamiltonian of the theory~\cite{DamNic2,DamNic1}. Such contributions behave as 
sharp wall potentials in the BKL limit, interrupting and reflecting the otherwise 
free-moving Kasner metric evolution. The latter can be represented mathematically by
the inertial dynamics of a vector in the Cartan subalgebra of $\eten$ undergoing Weyl-reflections when it reaches the boundary of a Weyl chamber. 
In the low energy $11D$ supergravity limit, these sharp walls terms can be regarded as fluxes, 
which are changed by integer amounts by instanton effects. This description, however, 
is valid only in a regime where all compactification radii can all become simultaneously
larger than the Planck length. 
In this case, the corresponding subset of positive real roots of $\eten$ can safely
be related to instanton effect. As shown in~\cite{MotlBanks2}, the regions of the moduli 
space of M-theory where this holds true are bounded by the (approximate) Kasner solution
mentioned above. A proper description of these regions calls for a modification 
of the Kasner evolution by introducing matter, which leads to a (possibly) 
non-chaotic behaviour of the system at late time (or large volume). The main contribution of~\cite{Gan1} was
to give evidence that these matter contributions have a natural description
in terms of imaginary roots of $\eten$. More precisely, these authors have
shown that extended objects such as Minkowskian Kaluza-Klein particles,
M2-branes, M5-branes, and Kaluza-Klein monopoles (KK7M-branes)
can be related to prime isotropic imaginary roots of $\e_{10}$ that, 
interestingly enough, are all Weyl-equivalent. These results, although derived 
in a compact setting, are amenable to the non-compact case~\cite{DamHenNic1,DamNic1}.

Ref.\cite{Gan2} only considers a certain class of orbifolds of M-theory, namely: 
$T^{10-q}\times T^{q}/\Z_{2}$ for $q=1,4,5,8,9$. 
After orbifold projection, the residual U-duality algebra $\ginv$ describing 
the untwisted sectors of all these orbifolds was
shown to possess a root lattice isomorphic to the root lattice of the
over-extended hyperbolic $\mathfrak{de}_{10|10}$. However, a careful root-space
analysis led the authors to the conclusion that $\ginv$ was
actually bigger than its hyperbolic counterpart, and contained 
$\mathfrak{de}_{10|10}$ as a proper subalgebra. Furthermore, in the absence of flux,
anomaly cancellation in such orbifolds of M-theory is known to require the insertion
of 16 M$(10-q)$-branes, Kaluza-Klein particles/monopoles or other BPS objects 
(the $S^{1}/\Z_{2}$ has to be treated from a type IA point of view,
where 16 D8-branes are required to compensate the charges of the two O8-planes)
extending in the directions transverse to the orbifold~\cite{Witten2,DasMuk1}.
In \cite{Gan2}, such brane configurations were shown to be nicely incorporated 
in the algebraic realization of the corresponding orbifolds. It was proven 
that the root lattice automorphism reproducing the $\Z_{2}$ action on the
metric and the three-form field of the low effective M-theory action could
always be rephrased in terms of a prime isotropic root, playing the r\^{o}le
of the orbifold shift vector and describing precisely the transverse
Minkowskian brane required for anomaly cancellation.

This construction in terms of automorphisms of the root lattice is however
limited to the $\Z_{2}$ case, where, in particular, the diagonal
components of the metric play no r{\^o}le. In order to treat the general 
$\Z_{n>2}$ orbifold case, we are in need of a more elaborate algebraic approach,
which operates directly at the level of the generators of the algebra. In
this regard, the works of Kac and Peterson on the classification of finite
order automorphisms of Lie algebras, have inspired a now standard procedure 
\cite{Koba,Koba2,Choi} to determine the residual invariant subalgebra of a
given finite dimensional Lie algebras, under a certain orbifold projection.
This has in particular been used to study systematically the breaking
patterns of the $E_{8}\times E_{8}$ gauge symmetry of the heterotic string 
\cite{Koba2,Nilles}. The method is based on choosing an eigenbasis in which
the orbifold charge operator can be rephrased as a Cartan preserving
automorphism Ad$e^{\frac{2i\pi }{n}H_{\La}}$, where $\La$ is an
element of the weight lattice having scalar product $(\La |\theta _{G})\leqslant n$
with the highest root of the algebra $\theta_{G}$, and $H_{\La}$ 
is its corresponding Cartan element. The shift vector $\La$ then determines 
by a standard procedure the invariant subalgebra $\ginv$ for all 
$\Z_{n} $ projections. However, the dimensionality and the precise set of
charges of the orbifold have to be established by other means. This is in particular
necessary to isolate possible degenerate cases. Finally, this method relies on the use of
extended (not affine) Dynkin diagrams, and it is not yet known how it can be
generalized to affine and hyperbolic Kac-Moody algebras.

Here we adopt a novel point of view based on the observation that the action
of an orbifold on the symmetry group of any theory that possesses at least
global Lorentz symmetry can be represented by the rigid action of a formal
rotation operator in any orbifolded plane. In algebraic terms, the orbifold
charge operator will be represented by a non-Cartan preserving, finite-order
automorphism acting on the appropriate complex combinations of generators.
These combinations are the components of tensors in the complex basis of 
the orbifolded torus which diagonalize the automorphism. Thus, they reproduce
the precise mapping between orbifolded generators in 
$\e_{10|10}/\k(\e_{10|10})$ and charged states 
in the moduli space of M-theory on $T^{10}$. It also enables one to keep track 
of the reality properties of the invariant subalgebra, provided we work with a Cartan 
decomposition of the original U-duality algebra. This is one reason which prompted us 
to choose the \textit{symmetric gauge} (in contrast to the \textit{triangular} Iwasawa 
gauge) to parametrize the physical fields of the theory. For this gauge choice, 
the orbifold charge operator is expressible as 
$\prod_{\a\in \Delta_{+}}\Ad e^{\frac{2\pi Q_{\a}}{n}(E_{\a}-F_{\a})}$, for 
$E_{\alpha }-F_{\alpha }\in \k(\e_{10|10})$, and $\Delta_{+}$ a set of positive
roots reproducing the correct orbifold charges $\{Q_{\a}\}$. The fixed
point subalgebra $\ginv$ is then obtained by truncating to the $Q_{\a}=kn$ sector, 
$k\in\Z$.

This method is general and can in principle be applied to the U-duality
symmetry of any orbifolded supergravities and their M-theory limits. In this
paper, we will restrict ourselves to the $\g^{U}=\e_{11-D|11-D}$
U-duality chain for $8\leqslant D\leqslant 1$. We will also limit our
detailed study to a few illustrative examples of orbifolds, namely: 
$T^{11-D-q}\times T^{q}/\Z_{n>2}$ for $q=2,4,6$ and 
$T^{11-D-q}\times T^{q}/\Z_{2}$ for $q=1,\ldots ,9$. In the 
${\mathbbm{Z}}_{2}$ case, we recover, for $q=1,4,5,8,9$, the results of~\cite{Gan2}. 
In the other cases, the results are original and lead, for $D=1$, to
several examples where we conjecture that $\ginv$ is
obtained by modding out either a Borcherds algebra or an indefinite (not
affine) Kac-Moody algebra, by its centres and derivations. As a first check
of this conjecture, we study in detail the $T^{8}\times T^{2}/\Z_{n}$ case, 
and verify its validity up to level $l=6$, investigating with
care the splitting of the multiplicities of the original $\e_{10}$
roots under the orbifold projection. We also show that the remaining cases
can be treated in a similar fashion. From a different perspective~\cite{HenryJulPaul1,HenryJulPaul2}, truncated real super-Borcherds 
algebras have been shown to arise already as more general symmetries of various 
supergravities expressed in the doubled formalism and compactified on square tori 
to $D=3$. Our work, on the other hand, gives other explicit examples of how Borcherds
algebras may appear as the fixed-point subalgebras of a hyperbolic Kac-Moody
algebra under a finite-order automorphism.

Subsequently, we engineer the relation between our orbifolding procedure which
relies on finite order non-Cartan preserving automorphism, and the formalism
of Kac-Peterson~\cite{KacPet}. We first show that there is a new \textit{primed} 
basis of the algebra in which one can derive a class of shift
vectors for each orbifold we have considered. Then, we prove that these
vectors are, for a given $n$, conjugate to the shift vector expected from 
the Kac-Peterson formalism. 
We show furthermore that, in the primed basis, every such class contains
a positive root of $\eten$, which can serve as class representative. 
This root has the form $\La'_{n}+n\tilde{\d}'$, where $\tilde{\d}'$ is in 
the same orbit as the null root $\d'$ of $\e_{9}$ under the Weyl group of $SL(10)$, 
and is minimal, in the sense that $\La_{n}'$ is the minimal weight 
leading to the required set of orbifold charges. We then list all such class 
representatives for all orbifolds of the type 
$T^{q_{1}}/\Z_{n_{1}}\times \cdots \times T^{q_{m}}/\Z_{n_{m}}$ 
with $\sum_{i=1}^{m}q_{i}\leqslant 10$, where the $\Z_{n_{i}}$ actions 
act independently on each $T^{2}$ subtori.

In particular, for the $T^{10-q}\times T^{q}/\Z_{2}$ orbifolds of
M-theory with $q=2,3,6,7$ that were not considered in \cite{Gan2}, we find 
that a consistent physical interpretation requires to consider them in the bosonic
M-theory that descend to type 0A strings. In such cases, we find class representatives 
that are either positive real roots of $\e_{10}$, 
or positive non-isotropic imaginary roots of norm $-2$. We then
show that these roots are related to the twisted sectors of some particular
non-supersymmetric type 0' orientifolds carrying magnetic fluxes. Performing 
the reduction to type 0A theory and T-dualizing appropriately, we actually 
find that these roots of $\eten$ descend to magnetized D9-branes in 
type 0' orientifolds carrying 
$(2\pi)^{-[q/2]}\int \underset{[q/2]}{\text{Tr}\underbrace{F\wedge ..\wedge F}}$
units of flux. This gives a partial characterization of open strings twisted sectors 
in non-supersymmetric orientifolds in terms of roots of $\e_{10}$. 
Moreover, the fact that these roots can be identified with Minkowskian D-branes 
even though none of them is prime isotropic, calls for a more general algebraic 
characterization of Minkowskian objects than the one propounded in~\cite{Gan2}. 
A new proposal supported by evidence from the $\Z_{n>2}$ case will be presented 
in Section~\ref{SecOrbi2}.

More precisely, the orbifolds of M-theory mentioned above descend to 
orientifolds of type 0A string theory by reducing on one direction
of the orbifolded torus for $q$ even, and on one direction outside the torus 
for $q$ odd. T-dualizing to type 0B, we find cases similar to those studied
in~\cite{BlumFontLust,BlumFontLust2}, where specific configurations 
of D-branes and D'-branes were used to cancel the two 10-form RR tadpoles. 
Here, in contrast, we consider a configuration in which the branes are tilted 
with respect to the orientifold planes in the 0A theory. This setup is T-dual 
to a type 0' orientifold with magnetic fluxes coupling to the electric charge 
of a D$(10-q)$-brane embedded in the space-filling D9-branes in the spirit 
of~\cite{BlumGoerKorLust,BlumKorsLust}.
In this perspective, the aforementioned roots of $\eten$ which 
determine the $\Z_{2}$ action also possess a dual description in 
terms of tilted D-branes of type 0A string theory. In the original 11-dimensional 
setting, these roots are related to exotic objects of M-theory and thereby 
provide a proposal for the M-theory origin of such configurations.

Finally, we will also comment on the structure of the $\eten$
roots that appear as class representatives for shift vectors of 
$\Z_{n}$ orbifolds of M-theory, and hint at the kind of flux configurations 
these roots could be associated to.

\section{Generalized Kac-Moody algebras}

In this section, we introduce recent mathematical constructions from the
theory of infinite-dimensional Lie algebras. Indeed, it is well-known in Lie
theory that fixed-point subalgebras of infinite-dimensional Lie algebras
under certain algebra automorphisms are often interesting mathematical
objects in their own right and might have quite different properties. Of
particular interest here is the fact that fixed-point subalgebras of
Kac-Moody algebras are not necessarily Kac-Moody algebras, but can belong to
various more general classes of algebras like extended affine Lie algebras~ 
\cite{AllBerPian,AllBerPian2,EALA}, generalized Kac-Moody algebras~\cite%
{Bor1,Bor2,Bor3}, Slodowy intersection matrices~\cite{Slo1} or Berman's
generalized intersection matrices~\cite{BerJurTan}. Indeed, invariant
U-duality symmetry subalgebras for orbifolds of M-theory are precisely
fixed-point subalgebras under a finite-order automorphism and can be
expected (at least in the hyperbolic and Lorentzian cases) to yield algebras
that are beyond the realm of Kac-Moody algebras.

\subsection{Central extensions of Borcherds algebras}
\label{GKMA}

Since they are particularly relevant to our results, we will focus here on
the so-called generalized Kac-Moody algebras, or GKMAs for short, introduced
by Borcherds in~\cite{Bor1} to extend the Kac-Moody algebras construction to
infinite-dimensional algebras with imaginary simple roots. We define here a
number of facts and notations about infinite-dimensional Lie algebras which
we will need in the rest of the paper, starting from very general
considerations and then moving to more particular properties. This will
eventually prompt us to refine the approach to GKMAs with a degenerate
Cartan matrix, by providing, in particular, a rigorous definition of how
scaling operators should by introduced in this case in accordance with the
general definition of GKMAs (see for instance Definitions \ref{Brel} and 
\ref{Defcomm} below). This has usually been overlooked in the literature, but
turns out to be crucial for our analysis of fixed point subalgebras of
infinite KMAs under a finite order automorphism, which occur, as we will see,
as hidden symmetries of the untwisted sector of M-theory under a given
orbifold.

In this perspective, we start by defining the necessary algebraic tools.
Let $\g$ be a (possibly infinite-dimensional) Lie algebra
possessing a Cartan subalgebra $\h$ (a complex nilpotent
subalgebra equal to its normalizer) which is \textit{splittable}, in other
words, the action of $\ad H$ on $\g$ is trigonalizable $\forall H\in \h$. 
The derived subalgebra $\g'=[\g,\g]$ possesses an $r$-dimensional Cartan
subalgebra $\h'=\g'\cap \h$ spanned by the basis $\Pi^{\vee}=\{H_i\}_{i\in I}$, 
with indices valued in the set $I=\{i_{1},..,i_{r}\}$.

We denote by $\h^{\prime\ast }$ the space dual to $\h'$.
It has a basis formed by $r$ linear functionals (or 1-forms) on 
$\h'$, the simple roots of $\g$: $\Pi=\{\a_i\}_{i\in I}$.
Suppose we can define an indefinite scalar product: $(\a_{i}|\a_{j})=a_{ij}$ for 
some real $r\times r$ matrix $a$, then:

\begin{definition}
\label{LieAlgebra} The matrix $a$ is called a generalized symmetrized Cartan
matrix, if it satisfies the conditions:

\begin{enumerate}
\item[i)] $a_{ij}=a_{ji}\,,$ $\forall i,j\in I$.

\item[ii)] $a$ has no zero column.

\item[iii)] $a_{ij}\leq 0\,,$ for $i\neq j$ and $\forall i,j\in I$.

\item[iv)] $\left. 
\begin{array}{cr}
\text{if }a_{ii}\neq 0\,: & 2\frac{a_{ij}}{a_{ii}} \\ 
\text{if }a_{ii}=0\,: & a_{ij}
\end{array}
\right\} $ $\in {\mathbbm{Z}}\,,$ for $i\neq j$ and $\forall i,j\in I$.
\end{enumerate}
\end{definition}

From integer linear combinations of simple roots, one constructs the root
lattice $Q=\sum_{i\in I}\Z\a_{i}$. The scalar product $(\,|\,)$ is then extended 
by linearity to the whole $Q\subset \h^{\prime \ast}$. Furthermore, by defining
fundamental weights $\{\La^{i}\}_{i\in I}$ satisfying $(\La^{i}|\a_{j})=
\d_{\phantom{i}j}^{i}\,,$ $\forall i,j\in I$, we introduce a duality relation 
with respect to the root scalar product. Then, from the set $\{\La^{i}\}_{i\in I}$ 
we define the lattice of integral weights $P=\sum_{i\in I}{\Z\La}^{i}$ dual to $Q$, 
such that $Q\subseteq P$.

Let us introduce the duality isomorphism $\nu$: $\h'\rightarrow \h'^{\ast}$ defined by 
\begin{equation}
\nu (H_{{i}})=\left\{ 
\begin{array}{lc}
2\frac{\a_{i}}{|{a}_{ii}|} & ,\text{ if } a_{ii}\neq 0\,. \\ 
{\alpha }_{i} & ,\text{ if }a_{ii}=0\,,
\end{array}
\right. \,.  \label{DuIs}
\end{equation}
We may now promote the scalar product $(\a_{i}|\a_{j})=a_{ij}$
to a symmetric bilinear form $B$ on $\h'$ through: 
\begin{equation*}
b_{ij}=B(H_{i},H_{j})=(\nu (H_{i})|\nu (H_{j}))\,,\text{ }\forall i,j\in I\,.
\end{equation*}
Suppose next that the operation $\mathrm{ad}(H)$ is diagonalizable $\forall H\in \h$, 
from which we define the following:

\begin{definition}
We call root space an eigenspace of $\mathrm{ad}(H)$ defined as 
\begin{equation}\label{RootSpace}
{\mathfrak{g}}_{\a}=\{X\in \g\,|\,\mathrm{ad}(H)X=\a(H)X,\,\forall H\in\h\}
\end{equation}
which defines the root system of $\g$ as 
$\Delta(\g,\h)=\{\a\neq 0\,|\,\g_{\a}\neq \{0\}\}$, depending on the choice of basis 
for $\h$.
\end{definition}

The multiplicities attached to a root $\a \in \Delta (\g,\h)$ are then given by 
$m_{\a}=\dim \g_{\a}$. As usual, the root system splits into a positive root system 
and a negative root system. The positive root system is defined as 
\begin{equation*}
\Delta_{+}(\g,\h)=\Big\{\a \in \Delta (\g,\h)\,|\, \a=\sum_{i\in I}n_{i}\a_{i},
\mathrm{ with}\;\,n_{i}\in \N\,,\forall i\in I\Big\}
\end{equation*}
and the negative root one as $\Delta_{-}(\g,\h)=-\Delta_{+}(\g,\h)$, so that 
$\Delta (\g,\h)=\Delta_{+}(\g,\h)\cup \Delta _{-}(\g,\h)$. We call 
ht$(\a)=\sum_{i\in I}n^{i}$ the height of $\a$. From now on, we shall write 
$\Delta \equiv \Delta(\g,\h)$ for economy, and restore the full notation 
$\Delta (\g,\h)$ or partial notation $\Delta (\g)$ when needed.

Finally, since $(\a|\a)$ is bounded above on $\Delta$, $\a$ is called real if 
$(\a|\a)>0$, isotropic imaginary if $(\a|\a)=0$ and (non-isotropic) imaginary 
if $(\a|\a)<0$. Real roots always have multiplicity one, as is the case for
finite-dimensional semi-simple Lie algebras, while (non-simple) isotropic
roots have a multiplicity equal to rk$a(\hat{\g})$ for some
affine subalgebra $\hat{\g}\subset \g$, while (non-simple) non-isotropic imaginary
roots can have very big multiplicities.

Generalized Kac-Moody algebras are usually defined with all 
\textit{imaginary simple} roots of multiplicity one, as well. One could in 
principle define a GKMA with simple roots of multiplicities bigger than one, 
but then the algebra would not be completely determined by its generalized 
Cartan matrix. In this case, one would need yet another matrix with coefficients 
specifying the commutation properties of all generators in the same simple root 
space. Here, we shall not consider this possibility further since it will turn 
out that all fixed point subalgebras we will be encountering in the framework 
of orbifold compactification of $11D$ supergravity and M-theory possess only 
isotropic simple roots of multiplicity one.

We now come to specifying the r\^{o}le of central elements and scaling
operators in the case of GKMAs with degenerate generalized Cartan matrix.

\begin{definition}
\label{Brel}If the matrix $a$ does not have maximal rank, define the centre
of $\g$ as $\z(\g)=\{c\in \h\,|\,B(H_{i},c)=0\,,$ $\forall i\in I\}$. In particular, 
if $l=\dim \z(\g)$, one can find $l$ linearly independent null
root lattice vectors $\{\d_{i}\}_{i=1,..,l}$ (possibly roots, but not
necessarily) satisfying $(\d_{i}|\nu (H_{j}))=0\,,$ $\forall i=1,\ldots,l$, 
$\forall j\in I$. One then defines $l$ linearly independent Cartan generators 
$\{{d}_{i}\}_{i=1,..,l}$ with ${d}_{i}\in \h/\h'$ thus extending the bilinear form 
$B$ to the whole Cartan algebra $\h$ as follows:

\begin{itemize}
\item $B({c}_{i},d_{j})={\delta }_{ij}\,,$ $\forall i,j=1,..,l$\thinspace .

\item $B(d_{i},d_{j})=0\,,$ $\forall i,j=1,..,l$\thinspace .

\item $B(H,d_{i})=0\,,$ $\forall i=1,..,l$ and for 
$H\in \h'/$Span$\{c_{i}\}_{i=1,..,l}$\thinspace .\end{itemize}

Then, we have rk$(a)=r-l$ and $\dim \h=r+l$.
\end{definition}

This definition univocally fixes the $i$-th level $k_{i}$ of all roots $%
\alpha \in \Delta $ to be $k_{i}=B(\nu^{-1}(\a),d_{i})$, using the
decomposition of $\nu^{-1}(\a)$ on orthogonal subspaces in 
$\h'=\left(\h'/\Span\{c_{1},\ldots,c_{l}\}\right) \oplus\Span\{c_{1}\}\oplus
\cdots\oplus \Span\{c_{n}\}$.

We are now ready to define a GKMA by its commutation relations. Definitions
of various levels of generality exist in the literature, but we choose one
that is both convenient (though seemingly complicated) and sufficient for
our purpose, neglecting the possibility that $[E_{i},F_{j}]\neq 0$ for 
$i\neq j$ (see, for example, \cite{Gebert1,BarGebGunNic} for such constructions), but
taking into account the possibility of degenerate Cartan matrices, a generic
feature of the type of GKMA we will be studying later on in this paper.

\begin{definition}
\label{Defcomm}The universal generalized Kac-Moody algebra associated to the
Cartan matrix $a$ is the Lie algebra defined by the following commutation
relations (Serre-Chevalley basis) for the simple root generators 
$\{E_{i},F_{i},H_{i}\}_{i\in I}$:

\begin{enumerate}
\item $[E_{i},F_{j}]=\d_{ij}H_{i}$, \hspace{1cm}
$[H_{i},H_{j}]=[H_{i},d_{k}]=0 \,,$ \hspace{1cm} 
$\forall i,j\in I$,\;\; 
$k=1,.. ,l\,.$

\item $[H_{i},E_{j}]=\left\{ 
\begin{array}{c}
\frac{2a_{ij}}{|a_{ii}|}E_{j}\,,\text{ if }a_{ii}\neq 0 \\ 
a_{ij}E_{j}\,,\text{ if }a_{ii}=0
\end{array}
\right.$, 
\hspace{0.5cm} 
$[H_{i},F_{j}]=\left\{ 
\begin{array}{c}
-\frac{2a_{ij}}{|a_{ii}|}F_{j}\,,\text{ if }a_{ii}\neq 0 \\ 
-a_{ij}F_{j}\,,\text{ if }a_{ii}=0
\end{array}
\right. \,,$ 
\hspace{0.5cm} 
$\forall i,j\in I\,.$

\item If ${a}_{ii}>0\,:$ 
\hspace{0.3cm}
$(\mathrm{ad}E_{i})^{1-2\frac{a_{ij}}{a_{ii}}}E_{j}=0$, 
\hspace{0.5cm} 
$(\mathrm{ad}F_{i})^{1-2\frac{a_{ij}}{a_{ii}}}F_{j}=0\,,$ 
\hspace{0.5cm} 
$\forall i,j\in I\,.$

\item $\forall i,j\in I$ such that $a_{ii}\leq 0$, $a_{jj}\leq 0$ 
and $a_{ij}=0\,:$ 
\footnote{Note that there is no a priori limit to the number of
times one can commute the generator $E_{i}$ for $\a_{i}$ imaginary
with any other generator $E_{j}$ in case $a_{ij}\neq 0$.} \hspace{0.5cm} 
$[E_{i},E_{j}]=0$, 
\hspace{0.5cm} $[F_{i},F_{j}]=0\,,$

\item $[d_{i},[E_{j_{1}},[E_{j_{2}},..,E_{j_{n}}]..]]=
k_{i}[E_{j_{1}},[E_{j_{2}},..,E_{j_{n}}]..]\,,$
\newline
$
[d_{i},[F_{j_{1}},[F_{j_{2}},..,F_{j_{n}}]..]]=
-k_{i}[F_{j_{1}},[F_{j_{2}},..,F_{j_{n}}]..]\,, 
$
\newline
where $k_{i}$ is the $i-th$ level of $\a=\a_{j_{1}}+\ldots+\a_{j_{n}}$, as defined above.
\end{enumerate}
\end{definition}

Since a generalized Kac-Moody algebra can be graded by its root system as: 
${\displaystyle \g=\h\oplus \bigoplus_{\a\in \Delta} \g_{\a}}$, the indefinite scalar 
product $B$ can be extended to an ad$(\g)$-invariant bilinear form satisfying:
$B(\g_{\a},\g_{\b})=0$ except if $\a+\b=0$, which we call the generalized Cartan-Killing 
form. It can be fixed uniquely by the normalization 
\begin{equation*}
B(E_{i},F_{j})=\left\{ 
\begin{array}{lr}
\frac{2}{|a_{ii}|}\d_{ij} & ,\text{ if }a_{ii}\neq 0 \\ 
\d_{ij} & ,\text{ if }a_{ii}=0
\end{array}
\right. \,,
\end{equation*}
on generators corresponding to simple roots. Then $\mathrm{ad}(\g)$-
invariance naturally implies: $B(H_{i},H_{j})=(\nu (H_{i})|\nu (H_{j}))$. 

The GKMA $\g$ can be endowed with an antilinear Chevalley
involution $\vartheta_{C}$ acting as $\vartheta_{C}(\g_{\a})
=\g_{-\a}$ and $\vartheta_{C}(H)=-H\,,$ $\forall H\in \h$, whose action on each simple 
root space $\g_{\a_i}$ is defined as as usual as $\vartheta_{C}(E_{i})=-F_{i}\,, $ 
$\forall i\in I$. The Chevalley involution extends naturally to the whole algebra 
$\g$ by linearity, for example:
\begin{equation}
\vartheta _{C}([E_{{i}},E_{{j}}])=[\vartheta _{C}(E_{{j}}),\vartheta _{C}(E_{{j}})]=
[F_{{i}},F_{j}]\,.  \label{ChevalMorph}
\end{equation}
This leads to the existence of an almost positive-definite contravariant
form $B_{\vartheta_{C}}(X,Y)=-B(\vartheta_{C}(X),Y)$. More precisely, it is
positive-definite everywhere outside $\h$.

Note that there is another standard normalization, the Cartan-Weyl basis,
given by:
\begin{equation*}
\begin{array}{l}
e_{{\alpha}_{i}}=\left\{ 
\begin{array}{lr}
\sqrt{\frac{|a_{ii}|}{2}}E_{i} & ,\text{ if }a_{ii}\neq 0 \\ 
E_{i} & ,\text{ if }a_{ii}=0
\end{array}
\right. \,, \hspace{1cm} f_{{\alpha }_{i}}=\left\{ 
\begin{array}{lr}
\sqrt{\frac{|a_{ii}|}{2}}F_{i} & ,\text{ if }a_{ii}\neq 0 \\ 
F_{i} & ,\text{ if }a_{ii}=0
\end{array}%
\right. , \\[4pt] 
h_{{\alpha }_{i}}=\left\{ 
\begin{array}{lr}
\frac{|a_{ii}|}{2}H_{i} & ,\text{ if }a_{ii}\neq 0 \\ 
H_{i} & ,\text{ if }a_{ii}=0
\end{array}
\right.\,,
\end{array}
\end{equation*}
and characterized by: $B(e_{\a},f_{\a})=1$, $\forall \a\in\Delta_{+}(\g)$.

We will not use this normalization here, but we will instead write the
Cartan-Weyl relations in a Chevalley-Serre basis, as follows:
\begin{definition}
\label{StructConst}For all $\a\in \Delta_{+}(\g)$ introduce $2m_{\a}$ generators: 
$E_{\a}^{a}$ and $F_{\a}^{a}$, $a=1,..,m_{\a}$. Generators corresponding to roots 
of height $\pm 2$ are defined as: 
\begin{equation*}
E_{\a_{i}+\a_{j}}=\mathcal{N}_{\a_{i},\a_{j}}[E_{i},E_{j}],\quad F_{\a_{i}+\a_{j}}=\mathcal{N}_{-\a_{i},-\a_{j}}[F_{i},F_{j}],\,\forall i,j\in I\,,
\end{equation*}
for a certain choice of structure constants $\mathcal{N}_{\a_{i},\a_{j}}$. Then, 
higher heights generators are defined recursively in the same way through:
\begin{equation}\label{DefComm}
[E_{\a}^{a},E_{\b}^{b}]=\sum_c (\mathcal{N}_{\a,\b})^{ab}_{\phantom{ab}c}E_{\a+\b}^{c}\, .
\end{equation}
\end{definition}

The liberty of choosing the structure constants is of course limited by the
anti-commutativity of the Lie bracket: 
$(\mathcal{N}_{\a,\b})^{ab}_{\phantom{ab}c}=-(\mathcal{N}_{\b,\a})^{ba}_{\phantom{ab}c}$ 
and the Jacobi identity, from which we can derive several relations. 
Among these, the following identity, valid for finite-dimensional Lie algebras, will be 
useful for our purposes: 
\begin{equation*}
\mathcal{N}_{\a,\b}\mathcal{N}_{-\a,-\b}=-(p+1)^{2},\,p\in \N\,,\text{ s.t. }\{\b-p\a,\ldots,\b+\a\}\subset \Delta (\g,\h)\,.
\end{equation*}
Note that this relation can be generalized to the infinite-dimensional case
if one chooses the bases of root spaces $\g_{\a}$ with $m_{\a}>1$ in a particular 
way such that there is no need for a sum in (\ref{DefComm}).
Imposing in addition 
$(\mathcal{N}_{\a,\b})^{ab}_{\phantom{ab}c}=-(\mathcal{N}_{-\a,-\b})^{ab}_{\phantom{ab}c}$
gives sign conventions compatible with 
$\vartheta_{C}(E_{\a}^a)=-F_{\a}^a\,$, $\forall \a\in \Delta_{+}(\g),a=1\ldots,m_{\a}$,
not only for simple roots. In the Serre-Chevalley normalization, this
furthermore ensures that: $\mathcal{N}_{\a,\b}\in \Z$, $\forall \a,\b\in \Delta$. 
Here lies our essential reason for sticking to this normalization, and we will 
follow this convention throughout the paper. In the particular case of {\it simply-laced}
semi-simple Lie algebras, we always have $p=0$, and we can choose 
$\mathcal{N}_{\a,\b}=\pm 1\,$, $\forall \a,\b\in \Delta$ (note, however,
that this is not true for infinite-dimensional simply-laced algebras).

Another important consequence of the Jacobi identity in the finite case,
which will turn out to be useful is the following relation 
\begin{equation*}
\mathcal{N}_{\a,-\b}=\mathcal{N}_{\b-\a,\a} \quad \forall \a,\b\in \Delta\,.
\end{equation*}

\subsection{Kac-Moody algebras as a special case of GKMA}

\label{KMAconv}

Standard symmetrized Kac-Moody algebras (KMA) can be recovered from the
preceding section by imposing $a_{ii}>0\,,$ $\,\forall i\in I$ in all the above
definitions. In addition, one usually rephrases the dual basis $\Pi^{\vee}$
in terms of coroots, by setting $\a_{i}^{\vee}\equiv H_{i}$. Their image
under the duality isomorphism reads
\begin{equation*}
\nu (\a_{i}^{\vee})=\frac{2}{(\a_{i}|\a_{i})}\a_{i}\,,\text{ }\forall i\in I\,,
\end{equation*}
so that instead of the symmetrized Cartan matrix $a$, one generally resorts
to the following non-symmetric version, defined as a realization of the
triple 
$\{\mathfrak{h},\Pi,\Pi^{\vee}\}$ with $\Pi^{\vee}=\{\a_i^{\vee}\}_{i\in I}
\subset \h^{\ast}$:
\begin{equation}
A_{ij}=\frac{2a_{ij}}{a_{ii}}\equiv \langle \a_{i}^{\vee},\a_{j}\rangle \,. \label{Asym}
\end{equation}
where the duality bracket on the RHS represents the standard action
of the one-form $\a_{j}$ on the vector $\a_{i}^{\vee}$.

The matrix $a$ is then called the symmetrization of the (integer) Cartan
matrix $A$. As a consequence of having introduced $l$ derivations in 
Definition~\ref{Brel}, the contravariant form $B_{\vartheta_{C}}(.,.)$ 
now becomes non-degenerate on the whole of $\g$, even in the case of central
extensions of multi-loop algebras, which are the simplest examples of
extended affine Lie algebra (EALA, for short).

For the following, we need to introduce the Weyl group of $\g$ as
\begin{definition}
\label{Weylgr} The Weyl group of $\g$, denoted $W(\g)$, is the group generated by 
all reflections mapping the root system into itself: 
\begin{equation*}
\begin{array}{rrcl}
r_{\a} \,: & \Delta(\g) & \rightarrow & \Delta(\g) \\[3pt] 
& \b & \mapsto & \b -\langle \a^{\vee },\b\rangle \a\,.
\end{array}
\end{equation*}
The set $\{r_{i_{1}},..,r_{i_{r}}\}$, where $r_{i}\doteq r_{\a_{i}}$
are called the fundamental reflections, is a basis of $W(\g)$.
Since $r_{i}^{-1}=r_{-\a_{i}}$, $W(\g)$ is indeed a group.
\end{definition}

The real roots of any finite Lie algebra or KMA can then be defined as being
Weyl conjugate to a simple root. In other words, $\a\in\Delta(\g,\h)$ is real 
if $\exists w\in W(\g)$ such that $\a=w(\a_{i})$ for $i\in I$ and $\g$ {\it is} a KMA.

A similar formulation exists for imaginary roots of a KMA, which usually turns
out to be useful for determining their multiplicities, namely (see \cite{Kac}):

\begin{theo}
\label{THim}\textit{Let }$\a=\sum_{i\in I}k_{i}\a _{i}\in Q\backslash \{0\}$
\textit{\ have compact support on the Dynkin diagram of }$\g$, \textit{ and set:}
\begin{equation*}
K=\{\a\,|\,\langle \a_{i}^{\vee},\a \rangle \leqslant 0\,,
\text{ }\forall i\in I\}\,,
\end{equation*}
\textit{Then denoting by }$\Delta_{im}$\textit{\ the set of imaginary roots
of }$\g$\textit{, we have:}
\begin{equation*}
\Delta _{im}({\mathfrak{g}})=\bigcup_{w\in W({\mathfrak{g}})}w(K)\,.
\end{equation*}
\end{theo}

It follows from Theorem~\ref{THim} that, in the affine case, every isotropic root $\a$ 
is Weyl-equivalent to $n\d$ (with $\d=\a_{0}+\theta$ the null root) for some 
$n\in \Z^{\ast }$, which is another way of showing
that all such roots have multiplicity $m_{\a}=r$. All isotropic roots
which are Weyl-equivalent to $\d$ are usually called \textit{prime}
isotropic. Note, finally, that statements similar to Theorem \ref{THim}
holds for non-isotropic imaginary roots of hyperbolic KMAs. For instance,
all positive roots with $(\a|\a)=-2$ can in this case be shown
to be Weyl-equivalent to $\La^{0}$, the weight dual to the extended root $\a_{0}$.

Intersection matrix algebras are even more general objects that allow for
positive non-diagonal elements in the Cartan matrix. Slodowy intersection
matrices allow such positive diagonal metric elements, while Berman
generalized intersection matrices give the most general framework by
allowing imaginary simple roots, as well, as in the case of Borcherds
algebras. Such more complicated algebras will not appear in the situations
considered in this paper, but it is not impossible that they could show up
in applications of the same methods to different situations.

\subsection{A comment on $\sl(10)$ representations in $\eten$ and their outer multiplicity}

Of particular significance for Kac-Moody algebras beyond the affine case are
of course the root system and the root multiplicities, which are often only
partially known. Fortunately, in the case of $\eten$, we can
rely on the work in~\cite{KleinNic1,Klein,Fisch2} to obtain information about a
large number of low-level roots, enough to study $\Z_{n}$ orbifolds up to $n=6$. 
These works rely on decomposing Lorentzian algebras in representations 
of a certain finite subalgebra. However, the set of representations is not 
exactly isomorphic to the root system (modulo Weyl reflections). Indeed, 
the multiplicity of a representation in the decomposition is in general smaller 
than the multiplicity of the root corresponding to its highest weight vector.
Typically, the $m$-dimensional vector space corresponding to a root of
(inner) multiplicity $m$ will be split into subspaces of several
representations of the finite subalgebra. Typically, a root ${\alpha}$ of
multiplicity $m_{\a}>1$ will appear $n_{\text{o}}(\a)$ times as
the highest weight vector of a representation, plus several times as a
weight of other representations. The number $n_{\text{o}}(\a)$ is
called the outer multiplicity, and can be 0. For a representation $\mathcal{R}$ of 
$\g$ it shall be denoted by a subscript as: $\mathcal{R}_{[n_{\text{o}}]}$ 
when needed. Even though the concept of outer multiplicity is of minor significance 
for our purpose, it is important to understand the mapping between the results 
of~\cite{NicFisch,Klein}, based on representation of finite subalgebras, and
ours, which focuses on tensorial representations with definite symmetry properties.

\section{Hidden symmetries in M-theory: the setup}

As a start, we first review some basic facts about hidden
symmetries of 11$D$ supergravity and ultimately M-theory, ranging from the early
non-linear realizations of toroidally compactified 11$D$ supergravity \cite%
{CremJulLuPope1,CremJulLuPope2} to the conjectured hyperbolic $\eten$
hidden symmetry of M-theory.

Then in Sections~\ref{ExE}-\ref{FluxBrane}, we do a synthesis of the algebraic approach
to U-duality symmetries of $11D$ supergravity on $T^q$ and the moduli space 
of M-theory on $T^{10}$, presenting in detail the physical material and mathematical
tools that we will need in the subsequent sections, and justify our choice
of parametrization for the coset element (algebraic field strength)
describing the physical fields of the theory. The reader familiar with these topics
may of course skip the parts of this presentation he will judge too detailed.

The global $E_{11-D|11-D}$ symmetry of classical 11$D$ supergravity reduced
on $T^{11-D}$ for $10\leqslant D\leqslant 3$ can be best understood as
arising from a simultaneous realization of the linear non-perturbative
symmetry of the supergravity Lagrangian where no fields are dualized and the
perturbative T-duality symmetry of type IIA string theory appearing in $D=10$
and below. Actually, the full $E_{11-D|11-D}$ symmetry has a natural
interpretation as the closure of both these groups, up to shift symmetries
in the axionic fields.

Type IIA string theory compactified on $T^{10-D}$ enjoys a $SO(10-D,10-D,\Z)$
symmetry\footnote{We consider $SO(10-D,10-D)$ instead of $O(10-D,10-D)$, as is 
sometimes done, because the elements of $O(10-D,10-D)$ connected to $-\Id$ flip 
the chirality of spinors in the type IIA/B theories. As such, this subset of elements 
is not a symmetry of the R-R sector of the type IIA/B supergravity actions, but 
dualities which exchange both theories.} which is valid order by order in 
perturbation theory. So, restricting to massless scalars arising from T-duality 
in $D\leqslant 8$, all inequivalent quantum configurations of the scalar sector 
of the bosonic theory are described by the moduli space 
\begin{equation*}
\mathcal{M}_{D}=SO(10-D,10-D,{\mathbbm{Z}})\backslash
SO(10-D,10-D)/(SO(10-D)\times SO(10-D))\,,
\end{equation*}%
where the left quotient by the arithmetic subgroup corrects the
over-counting of perturbative string compactifications. In contrast to the
NS-NS fields $B_{2}$ and $g_{\mu \nu }$ which, at the perturbative level,
couple to the string worldsheet, the R-R fields do so only via their field
strength. So a step towards U-duality can be achieved by dualizing the R-R
fields while keeping the NS-NS ones untouched. It should however be borne in
mind that such a procedure enhances the T-duality symmetry only when
dualizing a field strength to an equal or lower rank one. Thus, Hodge-duals
of R-R fields start playing a r\^{o}le when $D\leqslant 8$, those of NS-NS
fields when $D\leqslant 6$. However, when perturbative symmetries are concerned, 
we will not dualize NS-NS fields.

This enlarged T-duality symmetry can be determined by identifying its
discrete Weyl group $W(D_{10-D})$~\cite{LuPope1}, which implements the
permutation of field strengths required by electric-magnetic duality. In 
$D\leqslant 8$ it becomes now necessary to dualize R-R field strengths in
order to form Weyl-group multiplets. This results in $2^{9-D}$ R-R axions,
all exhibiting a shift symmetry, that enhances the T-duality group to: 
\begin{equation}
\widetilde{G}=SO(10-D,10-D)\ltimes {\mathbbm{R}}^{2^{9-D}}\,,  \label{Tsymm}
\end{equation}
the semi-direct product resulting from the fact that the R-R axions are now
linearly rotated into one another under T-duality. The (continuous) scalar
manifold is now described by the coset $\widetilde{G}/SO(10-D)\times SO(10-D)$, 
whose dimension matches the total number of scalars if we include the
duals of R-R fields only. The symmetry (\ref{Tsymm}) can now be enlarged to
accomodate non-perturbative generators, leading to the full global
symmetry $E_{11-D|11-D}$. However, this can only be
achieved without dualizing the NS-NS fields in the range 
$9\geqslant D\geqslant 7$. When descending to lower dimensions, indeed, 
the addition of non-perturbative generators rotating R-R and NS-NS fields 
into one another forces the latter to be dualized.

To evade this problem arising in low dimensions, we might wish to
concentrate instead on the global symmetry of the $11D$ supergravity
Lagrangian for $D\leqslant 9$, whose scalar manifold is described by the
coset
\begin{equation*}
GL(11-D)\ltimes {\mathbbm{R}}^{(11-D)!/((8-D)!3!)}/O(11-D)\,.
\end{equation*}
The corresponding group 
$G_{SG}=GL(11-D)\ltimes {\mathbbm{R}}^{(11-D)(10-D)(9-D)/6}$ encodes 
the symmetry of the totally undualized theory including the 
$(11-D)(10-D)(9-D)/6$ shift symmetries coming from the
axions produced by toroidal compactification of the three-form $C_{3}$.
Again, the semi-direct product reflects the fact that these axions combine
in a totally antisymmetric rank three representation of $GL(11-D)$. Since
NS-NS and R-R fields can be interchanged by $GL(11-D)$, the arithmetic subgroup
of $GL(11-D)\ltimes {\R}^{(11-D)(10-D)(9-D)/6}$ constitutes an
acceptable non-perturbative symmetry of type II superstring theory in 
$D\leqslant 9$. The price to pay in this case is to sacrifice T-duality,
since the subgroup of the linear group preserving the NS-NS and R-R sectors
separately is never big enough to accomodate $SO(10-D,10-D)$.

Eventually, the full non-perturbative symmetry $E_{11-D|11-D}$ can only be
achieved when both NS-NS and R-R fields are dualized, and may be viewed as
the closure of its $GL(11-D)$ and $SO(10-D,10-D)$ subgroups. However, the
number of shift symmetries in this fully dualized version of the theory is
given by $\{3,6,10,16,27,44\}$ for $8\geqslant D\geqslant 3$. Since, in 
$D\leqslant 5$, these numbers are always smaller or equal to
$(11-D)(10-D)(9-D)/6$ and $2^{9-D}$, neither $\widetilde{G}$ nor $G_{SG}$ are
subgroups of $E_{11-D|11-D}$ in low dimensions.

This exceptional symmetry is argued to carry over, in its discrete version,
to the full quantum theory. Typically, the conjectured U-duality group of
M-theory on $T^{11-D}$ can be rephrased as the closure 
\begin{equation}
E_{11-D|11-D}({\mathbbm{Z}})=SO(10-D,10-D,{\mathbbm{Z}})\bar{\times}SL(11-D,{%
\mathbbm{Z}})\,,  \label{Esymm}
\end{equation}%
where the first factor can be viewed as the perturbative T-duality symmetry
of IIA string theory, while the second one is the modular group of the torus 
$T^{11-D}$. In $D=9$, the latter can be rephrased in type IIB language as
the expected S-duality symmetry.


The moduli space of M-theory on $T^{11-D}$ is then postulated to be 
\begin{equation}
\mathcal{M}_{D}=E_{11-D|11-D}({\mathbbm{Z}})\backslash
E_{11-D|11-D}/K(E_{11-D|11-D})\,.  \label{modM}
\end{equation}%
It encodes both the perturbative NS-NS electric $p$-brane spectrum and the
spectrum of non-perturbative states, composed of the magnetic dual NS-NS $%
(9-D-p)$-branes and the R-R D-branes of IIA theory for $10\leqslant
D\leqslant 3$.

In dimensions $D<3$, scalars are dual to themselves, so no more enhancement
of the U-duality group is expected from dualization. However, an enlargement
of the hidden symmetry of the theory is nevertheless believed to occur
through the affine extension $E_{9|10}(\Z)$ in $D=2$, the
over-extended $E_{10|10}(\Z)$, generated by the corresponding
hyperbolic KMA, in $D=1$, and eventually the very-extended $E_{11}(\Z)$ 
in the split form, whose KMA is Lorentzian, for the totally compactified theory.

Furthermore, there is evidence that the latter two infinite-dimensional KMAs are also symmetries of unreduced $11D$ supergravity~\cite{KleinNic2,West}, viewed as a non-linear
realization (in the same spirit as the Monstruous Moonshine~\cite{Moore} has
been conjectured to be a symmetry of uncompactified string theory) and are believed
to be more generally symmetries of uncompactified M-theory itself~\cite{Nicolai97}.

\subsection{The exceptional $E_{r}$ serie: conventions and useful formul\ae}
\label{ExE}

Before going into more physical details, we need to introduce a few mathematical properties of the
U-duality groups and their related algebras. To make short, we will denote 
$G^{U}\doteq \text{Split}(E_{11-D})$, for $D=1,\ldots ,10$, and their
defining Lie or Kac-Moody algebras 
${\mathfrak{g}}^{U}\doteq \text{Split}({\mathfrak{e}}_{11-D})$.

Except in $D=10,9$, the exceptional serie $E_{r}$, with $r=11-D$, possesses
a physical basis for roots and dual Cartan generators:

\begin{definition}
\label{Phiz} Let the index set of Definition~\ref{LieAlgebra} be chosen as 
$I={9-r,\ldots ,8}$, for $3\leq r\leq 10$~\footnote{%
The physical basis makes sense only in cases where there are scalars coming
from compactification of the 3-form, which excludes the first two algebras
of the serie.}, then in the \textit{physical basis} of ${\mathfrak{h}}%
^{\prime \ast }$: $\{\varepsilon _{11-r}\doteq (1,0,\ldots ,0),\ldots
,\varepsilon _{10}\doteq (0,\ldots ,0,1)\}$, the set $\Pi =\{{\alpha }%
_{9-r},\ldots ,{\alpha }_{8}\}$ of simple roots of the semi-simple KMAs ${%
\mathfrak{e}}_{r}$ reads: 
\begin{equation}
\begin{array}{rcl}
{\alpha }_{9-r} & = & \varepsilon _{11-r}-\varepsilon
_{12-r}\;=\;(1,-1,0,\ldots ,0)\,, \\[3pt] 
& \vdots &  \\ 
{\alpha }_{7} & = & \varepsilon _{9}-\varepsilon _{10}\;=\;(0,\ldots
,0,1,-1)\,, \\[3pt] 
{\alpha }_{8} & = & \varepsilon _{8}+\varepsilon _{9}+\varepsilon
_{10}\;=\;(0,\ldots ,0,1,1,1)\,.
\end{array}
\label{PhysicalBasis}
\end{equation}
\end{definition}

The advantage of such a basis is to give a rank-, and hence dimension-,
independent formulation of $\Pi $, which is not the case for an orthogonal
basis $e_{i}$. Preserving the scalar product on the root system requires the
physical basis to be endowed with the following scalar product: 
\begin{equation}
(\a|\b)=\sum_{i=11-r}^{10}n^{i}m^{i}-\frac{1}{9}\sum_{i,j=11-r}^{10}n^{i}m^{j}\label{ScalarProduct}
\end{equation}%
for ${\alpha }=\sum_{i=11-r}^{10}n^{i}\varepsilon _{i}$ and $\beta
=\sum_{i=11-r}^{10}m^{i}\varepsilon _{i}$ (note that the basis elements
satisfy $(\varepsilon _{i}|\varepsilon _{j})=\delta _{ij}-(1/9)$ and have
length $2\sqrt{2}/3$).

In fact, writing this change of basis as
${\alpha }_{i}=(R^{-1})_{i}^{\phantom{i}j}\varepsilon _{j}$, we can invert this 
relation (which leads to the matrix $R$ given in Appendix A.i)and obtain the metric corresponding to the scalar product~(\ref{ScalarProduct}),
given in terms of the Cartan matrix as: 
\begin{equation}
\label{metrepsi}
g_{\varepsilon }=RAR^{\top }\,,
\end{equation}
in the simply-laced case we are interested in.

As seen in Section \ref{KMAconv}, the Cartan matrix of the ${\mathfrak{e}}_{r}$ 
serie is a realization of $(\mathfrak{h},\Pi,\Pi^{\vee })$, where
now $\Pi^{\vee}=\{{\alpha }_{9-r}^{\vee },\ldots ,{\alpha }_{8}^{\vee
}\}\cong \Pi $. Then, from $A_{ij}=\langle {\alpha }_{i}^{\vee },
{\alpha }_{j}\rangle =({\alpha }_{i}|{\alpha }_{j})$ we have:
\begin{equation}
A=\left( 
\begin{array}{rrrrrrrr}
2 & -1 & 0 & 0 & 0 & \cdots & 0 & 0 \\ 
-1 & 2 & -1 & 0 & 0 & \cdots & 0 & 0 \\ 
0 & -1 & 2 & -1 & 0 & \cdots & 0 & 0 \\ 
0 & \ddots & \ddots & \ddots & \ddots & \ddots & \vdots & \vdots \\ 
0 & \cdots & -1 & 2 & -1 & 0 & 0 & 0 \\ 
0 & \cdots & 0 & -1 & 2 & -1 & 0 & -1 \\ 
0 & \cdots & 0 & 0 & -1 & 2 & -1 & 0 \\ 
0 & \cdots & 0 & 0 & 0 & -1 & 2 & 0 \\ 
0 & \cdots & 0 & 0 & -1 & 0 & 0 & 2%
\end{array}%
\right)  \label{CartanMatrix}
\end{equation}%
and $A$ corresponds to a Dynkin diagram of the type: 
\begin{figure}[!h]
\begin{picture}(250,50)(-100,0)
\thicklines
\multiput(20,15)(30,0){3}{\circle{6}}
\multiput(100,15)(10,0){3}{\circle{2}}
\multiput(140,15)(30,0){4}{\circle{6}}
\multiput(23,15)(30,0){2}{\line(1,0){24}}
\put(83,15){\line(1,0){12}}
\put(125,15){\line(1,0){12}}
\multiput(143,15)(30,0){3}{\line(1,0){24}}
\put(170,18){\line(0,1){24}}
\put(170,45){\circle{6}}
\put( 13,0){$\a_{9-r}$}
\put( 43,0){$\a_{10-r}$}
\put( 73,0){$\a_{11-r}$}
\put(136,0){$\a_4$}
\put(166,0){$\a_5$}
\put(196,0){$\a_6$}
\put(226,0){$\a_7$}
\put(180,40){$\a_8$}
\end{picture}
\caption{Dynkin diagram of the $E_{r}$ serie}
\label{Fig:DynkinEr}
\end{figure}
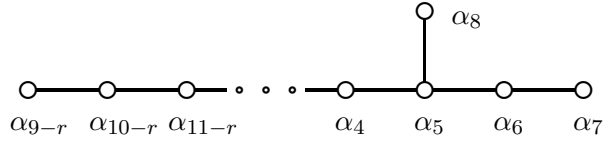
\vskip10pt

The simple coroots, in turn, form a basis of the derived Cartan subalgebra $%
\mathfrak{h}^{\prime }$, and we may choose (or alternatively define) $%
H_{i}\doteq {\alpha }_{i}^{\vee }$, $\forall i=9-r,\ldots ,8$. Since we
consider simply laced-cases only, the relation $\alpha _{i}^{\vee
}=(AR^{\top })_{ij}{\varepsilon }^{\vee j}$ determines the dual physical
basis for $r\neq 9$, \textit{i.e.} $D\neq 2$:%
\begin{equation}
\begin{array}{rcl}
H_{9-r} & = & \varepsilon ^{\vee \,11-r}-\varepsilon ^{\vee
\,12-r}=(1,-1,0,\ldots ,0) \\[3pt] 
\vdots &  &  \\ 
H_{7} & = & \varepsilon ^{\vee \,9}-\varepsilon ^{\vee \,10}=(0,\ldots
,0,1,-1)\,, \\[3pt] 
{\displaystyle H_{8}} & = & {\displaystyle-\frac{1}{3}\sum_{i=11-r}^{7}%
\varepsilon ^{\vee \,i}+\frac{2}{3}\sum_{i=8}^{10}\varepsilon ^{\vee \,i}=%
\frac{1}{3}(-1,\ldots ,-1,2,2,2)\,.}%
\end{array}
\label{Dualbasis}
\end{equation}%
In the same spirit as before, this dual basis is equipped with a scalar
product given in terms of the metric 
$g_{\varepsilon }^{\vee}=(g_{\varepsilon})^{-1}$ as: 
\begin{equation}
B(H,H^{\prime })=\sum_{i=11-r}^{10}h_{i}h_{i}^{\prime }+\frac{1}{9-r}
\sum_{i,j=11-r}^{10}h_{i}h_{j}^{\prime }\,,\text{ \ \ \ \ \ for }r\neq 9
\label{PhysicalKilling}
\end{equation}
for two elements $H=\sum_{i=11-r}^{10}h_{i}\varepsilon^{\vee \,i}$ and 
$H^{\prime}=\sum_{i=11-r}^{10}h_{i}^{\prime}\varepsilon^{\vee \,i}$. In
the affine case $r=9$, the Cartan matrix is degenerate. In order to determine 
$B(H,H^{\prime })$, one has to work in the whole Cartan subalgebra ${\mathfrak{h}}$, 
and not only in the derived one, and include a basis element related
to the scaling operator $d$. Consequently, there is no meaningful physical basis in 
this case. 

Not surprisingly, we recognize in~(\ref{PhysicalKilling}) the Killing form 
of ${\mathfrak{e}}_{r}$ restricted 
to ${\mathfrak{h}}({\mathfrak{e}}_{r})$. Since the dual metric is the inverse of $%
g_{\varepsilon }^{\vee }$, the duality bracket is defined as usual as $%
\langle \varepsilon _{i},\varepsilon ^{\vee \,j}\rangle =\varepsilon
_{i}(\varepsilon ^{\vee \,j})={\delta }_{i}^{\phantom{i}j}$, so that
consistently: 
\begin{equation}
\langle {\alpha }^{\vee },{\beta }\rangle =(\nu ({\alpha }^{\vee })|{\beta })={%
\beta }(H_{{\alpha }})\,.  \label{duabra}
\end{equation}%
Since ${\mathfrak{e}}_{r}$ is simply-laced, $\nu (\a^{\vee})=\a$, and
we then have various ways of expressing the Cartan matrix: 
\begin{equation}\label{Cartan-Killing}
A_{ij}\doteq \a_{i}(H_{j})\equiv (\a_{i}|\a_{j})\equiv B(H_{i},H_{j})\,.
\end{equation}

\subsubsection{A choice for structure constants}

We now fix the conventions for the $E_{r}$ serie that will hold throughout
the paper. For obvious reasons of economy, we introduce the following
compact notation to characterize $\e_{r}$ generators:

\begin{notation}
\label{GeneratorsIndices} Let $X_{\a}$ be a generator of the root
space $(\e_{r})_{\a}$, or of the dual subspace $\h_{\a}\subset \h$ for some root
$\a=\sum_{i=9-r}^{8}k^{i}\a_{i}\in \Delta (\e_{r})$. We write the corresponding 
generator as 
\begin{equation*}
X_{(9-r)^{k^{9-r}}\ldots \,8^{k^{8}}}\quad \mbox{ instead of }\quad
X_{k^{9-r}\a_{9-r}+\ldots +k^{8}\a_{8}}\,.
\end{equation*}
For example we will write: 
\begin{equation*}
E_{45^{2}678}\quad \mbox{ instead of }\quad E_{\a_{4}+2\a_{5}+\a_{6}+\a_{7}+\a_{8}}\,,
\end{equation*}
and similarly for $F$ and $H$. Sometimes, we will also write 
$\a_{(9-r)^{k^{9-r}}\ldots \,8^{k^{8}}}$ instead of $\sum_{i=9-r}^{8}k^{i}\a_{i}$.
\end{notation}

Furthermore, $\d$ always refers to the isotropic root of $\e_{9}$, namely 
$\d=\d_{E_{9}}=\a_{01^{2}2^{3}3^{4}4^{5}5^{6}6^{4}7^{2}8^{3}}$, $c=H_{\d}$ to its
center, and $d$ to its usual derivation operator $d=d_{E_{9}}$. Possible
subscripts added to $\d$, $c$ and $d$ will be used to discriminate $\e_{9}$ objects 
from objects belonging to its subalgebras.

Moreover, we use for $\e_{9}$ the usual construction based on
the loop algebra $\mathcal{L}(\e_{8})\doteq \C [z,z^{-1}]\otimes \e_{8}$, $z\in\C$. 
The affine KMA $\e_{9}$ is then obtained as a central extension thereof:
\begin{equation*}
\e_{9}=\mathcal{L}(\e_{8})\oplus \C c\oplus \C d\,.  
\end{equation*}
and is spanned by the basis of vertex operators satisfying
\begin{eqnarray}
[z^{m}\otimes H_{i},z^{n}\otimes H_{j}] &=&m\d_{ij}\d_{m+n,0}\,c\,, \notag \\[2pt]
[z^{m}\otimes H_{i},z^{n}\otimes E_{\a}] &=&\langle \a_{i}^{\vee},\a \rangle 
z^{m+n}\otimes E_{\a}\,, \label{affAlg} \\[3pt]
[z^{m}\otimes E_{\a},z^{n}\otimes E_{\b}] &=&\left\{ 
\begin{array}{ll}
\mathcal{N}_{\a,\b}z^{m+n}\otimes E_{\a+\b}\,, & \text{if }\a+\b \in \Delta (\e_{9}) \\ 
z^{m+n}\otimes H_{\a}+m\d_{m+n,0}\,c\,, & \text{if }\a=-\b \\ 
0\,, & \text{otherwise}
\end{array}
\right. \,.  \notag
\end{eqnarray}
In addition, the Hermitian scaling operators $d=z\frac{d}{dz}$
defined from $z\in S^{1}$ normalizes $\mathcal{L}(\e_{8})$: 
$[d,z^{n}\otimes X]=nz^{n}\otimes X$, $\forall X\in \e_{8}$.

In $\eten$, for which there is no known vertex operator construction yet, 
we rewrite the $\e_{9}$ subalgebra according to the usual prescriptions for 
KMAs by setting: $d=-H_{-1}$, $E_{n\d}^{a}=z^{n}\otimes H_{a}$, with $a=1,..,8$ 
its multiplicity, $E_{\a+n\d}=z^{n}\otimes E_{\a}$ and 
$E_{-\a+n\d}=z^{n}\otimes F_{\a}$, and similarly for negative-root generators.

Finally, there is a large number of mathematically acceptable sign
conventions for the structure constants $\mathcal{N}_{\a,\b}$,
as long as one satisfies the anti-commutativity and Jacobi identity of the
Lie bracket, as explained in Definition~\ref{StructConst}. If one decides to
map physical fields to generators of a KMA, which will eventually be done in
this paper, one has to make sure that the adjoint action of a rotation with
positive angle leads to a positive rotation of all physical tensors carrying
a covariant index affected by it. This physical requirement imposes more
stringent constraints on the structure constants. Though perhaps not the
most natural choice from a mathematical point of view, we fix signs
according to a lexicographical ordering for level 0 ($\mathfrak{sl}(r,\R)-$) 
roots, but according to an ordering based on their height for roots of higher 
level in $\a_{8}$. More concretely, if $\a=\a_{j\cdots k}$ 
has level 0, we set: 
\begin{equation*}
\mathcal{N}_{\a_{i},\a}=\left\{ 
\begin{array}{l}
1\text{ if }i<j \\ 
-1\text{ if }k<i
\end{array}
\right. \,.
\end{equation*}
On the other hand, we fix $\mathcal{N}_{5,8}=+1$, and always take the
positive sign when we lengthen a chain of simple roots of level $l>0$ by
acting with a positive simple root generator from the left, i.e.: 
\begin{equation*}
\mathcal{N}_{\a_{i},\a}=1\,,\text{ }\forall \a\text{ s.t. }l(\a)>0\,.
\end{equation*}
Structure constants for two non-simple and/or negative roots are then
automatically fixed by these choices.

\subsection{Toroidally reduced $11D$ supergravity: scalar fields and roots of $E_{11-D}$}

\label{NonLinSugra}

In this section, we rephrase the mapping between scalar fields of 11D supergravity on 
$T^q$, $q\geqslant 3$, and the roots of its finite U-duality algebras, in a way that 
will make clear the extension to the infinite-dimensional case. We start with 
$\mathcal{N}=1$ classical $11D$ supergravity, whose bosonic sector is described by the 
Lagrangian: 
\begin{equation}
S_{11}=\frac{1}{l_{P}^{9}}\int d^{11}x\,e\Big( R-\frac{l_{P}^{6}}{2\cdot 4!}
(G_{4})^{2}\Big) +\frac{1}{2\cdot 3!}\int C_{3}\wedge G_{4}\wedge G_{4}
\label{SugLag}
\end{equation}
where the four-form field strength is exact: $G_{4}=dC_{3}$. There are various
conventions for the coefficients of the three terms in the Lagrangian~(\ref{SugLag}), 
which depend on how one defines the fermionic sector of the theory. In any case, 
the factors of Planck length can be fixed by dimensional analysis. Here, we adopt 
the conventions of~\cite{OberPiol1}, where we have, in units of length
\begin{equation*}
[g_{AB}]=2\,,\quad [C_{ABC}]=0\,,\quad [d]=[dx]=0\,.
\end{equation*}
As a consequence of the above: $[R]=-2$.

The action (\ref{SugLag}) rescales homogeneously under:
\begin{equation}
g_{AB}\rightarrow M_{P}^{2}g_{AB}\,,\quad C_{ABC}\rightarrow M_{P}^{3}C_{ABC}
\label{rescal}
\end{equation}
which eliminates all $l_{P}$ terms from the Einstein-Hilbert and gauge
Lagrangian while rescaling the Chern-Simons (CS) part by $l_{P}^{-9}$, and renders, in
turn, $g_{AB}$ dimensionless. This is the convention we will adopt in the
following which will fix the mapping between the $E_{11-D}$ root system
and the fields parametrizing the scalar manifold of the reduced theory for 
$D\geqslant 3$. How to extend this analysis to the conjectured affine and hyperbolic 
U-duality groups $E_{9}$ and $E_{10}$ will be treated in the next section.

The Kaluza-Klein reduction of the theory to $D\geqslant 3$ dimensions is
performed according to the prescription 
\begin{equation}
ds_{11}^{2}=e^{\frac{\sqrt{2}}{D-2}(\rho_{D}|\varphi)}ds_{D}^{2}+
\sum_{i=D}^{10}e^{-\sqrt{2}(\varepsilon_{i}|\varphi )}
\left(\tilde{\c}_{\phantom{i}j}^{i}dx^{j}+\mathcal{A}_{1}^{i}\right) ^{2}\,,
\label{metrcompa}
\end{equation}
with $\tilde{\c}_{\phantom{i}j}^{i}=(\d_{\phantom{i}j}^{i}+\mathcal{A}_{0\,j}^{i})$ 
with $i<j$ for $\mathcal{A}_{0\,j}^{i}$. The compactification vectors $\varepsilon_{i}$ 
are the ones defining the physical basis of Definition~\ref{Phiz}, and can be expressed 
in the orthonormal basis $\{e_{i}\}_{i=1}^{11-D}$, $e_{i}\cdot e_{j}=\d_{ij}$, as
\begin{equation*}
\varepsilon _{k}=-\sum_{i=1}^{10-k}\frac{1}{\sqrt{(10-i)(9-i)}}e_{i}+\sqrt{%
\frac{k-2}{k-1}}e_{11-k}\,.\text{ \ \ \ for }k\leqslant 8\,.
\end{equation*}%
In the $D=2,1$ cases, the additional vectors completing the physical basis
are defined formally, without reference to the compactification procedure.

Accordingly, the vector of dilatonic scalars can be expanded as 
$\varphi=\sum_{i=1}^{11-D}\varphi_{i}e_{i}$. We will however choose to stick to the
physical basis. The expression of $\varepsilon _{k}$ in terms of the
orthonormal basis will help to connect back to the prescription of~\cite{CremJulLuPope1} 
and~\cite{Keuren1,Keuren2}. In this respect, the scalar product $(\,|\,)$ used in 
expression~(\ref{metrcompa}) is precisely the product on the root 
system~(\ref{ScalarProduct}). Finally, we also introduce the "threshold" vector
\begin{equation}
\rho_{D}=\sum_{i=D}^{10}\varepsilon_{i}  \label{thresh}
\end{equation}
which will be crucial later on when studying the structure of Minkowskian
objects in $E_{10}$.

From expression (\ref{metrcompa}), we see that the elfbein produces $(11-D)$
one-forms $\mathcal{A}_{1}^{i}$ and $(11-D)(12-D)/2$ scalars $\mathcal{A}_{0\,j}^{i}$, 
whereas the three form generates the following two-, one- and zero-form potentials: 
$(11-D)$ $C_{2\,i}$, $(11-D)(10-D)/2$ $C_{1\,ij}$ and $(11-D)(10-D)(9-D)/6$ $C_{0\,ijk}$. 
The reduction of the $11D$ action (\ref{SugLag}) to any dimension greater than two reads: 
\begin{equation}
\label{accomp}
\begin{array}{ccl}
e^{-1}\mathcal{L}\! & \!\!=\!\! & \!{\displaystyle R-\frac{1}{2}(\partial \varphi )^{2}-
\frac{1}{2\cdot 4!}e^{\sqrt{2}(\kappa |\varphi )}(\underline{G}_{4})^{2}-\frac{1}{2\cdot 3!}
\sum_{i}e^{\sqrt{2}(\kappa_{i}|\varphi )}(\underline{G}_{3\,i})^{2}}
\\[3pt] 
&&
{\ds -\frac{1}{2\cdot 2!}\sum_{i<j}e^{\sqrt{2}(\kappa _{ij}|\varphi )}(\underline{G}_{2\,ij})^{2}} 
{\displaystyle-\frac{1}{2\cdot 2!}\sum_{i}e^{\sqrt{2}(\lambda_{i}|\varphi )}(\underline{\mathcal{F}}_{2}^{i})^{2} }
\\[3pt]
&&
{\ds -\frac{1}{2}
\sum_{i<j<k}e^{\sqrt{2}(\kappa _{ijk}|\varphi )}(\underline{G}_{1\,ijk})^{2}-
\frac{1}{2}\sum_{i<j}e^{\sqrt{2}(\lambda _{ij}|\varphi )}
(\underline{\mathcal{F}}_{1\,j}^{i})^{2}+e^{-1}\mathcal{L}_{CS}}
\end{array}
\end{equation}
where $\mathcal{L}_{CS}$ is the reduction of the CS-term $C_{3}\wedge
G_{4}\wedge G_{4}$, and again indices run according to $i,j,k=D,..,10$. The
field strengths appearing in the above kinetic term exhibit the exterior
derivative of the corresponding potentials as leading term, but contain
additional non-linear Kaluza-Klein modifications. For instance:
\begin{equation}
\begin{array}{ll}
\underline{G}_{4}=G_{4}-\c_{\phantom{i}j}^{i}G_{3\,i}\wedge \mathcal{A}_{1}^{j}+
\c_{\phantom{i}k}^{i}\c_{\phantom{k}l}^{j}G_{2\,ij}\wedge \mathcal{A}_{1}^{k}\wedge \mathcal{A}_{1}^{l}+\ldots \,, & 
\underline{\mathcal{F}}_{2}^{i}=\mathcal{F}_{2}^{i}-\c_{\phantom{j}k}^{j}\mathcal{F}_{1\,j}^{i}\wedge \mathcal{A}_{1}^{k}\,, \\ 
\vdots & \underline{\mathcal{F}}_{1\,j}^{i}=\c_{\phantom{j}j}^{k}\mathcal{F}_{1\,k}^{i}\,. \\ 
\underline{G}_{2\,ij}=\c_{\phantom{i}i}^{m}\c_{\phantom{i}j}^{n}G_{2\,mn}- 
\c_{\phantom{i}i}^{l}\c_{\phantom{i}j}^{m}\c_{\phantom{k}k}^{n}G_{1\,lmn}\wedge \mathcal{A}_{1}^{k} \,,& 
\\[3pt]
\underline{G}_{1\,ijk}=\c_{\phantom{i}i}^{l}\c_{\phantom{i}j}^{m}\c_{\phantom{k}k}^{n}G_{1\,lmn}\,, & 
\end{array}
\label{fields}
\end{equation}
with $\c_{\phantom{i}j}^{i}=(\tilde{\c}^{-1})_{\phantom{i}j}^{i}$,
the not-underlined field strengths being total derivatives: 
$G_{(n)\,i_{1}\cdots i_{l}}=dC_{(n-1)\,i_{1}\cdots i_{l}}$ and 
$\mathcal{F}_{(n)\,i_{1}\cdots i_{l}}^{i}=d\mathcal{A}^{i}_{(n-1)\,i_{1}\cdots i_{l}}$,
where $n$ is the rank of the form. The whole set of field strengths and the details 
of the reduction of the CS term are well known and can be found in~\cite{LuPope2}.

The global symmetry of the scalar manifold which, upon quantization, is
conjectured to become the discrete U-duality symmetry of the theory is
encoded in the compactification vectors 
$\overline{\Delta}=\{\kappa;\kappa_{i};\kappa_{ij};\kappa_{ijk};\lambda_{i};\lambda_{ij}\}$
appearing in the Lagrangian (\ref{accomp}). As pointed out previously, this global
symmetry will only be manifest if potentials of rank $D-2$ are dualized to
scalars, thereby allowing gauge symmetries to be replaced by internal ones.
In each dimension $D$, this will select a subset of $\overline{\Delta}$ to
form the positive root system of $E_{11-D}$. One has to keep in mind,
however, that in even space-time dimensions, this rigid symmetry is usually
only realized on the field strengths themselves, and not on the potentials.
This is attributable to the customary difficulty of writing a covariant lagrangian for
self-dual fields \footnote{In some cases -for the $11D$
five-brane for instance-, this can be achieved by resorting to the Pasti-Sorokin-Tonin
formalism.}. In even dimensional cases then, the $E_{11-D}$ symmetry
appears as a local field transformation on the solution of the equations of
motion.

We now give the whole set $\overline{\Delta}$ in the physical basis. One has to
bear in mind that some of these vectors become roots only in particular
dimensions, and thus do not, in general, have squared length equal to 2. In
constrast, $\lambda_{ij}$ and $\kappa_{ijk}$ are always symmetries of the
scalar manifold, and can therefore be directly translated into positive
roots of $E_{11-D}$ for the first two levels $l=0,1$ in $\a_{8}$. We
have for $i<j<k$: 
\begin{equation*}
\begin{array}{lrcl}
l=0: & W_{\text{KKp}} & \ni & \lambda_{ij}=\varepsilon_{i}-\varepsilon_{j}\\[3pt] 
l=1: & W_{\text{M2}} & \ni & \kappa_{ijk}=\varepsilon_{i}+\varepsilon_{j}+\varepsilon_{k}
\end{array}
\end{equation*}
In fact, they build orbits of $E_{11-D}$ under the Weyl group of $SL(11-D,\R)$, 
which we denote by $W_{\text{KKp}}$ and $W_{\text{M2}}$, anticipating results from 
M-theory on $T^{10}$ which associates $\lambda_{ij}$ with Euclidean KK particles and 
$\kappa_{ijk}$ with Euclidean M2-branes. So, in $10\geqslant D\geqslant 6$, since no 
dualization occurs, the root system of the U-duality algebra is completely covered by 
$W_{\text{KKp}}$ and $W_{\text{M2}}$, with the well known results 
$G^{U}=\{SO(1,1);GL(2,\R);SL(3,\R)\times SL(2,\R);SL(5,\R);SO(5,5)\}$, for 
$D=\{10;9;8;7;6\}$.

For $D=5$, we dualize $\underline{G}_{4}=e^{-\kappa \cdot \varphi }\ast 
\underline{\widetilde{G}}_{1}$, with $-\kappa =\theta_{E_{6}}=((1)^{6})$
the highest root of $E_{6}$, which constitutes a Weyl orbit all by itself.
For highest roots of the Lie algebra relevant to our purpose, we refer the
reader to Appendix~\ref{AppendixA}~ii). In $D=4$, we dualize 
$(\underline{G}_{3\,i})=e^{-\kappa_{i}\cdot \varphi}\ast \underline{\widetilde{G}}_{1\,i}$, 
with $-\kappa_{i}=((1)^{i-1},0_{i},(1)^{7-i})$ forming the Weyl orbit of $\theta _{E_{7}}$
(which contains $\theta_{E_{6}}$). Finally, in $D=3$, dualizing 
$\underline{G}_{2\,ij}=e^{-\kappa_{ij}\cdot \varphi}\ast \underline{\widetilde{G}}_{1\,ij}$ 
and $\underline{\mathcal{F}}_{2}^{i}=e^{-\lambda_{i}\cdot \varphi}\ast 
\widetilde{\underline{\mathcal{F}}}_{1}^{i}$ increases the size of the former
$\theta_{E_{7}}$ Weyl orbit and creates the remaining $\theta_{E_{8}}$ orbit:
\begin{equation}
\begin{array}{ccccl}
D=5 & l=2: & W_{\text{M5}} & \ni & -\kappa=\frac{3}{D-2}\rho _{D} \\ 
D=4 & l=2: & W_{\text{M5}} & \ni & -\kappa_{i}=\frac{2}{D-2}\rho_{D}-\varepsilon_{i} \\ 
D=3 & l=2: & W_{\text{M5}} & \ni & -\kappa_{ij}=\frac{1}{D-2}\rho_{D}-\varepsilon_{i}-\varepsilon_{j} \\ 
& l=3: & W_{\text{KK7M}} & \ni & -\lambda_{i}=\frac{1}{D-2}\rho_{D}+\varepsilon_{i}
\end{array}
\label{W}
\end{equation}

For the same reason as before, we denote these two additional orbits $W_{%
\text{M5}}$ and $W_{\text{KK7M}}$ since they will be shown to describe
 totally wrapped Euclidean M5-branes and KK monopoles. For $D=3$,
for instance, we can check that $\dim W_{\text{KKp}}=\dim W_{\text{M5}}=28$, 
$\dim W_{\text{M2}}=56$ and $\dim W_{\text{KK7M}}=8$, which reproduces the
respective number of scalars coming from the KK gauge fields, 3-form, and their
magnetic duals, and verifies $\dim \Delta_{+}(E_{8})=\sum_{i}\dim W_{i}$.

For what follows, it will turn out useful to take advantage of the
dimensionless character of the vielbein, resulting from the rescaling 
(\ref{rescal}), to rewrite the internal metric in terms of the duality bracket 
(\ref{duabra}): 
\begin{equation}
ds_{11-D}^{2}=\sum_{i=D}^{10}e^{2\langle H_{R},\varepsilon _{i}\rangle
}\delta _{ij}\tilde{\gamma}_{\phantom{i}k}^{i}\tilde{\gamma}_{\phantom{i}%
l}^{j}dx^{k}\otimes dx^{l}  \label{metcomp}
\end{equation}%
with $H_{R}=\sum_{i=D}^{10}\ln (M_{P}R_{i})\varepsilon ^{\vee i}$. Thus in
particular: $e^{-\sqrt{2}(\varepsilon _{i}|\varphi )}=(M_{P}R_{i})^{2}$. In
this convention, the scalar Lagrangian for $D=3$ reads 
\begin{equation}
\begin{array}{l}
{\displaystyle-eg^{AB}(g_{\varepsilon }^{\vee })^{ij}\left( \frac{\partial
_{A}R_{i}}{R_{i}}\right) \left( \frac{\partial _{B}R_{j}}{R_{j}}\right) -%
\frac{1}{2}e\sum_{i<j<k}\frac{1}{(M_{P}^{3}R_{i}R_{j}R_{k})^{2}}(\underline{G%
}_{1\,ijk})^{2}} \\[3pt]
{\displaystyle-\frac{1}{2}e\sum_{i<j}\left( \frac{R_{j}}{R_{i}}\right) ^{2}(%
\underline{\mathcal{F}}_{1\,j}^{i})^{2}-\frac{1}{4}e\sum_{i<j}\left( \frac{%
R_{i}R_{j}}{M_{P}^{6}V_{8}}\right) ^{2}(\widetilde{\underline{G}}%
_{1\,ij})^{2}-\frac{1}{4}e\sum_{i}\left( \frac{1}{M_{P}^{9}R_{i}V_{8}}%
\right) ^{2}(\widetilde{\underline{\mathcal{F}}}_{1}^{i})^{2}}
\end{array}
\label{scalLagr}
\end{equation}
with the dual metric 
$(g_{\varepsilon}^{\vee})^{ij}=\d^{ij}+(D-2)^{-1}\sum_{k,l}\d^{ik}\d^{lj}$ (\ref{PhysicalKilling})
and the internal volume $V_{8}=\prod_{i=3}^{10}R_{i}$. Clearly, the
coefficients\footnote{Note that the interpretation of these coefficients is 
somewhat different than in \cite{OberPiol1}, since here we are working in the conformal
Einstein frame.} in front of the one-form kinetic terms reproduce the
inverse squared tensions for totally wrapped Euclidean KK particles,
M2-branes, M5-branes and KK monopoles (KK7M-branes). This will be the touchstone of our
analysis, and in the next section we will present, following \cite{Gan1}, how
the corresponding Minkowskian branes arise in $D=1$ in the framework of 
$E_{10|10}/K(E_{10|10})$ hyperbolic billiards.

\subsection{Non-linear realization of supergravity: the triangular and symmetric gauges}

The final step towards unfolding the hidden symmetry of the scalar manifold of the reduced 
theory consists in showing that one can construct its Langrangian density as a coset 
$\sigma$-model from a non-linear realization. Here, we rederive, in the formalism we 
use later on, only the most symmetric $D=3$ case, since the $D\geqslant 4$ constructions 
are obtainable as restriction thereof by referring to table~(\ref{W}). For the detailed 
study of the $D>3$ cases, see~\cite{CremJulLuPope1}.

Furthemore, the use of a parametrization of the coset sigma-model based on the 
Borel subalgebra of the U-duality algebra, called triangular gauge, is crucial to 
this type of non-linear realization. In contrast, we will show in the second part of 
this section, that the most natural setup to treat orbifolds of the corresponding
supergravities is given by a parametrization of the coset based on the Cartan decomposition 
of the U-duality algebra. We refer to this choice as the symmetric gauge.

In $D=3$, the non-linear realization of the scalar manifold is based on the
group element: 
\begin{equation}
\begin{array}{rcl}
g & = & {\displaystyle{\exp}\Big[-\frac{1}{\sqrt{2}}\sum_{i}\ln(M_{P}R_{i})
\varepsilon^{\vee i}\Big] \cdot \Big( \prod_{i<j}e^{\mathcal{A}_{\phantom{i}j}^{i}K_{i}^{+\,j}}\Big) \cdot 
\exp\Big[ \sum_{i<j<k}C_{ijk}Z^{+\,ijk}\Big] \cdot} \\[3pt] 
&  & {\displaystyle\cdot \exp \Big[ \sum_{i_{1}<..<i_{6}}\widetilde{C}_{i_{1}..i_{6}}\widetilde{Z}^{+\,i_{1}..i_{6}}\Big] \cdot 
\exp \Big[ \sum_{i_{1}<..<i_{8},j}\widetilde{\mathcal{A}}_{\phantom{j}i_{1}..i_{8}}^{j}\widetilde{K}_{j}^{+\,i_{1}..i_{8}}\Big] {\,,}}
\end{array}
\label{gelement}
\end{equation}
where the dual potentials are reformulated to exhibit a tensorial rank
that would generalize to $D<3$ (we drop all 0 subscripts since we are only
dealing with scalars). In particular, they are related to the $D=3$ dual potentials as 
$\widetilde{C}_{i_{1}..i_{6}}=\frac{1}{2!}\epsilon_{i_{1}..i_{6}kl}\widetilde{C}_{0}^{\phantom{0}kl}$
and $\widetilde{\mathcal{A}}_{\phantom{j}i_{1}..i_{8}}^{j}=\epsilon_{i_{1}.i_{8}}\widetilde{\mathcal{A}}_{0}^{j}$, 
by the totally antisymmetric rank 8 tensor of $SL(8,\R)$, $\epsilon_{i_{1}..i_{8}}$. 
These entered expression~(\ref{scalLagr}) as $\widetilde{G}_{1\,ij}=d\widetilde{C}_{0\,ij}$
and $\widetilde{\mathcal{F}}_{1}^{i}=d\widetilde{\mathcal{A}}_{0}^{i}$.

The group element~(\ref{gelement}) is built out of the Borel subalgebra of 
$E_{8}$, which is spanned by the following raising operators 
\begin{equation}
\begin{array}{rclcrcl}
[K_{i}^{+\,j},K_{k}^{+\,l}] & = & \d_{k}^{j}K_{i}^{+\,l}-\d_{i}^{l}K_{k}^{+\,j}\,, &  &  &  &  \\[3pt] 
[K_{i}^{+\,j},Z^{+\,k_{1}k_{2}k_{3}}] & = & -3\d_{i}^{[k_{1}}Z^{+\,|j|k_{2}k_{3}]}\,, &  & 
[K_{i}^{+\,j},\widetilde{Z}^{+\,k_{1}..k_{6}}] & = & -6\d_{i}^{[k_{1}}\widetilde{Z}^{+\,|j|k_{2}..k_{6}]}\,, \\[3pt] 
[K_{i}^{+\,j},\widetilde{K}_{k}^{+\,k_{1}..k_{8}}] & = & \d_{k}^{j}\widetilde{K}_{i}^{+\,k_{1}..k_{8}}\,, &  &  &  &  \\[3pt]
[Z^{+\,i_{1}i_{2}i_{3}},Z^{+\,i_{4}i_{5}i_{6}}] & = & -\widetilde{Z}^{+\,i_{1}..i_{6}}\,, &  & 
[Z^{+\,i_{1}i_{2}i_{3}},\widetilde{Z}^{+\,i_{4}..i_{9}}] & = & -3\widetilde{K}^{+\,[i_{1}|i_{2}i_{3}]i_{4}..i_{9}}\,,
\end{array}
\label{algE8}
\end{equation}
and the Cartan subalgebra, acting on the former as (without implicit summations 
on repeated indices) 
\begin{equation*}
\begin{array}{rclcrcl}
[\varepsilon^{\vee i},K_{j}^{+\,k}]\! & \!=\! & \!
\lambda_{jk}(\varepsilon^{\vee i})\,K_{j}^{+\,k}\,, &  & 
[\varepsilon^{\vee i},Z_{jkl}^{+}]\! & \!=\! & \!\kappa_{jkl}(\varepsilon^{\vee i})\,Z_{jkl}^{+}\,, \\[3pt] 
[\varepsilon^{\vee i},\widetilde{Z}_{j_{1}..j_{6}}^{+}]\! & \!=\! & 
\!{\displaystyle -\sum_{l<m}\epsilon_{j_{1}..j_{6}}^{\phantom{j_1..j_6}lm}\,
\kappa _{lm}(\varepsilon^{\vee i})\widetilde{Z}_{j_{1}..j_{6}}^{+}\,,} &  & 
[\varepsilon^{\vee i},\widetilde{K}_{j}^{+\,k_{1}..k_{8}}]\! & \!=\! & 
\! -\kappa_{j}(\varepsilon^{\vee i})\,\widetilde{K}_{j}^{+\,k_{1}..k_{8}}\,.
\end{array}
\end{equation*}
Anticipating the extension to $D=2,1$, we redefine positive roots 
$\kappa_{i_{1}..i_{6}}\doteq \sum_{l<m}\epsilon _{i_{1}..i_{6}}^{\phantom{j_1..j_6}lm}\kappa _{lm}$ 
and $\lambda_{j|i_{1}..i_{8}}\doteq \epsilon_{i_{1}..i_{8}}\kappa_{j}$ 
corresponding to the generators $\widetilde{Z}$ and $\widetilde{K}$. The scalar 
Lagrangian in $D$ dimensions is then expressible as a coset sigma-model, obtained
from the algebraic field strength $\mathcal{G}\doteq g^{-1}dg$ 
\begin{equation}  \label{FS}
\begin{array}{rcl}
\mathcal{G} \! & \!=\! & \!{\displaystyle -\frac{1}{\sqrt{2}}\sum_{i}\ln(M_{P}R_{i})
\varepsilon^{\vee i}+\sum_{i<j<k}e^{\langle H_{R},\kappa_{ijk}\rangle}
\underline{G}_{1\,ijk}Z^{+\,ijk}+\sum_{i<j}e^{\langle H_{R},\lambda _{ij}\rangle }
\underline{\mathcal{F}}_{1\,j}^{i}K_{i}^{+\,j}}\\ 
&  & {\displaystyle +\sum_{i_{1}<..<i_{6}}e^{-\langle H_{R},\kappa_{i_{1}..i_{6}}\rangle }
\widetilde{\underline{G}}_{1\,i_{1}..i_{6}\,}\widetilde{Z}^{+\,i_{1}..i_{6}}+
\sum_{i_{1}<..<i_{8},j}e^{-\langle H_{R},\lambda_{j|i_{1}..i_{8}}\rangle }
\widetilde{\underline{\mathcal{F}}}_{1\,\phantom{j}i_{1}..i_{8}}^{\,j}
\widetilde{K}_{j}^{+\,i_{1}..i_{8}}\,. }
\end{array}
\end{equation}
This particular parametrization of the coset $\e_{8|8}/\k(\e_{8|8})$ is known 
as the Iwasawa decomposition. In other words, the split form $\g^{U}=\e_{11-D|11-D}$ 
decomposes as a sum of closed factors $\g^{U}=\k\oplus \an\oplus \mathfrak{n}$, 
where $\k$ is its maximal compact subalgebra. Then, the coset 
$\g^{U}/\k=\an\oplus \mathfrak{n}$ is parametrized by the direct sum of an abelian 
and a nilpotent subalgebra. This can be interpreted as a "gauge" choice, where 
the coset elements are either diagonal (Cartan generators) or upper triangular
(Borel, or positive root, generators). In the following, this choice will be
referred to as the \textit{triangular gauge}.

The negative root generators can be retrieved from the Borel subalgebra by
defining the appropriate transposition operation. Since we want it to be
applicable to $\g^{U}=\e_{11-D|11-D}$ $\forall D$, and not only to U-duality 
algebras with orthogonal maximal compact subalgebra, we construct it as 
in~\cite{CremJulLuPope1} out of the Cartan involution $\vartheta$ as 
$T(X)=-\vartheta (X)$, $\forall X\in \g^{U}$. This induces a corresponding 
generalized transposition~\cite{CremJulLuPope1,CremJulLuPope2} on the group level
denoted by: $T(\mathcal{X})=\Theta (\mathcal{X}^{-1})$, $\forall \mathcal{X}\in G^{U}$. 
In the present case, since $\g^{U}$ is the split form, we have 
$\vartheta =\vartheta_{C}$, the latter being the Chevalley involution.

By requiring the following normalizations:
\begin{equation*}
\begin{array}{l}
\Tr(\varepsilon^{\vee i}\varepsilon^{\vee j})=2(g_{\varepsilon}^{\vee })^{ij}\,,\qquad \qquad 
\Tr\Big(K_{i}^{\phantom{i}j}T(K_{k}^{\phantom{i}l})\Big)=\d_{ik}\d^{jl}\,, \\[3pt] 
\Tr\Big(X^{i_{1}..i_{p}|j_{1}..j_{q}}T(X^{k_{1}..k_{p}|l_{1}..l_{q}})\Big)=
p!q!\d_{k_{1}}^{[i_{1}}\cdots \d_{k_{p}}^{i_{p}]}\d_{l_{1}}^{[j_{1}}\cdots\d_{l_{q}}^{j_{q}]}\,,
\end{array}
\end{equation*}
where $X^{i_{1}..i_{p}|j_{1}..j_{q}}$ stand for the remaining generators of (\ref{algE8}), 
the bosonic scalar Lagrangian (\ref{scalLagr}) is readily obtained from the coset sigma-model
\begin{equation}
\mathcal{L}=-\frac{1}{2}e\text{Tr}\left[ g^{-1}\partial g(\Id+T)g^{-1}\partial g\right] 
\equiv \frac{1}{4}e\text{Tr}(\partial \mathcal{M}^{-1}\partial \mathcal{M})  \label{LagNL}
\end{equation}
where $\mathcal{M}=gT(g)$ is the internal $\sigma$-model metric. The equations of motion
for the moduli of the theory are then summarized in the Maurer-Cartan
equation: $d\mathcal{G=G}\wedge \mathcal{G}$. By adding the negative root
generators, we restore the $K(E_{11-D})$ local gauge invariance, and enhance
the coset to the full continuous U-duality group. In $D=3$, for instance, we
thus recover the dimension of $E_{8}$ as:%
\begin{eqnarray*}
248 &=&8(H_{i})+28(K_{i}^{+j})+56(Z^{+ijk})+28(\widetilde{Z}^{+i_{1}..i_{6}})+
8(\widetilde{K}_{j}^{+i_{1}..i_{8}}) \\
&&+\overline{28}\left( T(K_{i}^{+j})\right) +\overline{56}\left(
T(Z^{+ijk})\right) +\overline{28}\left( T(\widetilde{Z}^{+i_{1}..i_{6}})\right)
+\overline{8}\left( T(\widetilde{K}_{j}^{+i_{1}..i_{8}})\right)
\end{eqnarray*}
Note that the triangular gauge is not preserved by a rigid left
transformation $U$ from the symmetry group $G^{U}$: 
$g(x)\rightarrow Ug(x)$ for $g\in G^{U}/K(G^{U})$. This leaves $\mathcal{G}$ 
invariant but will generally send $g$ out of the positive root
gauge. We will then usually need a local compensator $h(x)\in K(G^{U})$ to
bring it back to the original gauge. So the Lagrangian (\ref{LagNL}) is kept
invariant by the compensated transformation $g(x)\rightarrow Ug(x)h(x)^{-1}$
which sends: $\mathcal{M}\rightarrow U\mathcal{M}T(U)$, provided $hT(h)=%
\mathrm{1\kern-.28emI}$\texttt{.}

If the triangular gauge is the natural choice to obtain a closed non-linear
realization of a coset sigma-model, it will show to be quite unhandy when
trying to treat orbifolds of reduced $11D$ supergravity and M-theory. In
this case, a parametrization of the coset based on the Cartan decomposition
into eigenspace of the Chevalley involution $\g^{U}=\k\oplus \p$ is more 
appropriate. In other words, one starts from an algebraic field strength 
valued in $\g^{U}$: 
\begin{equation}
\widetilde{g}^{-1}d\widetilde{g}=\mathcal{P}+\mathcal{Q} \quad, \label{PQ}
\end{equation}
so that the coset is parametrized by:
\begin{equation}
\mathcal{P}=\frac{1}{2}(\Id+T)\,g^{-1}dg\,.\label{P}
\end{equation}
and $\mathcal{Q}$ ensures local (now unbroken) $K(G^{U})$ invariance of the model. 
Note that the Lagrangian (\ref{LagNL}) is, as expected, insensitive to this 
different parametrization since 
$\frac{1}{4}e\Tr(\partial \mathcal{M}^{-1}\partial \mathcal{M})\equiv 
-e\Tr\left[ \mathcal{P}T(\mathcal{P})\right]$.

We then associate symmetry generators to the moduli of compactified $11D$
supergravity / M-theory in the following fashion: for economy, we will
denote all the Borel generators of $\sl(11-D,\R)\subset \g^{U}$ by $K_{i}^{+\,j}$ for 
$i\leqslant j$, by setting in particular 
$K_{i}^{\pm\,i}=K_{i}^{\phantom{i}i}\doteq \varepsilon^{\vee \,i}$.
Using relation (\ref{Dualbasis}), the Cartan generators can now be
reexpressed as 
\begin{equation*}
H_{i}=K_{i+2}^{\phantom{A}i+2}-K_{i+3}^{\phantom{A}i+3}\,,\quad i=1,..,7\,,\qquad 
H_{8}=-\frac{1}{3}\sum_{i=1}^{5}K_{i+2}^{\phantom{A}i+2}+
\frac{2}{3}\left( K_{8}^{\phantom{8}8}+K_{9}^{\phantom{9}9}+K_{10}^{\phantom{10}10}\right) \,.
\end{equation*}
The dictionary relating physical moduli and coset generators can then be
established for all moduli fields corresponding to real roots of level 
$l=0,1,2,3$, generalizing to $D=2,1$ the previous result (\ref{W}). We will
denote the generalized transpose of a Borel generator $X^{+}$ as 
$X^{-}\doteq T(X^{+})$.
\begin{equation}
\begin{array}{c|c|c}
\text{modulus} & \text{generator} & \text{physical basis} \\ \hline
\ln (M_{P}R_{i}) & K_{i}^{\phantom{i}i} & \varepsilon ^{\vee \,i} \\[3pt] 
\mathcal{A}_{\phantom{i}j}^{i} & K_{i}^{\phantom{i}j}=
\frac{1}{2}\left(K_{i}^{+\,j}+K_{i}^{-\,j}\right) & \varepsilon_{i}-\varepsilon_{j}\\[3pt] 
C_{ijk} & Z^{ijk}=\frac{1}{2}\left( Z^{+\,ijk}+Z^{-\,ijk}\right) & 
\varepsilon^{i}+\varepsilon^{j}+\varepsilon^{k} \\[3pt] 
\widetilde{C}_{i_{1}..i_{6}} & Z^{i_{1}..i_{6}}=
\frac{1}{2}\left(Z^{+\,i_{1}..i_{6}}+Z^{-\,i_{1}..i_{6}}\right) & 
\sum_{l=1}^{6}\varepsilon_{i_{l}} \\[3pt] 
\widetilde{\mathcal{A}}_{\phantom{j}i_{1}..i_{8}}^{j},\;\text{{\footnotesize 
{$j\in \{i_{1},..,i_{8}\}$}}} & \widetilde{K}_{j}^{\phantom{j}i_{1}..i_{8}}=
\frac{1}{2}\left( \widetilde{K}_{j}^{+\,i_{1}..i_{8}}+
\widetilde{K}_{j}^{-\,i_{1}..i_{8}}\right) & 
\sum_{l=1}^{8}(1+\d_{j}^{i_{l}})\varepsilon _{i_{l}}
\end{array}
\label{tablist}
\end{equation}
This list exhausts all highest weight $\sl(11-D,\R)$ representations present 
for $D=3$. In the infinite-dimensional case, there is an infinite number of 
other $\sl(11-D,\R)$ representations. The question of their identification is 
still a largely open question. Progresses have been made lately in identifying 
some roots of $E_{10}$ as one-loop corrections to $11D$ supergravity~\cite{DamNic1} 
or as Minkowskian M-branes and additional solitonic objects of M-theory~\cite{Gan1}. 
These questions will be introduced in the next section, and will become one of 
the main topics of the last part of this paper.

However, it is worth noting that for $D\leqslant 2$, the 8-form generator is
now subject to the Jacobi identity
\begin{equation}
\widetilde{K}^{[i_{1}|i_{2}..i_{9}]}=0\,.  \label{Jac}
\end{equation}%
In $D=2$, this reflects the fact that the would-be totally antisymmetric
generator $\widetilde{K}^{[i_{1}i_{2}..i_{9}]}$ attached to the null root $%
\delta $ is not the dual of a supergravity scalar, but corresponds to the
root space $\{z\otimes H_{i}\}_{i=1,..,8}$, and reflects the localization of
the U-duality symmetry.

In addition, we denote the compact generators by 
$\mathcal{K}_{i}^{\phantom{i}j}=K_{i}^{+\,j}-K_{i}^{-\,j}$, 
$\mathcal{Z}^{ijk}=Z^{+\,ijk}-Z^{-\,ijk}$ and
similarly for $\mathcal{Z}^{i_{1}..i_{6}}$ and 
$\widetilde{\mathcal{K}}_{j}^{i_{1}..i_{8}}$. Then:
\begin{equation*}
K(G^{U})=\text{Span}\left\{ \mathcal{K}_{i}^{\phantom{i}j};\mathcal{Z}^{ijk};
\mathcal{Z}^{i_{1}..i_{6}};\widetilde{\mathcal{K}}_{j}^{\phantom{j}
i_{1}..i_{8}}\right\}
\end{equation*}
Fixing the normalization of the compact generators to 1 has been motivated
by the algebraic orbifolding procedure we will use in the next sections, and
ensures that automorphism generators and the orbifold charges they induce
have the same normalization.

In particular, the compact Lorentz generators 
$\mathcal{K}_{i}^{\phantom{i}i+1}\equiv E_{\a_{i-2}}-F_{\a_{i-2}}$, 
$\forall i=D,\ldots,9$ clearly generate rotations in the $(i,i+1)$-planes, so 
that a general rotation in the $(i,j)$-plane is induced by 
$\mathcal{K}_{i}^{\phantom{i}j}$. One can check that, as expected,
$[\mathcal{K}_{k}^{\phantom{k}j},X_{i_{1}..j..i_{p}}]=X_{i_{1}..k..i_{p}}$ 
$\forall X\in \g^{U}/\k(\g^{U})$ in Table~(\ref{tablist}).
For instance, the commutator $[\mathcal{K}_{i}^{\phantom{i}i+1},Z_{i+1\,jk}]$ 
for $i+1<j<k$ belongs to the root space of:
\begin{equation*}
\a=\a_{i-2}+..+\a_{j-3}+2(\a_{j-2}+..+\a_{k-3})+3(\a_{k-2}+..+\a_{5})+
2\a_{6}+\a_{7}+\a_{8}
\end{equation*}
that defines $Z_{ijk}=(1/2)(E_{{\alpha }}+F_{{\alpha }})$.

As a final remark, note that the group element $\widetilde{g}$ with value in 
$G^{U}$ can be used to reinstate local $K(G^{U})$-invariance of the algebraic 
field strength $\widetilde{g}^{-1}d\widetilde{g}=\mathcal{P}+\mathcal{Q}$ under
the transformation as $\widetilde{g}(x)\rightarrow U\widetilde{g}%
(x)h(x)^{-1} $ for $h(x)\in K(G^{U})$ and a rigid U-duality element $U\in
G^{U}$. In this case, $\mathcal{Q}$ transforms as a generalized connection:
\begin{equation*}
\mathcal{Q}\rightarrow h(x)\mathcal{Q}h(x)^{-1}-h(x)\,dh(x)^{-1}\,,
\end{equation*}%
and $\mathcal{P}+\mathcal{Q}$ as a generalized field-strength: 
$\mathcal{P}+\mathcal{Q}\rightarrow U^{-1}(\mathcal{P}+\mathcal{Q})U$. 
Performing a level expansion of $\mathcal{Q}$:
\begin{equation*}
\mathcal{Q}=\frac{1}{2}dx^{A}\,\big(\omega_{Aj}^{\phantom{Aj}i}
\mathcal{K}_{i}^{\phantom{i}j}+\omega_{A}^{\phantom{A}ijk}\mathcal{Z}_{ijk}\big)+\ldots
\end{equation*}
we recognize for $l=0$ the Lorentz connection, for $l=1$ the 3-form gauge
connection, etc.

Actually our motivation for working in the symmetric gauge comes from the
fact that, at the level of the algebra, the orbifold charge operator acting as 
Ad$h$ preserves this choice. Indeed $h\in K(G^{U})$ is in this case a rigid
transformation, so that one can drop the connection part $\mathcal{Q}$ in 
expression (\ref{PQ}), and Ad$h$ normalizes $\mathcal{P}$.

Along this line, a non-linear realization where only local Lorentz invariance 
is implemented has been used extensively in \cite{West,West2,West3} to uncover
very-extended Kac-Moody hidden symmetries of various supergravity theories.
This has led to the conjecture that $\mathfrak{e}_{11}$ is a symmetry of $11D$ 
supergravity, and possibly M-theory, as this very-extended algebra can be obtained 
as the closure of the finite Borel algebra of a non-linear realization similiar
to the one we have seen above, with the $11D$ conformal algebra.

\subsection{M-theory near a space-like singularity as a 
$E_{10|10}/K(E_{10|10})$ $\protect\sigma $-model}
\label{sec:sing}

In the preceding section, we have reviewed some basic material about $11D$
supergravity compactified on square tori, which we will need in this paper
to derive the residual U-duality symmetry of the untwisted sector of the
theory when certain compact directions are taken on a orbifold. The
extension of this analysis to the orbifolded theory in $D=2,1$ dimensions,
where KM hidden symmetries are expected to arise, will require a
generalization of the low-energy effective supergravity approach. The proper
framework to treat hidden symmetries in $D=1$ involves a $\sigma$-model
based on the infinite coset $E_{10|10}/K(E_{10|10})$. In the vicinity of a
space-like singularity, this type of model turns out to be a generalization
of a Kasner cosmology, leading to a null geodesic motion in the moduli space
of the theory, interrupted by successive reflections against potential
walls. This dynamics is usual referred to as a cosmological billiard, where
by billiard, we mean a convex polyhedron with finitely many vertices, some
of them at infinity.

In \cite{DamHenNic1,DamNic2} the classical dynamics of M-theory near a
spacelike singularity has been conjectured to possess a dual description in
terms of this chaotic hyperbolic cosmological billiard. In particular, these
authors have shown that, in a small tension limit $l_{p}\rightarrow 0$
corresponding to a formal BKL expansion, there is a mapping\footnote{%
Which has been worked out up to order $l=6$ and ht$(\alpha )=29$.} between
(possibly composite) operators\footnote{%
Constructed from the vielbein, electric and magnetic components of the
four-form field-strength, and their multiple spatial gradients.} of the
truncated equations of motion of $11D$ supergravity at a given spatial
point, and one-parameter quantities (coordinates) in a formal $\sigma $%
-model over the coset space $E_{10|10}/K(E_{10|10})$. More recently, \cite%
{DamNic1} has pushed the analysis even further, and shown how higher order
M-theory corrections to the low-energy $11D$ supergravity action (similar to 
$\alpha ^{\prime }$ corrections in string theory) are realized in the $%
\sigma $-model, giving an interpretation for certain negative imaginary
roots of $E_{10}$.

In particular, the regime in which this correspondence holds is reached when
at least one of the diagonal metric moduli is small, in the sense that $%
\exists i\mbox{ s.t. }R_{i}\ll l_{P}$. In this case, the contributions to
the Lagrangian of $11D$ supergravity (with possible higher order
corrections) coming from derivatives of the metric and $p$-form fields can
be approximated by an effective potential, with polynomial dependence on the
diagonal metric moduli. In the BKL limit, these potential terms become
increasingly steep, and can be replaced by sharp walls or cushions, which,
on the $E_{10|10}/K(E_{10|10})$ side of the correspondence, define a Weyl
chamber of $E_{10}$. The dynamics of the model then reduces to the time
evolution of the diagonal metric moduli which, in the coset, map to a null
geodesic in the Cartan subalgebra of $E_{10}$ deflected by successive
bounces against the billiard walls. In the leading order approximation, one
can restrict his attention to the dominant walls, \textit{i.e.} those given
by the simple roots of $E_{10}$, so that the billiard motion is confined to
the fundamental Weyl chamber of $E_{10}$. As mentioned before, \cite%
{DamHenNic1,DamNic2,DamNic1} have shown how to extend this analysis to other
Weyl chambers by considering higher level non-simple roots of $E_{10}$, and
how the latter can be related, on the supergravity side, to composite
operators containing multiple gradients of the supergravity fields and to
M-theory corrections. These higher order terms appear as one considers
smaller and smaller corrections in $l_{P}$ as we approach the singularity $%
x^{0}\rightarrow \infty $. These corrections are of two different kind: they
correspond either to taking into account higher and higher spatial gradients
of the supergravity fields in the truncated equations of motion of $11D$
supergravity at a given point of space, or to considering M-theory
corrections to the classical two-derivative Lagrangian.

In the following, since we ultimately want to make contact with \cite%
{Gan1,Gan2}, we will consider the more restrictive case in which the space
is chosen compact, and is in particular taken to be the ten-dimensional
torus $T^{10}$, with periodic coordinates $0\leqslant x^{i}<2\pi$, $\forall
i=1,\ldots,10$. This, in principle, does not change anything to the non-compact
setup of \cite{DamHenNic2}, since there the mapping relates algebraic
quantities to supergravity fields at a \textit{given point} in space,
regardless in principle of the global properties of the manifold.

Before tackling the full-fledged hyperbolic $E_{10}$ billiard and the
effective Hamiltonian description of 11D supergravity dual to it, it is
instructive to consider the toy model obtained by setting all fields to zero
except the dilatons. This leads to a simple cosmological model characterized
by a space-like singularity at constant time slices $t$. This suggests to
introduce a lapse function $N(t)$. The proper time $\sigma $ is then defined
as $d\sigma =-N(t)dt$, and degenerates ($\sigma \rightarrow 0^{+}$) at the
singularity $N(t)=0$. This particular limit is referred to as the BKL limit,
from the work of Belinskii, Khalatnikov and Lifshitz~\cite{BKL1,BKL2}. As
one approaches the singularity, the spatial points become causally
disconnected since the horizon scale is smaller than their spacelike
distance.

In this simplified picture, the metric (\ref{metrcompa}) reduces to a Kasner
one, all non-zero fields can be taken to depend only on time (since the
space points are fixed): 
\begin{equation}
ds^{2}=-(N(t)dt)^{2}+\sum_{i,j=1}^{10}e^{2\langle H_{R}(t),\varepsilon
_{i}\rangle }\delta _{ij}dx^{i}\otimes dx^{j}\,.  \label{metplat}
\end{equation}%
In addition to proper time $\sigma $, we introduce an "intermediate time"
coordinate $u$ defined as%
\begin{equation}
du=-\frac{1}{\sqrt{\bar{g}}}d\sigma =\frac{N(t)}{M_{P}^{10}V(t)}dt\,,
\label{du}
\end{equation}%
where $\sqrt{\bar{g}}=\sqrt{\det g_{ij}}=e^{\langle H_{R},\rho _{1}\rangle }$
and $V(t)=\prod_{i=1}^{10}R_{i}(t)$, where $\rho _{1}$ is the "threshold"
vector (\ref{thresh}). In this frame, one approaches the singularity as
$u\rightarrow +\infty $.

Extremizing $\int eR$ with respect to the $R_{i}$ and $N/\sqrt{\bar{g}}$, we
get the equations of motion for the compactification radii and the zero mass
condition:%
\begin{equation}
\frac{d}{du}\left( \frac{1}{R_{i}}\frac{dR_{i}}{du}\right) =0\,,\ \ \ \ \ \
\ \ \ \ \ \ \ \ \ \ \sum_{i}\left( \frac{\dot{R}_{i}}{R_{i}}\right)
^{2}-\left( \sum_{i}\frac{\dot{R}_{i}}{R_{i}}\right) ^{2}=0\,,
\label{zeromass}
\end{equation}%
where the dot denotes $\frac{d}{dt}$. Setting $%
R_{i}(u_{0})=R_{i}(s_{0})=M_{P}^{-1}$, one obtains $R_{i}$ in terms either
of $u$ or of $\sigma$
\begin{equation}
M_{P}R_{i}=e^{-v_{i}\left( u-u_{0}\right) }=\left( \frac{\sigma }{\sigma _{0}%
}\right) ^{\frac{v_{i}}{\sum_{j}v_{j}}}  \label{Kasnsol}
\end{equation}%
since $u=-\frac{1}{\sum_{j}v_{j}}\ln (\sigma +const)+const^{\prime }$. Then,
the evolution of the system reduces to a null geodesic in $\mathfrak{h}%
(E_{10})$. In the $u$-frame in particular, the vector $H_{R}(u)=\sum_{i}\ln
(M_{P}R_{i}(u)){\varepsilon ^{\vee i}}$ can be regarded as a particle moving
along a straight line at constant velocity $-\vec{v}$. In the $u$-frame, it
is convenient to define $\vec{p}=(\sum_{j}v_{j})^{-1}\vec{v}$, whose components are
called Kasner exponents. These satisfy in particular: 
\begin{equation}
\sum_{i}p_{i}^{2}-\Big(\sum_{i}p_{i}\Big)^{2}=0\,,\text{ \ \ \ \ \ \ \ \ \ \
\ \ \ \ \ \ }\sum_{i}p_{i}=1\,.  \label{pow}
\end{equation}%
The first constraint originates from the zero mass condition (\ref{zeromass}%
) and implies $\vec{p}\in \mathfrak{h}^{\ast }(E_{10})$, while the second
one comes from the very definition of the $p_{i}$'s. These two conditions
result in at least one of the $p_{i}$'s being positive and at least another
one being negative, which leads, as expected, to a Schwartzschild type
singularity in the far past and far future.

In the general case, we reinstate off-diagonal metric elements in the
line-element (\ref{metplat}) by introducing the vielbein (\ref{metcomp}) in
triangular gauge:%
\begin{equation}
\delta _{ij}dx^{i}\otimes dx^{j}\rightarrow \delta _{ij}\tilde{\gamma}_{%
\phantom{i}p}^{i}\tilde{\gamma}_{\phantom{i}q}^{j}dx^{p}\otimes dx^{q}\,,
\label{Iwa}
\end{equation}%
with $\tilde{\gamma}_{\phantom{i}p}^{i}=(\delta _{\phantom{i}p}^{i}+\mathcal{%
A}_{\phantom{i}\,p}^{i})$, and $\mathcal{A}_{\phantom{i}\,p}^{i}$ defined
for $i<p$. For reasons of clarity, we discriminate this time the flat
indices $(i,j,k,l)$ from the curved ones $(p,q,r,s)$.

In this more general case, it can be shown \cite{DamHenNic2}, that
asymptotically (when approaching the singularity), the log of the scale
factors $\ln M_{P}R_{i}$ are still linear functions of $u$, while the
off-diagonal terms $\mathcal{A}_{\phantom{i}j}^{i}$ tend to constants: in
billiard language, they \textit{freeze} asymptotically.

To get the full supergravity picture, one will in addition turn on electric
3-form and magnetic 6-form fields and the duals to the Kaluza-Klein vectors,
and possibly other higher order corrective terms. Provided we work in the
Iwasawa decomposition (\ref{metrcompa}), one can show that, similarly to the
off-diagonal metric components, these additional fields and their multiple
derivatives also freeze as one approaches the singularity. In particular,
all $(p+1)$-form field strengths will tend to constants in this regime, and
therefore behave like potential terms for the dynamical scale factors.

An effective Hamiltonian description of such a system has been proposed \cite%
{DamHenNic2,DamNic1}:%
\begin{equation}
\mathcal{H}(H_{R},\partial _{u}H_{R},F)=B(\partial _{u}H_{R},\partial
_{u}H_{R})+\frac{1}{2}\sum_{A}e^{2w_{A}(H_{R})}c_{A}(F)\,,  \label{ham1}
\end{equation}%
For later convenience, we want to keep the dependence on conformal time
apparent, so that we use $\partial _{u}H_{R}$ to represent the canonical
momenta given by $\pi ^{i}=2(g_{\varepsilon }^{\vee })^{ij}\,\partial
_{u}\ln M_{P}R_{i}$. In units of proper time (\ref{conf}), the Hamiltonian
is then given by the integral: 
\begin{equation}
H=\int d^{10}x\,\frac{N}{\sqrt{\bar{g}}}\,\mathcal{H}\,.  \label{hamresc}
\end{equation}%
Let us now discuss the structure of $\mathcal{H}$ (\ref{ham1}) in more
details. First, the Killing form $B$ is defined as in eqn.~(\ref%
{PhysicalKilling}) and is alternatively given by the metric ${g_{\varepsilon
}^{\vee }}$. It determines the kinetic energy of the scale factors. The
second term in expression (\ref{ham1}) is the effective potential generated
by the frozen off-diagonal metric components, the $p$-form fields, and
multiple derivatives of all of them, which are collectively denoted by $F$.
The (possibly) infinite sum over $A$ includes the basic contributions from
classical $11D$ supergravity (\ref{scalLagr}), plus higher order terms
related to quantum corrections coming from M-theory. In the vicinity of a
spacelike singularity, the dependence on the diagonal metric elements
factorizes, so that these contributions split into an exponential of the
scale factors, $e^{2w_{A}(H_{R})}$, and a part that freezes in this BKL
limit, generically denoted by $c_{A}(F)$.

These exponential factors $e^{2w_{A}(H_{R})}$ behave as sharp wall
potentials, now interrupting the former straight line null geodesics $%
H_{R}(u)$ and reflecting its trajectory, while conserving the energy of the
corresponding virtual particle and the components of its momentum parallel
to the wall. In contrast, the perpendicular components change sign. Despite
these reflections, the dynamics remains integrable and leads to a chaotic
billiard motion. The reflections off the walls happen to be
Weyl-reflections in $\mathfrak{h}(E_{10})$, and therefore conserve the
kinetic term in $\mathcal{H}$ (\ref{ham1}). However, since the Weyl group of 
$E_{10}$ is a subgroup of the U-duality group, it acts non-trivially on the
individual potential terms of $\mathcal{H}$. As the walls represent
themselves Weyl reflections, they will be exchanged under conjugation by the
Weyl group. More details on the action of the U-duality group in the general
case, and in relation with hyperbolic billiard dynamics can be found in
Appendix \ref{AppendixB}.

In the BKL limit then, the potential terms $e^{2w_{A}(H_{R})}$ can be
mimicked by theta-functions: $\Theta \left( w_{A}(H_{R})\right) $ so that
the dynamics is confined to a billiard table defined by the inequalities $%
w_{A}(H_{R})\leqslant 0$. If one can isolate, among them, a finite set of
inequalities $I=\{A_{1},..,A_{n}\}$, $n<\infty $, which imply all the
others, the walls they are related to are called \textit{dominant}.

The contributions to the effective potential in $\mathcal{H}$ (\ref{ham1})
arising from classical supergravity can be described concretely, and we can
give to the corresponding walls an interpretation in terms of roots of $%
E_{10}$. As a first example, we give the reduction on $T^{10}$ of the
kinetic energy for the 3-form potential, and write it in terms of the
momenta conjugate to the $C_{ijk}$:%
\begin{equation}
\frac{1}{2}{\sum_{i<j<k}e^{2w_{ijk}(H_{R})}(\underline{\pi }}^{ijk}{)^{2}=}%
\frac{1}{2}{\sum_{i<j<k}(M_{P}^{3}R_{i}R_{j}R_{k})^{2}}\left[ \tilde{\gamma}%
_{\phantom{i}p}^{i}\tilde{\gamma}_{\phantom{i}q}^{j}\tilde{\gamma}_{%
\phantom{k}r}^{k}\,{\pi ^{pqr}}\right] ^{2}\,.  \label{potM2}
\end{equation}%
As pointed out above, the momenta ${\underline{\pi }}^{ijk}$ freeze in the
BKL limit. Their version in curved space can be computed to be%
\begin{equation*}
\pi ^{p_{1}p_{2}p_{3}}={\sum_{i<j<k}e^{-2w_{ijk}(H_{R})}}\gamma _{%
\phantom{p_1}i}^{p_{1}}\gamma _{\phantom{p_1}j}^{p_{2}}\gamma _{\phantom{p_1}%
k}^{p_{3}}\gamma _{\phantom{p_1}i}^{q_{1}}\gamma _{\phantom{p_1}%
j}^{q_{2}}\gamma _{\phantom{p_1}k}^{q_{3}}\partial _{u}C_{q_{1}q_{2}q_{3}}\,,
\end{equation*}%
with $\partial _{u}C_{q_{1}q_{2}q_{3}}=\sqrt{\bar{g}}\,G_{0q_{1}q_{2}q_{3}}$%
, since the flat time-index is defined by: $dx^{0}=N(t)dt$. From expression (%
\ref{potM2}), one identifies the walls related to the three-form effective
potential, and referred to as "electric" in \cite{DamHenNic2}, with $l=1$
positive roots of $E_{10}$, namely: $w_{ijk}=\varepsilon _{i}+\varepsilon
_{j}+\varepsilon _{k}\in W_{\text{M2}}(E_{10})$.

Note, in passing, that the exponential in eqn.(\ref{potM2}) has the opposite
sign compared to the reduced Lagrangian (\ref{scalLagr}) for $D\geqslant 3$.
This is a consequence of opting for the Hamiltonian formalism, where the
Legendre transform inverts the sign of the phase factor $e^{2w_{A}(H_{R})}$
for the momenta ${\pi ^{pqr}}$. In this respect, the latter are defined with
upper curved indices (flattened by $\tilde{\gamma}_{\phantom{i}p}^{i}$), as
in expression (\ref{potM2}), while their conjugate fields carry lower curved
ones (flattened by the inverse $\gamma _{\phantom{i}p}^{i}\doteq (\tilde{%
\gamma}^{-1})_{\phantom{i}p}^{i}$, see (\ref{potM5}) below). For more
details, see \cite{DamHenNic2}. In any case, one can simultaneously flip all
signs in the wall factors for both the Lagrange and Hamiltonian formalisms,
by choosing a lower triangular parametrization for the vielbein (\ref{Iwa}),
which corresponds to an Iwasawa decomposition with respect to the set of
negative roots of the U-duality group.

Similarly, there will be a potential term resulting from the dual six-form $%
\widetilde{C}_{6}$ kinetic term (the second term in the second line of
expression (\ref{scalLagr})). In contrast to eqn.~(\ref{scalLagr}), we
rewrite the electric field energy for $\widetilde{C}_{6}$ as the magnetic
field energy for $C_{3}$: 
\begin{equation}
\begin{array}{l}
{\displaystyle
\frac{1}{2}\sum_{i<j<k<l}e^{2w_{ijkl}(H_{R})}(\underline{G}_{ijkl})^{2}=} \\[2pt] 
{\displaystyle\qquad \frac{1}{2}\sum_{i_{1}<\ldots <i_{6}}\sum_{i_{7}<\ldots
<i_{10}}{(M_{P}^{6}R_{i_{1}}}\cdots {R_{i_{6}})^{2}}\left[ \gamma _{%
\phantom{p}i_{7}}^{p}\gamma _{\phantom{q}i_{8}}^{q}\gamma _{\phantom{r}%
i_{9}}^{r}\gamma _{\phantom{s}i_{10}}^{s}G_{pqrs}\,\epsilon ^{i_{1}\ldots
i_{10}}\right] ^{2}}\label{potM5}%
\end{array}%
\end{equation}%
Again, the components ${\underline{G}_{ijkl}}$ freeze in the BKL limit,
leaving a dependence on the "magnetic" walls given by $l=2$ roots of $E_{10}$%
: $w_{ijkl}=\sum_{m\notin \{i,j,k,l\}}\varepsilon _{m}\in W_{\text{M5}%
}(E_{10})$. Dualizing this expression with respect to the ten compact
directions, we can generate Chern-Simons terms resulting from the
topological couplings appearing in the definition of $\underline{G}_{2\,ij}$
in eqn.~(\ref{fields}), namely: $2\gamma _{\phantom{i}i_{1}}^{p_{1}}\gamma _{%
\phantom{i}i_{2}}^{p_{2}}\gamma _{\phantom{m}i_{3}}^{p_{3}}\gamma _{%
\phantom{k}i_{4}}^{p_{4}}\gamma _{\phantom{k}%
i_{5}}^{p_{5}}G_{p_{3}p_{4}p_{5}[p_{1}}\mathcal{A}_{\phantom{i_5}%
p_{2}]}^{i_5}$. However, such contributions are characterized by the same
walls as expression (\ref{potM5}), and thus have no influence on the
asymptotic billiard dynamics, but only modify the constraints.

The off-diagonal components of the metric $\mathcal{A}_{\phantom{i}\,j}^{i}$
will also contribute a potential term in $\mathcal{H}$ (\ref{ham1}).
Inspecting the second line of expression (\ref{scalLagr}), we recognize it
as the frozen kinetic part of the first term on this second line: 
\begin{equation}
\frac{1}{2}\sum_{i<j}e^{2w_{ij}(H_{R})}(\underline{\pi }_{\phantom{i}%
j}^{i})^{2}=\frac{1}{2}\sum_{i<j}\left( \frac{R_{i}}{R_{j}}\right) ^{2}\big[%
\tilde{\gamma}_{\phantom{k}p}^{i}\pi _{\phantom{i}j}^{p}\big]^{2}
\label{potmetric}
\end{equation}%
where the momentum with curved indices is defined as%
\begin{equation*}
\pi _{\phantom{i}j}^{p}=\sum_{k}e^{-2w_{kj}(H_{R})}\gamma _{\phantom{k}%
k}^{p}\gamma _{\phantom{k}j}^{r}\partial _{u}\mathcal{A}_{\phantom{i}%
r}^{k}\,,\text{ \ \ \ \ with }k<j\,.
\end{equation*}%
The sharp walls appearing in this case are usually called \textit{symmetry}
(or \textit{centrifugal}) walls and correspond to $l=0$ roots of $E_{10}$,
namely: $w_{ij}=\varepsilon _{i}-\varepsilon _{j}\in W_{\text{KKp}}(E_{10})$.

Finally, the curvature contribution to the potential in $\mathcal{H}$ (\ref%
{ham1}) produces two terms:
\begin{equation}
\label{KK7M}
\begin{array}{l}
{\ds \frac{1}{2}\sum_{j<k}\sum_{i\neq \{j,k\}}e^{2\widetilde{w}_{ijk}(H_{R})}(%
\underline{\mathcal{F}}_{\phantom{i}jk}^{i})^{2}-
\sum_{i}e^{2w_{i}(H_{R})}(\underline{\mathcal{F}}_{i})^{2} } \\[2pt]
\quad{\ds =2\sum_{i_{1}<\ldots<i_{7},i_{8}}\sum_{i_{9}<i_{10}}(M_{P}^{9}R_{i_{1}}\cdots
R_{i_{7}}R_{i_{8}}^{2})^{2}(\gamma _{\phantom{m}i_{9}}^{p}\gamma _{%
\phantom{n}i_{10}}^{q}\partial _{\lbrack p}\mathcal{A}_{\phantom{i}%
q]}^{i_{8}}\epsilon ^{i_{1}\ldots i_{10}})^{2} }  \\[2pt]
\qquad{\ds-\sum_{i_{1}<\ldots <i_{9},i_{10}}(M_{P}^{9}R_{i_{1}}\cdots R_{i_{9}})^{2}(%
\underline{\mathcal{F}}_{i_{10}}\epsilon ^{i_{1}\ldots i_{10}})^{2}\,.}
\end{array}
\end{equation}
The first one is already present as the third term on the second line of
expression (\ref{scalLagr}), the ${\underline{\mathcal{F}}_{\phantom{i}%
jk}^{i}}$ being related to the spatial gradients of the metric, or,
alternatively, to the structure functions of the Maurer-Cartan equation for
the vielbein (\ref{Iwa}):%
\begin{equation}
{\underline{\mathcal{F}}_{\phantom{i}jk}^{i}=2}\gamma _{\phantom{i}%
j}^{p}\gamma _{\phantom{m}k}^{q}\partial _{\lbrack p}\mathcal{A}_{\phantom{i}%
q]}^{i}  \label{FKK}
\end{equation}%
As for expression (\ref{potM5}), one can generate Chern-Simons couplings $%
\gamma _{\phantom{i}j}^{p}\gamma _{\phantom{m}k}^{q}\gamma _{\phantom{i}%
l}^{r}\mathcal{F}_{\phantom{i}r[p}^{i}\mathcal{A}_{\phantom{i}q]}^{l}$ by
dualizing the above expression in the ten compact directions. This again
will not generate a new wall, and, as for expression (\ref{FKK}),
corresponds to $l=3$ roots of $E_{10}$ given by $\widetilde{w}%
_{ijk}=\sum_{l\notin \{i,j,k\}}\varepsilon _{l}+2\varepsilon _{i}\in W_{%
\text{KK7M}}(E_{10})$.

The ${\underline{\mathcal{F}}}_{i}$ on the other hand are some involved
expressions depending on the fields $R_{i}$, $\partial R_{i}$, ${\underline{%
\mathcal{F}}_{\phantom{i}jk}^{i}}$ and $\partial {\underline{\mathcal{F}}_{%
\phantom{i}jk}^{i}}$. In eqn.(\ref{KK7M}), they are related to lightlike
walls $w_{i}=\sum_{k\neq i}\varepsilon _{k}$ given by all permutations of
the null root ${\delta }=(0,(1)^{9})$. These prime isotropic roots are
precisely the ones at the origin of the identity (\ref{Jac}). Since they can
be rewritten as $w_{i}=(1/2)(\widetilde{w}_{jki}+\widetilde{w}_{kij})$, they
are subdominant with respect to the $\widetilde{w}_{ijk}$, and will not
affect the dynamics of $H_{R}$ even for $\vec{p}$ close to the lightlike
direction they define. So they are usually neglected in the standard BKL
approach. In the next section, we will see that these walls have a natural
interpretation as Minkowskian KK-particles~\cite{Gan1}, and contribute
matter terms to the theory.

All the roots describing the billiard walls we have just listed are, except
for $w_{i}$, real $l\leqslant 3$ roots of $E_{10}$, and the billiard
dynamics constrains the motion of $H_{R}$ to a polywedge bounded by the
hyperplanes: $\langle H_{R}(t),w_{A}\rangle =0$, with $A$ spanning the
indices of the walls mentioned above. The dominant walls are then the simple
roots of $E_{10}$. In this respect, the orbits $W_{\text{M5}}(E_{10})$ and $%
W_{\text{KK7M}}(E_{10})$ contains only subdominant walls, which are \textit{%
hidden} behind the dominant ones, and can, in a first and coarse
approximation, be neglected. The condition $\alpha _{i}(H_{R}(t))\leqslant 0$%
, with $\alpha _{i}\in \Pi (E_{10})\subset W_{\text{KKp}}(E_{10})\cup W_{%
\text{M2}}(E_{10})$, leads to the constraints:%
\begin{equation}
R_{1}\leqslant R_{2}\,,\text{ \ }R_{2}\leqslant R_{3}\,,\ldots ,\text{ }%
R_{9}\leqslant R_{10}\,,\text{ \ and }R_{8}R_{9}R_{10}\leqslant l_{P}^{3}
\label{ineq2}
\end{equation}%
and the motion on the billiard is indeed confined to the fundamental Weyl
chamber of $E_{10}$. The order in expression (\ref{ineq2}) depends on the
choice of triangular gauge for the metric (\ref{Iwa}), and does not hold for
an arbitrary vielbein. In the latter case, the formal $E_{10}$ coset $\sigma 
$-model is more complicated than expression (\ref{ham2}) below.

At this stage, we can rederive the mapping between geometrical objects of
M-theory on $T^{10}$ and the formal coset $\sigma $-model on $%
E_{10|10}/K(E_{10|10})$ proposed by \cite{DamHenNic1}, for the first $%
l=0,1,2,3$ real positive roots of $E_{10}$. This geodesic $\sigma $-model is
governed by the evolution parameter $t$, which will be identified with the
time parameter (\ref{du}). To guarantee reparametrization invariance of the
latter, we introduce the lapse function $n$, different from $N$. Then, in
terms of the rescaled evolution parameter $d\tau =ndt$, the formal $\sigma $%
-model Hamiltonian reads \cite{DamHenNic2}:%
\begin{equation}
\mathcal{H}(H_{R},\partial _{\tau }H_{R},\nu ,\partial _{\tau }\nu )=n\Big(%
B(\partial _{\tau }H_{R}|\partial _{\tau }H_{R})+\frac{1}{2}\sum_{{\alpha }%
\in \Delta _{+}(E_{10})}\sum_{a=1}^{m_{{\alpha }}}e^{2\langle H_{R},\alpha
\rangle }[P_{\alpha ,a}(\nu ,\partial _{\tau }\nu )]^{2}\Big)  \label{ham2}
\end{equation}%
where $(\nu ,\partial_{\tau}\nu)$ denotes the infinitely-many canonical variables of
the system. We again use $\partial _{\tau }H_{R}$ to represent the momenta $%
\pi ^{i}=2(g_{\varepsilon }^{\vee })^{ij}\,(R_{i})^{-1}\partial _{\tau }%
R_{i}$ conjugate to $\ln M_{P}R_{i}$. The metric entering the kinetic
term is chosen to be ${g_{\varepsilon }^{\vee }}$, which is dictated by
comparison with the bosonic sector of toroidally reduced classical 11D
supergravity.

Expression (\ref{ham2}) is obtained by computing the formal Lagrangian
density from the algebraic field strength valued in $\mathfrak{a}%
(E_{10|10})\oplus \mathfrak{n}(E_{10|10})$ as: 
\begin{equation}
g^{-1}\frac{d}{dt}g={-\frac{1}{\sqrt{2}}}\sum_{i}\frac{\dot{R}_{i}}{R_{i}}%
\,\varepsilon ^{\vee i}+\sum_{{\alpha }\in \Delta
_{+}(E_{10})}\sum_{a=1}^{m_{{\alpha }}}Y_{{\alpha },a}(\nu ,\dot{\nu}%
)\,e^{-\langle H_{R},\alpha \rangle }E_{{\alpha }}^{a}\,,  \label{coselmn}
\end{equation}%
As in eq.(\ref{LagNL}), one starts by calculating $\mathcal{L}=n^{-1}$Tr$(%
\mathcal{P}T(\mathcal{P}))$ with $\mathcal{P}$ given in expression (\ref{P}%
). One then switches to the Hamiltonian formalism, with momentum-like
variables given by the Legendre transform $P_{\alpha ,a}(\nu ,\dot{\nu})=\frac{1%
}{n}e^{-2\langle H_{R},\alpha \rangle }Y_{{\alpha },a}(\nu ,\dot{\nu})$,
eventually leading to expression (\ref{ham2}).

In the BKL limit, the (non-canonical) momenta tend to constant values $%
P_{\alpha ,a}(\nu ,\dot{\nu})\rightarrow C_{\alpha ,a}$, and the potential
terms in expression (\ref{ham2}) exhibit the expected sharp wall behaviour.
One can now try and identifiy the roots ${\alpha }\in \Delta _{+}(E_{10})$
of the formal Hamiltonian (\ref{ham2}) with the wall factors $w_{A}$ in the
effective supergravity Hamiltonian (\ref{ham1}). With a consistent
truncation to $l=3$, for instance, one recovers the supergravity sector (\ref%
{FS}) on $T^{10}$. This corresponds to the mapping we have established
between real simple roots of $E_{10}$ and the symmetry, electric, magnetic
and curvature walls $w_{ij}$ $w_{ijk}$, $w_{ijkl}$ and $\widetilde{w}_{ijk}$%
. which are all in $\Delta _{+}(E_{10})$. The identification of the
algebraic coordinates $C_{\alpha ,a}$ with geometrical objects in the low
energy limit of M-theory given by $c_{A}(F)$ (as defined in (\ref{ham1}) and below)
 can then be carried out.

Proceeding further to $l=6$, one would get terms related to multiple spatial
gradients of supergravity fields appearing in the truncated equations of
motion of $11D$ supergravity~\cite{DamHenNic1,DamNic1} at a given point.
Finally, considering a more general version of the Hamiltonian (\ref{ham2})
by extending the second sum in the coset element (\ref{coselmn}) to negative
roots,\textit{\ i.e}.: ${\alpha }\in \Delta _{+}(E_{10})\cup \Delta
_{-}(E_{10}),$ and pushing the level truncation to the range $l=10$ to $28$,
one eventually identifies terms corresponding to 8th order derivative
corrections to classical supergravity \cite{DamNic1} of the form $%
R_{2}^{m}(DG_{4})^{n}$, where $R_{2}$ is the curvature two-form and $D$ is the
Lorentz covariant derivative. At eighth order in the derivative, \textit{i.e}%
. for $(m,n)\in \{(4,0),(2,2),(1,3),(0,4)\}$, they are typically related to $%
\mathcal{O}(\alpha ^{\prime 3})$ corrections in $10D$ type IIA string
theory, at tree level. In this case however, it may happen that the
corresponding subleading sharp walls $w_{A}$ are negative, which means that
they can only be obtained for a non-Borel parametrization of the coset. In
addition, they may not even be roots of $E_{10}$. However, these walls
usually decompose into $w_{A}=-(n+m-1)\rho _{1}+\zeta $, for $\zeta \in
\Delta _{+}(E_{10})$, where the first term on the RHS represents the leading 
$R^{n+m}$ correction. If $n+m=3{\mathbbm{N}}+1$, the $R^{m}(DF)^{n}$
correction under consideration is compatible with $E_{10}$, and $\zeta $ is
regarded as the relative positive root associated to it.

This means that the $w_{A}$ are not necessarily always roots of ${\mathfrak{e%
}_{10}}$, and when this is not the case, a (possibly infinite) subset of
them can still be mapped to roots of $E_{10}$, by following a certain
regular rescaling scheme.

\subsection{Instantons, fluxes and branes in M-theory on $T^{10}$: an
algebraic approach}
\label{FluxBrane}

If we now consider the hyperbolic U-duality symmetry $E_{10}$ to be a
symmetry not only of $11D$ supergravity, but also of the moduli space space
of M-theory on $T^{10}$, which is conjectured to be the extension of
expression (\ref{modM}) to $D=1$:%
\begin{equation}
\mathcal{M}_{10}=E_{10|10}({\mathbbm{Z}})\backslash E_{10|10}/K(E_{10|10})
\label{modM10}
\end{equation}%
the real roots appearing in the definition of the cosmological billiard are
now mapped to totally wrapped Euclidean objects of M-theory, and can be
identified by computing the action:%
\begin{equation}
S_{\alpha }[M_{P}R_{i}]=2\pi e^{\langle H_{R},\alpha \rangle }\,,\qquad
\alpha \in \Delta _{+}(E_{10})\,.  \label{actinst}
\end{equation}%
Thus, the roots of $E_{10}$ found in the preceding section, namely: $%
w_{ij}=\varepsilon _{i}-\varepsilon _{j}\in W_{\text{KKp}}(E_{10})$, $%
w_{ijk}=\varepsilon _{i}+\varepsilon _{j}+\varepsilon _{k}\in W_{\text{M2}%
}(E_{10})$, $w_{i_{1}..i_{6}}=(\epsilon _{i_{1}}+..+\epsilon _{i_{6}})\in W_{%
\text{M5}}(E_{10})$ and $\widetilde{w}_{ijk}=\sum_{l\notin
\{i,j,k\}}\varepsilon _{l}+2\varepsilon _{i}\in W_{\text{KK7M}}(E_{10})$
describe totally wrapped Euclidean Kaluza-Klein particles, M2-branes,
M5-branes and Kaluza-Klein monopoles. The dictionary relating these roots of 
$E_{10}$ to the action of extended objects of M-theory can be found in Table~%
\ref{ListBranes}, for the highest weight of the corresponding representation
of ${\mathfrak{sl}}(10,{\mathbbm{R}})$ in ${\mathfrak{e}_{10}}$.

Now, as pointed out in \cite{MotlBanks2}, the (approximated) Kasner solution
defines a past and future spacelike singularity. On the other hand, the
low-energy limit in which $11D$ supergravity becomes valid requires all
eleven compactification radii to be larger than $l_{P}$, and consequently
the Kasner exponents to satisfy (for a certain choice of basis for $%
\mathfrak{h}(E_{10})$, which can always be made):%
\begin{equation}
0<p_{10}\leqslant p_{9}\leqslant \ldots \leqslant p_{1}  \label{insugr}
\end{equation}%
so that the vector $\overrightarrow{p}$ is timelike with respect to the
metric $|\overrightarrow{p}|^{2}=\sum_{i}p_{i}^{2}-\left(
\sum_{i}p_{i}\right) ^{2}$ (\ref{PhysicalKilling}). Clearly, this does not
satisfy the constraints (\ref{pow}) which require $\overrightarrow{p}$ to be
lightlike. Such a modification of the Kasner solution (\ref{Kasnsol}) has
been argued in \cite{MotlBanks2} to be achieved if one includes matter,
which dominates the evolution of the system in the infinite volume limit and
thereby changes the solution for the geometry. This does not invalidate the
Kasner regime prevailing close to the initial spacelike singularity, since,
as pointed out in \cite{MotlBanks2}, matter and radiation become negligible
when the volume of space tends to zero (even though their density becomes
infinite). In the following, we will see how matter, in the form of
Minkowskian particles and branes, have a natural interpretation in terms of
some class of imaginary roots of ${\mathfrak{e}_{10}}$, and can thus be
incorporated in the hyperbolic billiard approach.

In particular, the inequality (\ref{insugr}) is satisfied if at late time we
have%
\begin{equation}
R_{1}\gg R_{2}\,,\text{ \ }R_{2}\gg R_{3}\,,\ldots ,\text{ }R_{9}\gg
R_{10}\,,\text{ \ and }R_{8}R_{9}R_{10}\gg l_{P}^{3}  \label{regR}
\end{equation}%
which can be rephrased as: $\langle H_{R},\alpha _{i}\rangle \gg 0$, $%
\forall \alpha _{i}\in \Pi ({\mathfrak{e}_{10}})$. The action (\ref{actinst}%
) related to such roots is then large at late time, and the corresponding
Euclidean objects of Table \ref{ListBranes} can then be used to induce
fluxes in the background, and thus be related to an instanton effect. This
is in phase with the analysis in \cite{MotlBanks2}, which states that at
large volume, the moduli of the theory become slow variables (in the sense
of a Born-Oppenheimer approximation) and can be treated semi-classically.

Let us now make a few remarks on the two different regimes encountered so far,
the billiard and semi-classical dynamics. In the semi-classical regime of~\ref{regR}, 
we are well outside the fundamental Weyl chamber (\ref{ineq2}) and higher level 
roots of ${\mathfrak{e}_{10}}$ have to be taken into account and given a physical 
interpretation. In this limit of large radii, the dominance of matter and radiation 
will eventually render the dynamics non-chaotic at late times, but the vacuum 
of the theory can be extremely complicated, because of the presence of instanton 
effects and solitonic backgrounds. In contrast, in the vicinity of the spacelike 
singularity, matter and radiation play a negligible r\^{o}le, leading to the chaotic
dynamics of billiard cosmology. On the other hand, the structure of the vacuum is 
simple in the BKL regime, in which the potential walls appear to be extremely sharp. 
It is characterized by ten flat directions bounded by infinite potential walls,
the dominant walls of the fundamental Weyl chamber of $\eten$. 

Finally, when $\overrightarrow{p}$ is timelike, it has been shown in~\cite{MotlBanks2}
that the domain (\ref{insugr}) where $11D$ supergravity is valid can be mapped, 
after dimensional reduction, to weakly coupled type IIA or IIB supergravity.
For instance, the safe domain for type IIA string theory (where all the nine radii 
are large compared to $l_{s}$ and $g_{IIA}<1$, two parameters given in terms of 
$11D$ quantities in eqns.(\ref{compform})) is given by:
\begin{equation}
\label{inIIA}
p_{10}<0<p_{10}+2p_{9}\,,\text{ \ and \ \ }p_{9}\leqslant p_{8}\leqslant
\ldots \leqslant p_{1}\,.
\end{equation}
The two domains (\ref{insugr}) and (\ref{inIIA}) are then related by
U-duality transformations (cf. Appendix \ref{AppendixB}).

Let us now discuss the issue of fluxes in this setup. From now on and
without any further specification, we assume that the conditions (\ref{regR}%
) are met. Then, in addition to the instanton effects we have just
mentioned, one can consider more complicated configurations by turning on
some components of the $p$-form potentials of the theory. In this case, the
action (\ref{actinst}) receives an additional contribution due to the
Wess-Zumino coupling of the $p$-form potential to the world-volume of the
corresponding brane-like object. The action (\ref{actinst}) will now receive
a flux contribution which can be rephrased in algebraic terms as~\cite{Gan,Gan1}%
:
\begin{equation}
S_{\alpha _{(p)}}[M_{P}R_{i};\mathcal{C}_{\alpha _{(p)}}]=2\pi e^{\langle
H_{R},\alpha _{(p)}\rangle }+i\mathcal{C}_{\alpha _{(p)}}=\frac{M_{P}^{p+1}}{%
(2\pi )^{p}}\int_{\mathcal{W}_{p+1}}e\,d^{p+1}x+i\int_{\mathcal{W}%
_{p+1}}C_{p+1}\,,  \label{actInst}
\end{equation}%
where the $\alpha _{(p)}$ are positive real roots of ${\mathfrak{e}_{10}}$
given by the second column of Table \ref{ListBranes}, for all possible
permutations of components in the physical basis. In particular, we will
have three-form and six-form fluxes for non-zero potentials $C_{3}$ and $%
\widetilde{C}_{6}$ coupling to the Euclidean world-volumes $\mathcal{W}_{3}$/%
$\mathcal{W}_{6}$ of M2-/M5-branes respectively. For fluxes associated to
Kaluza-Klein particle, we have the couplings $\mathcal{C}_{\alpha
_{i-2}}=\int_{\c} g_{i\,i+1}g_{i+1\,i+1}^{-1}\,dx^{i}$, $i=1,..,9$, where $\gamma $ is the KK-particle
world-line, and the internal metric $g$ can be written in terms of our variables $R_i$
and $\mathcal{A}_{\phantom{i}8}^i$ using eqs (\ref{metrcompa}) and (\ref{metcomp}).
There is also a similar coupling of the dual potential
$\widetilde{\mathcal{A}}_{\phantom{i}8}^{i}$ to the eight-dimensional KK7M world-volume.

The moduli $M_{P}R_{i}$, $i=1,..,10$, together with the fluxes from $p$-form
potentials (\ref{actInst}) parametrize the moduli space (\ref{modM10}).
Furthermore, on can define the following function which is harmonic under
the action of a certain Laplace operator defined on the variables $%
\{M_{P}R_{i};\mathcal{C}_{\alpha }\}_{\alpha \in \Delta _{+}({\mathfrak{e}%
_{10}})}^{i=1,..,10}$ in the Borel gauge of ${\mathfrak{e}_{10|10}}$, and
which is left-invariant under $E_{10|10}$:%
\begin{equation*}
\sqrt{N_{p}}\exp \left[ -2\pi N_{p}\left( e^{\langle H_{R},\alpha
_{(p)}\rangle }\pm \frac{i}{2\pi }\mathcal{C}_{\alpha _{p}}\right) \right]
\,.
\end{equation*}%
In the limit of large radii, $N_{p}$ is the instanton number and this
expression  is an extension to ${\mathfrak{e}_{10}}$ of the usual instanton
corrections to string thresholds appearing in the low-energy effective
theory. As such, it is conjectured to capture some of the non-perturbative
aspects of M-theory \cite{Gan}.

Another kind of fluxes arise from non-zero expectation values of $(p+1)$%
-form field strengths. If we reconsider the effective Hamiltonian (\ref{ham1}%
) in the region (\ref{regR}) where instanton effects are present, we notice
that the action (\ref{actinst}) appears in the effective potential as $\frac{1}{2\pi}S_{\a}
=e^{2\a(H_{R})}$. On the
other hand, since their coefficients $c_{A}(F)$ freeze in the BKL limit, we
may regard them as fluxes or topology changes provided the $(p+1)$-form
field strengths appearing in eqns.(\ref{potM2}), (\ref{potM5}), (\ref%
{potmetric}) and (\ref{FKK}) have, in this limit, integral background value: 
\begin{eqnarray*}
{\underline{\pi }_{ijk}} &{\rightarrow }&{(2\pi )}^{6}\langle (\ast _{10}{%
\underline{\pi })_{i_{1}..i_{7}}\rangle =}\frac{1}{2\pi }\int_{\mathcal{C}%
_{i_{1}..i_{7}}}\ast _{10}{\underline{\pi }_{3}\in {\mathbbm{Z}\,,}} \\
{\underline{G}_{ijkl}} &{\rightarrow }&{(2\pi )}^{3}{\langle \underline{G}%
_{ijkl}\rangle =}\frac{1}{2\pi }\int_{\mathcal{C}_{ijkl}}{\underline{G}%
_{4}\in {\mathbbm{Z}\,,}} \\
\underline{\pi }_{\phantom{i}j}^{i} &\rightarrow &{(2\pi )}^{8}\langle (\ast
_{10}\underline{\pi }^{i}{)_{j_{1}..j_{9}}\rangle =}\frac{1}{2\pi }\int_{%
\mathcal{C}_{j_{1}..j_{9}}}\ast _{10}{\underline{\pi }_{\phantom{i}1}^{i}\in 
{\mathbbm{Z}\,,}} \\
\underline{\mathcal{F}}_{\phantom{i}jk}^{i} &\rightarrow &{2\pi \langle 
\underline{\mathcal{F}}_{\phantom{i}jk}^{i}\rangle =}\int_{\mathcal{C}_{jk}}%
\text{ch}_{1}\left( {\underline{\mathcal{F}}_{\phantom{i}2}^{i}}\right)
\equiv \text{Ch}_{1}({\underline{\mathcal{F}}_{\phantom{i}2}^{i};}\mathcal{C}%
_{jk}){\in {\mathbbm{Z}\,,}}
\end{eqnarray*}%
Where $\mathcal{C}_{i_{1}..i_{p+1}}$ is a $(p+1)$-cycle chosen along the
appropriate spatial directions.

In particular, the coefficients $c_{A}(F)$ appearing in the potential terms (%
\ref{potM2}) and (\ref{potM5}) are now restricted to be integers: $%
c_{A}(F)\rightarrow \big[{(2\pi )}^{6}\langle (\ast _{10}{\underline{\pi }%
)_{i_{1}..i_{7}}\rangle \big]}^{2}$ and $\big[{(2\pi )}^{3}{\langle 
\underline{G}_{ijkl}\rangle \big]}^{2}$ and generate respectively seven-form
and four-form fluxes. In this perspective, the instantons encoded in the
exponential term $e^{2w_{A}(H_{R})}\equiv e^{2\langle H_{R},\alpha
_{(p)}\rangle }$ for $\alpha _{\text{M5}}=\sum_{m\notin
\{i,j,k,l\}}\varepsilon _{m}$ and $\alpha _{\text{M2}}=\sum_{n\notin
\{i_{1},..,i_{7}\}}\varepsilon _{n}$ are understood as the process that
changes the fluxes by an integral amount.

The wall coefficient $c_{A}(F)=\big[{2\pi \langle \underline{\mathcal{F}}_{%
\phantom{i}jk}^{i}\rangle \big]}^{2}$ (\ref{FKK}), on the other hand,
corresponds to a deformation of the basic torus $T^{10}$ to an $S^{1}$
fibration of the $i$th direction over the two-torus $T^{2}=\{x^{j},x^{k}\}$,
in other words to the metric:%
\begin{equation*}
ds^{2}=-(Ndt)^{2}+\sum_{m\neq i}(M_{P}R_{m})^{2}(dx^{m})^{2}+(M_{P}R_{i})^{2}%
\Big[dx^{i}-\frac{1}{2\pi }\text{Ch}_{1}({\underline{\mathcal{F}}_{%
\phantom{i}2}^{i};}\mathcal{C}_{jk})x^{k}dx^{j}\Big]^{2}\,.
\end{equation*}%
where the periodicity of $x^{k}$ implies $x^{i}\rightarrow x^{i}+$Ch$_{1}
({\underline{\mathcal{F}}_{\phantom{i}2}^{i};}\mathcal{C}_{jk})x^{j}$ for the
fiber coordinate, all other coordinates retaining their usual $2\pi$-periodicity. 
The value of $c_{A}(F)$ determines the first Chern character
(or Chern class, since $\mbox{ch}_1=\mbox{c}_1$) of the fibration, and the
instanton effect associated to the root 
$\a_{\text{KK7M}}=\sum_{l\notin\{i,j,k\}}\varepsilon_{l}+2\varepsilon_{i}$ 
creates an integral jump in this first Chern number.

\subsubsection{Minkowskian branes from prime isotropic roots of $\eten$}

As mentioned in the preceding section, when considering the large volume
limit (\ref{regR}) in the domain (\ref{insugr}) where $11D$ supergravity is
valid, one should in principle start considering higher level roots of $\eten$ 
in other Weyl chambers than the fundamental one. These roots, which, in the strict
BKL limit, appear as subdominant walls and can be neglected in a first
approximation, should now be taken into account as corrective or mass terms.
In \cite{Gan1,Gan2}, a program has been proposed to determine the physical
interpretation of a class of null roots of $\eten$. These
authors have, in particular, shown the correspondence between prime
isotropic roots and Minkowskian extended objects of M-theory, for the first
levels $l=3,5,6,7,8$. On the other hand, as seen in Section \ref{sec:sing},
the autors of \cite{DamNic1} have developed a different program where they
identify imaginary (both isotropic and non-isotropic) bur also real roots,
with $R^{m}(DF)^{n}$ type M-theory corrections to classical supergravity.
However, these results have been obtained in an intermediate domain of the
dynamical evolution, where not only negative roots become leading
(accounting for the fact that these higher order corrections are described
by negative roots), but where we expect quantum corrections to be visible.
In the approach of \cite{Gan1}, in contrast, the regime (\ref{regR}) allows
for effects related to extended objects to become important, pointing, in
the line of Section \ref{FluxBrane}, at an interpretation for certain higher
level roots in terms of branes and particles.

In the following, we give a condensed version of the correspondence between
prime isotropic roots of $\eten$ and Minkowskian extended objects of
M-theory, which can be found in a much more detailed and ample version in
\cite{Gan1}, which we follow closely until the end of this section.

First of all, since we now restrict to the region (\ref{insugr}), we are
sufficiently far from the singularity for the lapse function $N(t)$ to have
any non-zero value. In particular, we can gauge-fix to $N(t)=M_{P}$ in
expression (\ref{du}), which defines the conformal time:
\begin{equation}
d\tilde{u}=\frac{dt}{M_{P}^{9}V(t)}\,.  \label{conf}
\end{equation}
As we will see below, these are the "natural" units to work out the relation
between prime isotropic roots of $\eten$ and Minkowskian
particles and branes in M-theory.

Consider, for instance two M5 instantons at times $t_{\b_{\text{M5}}}\ll
t_{\a_{\text{M5}}}$ encoded algebraicly in 
\begin{equation*}
\a_{\text{M5}}=((1)^{4},(0)^{4},(1)^{2})\text{, \ \ \ \ \ \ \ }
\b_{\text{M5}}=((0)^{4},(1)^{6})
\end{equation*}
Since each of them creates a jump in their associated flux, inverting the
time order to $t_{\b_{\text{M5}}}\gg t_{\a_{\text{M5}}}$ will
pass one instanton through the other, thereby creating a Minkowskian
M2-brane stretched between them in the interval $[t_{\a_{\text{M5}}},t_{\b_{\text{M5}}}]$, 
where their respective fluxes overlap. This M2-brane will be associated to the root 
$\c_{\text{M2}}=\a_{\text{M5}}+\b_{\text{M5}}=((1)^{8},(2)^{2})$. Recalling that we
gauge-fixed to conformal time (\ref{conf}), the action for such an object
has to be expressed in unit of conformal time, then: 
\begin{equation}\label{formmass}
\frac{d}{d\tilde{u}}\widetilde{S}_{\a}=M_{P}^{9}V\frac{d}{dt}S_{\a}=
2\pi e^{\langle H_{R},\a\rangle}\text{ }\longrightarrow \text{\ }
M_{\a}=\frac{1}{2\pi }\frac{d}{dt}S_{\a}=M_{P}^{-9}V^{-1}e^{\langle H_{R},\a\rangle }
\end{equation}
This expression for the mass of the object could also be deduced from the
rescaling (\ref{hamresc}). Thus, in particular: $M_{\gamma _{\text{M2}%
}}=M_{P}^{3}R_{9}R_{10}$ as expected from a membrane wrapped around
directions $x^{9}$ and $x^{10}$.

From the supergravity perspective, the instanton described by $\a_{\text{M5}}$
will create a jump in the flux: $(2\pi)^{3}\langle \underline{G}_{5678}\rangle
\rightarrow (2\pi)^{3}\langle\underline{G}_{5678}\rangle +1$
when going from $t<t_{\a_{\text{M5}}}$ to $t_{\a_{\text{M5}}}<t $, while instanton 
$\b_{\text{M5}}$ induces $(2\pi)^{3}\langle \underline{G}_{1234}\rangle\rightarrow
(2\pi)^{3}\langle \underline{G}_{1234}\rangle-1$ when $t$ passes $t_{\b_{\text{M5}}}$.
Now the M2-brane flux sourced by $(2\pi)^{6}\langle \underline{G}_{1234}\rangle\langle
\underline{G}_{5678}\rangle$, via the topological term $\int C_{3}\wedge G_{4}\wedge G_{4}$
of $11D$ supergravity, has to be counterbalanced by an equal number of anti M2-branes.
Thus, going from configuration $t_{\b_{\text{M5}}}\ll t_{\a_{\text{M5}}}$ to
configuration $t_{\b_{\text{M5}}}\gg t_{\a_{\text{M5}}}$ after
setting both initial fluxes to zero produces one unit of Minkowskian
anti-M2-brane flux which must be compensated by the creation of a M2-brane
in the same directions, as expected.

Another process involves passing an M2-instanton $\a_{\text{M2}}=((1)^{2},(0)^{7},1)$ 
through a KK-monopole $\b_{\text{KK7M}}=((0)^{2},(1)^{6},2,1)$. Since the KK-monopole 
shifts by one unit the first Chern class of the circular $x^{9}$-fibration over the 
$(x^{1},x^{2})$-torus: Ch$_{1}(\underline{\mathcal{F}}_{\phantom{i}(2)}^{9};
\mathcal{C}_{12})\rightarrow $Ch$_{1}(\underline{\mathcal{F}}_{\phantom{i}(2)}^{9};
\mathcal{C}_{12})+1$, it creates an obstruction that blocks the M2-instanton
somewhere in the $(x^{1},x^{2})$ plane, and, by means of the fibration,
produces an object at least wrapped along the $x^{9}$ direction. It is not
hard to see that the Minkowskian object resulting from this process is a
M2-brane wrapped around $x^{9}$ and the original $x^{10}$. One can check
that the root $\a_{\text{M2}}+\b_{\text{KK7M}}=\c_{\text{M2}}=((1)^{8},(2)^{2})$, 
recovering the same object as before.

Furthermore, by combining an M5-instanton $\a_{\text{M5}}=(0,(1)^{6},(0)^{3})$ 
shifting the magnetic flux $(2\pi)^{3}\langle \underline{G}_{189\,10}\rangle$ 
by one unit with an M2-instanton $\b_{\text{M2}}=((0)^{7},(1)^{3})$ shifting 
the electric flux $(2\pi)^{6}\langle (\ast_{10}{\underline{\pi})_{1\cdots 7}\rangle}$ 
accordingly, one creates a Minkowskian KK-particle $\a_{\text{KKp}}=(0,(1)^{9})$
corresponding to the following contribution the momentum in the $x^{1}$
direction: $\mathcal{P}^{1}=\int d^{10}x\,G^{189\,10}\pi _{89\,10}$. The
mass (\ref{formmass}) of this object $M_{\c_{\text{KKp}}}=R_{1}^{-1}$
is in accordance with this interpretation.

\begin{table}[h]
\label{TabBranes}
\par
\begin{center}
$\hspace{-2.5mm}
\begin{array}{|c||c|c||c|c|}
\hline
\text{object} & \text{real root} & S_{\a} & \text{prime isotropic} & M_{\a} \\ \hline
\text{KKp} & (0,0,0,0,0,0,0,0,1,-1) & 2\pi R_{9}R_{10}^{-1} & 
(0,1,1,1,1,1,1,1,1,1) & R_{1}^{-1} \\ 
\text{M2} & (0,0,0,0,0,0,0,1,1,1) & 2\pi M_{P}^{3}R_{8}R_{9}R_{10} & 
(1,1,1,1,1,1,1,1,2,2) & M_{P}^{3}R_{9}R_{10} \\ 
\text{M5} & (0,0,0,0,1,1,1,1,1,1) & 2\pi M_{P}^{6}R_{5}\cdots R_{10} & 
(1,1,1,1,1,2,2,2,2,2) & M_{P}^{6}R_{5}\cdots R_{10} \\ 
\text{KK7M} & (0,0,1,1,1,1,1,1,1,2) & 2\pi M_{P}^{9}R_{3}\cdots R_{9}R_{10}^{2} & 
(1,1,1,2,2,2,2,2,2,3) & M_{P}^{9}R_{3}\cdots R_{9}R_{10}^{2} \\ 
\text{KK9M} & (1,1,1,1,1,1,1,1,1,3) & 2\pi M_{P}^{12}R_{1}\cdots R_{9}R_{10}^{3} & 
(1,2,2,2,2,2,2,2,2,4) & M_{P}^{12}R_{2}\cdots R_{9}R_{10}^{3} \\ \hline
\end{array}
$
\end{center}
\caption{Euclidean and Minkowskian branes of M-theory on $T^{10}$ and
positive roots of $E_{10}$}
\label{ListBranes}
\end{table}

Similar combinations of Euclidean objects can be shown, by various brane
creation processes, to produce Minkowskian M5-branes and KK7M-branes. To
conclude, all four types of time-extended matter fields are summarized in
Table~\ref{ListBranes} by their highest weight representative with its mass
formula. At present, it is still unclear how matter terms produced by the
prime isotropic roots of Table~\ref{ListBranes} should be introduced in the
effective Hamiltonian~(\ref{ham1}). Since the corresponding Minkowskian
branes originate from creation processes involving two instantons, as
explained above, we expect such a contribution to be~\cite{Gan1}: 
$2\pi n_{\c_{m}}e^{\langle H_{R},\c_{m}\rangle}$, where the isotropic
root $\c_{m}$ describing the Minkowskian brane decomposes into two real
roots related to instantons $\c_{m}=\a_{e}+\b_{e}$, and turns
on $n_{\c_{m}}=n_{\a_{e}}n_{\b_{e}}$ units of flux which
compensates for the original $n_{\a_{e}}$ and $n_{\b_{e}}$ units of
flux produced by the two instantons. Since $\c_{m}$ is a root, such a
term will never arise as a term in the serie (\ref{ham2}). We then expect
the Hamiltonian (\ref{ham2}) to be modified in the presence of matter. In
this respect, a proposal for a corrective term has been made in \cite{Gan1},
which reproduces the energy of the Minkowskian brane only up to a 
$2\pi$-factor. Moreover, it generates additional unwanted contributions for 
which one should find a cancelling mechanism.

From Table \ref{ListBranes}, one readily obtains the spectrum of BPS objects
of\ type IIA string theory, by compactifying along $x^{10}$, and taking the
limit $M_{P}R_{10}\rightarrow 0$, thereby identifying: 
\begin{equation}
R_{10}=\frac{g_{A}}{M_{s}}\,,\text{ \ \ \ \ \ \ \ \ \ \ \ \ \ \ }
M_{P}=\frac{M_{s}}{\sqrt[3]{g_{A}}}\,.  \label{compform}
\end{equation}
In this respect, we have included in Table \ref{ListBranes} the conjectured
KK9M-brane as the putative M-theory ascendant of the D8, KK8A and
KK9A-branes of IIA string theory, the latter two being highly
non-perturbative objects which map, under T-duality, to the IIB S7 and
S9-branes.

Since we mention type IIB string theory, we obtain its spectrum after
compactification (\ref{compform}) by T-dualizing along the toroidal $x^{9}$
direction, which maps: 
\begin{equation*}
R_{9,B}=\frac{1}{M_{s}^{2}R_{9,A}}\,,\text{ \ \ \ \ \ \ \ \ \ \ \ \ \ }
g_{B}=\frac{g_{A}}{M_{s}R_{9,A}}\,.
\end{equation*}%
From the $E_{10}$ viewpoint, T-dualizing from type IIA to IIB string theory
corresponds to different embeddings of $\mathfrak{sl}(9,\R)$ in $\e_{10|10}$. 
Going back to the Dynkin diagram~\ref{Fig:DynkinEr} for $r=10$, type IIA theory 
corresponds to the standard embedding of $\sl(9,\R)$ along the preferred subalgebra 
$\sl(10,\R)\subset \e_{10|10}$ (the gravity line), while type IIB is obtained 
by choosing the the Dynkin diagram of $\sl(9,\R)$ to extend in the $\a_{8}$ direction. 
This two embeddings consist in the two following choices of basis of simple roots 
for the Dynkin diagram of $\sl(9,\R)$:
\begin{equation*}
\Pi_{A}=\{\a_{-1},\a_{0},\ldots,\a_{5},\a_{6}\}\,,
\text{ \ \ \ \ \ \ \ \ \ \ \ \ \ \ \ \ \ \ \ }
\Pi _{B}=\{\a_{-1},\a_{0},\ldots ,\a_{5},\a_{8}\}\,,
\end{equation*}
which results in two different identifications of the NS-NS sector of both
theories. Then, a general T-duality on $x^{i}$, $i\neq 10$, can be expressed
in purely algebraic language, as the following mapping:
\begin{eqnarray*}
\mathcal{T}_{i} &:&\mathfrak{h}(E_{r})\rightarrow \mathfrak{h}(E_{r}) \\
&&\varepsilon_{i}\mapsto -\varepsilon_{i}
\end{eqnarray*}%
From the $\sigma$-model point of view, the type IIA and IIB theories correspond
to two different level truncations of the algebraic field strength (\ref{coselmn}), 
namely a level decomposition with respect to $(l_{7},l_{8})$ for type IIA, and 
$(l_{6},l_{7})$ for type IIB, $l_{i}$ being the level in the simple root 
$\a_{i}$ of $E_{10}$. The NS-NS and RR sectors of both supergravity
theories are then obtained by pushing the decomposition up to level 
$(l_{7},l_{8})=(2,3)$ for IIA, and up to $(l_{6},l_{7})=(4,2)$ for IIB. See
for instance \cite{KleinSchnakWest}, where the results are directly
transposable to $E_{10}$ (all roots considered there are in fact $E_{10}$
roots).

\subsubsection{Minkowskian objects from threshold-one roots of $\eten$}

\label{Thresh}

By inspecting the second column of Table~\ref{ListBranes}, we observe that
all Minkowskian objects extended in $p$ spatial directions, are
characterized, on the algebraic side, by adding $\sum_{i=1}^{p}\varepsilon_{k_{i}}$ 
to the threshold vector $\rho_{1}=((1)^{10})$. For the Minkowskian KK particle, 
the corresponding root of $\eten$ is related to its quantized momentum, and one 
needs therefore to substract a factor of $\varepsilon_{j}$ to the threshold vector.

The Minkowskian world-volume of these objects naturally couples to the
respective $(p+1)$-form potentials (\ref{actinst}). So, in contrast to the
Hamiltonian formalism (\ref{ham1}), which treats, for a different purpose,
the temporal components of the field-strengths as conjugate momenta, one now
needs to keep the time index of the tensor potentials apparent, thereby
working in the Lagrange formalism. This is similar to what is done in \cite%
{DamNic1} where the authors perform a component analysis of one-loop
corrections to classical $11D$ supergravity.

As pointed out in Section \ref{sec:sing}, the $(p+1)$-form components
separate into an oscillating part and a part that freezes as 
$u\rightarrow +\infty $, so that we have:
\begin{equation}\label{pottens}
C_{0i_{1}..i_{q}}=\frac{1}{N(t)}e_{\phantom{j_2}i_{1}}^{p_{1}}\cdots 
e_{\phantom{j_p}i_{q}}^{p_{q}}C_{tp_{1}\cdots p_{q}}=
e^{-\langle \rho_{1}+\sum_{n=1}^{q}\varepsilon_{i_{n}},H_{R}\rangle }
\c_{\phantom{j_2}i_{2}}^{p_{1}}\cdots \c_{\phantom{j_p}i_{q}}^{p_{q}}C_{up_{1}\cdots p_{q}}\,.
\end{equation}
where we have used $\sqrt{\bar{g}}=e^{\langle \rho _{1},H_{R}\rangle }$, and
the index $0$ stands for the flat time coordinate $dx^{0}=N(t)dt$. Following
the analysis of Section \ref{sec:sing}, the component $C_{up_{1}\cdots p_{q}}$ 
can be shown to freeze. Now, by selecting the appropriate basis vectors 
$\varepsilon_{j_{n}}$, we observe that all imaginary roots in the second
column of Table~\ref{ListBranes} are related to a tensor component of the
form (\ref{pottens}) with the expected value of $q$. As a side remark, the
minus sign appearing in the exponential wall factor in eqn.(\ref{pottens})
comes from working in the Lagrange formalism, as discussed in Section~\ref{FluxBrane}.

For Minkowskian KK particles and M2-branes, this approach is related to
performing the $\eten$ extension of the last two sets of roots
in eqns.(\ref{W}) by setting $D=1$. When restricting to the roots obtained
by this procedure which are highest weights under the Weyl group of 
$\sl(10,{\mathbbm{R}})$, one again recovers the two first terms in
the second column of Table~\ref{ListBranes}. Since we have not performed any
Hodge duality in this case, we obtain, as expected, roots characterizing KK
particles and M2-branes.

It results from this simple analysis that it is the presence of the
threshold vector $\rho_{1}=((1)^{10})$ which determines if an object is
time-extended, and not necessarily the fact that the corresponding root is
isotropic. We shall see in fact when working out 0B' orientifolds that
certain types of magnetized Minkowskian D-branes can be associated to real
roots and non-isotropic imaginary roots, provided they decompose as 
$\a=\rho_{1}+\vec{q}\in \Delta_{+}(E_{10})$, where $\vec{q}$ is a positive
vector (never a root) of threshold $0$, \textit{i.e.} that can never be
written as $\vec{q}=n\rho_{1}+\vec{q}'$ for $n\neq 0$.

\section{Orbifolding in a KMA with non-Cartan preserving automorphisms}

\label{OrbiLie}

In this section, we expose the method we use to treat physical orbifolds
algebraicly. It is based on the simple idea that orbifolding a torus by 
$\Z_{n}$ is geometrically equivalent to a formal $2\pi /n$
rotation. Using the mapping between physical and algebraic objects, one can
then translate the geometrical rotation of tensors into purely algebraic
language as a formal rotation in the KM algebra. This is given by an adjoint
action of the group on its KM algebra given by a finite-order inner
automorphism.

More concretely, let us consider an even orbifold $T^{q}/Z_{n}$, acting as a
simultaneous rotation of angle $2\pi Q_{a}/n$, $a=1,\ldots ,q/2$ in each
pair of affected dimensions determined by the charges $Q_{a}\in \{1,\ldots,n-1\}$. 
A rotation in the $(x^i,x^j)$ plane is naturally generated by
the adjoint action of the compact group element 
$\exp(\frac{2\pi}{n}Q_{a}\mathcal{K}_{i}^{\phantom{i}j})\equiv 
\exp(\frac{2\pi }{n}Q_{a}(E_{\a}-F_{\a}))$ for $\a=\a_{i-2}+\ldots +\a%
_{j-3}$. In particular, rotations on successive dimensions $(i+2,i+3)$
are generated by $E_{\a_{i}}-F_{\a_{i}}$. We will restrict
ourselves to orbifolds acting only on successive pair of dimensions in the
following, although everything can be easily extended to the general case.
In particular, physically meaningful results should not depend on that
choice, since it only amounts to a renumbering of space-time dimensions. For
the same reason, we can restrict our attention to orbifolds that are taken
on the last $q$ spatial dimensions of space-time $\{x^{11-q},\ldots ,x^{10}\}$.
In that case, we have the $q/2$ rotation operators 
\begin{equation}
\begin{array}{rcl}
V_{1}\doteq & e^{\frac{2\pi }{n}Q_{1}\,\mathcal{K}_{11-q\,12-q}} & 
=e^{\frac{2\pi}{n}Q_{1}\,(E_{\a_{9-q}}-F_{\a_{9-q}})} \\ 
& \vdots &  \\ 
V_{\frac{q}{2}}\doteq & e^{\frac{2\pi }{n}Q_{\frac{q}{2}}\,\mathcal{K}
_{9\,10}} & =e^{\frac{2\pi}{n}Q_{q/2}\,(E_{\a_{7}}-F_{\a_{7}})}
\end{array}
\label{RotationOperators}
\end{equation}
that all mutually commute, so that the complete orbifold action is given by: 
\begin{equation}
\mathcal{U}_{q}^{\Z_{n}}\doteq \prod_{a=1}^{q/2}V_{a}\,.
\label{OrbifoldAction}
\end{equation}

Note that $\mathcal{U}_{q}^{\Z_n}$ generically acts non-trivially
only on generators $\in\g_{\pm\a}$ (and the corresponding $%
H_{\a}\in\h$) for which the decomposition of $\a$ in
simple roots contains at least one of the root $\a_{8-q},\ldots,\a_{8}$ 
for $q>2$ ($\a_6$ and $\a_7$ for $q=2$).

In the particular case of $\Z_{2}$ orbifolds, the orbifold action
leaves the Cartan subalgebra invariant, so that it can be expressed as a
chief inner automorphism by some adjoint action $\exp (i\pi H),\,H\in {%
\mathfrak{h}}_{\Q}$ on $\g$. It indeed turns out that
the action of such an automorphism on $\g/\h%
=\bigoplus_{\a\in \Delta (\g)}\g_{\a}$
depends linearly on the root $\a$ grading $\g_{\a}$
and can thus be simply expressed as: 
\begin{equation}
\label{OddOrbifolding}
\begin{array}{l}
{\ds \mathrm{Ad}(e^{i\pi H})\g_{\a}=(-1)^{\a(H)}\g_{\a}=
(-1)^{\sum_{i=9-r}^{8}k^{i}\a_{i}(H)}\g_{\a}\,,}\\[4pt] \hspace{5cm}
\mathrm{for}\,H\in {{
{\ds \h}}_{\Q}\;\text{ and }\;\a=\sum_{i=9-r}^{8}k^{i}\a_{i}\,.}
\end{array}
\end{equation}
If $q$ is even, both methods are completely equivalent, while, if $q$ is odd,
the determinant is negative and it cannot be described by a pure $SO(r)$
rotation. If one combines the action of $E_{\a}-F_{\a}$ with
some mirror symmetry, however, one can of course reproduce the 
action~(\ref{OddOrbifolding}). Indeed, the last form of the orbifold action 
in expression~(\ref{OddOrbifolding}) is the one given in~\cite{Gan2}. The
results of~\cite{Gan2} are thus a subset of those obtainable by our more
general method.

\subsection{Cartan involution and conjugation of real forms}

\label{realform}

In the preceding sections, we have already been acquainted with the
Chevalley involution $\vartheta _{C}$. Here, we shall introduce just a few
more tools we shall need later in this work to deal with real forms in the
general sense. Let $\g$ be a complex semisimple Lie algebra. If
it is related to a real Lie algebra $\g_{0}$ as 
$\g=(\g_{0})^{\C}\doteq \g_{0}\otimes _{\R}\C$, $\g$ will be called a
complexification thereof. Reciprocally, $\g_{0}$ is a real form of 
$\g$ with $\g=\g_{0}\oplus i\g_{0}$. Next, a semisimple real Lie algebra 
is called compact if it can be endowed
with a Killing form satisfying 
\begin{equation}
B(X,X)<0\text{ , \ }\forall X\in \g_{0}\text{ \ \ }(X\neq 0)
\text{ ,}  \label{compcond}
\end{equation}
and non-compact otherwise.

Thus, a non-compact real form can in general be obtained from its
complexification $\g$ by specifying an involutive automorphism
$\vartheta$ defined on $\g_{0}$, such that $B(\vartheta X,\vartheta Y)=B(X,Y)$
and $B_{\vartheta}(X, Y)=-B(X,\vartheta Y)$, $\forall X,Y\in
\g_{0}$, is a symmetric and positive definite form. $\vartheta$ is
called a Cartan involution (the argument here is a generalization of the
construction of the almost-positive definite covariant form based on the
Chevalley involution $\vartheta _{C}$ in Section \ref{GKMA}). It can be
shown that every real semisimple Lie algebra possesses such an involution,
and that the latter is unique up to inner automorphisms. This is a corollary
of the following theorem:

\begin{theorem}\label{ConjugationThm}
\textit{Every automorphism $\psi$ of $\g$ is conjugate to a chief
automorphism $\vartheta$ of $\g$ through an inner automorphism $\phi$, ie:} 
\begin{equation}
\psi =\phi^{-1}\circ \vartheta \circ \phi ,\text{ \ \ \ \ \ }
\phi \in \text{Int}(\g)  \label{conjug}
\end{equation}
\end{theorem}

Then, it is clear that $\psi$ is involutive iff $\vartheta$ is involutive.
In this case, the two real forms of $\g$ generated by $\psi$ and $\vartheta$ 
are isomorphic, so that for every conjugacy class of involutive
automorphisms, one needs only consider the chief involutive automorphism (as
class representative), which can in turn be identified with the Cartan
involution.

The Cartan involution induces an orthogonal $(\pm 1)-$eigenspace
decomposition into the direct sum $\g_{0}=\k\oplus^{\perp}\p$, called Cartan decomposition of $\g_{0}$,
with property 
\begin{equation}
\left. \vartheta \right\vert_{\k}=1\text{ and }\left. \vartheta
\right\vert_{\p}=-1\text{ .}  \label{chief}
\end{equation}
More specifically, $\k$ is a subalgebra of $\g_{0}$
while $\p$ is a representation of $\k$, since: $[\k,\k]\subseteq \k$, 
$[\k,\p]\subseteq \p$ and $[\p,\p]\subseteq \k$. Finally, as our notation
for the Cartan decomposition suggested, $\k$ and $\p$ are orthogonal with respect to
the Killing form and $B_{\vartheta}$.

Alternatively, it is sometimes more convenient to define a real form $\g_{0}$ of 
$\g$ as the fixed point subalgebra of $\g$ under an involutive automorphism 
called conjugation $\tau$ such that 
\begin{equation}  \label{conj}
\tau(X)=X\,,\qquad \tau(iX)=-iX\,,\quad \forall X\in \g^{U}
\end{equation}
then: $\g_{0}=\left\{X\in \g\,|\,\tau (X)=X\right\} $.

Finally, by Wick-rotating $\p$ in the Cartan decomposition of $\g_{0}$ one obtains 
the compact Lie algebra $\g_{c}=\k\oplus^{\perp}i\p$ which is a compact real form of 
$\g=(\g_{0})^{\C}$.

Because of Theorem~\ref{ConjugationThm}, one needs an invariant quantity sorting out 
involutive automorphisms leading to isomorphic real forms. This invariant is the
signature (or character of the real Lie algebra) $\sigma$, defined as the
difference between the number $d_{-}=\dim \k$ of compact
generators and the number $d_{+}=\dim \p$ of non-compact
generators (the $\pm $-sign recalling the sign of the Killing form): 
\begin{equation*}
\sigma =d_{+}-d_{-}\,.
\end{equation*}
For \textit{simple} real Lie algebras, $\sigma $ \textit{uniquely} specifies 
$\g_{0}$. The signature varies between its maximal value for the
split form $\sigma=r$ and its minimal one for the compact form $\sigma=-\dim \g$.

Defining the following linear operator constructed from $\vartheta$
(see~\cite{GTP2}, p.543)
\begin{equation}\label{sqr}
\sqrt{\vartheta}=\frac{1}{2}(1+i)\vartheta +\frac{1}{2}(1-i)\Id\text{ ,}  
\end{equation}
satisfying $\sqrt{\vartheta}\circ \sqrt{\vartheta}\,X=\vartheta X$, 
$\forall X\in \g$, all non-compact real forms of $\g$ will be obtained through 
\begin{equation}
\g_{0}=\sqrt{\vartheta }\,\g_{c}  \label{sqr2}
\end{equation}
by selecting the appropriate \textit{chief} involutive automorphism 
$\vartheta$.

\subsection{Determining the real invariant subalgebra from its complexification}
\label{defin}

For a given orbifold $T^{11-q-D}\times T^{q}/\Z_{n}$ of
eleven-dimensional supergravity/M-theory, the orbifold action on the
corresponding U-duality algebra in $D$ dimensions is given by the
inner automorphism $\mathcal{U}_{q}^{\Z_{n}}$, $\forall D$. This
automorphism has a natural extension to the complexification $(\g^{U})^{\C}$ of 
the split form $\g^{U}$, where the appropriate set of generators describing physical 
fields and duality transformations on a complex space can be properly defined.

The requirement that these new generators diagonalize $\mathcal{U}_{q}^{\Z_{n}}$ and 
are charged according to the index structure of their corresponding physical objects 
will select a particular complex basis of $(\g^{U})^{\C}$. We will henceforth refer 
to this algebraic procedure as "orbifolding the theory".

Projecting out all charged states under $\mathcal{U}_{q}^{\Z_{n}}$ is then equivalent 
to an orbifold projection in the U-duality algebra, resulting in the invariant subalgebra
$(\ginv)^{\C}=\mbox{Fix}_{\mathcal{U}_{q}^{\Z_{n}}}(\g^{U})^{\C}$ (the notation 
$\mbox{Fix}_{V}\g$ stands for the fixed-point subalgebra of $\g$ under the automorphism $V$).

Since we expect the untwisted sector of the theory to be expressible from 
the non-linear realization of $G_{\text{inv}}/K(G_{\text{inv}})$ as a
coset sigma-model, we are particularly interested in determining the reality properties 
of $\ginv$, the algebra that describes the residual U-duality symmetry of the theory. 

Retrieving the real form $\ginv$ from its
complexification $(\ginv)^{\C}$ can be achieved by
restricting the conjugation (\ref{conj}) to $(\ginv)^{\C}$. Denoting such a restriction 
$\tau_{0}\doteq \tau|_{\ginv}$, the real form we
are looking for is given by
\begin{equation*}
\ginv=\text{Fix}_{\tau_{0}}
(\ginv^{\C})\,.
\end{equation*}
Since $\g^{U}$ is naturally endowed with the Chevalley
involution $\vartheta_{C}$, the Cartan involution associated to the real
form $\ginv$ is then the restriction of $\vartheta_{C}$
to the untwisted sector of the U-duality algebra, which we denote 
$\phi=\vartheta_{C}|_{\ginv}$. Consequently, the real form $\ginv$ possesses 
a Cartan decomposition $\ginv=\k_{\text{inv}}\oplus \p_{\text{inv}}$, with eigenspaces 
$\phi (\k_{\text{inv}})=\k_{\text{inv}}$ and $\phi (\p_{\text{inv}})=-\p_{\text{inv}}$. 
The whole procedure outlined in this section can be summarized by the following sequence:
\begin{equation}
\begin{CD} (\g^{U},\vartheta _{C})
@>{\mathcal{U}_{q}^{\Z_{n}}}>> {(\g}^{U})^{\C}
@>{\text{Fix}_{\mathcal{U}_{q}^{\Z_{n}}}}>>
(\ginv)^{\C} @>{\text{Fix}_{\tau_0}}>>
(\ginv,\phi) \,.
\end{CD}  
\label{chainFix}
\end{equation}

\subsection{Non-compact real forms from Satake diagrams}

\label{secSatak}

As we have seen before, real forms are described by classes of involutive
automorphisms, rather than by the automorphisms themselves. As such, the
Cartan involution, which we will refer to as $\vartheta$, can be regarded
as some kind of preferred involutive automorphism, and is encoded in the
so-called \textit{Satake} diagram of the real form it determines. The Cartan
involution splits the set of simple roots $\Pi$ into a subset of black
(invariant) roots ($\vartheta(\a_{i})=\a_{i}$) we call $\Pi_{c}$, and the subset
$\Pi_{d}=\Pi \,/\,\Pi_{c}$ of white roots, such as
\begin{equation*}
\vartheta (\a_{i})=-\a_{p(i)}+\sum_{k}\eta_{ik}\a_{k}\text{, with }
\a_{p(i)}\in \Pi_{d}\text{ and }\a_{k}\in \Pi_{c}
\end{equation*}
where $p$ is an involutive permutation rotating white simple roots into
themselves and $\eta_{ik}$ is a matrix of non-negative integers. A Satake
diagram consists in the Dynkin diagram of the complex form of the algebra
with nodes painted in white or in black according to the above prescription.
Moreover, if two white roots are exchanged under $p$, they will be joined on
the Satake diagram by an arrow.

From the action of $\vartheta$ on the root system, one can furthermore
determine the Dynkin diagram and multiplicities of the so-called restricted
roots, which are defined as follows: for a Cartan decomposition 
$\g_{0}=\k\oplus \p$, let $\mathfrak{a}\subset \p$ be maximal abelian. 
Then, one can define the partition under $\mathfrak{a}$ into simultaneous 
orthogonal eigenspaces (see \cite{Knapp} for a detailed discussion):
\begin{equation}
(\g_{0})_{\bar{\a}}=\{X\in \g_{0}\,|\,\text{ad}(H_{\mathfrak{a}})X=
\bar{\a}(H_{\mathfrak{a}})X,\text{ }\forall H_{\mathfrak{a}}\in \mathfrak{a}\}
\text{ .}  \label{restric}
\end{equation}
This defines the restricted roots $\bar{\a}\in \mathfrak{a}^{\ast}$ as
the simultaneous eigenvalues under the commuting family of self-adjoint
transformations $\{$ad$(H_{\mathfrak{a}})|\forall H_{\mathfrak{a}}\in 
\mathfrak{a}\}$. Then, we can choose a basis such that 
$\h_{0}=\mathfrak{t}\oplus \mathfrak{a}$, where $\mathfrak{t}$ is the maximal
abelian subalgebra centralizing $\mathfrak{a}$ in $\k$. The Cartan
subalgebra can be viewed as a torus with topology $(S^{1})^{n_{c}}\times (\R)^{n_{s}}$ 
where $n_{s}=\dim \mathfrak{a}$ is called the $\R$-rank. 
Restricted-root spaces are the basic ingredient of the Iwasawa decomposition, 
so we shall return to them when discussing non-linear realizations 
(see Section \ref{NonLinSugra}) of orbifolded $11D$ supergravity/M-theory models.

We denote by $\Sigma$ the set of roots not restricting to zero on $\mathfrak{a}^{\ast}$. 
As an example, one can choose a basis where such a set $\Sigma$ reads:
\begin{equation}
\Sigma =\left\{\bar{\a}=\frac{1}{2}(\a-\vartheta(\a))\in 
\mathfrak{a}^{\ast}\,|\,\bar{\a}\neq 0\right\} \text{ .}
\label{restSigm}
\end{equation}
Then, a real form can be encoded in the triple 
$\left(\mathfrak{a},\Sigma,m_{\bar{\a}}\right)$ and $m_{\bar{\a}}$ is the function 
giving the multiplicity of each restricted root, in other words 
$m_{\bar{\a}}=\dim(\g_{0})_{\bar{\a}}$. If we denote by $\overline{\Pi}$ a basis
of $\Sigma$, all non-compact real forms of $\g$ can be encoded graphically in
\begin{itemize}
\item[I)] the Satake diagram of $(\Pi,\vartheta)$;

\item[II)] the Dynkin diagram of $\overline{\Pi}$;

\item[III)] the multiplicities $m_{\bar{\a}_{i}}$ and $m_{2\bar{\a}_{i}}$ for 
$\bar{\a}_{i}\in \bar{B}$.
\end{itemize}

On the other hand, given a Satake diagram, we can determine the real form
associated to it as a fixed point algebra under $\tau$. Indeed from the
Satake diagram one readily determines $\vartheta$, and since it can be
shown that there always exists a basis of $\h$ such that the
"compact" conjugation $\tau ^{c}=\vartheta \circ \tau$ acts as 
$\tau^{c}(\a)=-\a$, $\forall \a \in \Delta$, then the conjugation
is determined by $\tau =-\vartheta$ on the root lattice.

Finally, in the finite case, the $\R$-rank $n_{s}$ is given in
the Satake diagram by the number of white roots minus the number of arrows,
and $n_{c}$ by the number of black roots plus the number of arrows.

\section{The orbifolds $T^{2}/\Z_{n>2}$}
\label{SecT2}

From the algebraic method presented in Section \ref{defin}, it 
is evident that a $T^{2}/\Z_{n}$ orbifold on the pair of
spatial dimensions $\{x^{9},x^{10}\}$ is only expected to act non-trivially on
the root spaces $(\g^{U})_{\a}\subset \g^{U}$,
characterized by all roots ${\a}$ containing $\a_{6}$ and/or $\a_{7}$, as well 
as on the corresponding Cartan element $H_{\a}$. 

The basis of $(\g^U)^{\C}$ diagonalizing the orbifold automorphism
$\mathcal{U}_{2}^{\Z_{n}}$ with the appropriate set of charges
will be derived step by step for the chain of compactification ranging
from $D=8$ to $D=1$. This requires applying the machinery of Section \ref{defin}
to the generators of the root spaces $(\g^{U})_{\a}$ mentioned above and
selecting combinations thereof to form a basis of $(\g^{U})^{\C}$ 
with orbifold charges compatible with their tensorial properties. 
We will at the same time determine the real invariant
subalgebra $\ginv$ by insisting on always selecting
lowest-height invariant simple roots, which ensures that the resulting invariant
subalgebra is maximal. In $D=2,1$, subtleties connected with roots of multiplicities 
greater than one and the splitting of their corresponding root spaces will be 
adressed.

For a start, we will work out the $D=8$ case in detail, and then show how this construction
can be regularly extended down to the $D=3$ case. The affine and hyperbolic $D=2,1$ cases require 
more care and will be treated separately. In $D=8$, then, we consider eleven-dimensional supergravity 
on $T^{3}$, which possesses U-duality algebra 
$\g^{U}=\sl(3,\R)\oplus \sl(2,\R)$, whose complexification is described by the Dynkin diagram of
${\mathfrak{a}}_{2}\oplus{\mathfrak{a}}_{1}$. It has positive-root space 
$\Delta_{+}=\{\a_{6},\a_{7},\a_{6}+\a_{7},\a_{8}\}$, and its Cartan subalgebra is spanned by
$\big\{H_{6}=~\varepsilon_{8}^{\vee}-~\varepsilon_{9}^{\vee}\,;\,H_{7}=\varepsilon_{9}^{\vee}
-\varepsilon_{10}^{\vee}\,;\,H_{8}=(2/3)(\varepsilon_{8}^{\vee}+\varepsilon_{9}^{\vee}+
\varepsilon_{10}^{\vee})\big\}$. The $\an_{1}$ factor corresponds to transformations acting 
on the unique scalar $C_{89\,\,10}$ produced by dimensional reduction of the 3-form field on $T^{3}$.

The orbifold action on the two-torus
\begin{equation}
(z,\bar{z})\rightarrow (e^{2\pi i/n}z,e^{-2\pi i/n}\bar{z})
\label{GeometricOrbifoldAction}
\end{equation}
induces the following inner automorphism on the U-duality algebra 
\begin{equation*}
\mathcal{U}_{2}^{\Z_{n}}=\mathrm{Ad}(e^{\frac{2\pi}{n}\,(E_{7}-F_{7})})\doteq
\mathrm{Ad}(e^{-\frac{2\pi}{n}i\mathcal{K}_{z\bar{z}}})\text{ ,}\qquad \text{for }n>2\,.
\end{equation*}
This automorphism acts diagonally on the choice of basis for $(\g^U)^{\C}$ appearing in 
Table \ref{AlgebraicT2/ZnCharges}, where both compact and non-compact generators have the charge 
assignment expected from their physical counterparts.

\begin{table}[!h]
\begin{center}
$ \begin{array}{|c|c|}
\hline
Q_{A} & \text{generators} \\ \hline
0 & K_{z\bar{z}}=-\frac{1}{6}(2H_{6}+H_{7})+\frac{1}{2}H_{8} \\[4pt] 
& K_{88}=\frac{1}{3}(2H_{6}+H_{7})+\frac{1}{2}H_{8} \\[3pt] 
& Z_{8z\bar{z}}=\frac{i}{2}(E_{8}+F_{8})\,,\quad 
\mathcal{Z}_{8z\bar{z}}=i(E_{8}-F_{8}) \\[2pt] 
& \mathcal{K}_{z\bar{z}}=i(E_{7}-F_{7}) \\ \hline
\pm 1 & \left\{ 
\begin{array}{c}
K_{8\bar{z}} \\ 
K_{8z}
\end{array}
\right\} =\frac{1}{2\sqrt{2}}\left( E_{6}+F_{6}\pm i(E_{67}+F_{67})\right) 
\\[3pt] 
& \left\{ 
\begin{array}{c}
\mathcal{K}_{8\bar{z}} \\ 
\mathcal{K}_{8z}
\end{array}
\right\} =\frac{1}{\sqrt{2}}\left( E_{6}-F_{6}\pm i(E_{67}-F_{67})\right) 
\\ \hline
\pm 2& \left\{ 
\begin{array}{c}
K_{\bar{z}\bar{z}} \\ 
K_{zz}
\end{array}
\right\} =\frac{1}{2}(H_{7}\pm i(E_{7}+F_{7})) \\ \hline
\end{array}$
\end{center}
\caption{Algebraic charges for $S^1\times T^2/\Z_{n>2}$ orbifolds}
\label{AlgebraicT2/ZnCharges}
\end{table}

The invariant subalgebra $(\ginv)^{\C}$ can be directly read off 
Table \ref{AlgebraicT2/ZnCharges}, the uncharged objects building an
$\an_1\oplus \C^{\oplus^2}$ subalgebra, since the original $\an_2$
factor of $(\g^U)^{\C}$ now breaks into two abelian generators $H^{[2]}\doteq 2H_6+H_7$
and $\widetilde{H}^{[2]}\doteq-\mathcal{K}_{z\bar{z}}$. The total rank (here 3) is conserved,
which will appear to be a generic feature of $(\ginv)^{\C}$.

The real form $\ginv$ is then easily identified by
applying the procedure outlined in eq. (\ref{chainFix}).
Since $K_{z\bar{z}}$ and $K_{88}$ are already in Fix$_{\tau_0}(\ginv^{\C})$
while $\tau_0(Z_{8z\bar{z}})=-Z_{8z\bar{z}}$, 
$\tau_0(\mathcal{Z}_{8z\bar{z}})=-\mathcal{Z}_{8z\bar{z}}$
and $\tau_0(\mathcal{K}_{z\bar{z}})=-\mathcal{K}_{z\bar{z}}$, a proper basis of
the invariant real form is, in terms of $(\ginv)^{\C}$ generators:
$\ginv=\text{Span}\big\{2/3(K_{88}+2K_{z\bar{z}})\,;\,iZ_{8z\bar{z}}\,;
\,i\mathcal{Z}_{8z\bar{z}}\big\}\oplus\text{Span}\{2(K_{88}-K_{z\bar{z}})\}\oplus
\text{Span}\{-i\mathcal{K}_{z\bar{z}}\}$. 
From now on, the last two abelian factors will be replaced by $H^{[2]}$ and $i\widetilde{H}^{[2]}$.
Now, how such a basis behaves under the associated Cartan involution $\phi$ is clear from 
Section~\ref{defin}. This determines the invariant real form to be
$\sl(2,\R)\oplus \so(1,1)\oplus \u(1)$, with total signature $\sigma =1$. In general, 
the signature of the subalgebra kept invariant by a $T^{2n}/\Z_{n>2}$ orbifold will be 
given by $\sigma(\g^U)-2n$ (keeping in mind that some orbifolds are equivalent under
a trivial $2\pi$ rotation).

The coset defining the non-linear realization of the orbifolded supergravity is 
obtained in the usual way by modding out by the maximal compact subgroup: 
\begin{equation*}
\frac{SL(2,\R)}{SU(2)}\times \frac{SO(1,1)}{\Z_{2}}
\text{ }{.}
\end{equation*}

In $D=7$, there appears an additional simple root $\a_{5}$, which, in the purely toroidal 
compactification, enhances and reconnects the U-duality algebra into $\g^{U}=\sl(5,\R)$, 
following the well known $\e_{n|n}$ serie. The complexification $(\g^{U})^{\C}$ resulting 
from orbifolding the theory calls for six additional generators: \{$K_{7\bar{z}},Z_{7z\bar{z}},
\mathcal{K}_{7\bar{z}},\mathcal{Z}_{7z\bar{z}}\}$ and the 2 corresponding Hermitian conjugates, 
produced by acting with ad($E_{5}\pm F_{5}$) on the objects in Table \ref{AlgebraicT2/ZnCharges}, 
all of which, together with the Cartan element $K_{77}$, have the expected orbifold charges. 

Beside these natural combinations, we now have four new types of
objects with charge $\pm 1$, namely: 
\begin{equation}
\left\{ 
\begin{array}{c}
Z_{78\bar{z}}\,/\,\frac{1}{2}\mathcal{Z}_{78\bar{z}} \\ 
Z_{78z}\,/\,\frac{1}{2}\mathcal{Z}_{78z}
\end{array}\right\} 
=\frac{1}{2\sqrt{2}}\left(E_{5678}(+/-)F_{5678}\pm
i(E_{568}(+/-)F_{568})\right) \,,  \label{NewChargedFields}
\end{equation}
so that the invariant subalgebra $(\ginv)^{\C}$ is a
straightforward extension by $\a_5$ of the $D=8$ case, as can be seen in 
Table \ref{T2/ZnDynkin}. Its real form is obtained from the sequence (\ref{chainFix}) 
just as in $D=8$, yielding the expected $\ginv=\sl(3,\R)\oplus\so(1,1)\oplus\u(1)$,
where the non-compact abelian factor is now generated by the combination 
\begin{equation}
H^{[2]}=4H_{5}+6H_{6}+3H_{7}+2H_{8}=\frac{10}{3}\left(K_{7}^{\phantom{7}7}+
K_{8}^{\phantom{8}8}-K_{z}^{\phantom{z}z}\right) \,,  
\label{CommutingCartan}
\end{equation}
while, as before, $\u(1)=\R(E_{7}-F_{7})$. Thus $\sigma(\ginv)=2$, while the total rank 
is again conserved by the orbifold projection.

The above procedure can be carried out in $D=6$. In this case however, the
invariant combination $H^{[2]}$ which generated earlier the non-compact $\so(1,1)$ factor 
is now dual to a root of $\g^{U}=\so(5,5)$, namely: 
\begin{equation}\label{CommutingCartan2}
H^{[2]}=H_{\theta _{D_{5}}}=\frac{2}{3}\left( K_{6}^{\phantom{6}6}+
K_{7}^{\phantom{7}7}+K_{8}^{\phantom{8}8}-K_{z}^{\phantom{z}z}\right) \,.  
\end{equation}
The abelian factor is thus enhanced to a full $\sl(2,\R)$ subalgebra with root system 
$\{\pm \theta_{D_{5}}\}$, while the real invariant subalgebra clearly becomes 
$\sl(4,\R)\oplus \sl(2,\R)\oplus \u(1)$. In $D=5$, $\theta_{D_{5}}$ connects to $\a_{3}$ 
giving rise to $\ginv=\sl(6,\R)\oplus\u(1)$. The extension to $D=4,3$ is completely 
straightforward, yielding respectively $\ginv=\so(6,6)\oplus\u(1)$ 
and $\e_{7|7}\oplus\u(1)$. The whole serie of real invariant subalgebras appears in 
Table \ref{T2/ZnDynkin}, beside their Satake diagram, which encodes the set of simple 
invariant roots $\Pi_0$ and the Cartan involution $\phi$.

\subsection{Affine central product and the invariant subalgebra in $D=2$}
\label{affT2}

New algebraic features appear in $D=2$, since, in the purely toroidal case,
the U-duality algebra is now conjectured to be the affine 
${\e}_{9|10}\doteq \text{Split}(\hat{\e}_{8}).$\footnote{Since
$\vartheta_C(\delta )=-\delta $ implies $\vartheta_C(d)=-d$ and 
$\vartheta_C(c)=-c$, the split form of any KMA has signature $\sigma =\dim\h$.}

The invariant subalgebra $(\ginv)^{\C}$ consists in
the affine $\hat{\e}_7$ together with the Heisenberg algebra
$\hat{\u}(1)^{\C}$ spanned by 
$\{z^{n}\otimes (E_{7}-F_{7}),\forall n\in {\Z\,;c\,;d\}}$. Though both
terms commute at the level of loop-algebras, their affine extensions share
the same central charge $c=H_{\d}$ and scaling operator $d$. Now, a
product of two finite-dimensional algebras possessing (at least partially) a
common centre is called a \textit{central product} in the mathematical
literature. The present situation is a natural generalization of this
construction to the infinite-dimensional setting, where not only the central
charge but also the scaling element are in common. Since the latter is a
normalizer, we are not strictly dealing with a central product. We will
therefore refer to such an operation as an \textit{affine central product}
and denote it by the symbol $\Join$. Anticipating the
very-extended $D=1$ case, we can write the invariant subalgebra as the complexification
\begin{equation}
\label{kotien}
\hat{\e}_{7}\Join\hat\u(1)\equiv 
\big(\hat{\e}_{7}\oplus \hat{\u}(1)\big)
\,/\,\{\z,\bar{d}\}\text{ ,}
\end{equation}
where $\z=H_{\d_{E_7}}-c_{\hat{\u}(1)}$ is the centre of the algebra and 
$\bar{d}=d_{\d_{E_7}}-d_{\hat{\u}(1)}$ is the difference of scaling operators.

The real form $\ginv$ can be determined first by observing that the non-compact 
and compact generators $H_{\a}$ and $E_{\a}\pm F_{\a}$ (with $(\a|\a_7)=0$) 
of the $\e_{7|7}$ factor in $D=3$ naturally extend to the $(t^n\pm~t^{-n})
\otimes~H_{\a}$ and $(t^n\otimes E_{\a}\pm t^{-n}\otimes F_{\a})$ vertex operators
of an affine $\hat{\e}_{7|9}$ and second, by noting that the remaining factor in 
the central product $\hat{\e}_{7}\Join\hat{\u}(1)$ is in fact the loop 
algebra $\mathcal{L}(\u(1))$ whose tower of generators can be grouped in pairs 
of one compact and one non-compact generator, according to~\footnote{Note that
the combination used in eqn.~(\ref{invol}) is well-defined in $\e_{9|10}$, 
since it can be rewritten in the following form: 
$(z^{n}\otimes E_{7}\mp z^{-n}\otimes F_{7})\pm (z^{-n}\otimes E_{7}\mp z^{n}\otimes F_{7})$.} 
\begin{equation}
\label{invol}
\phi\big((z^{n}\pm z^{-n})\otimes (E_{7}-F_{7})\big)=\pm (z^{n}\pm z^{-n})\otimes (E_{7}-F_{7})\,,
\end{equation}
in addition to the former compact Cartan generator
$i\widetilde{H}^{[2]}=E_{7}-F_{7}$. In short, the $\mathcal{L}(\u(1))$ factor 
contributes $-1$ to the signature of $\ginv$, so that in total: $\sigma=8$. Restoring
the central charge and the scaling operator in $\mathcal{L}(\u(1))$ so as
to write $\ginv$ in the form (\ref{kotien}), we will denote the resulting
real Heisenberg algebra $\hat{\u}_{|1}(1)$, so as to render its signature apparent.

For the sake of clarity, we will represent $\ginv$ in Table \ref{T2/ZnDynkin} 
by the Dynkin diagram of $\hat{\e}_{7|9}\oplus\hat{\u}_{|1}(1)$ supplemented 
by the signature $\sigma(\ginv)$, but it should be kept in mind that $\ginv$ 
is really given by the quotient (\ref{kotien}). In Table \ref{T2/ZnDynkin}, 
we have separated  the $D=2,1$ cases from the rest, to stress that the Satake 
diagram of $(\Pi_0,\phi)$ describes $\ginv$ completely only in the finite case.

Finally, to get some insight into the structure of the algebra 
$\hat{\e}_{7}\Join\hat{\u}(1)$, it is worthwhile
noting that the null root of the original ${\e}_{9}$ is also the
null root of $\hat{\e}_{7}$, as
\begin{eqnarray*}
\delta &=&{\a}_{0}+2{\a}_{1}+3{\a}_{2}+4{\a}_{3}+5{\a}_{4}+6{\a}_{5}+
4{\a}_{6}+2{\a}_{7}+3{\a}_{8} \\
&=&{\a}_{0}+2{\a}_{1}+3{\a}_{2}+4{\a}_{3}+3{\a}_{4}+2{\a}_{5}+{\a}_{8}+
2\theta_{D_{5}}=\d_{E_{7}}\text{ .}
\end{eqnarray*}
Although the root space $\g_{\d}\subset\e_{9}$
is eight-dimensional, we have $\text{mult}(\d_{E_{7}})=7$, since the
eighth generator $z\otimes H_{7}$ of $\g_{\d}$ is projected out. The latter 
is now replaced by the invariant combination $z\otimes (E_{7}-F_{7})$ whose 
commutator with itself creates the central charge of the $\hat{\u}(1)$, whereas 
the seven remaining invariant generators 
$\{z\otimes H_{\theta_{D_{5}}}\,;z\otimes H_{i},\forall i=1,\cdots,5,8\}$ build 
up the root space $\g_{\d_{E_{7}}}$. In a sense that will become clearer in $D=1$, 
the multiplicity of $\d_{E_{9}}$ is thus preserved in 
$\hat{\e}_{7}\Join\hat{\u}(1)$.

\begin{table}[!t]
\begin{center}
\begin{tabular}{c|c|c|c}
$D$ & $(\Pi_0,\phi)$ & $\ginv$ & $\sigma(\ginv)$ \\ \hline\hline
$8$ & \quad 
\begin{picture}(260,30) \thicklines
\put(60,0){\circle{6}}
\put(54,-12){$\a_8$} 
\put(90,-3){ $\times$ } 
\put(120,-3){$H^{[2]}$}
\put(150,-3){ $\times$ } 
\put(180,-3){$i\widetilde{H}^{[2]}$} 
\end{picture}
\hspace{1.5mm} & $
\begin{array}{c}
\sl(2,\R)\oplus\so(1,1) \\[3pt] 
\oplus\;\u(1)
\end{array}
$ & 1 \\[10pt] 
$7$ & \quad 
\begin{picture}(260,30) \thicklines
\multiput(45,0)(30,0){2}{\circle{6}}
\multiput(48,0)(30,0){1}{\line(1,0){24}}
 \put(39,-12){$\a_5$}
\put(69,-12){$\a_8$} 
\put(105,-3){$\times$} 
\put(135,-3){$H^{[2]}$}
\put(165,-3){ $\times$ } 
\put(195,-3){$i\widetilde{H}^{[2]}$} 
\end{picture}
\hspace{1.5mm} & $
\begin{array}{c}
\sl(3,\R)\oplus\so(1,1) \\[3pt] 
\oplus\;\u(1)
\end{array}
$ & 2 \\[10pt] 
$6$ & \quad 
\begin{picture}(260,30) \thicklines
\multiput(30,0)(30,0){3}{\circle{6}} 
\put(135,0){\circle{6}}
\multiput(33,0)(30,0){2}{\line(1,0){24}} 
\put(24,-12){$\a_4$}
\put(54,-12){$\a_5$} \put(84,-12){$\a_8$} 
\put(129,-14){$\theta_{D_5}$}
\put(165,-3){ $\times$ } 
\put(195,-3){$i\widetilde{H}^{[2]}$}
\end{picture}
\hspace{1.5mm} & $
\begin{array}{c}
\sl(4,\R)\oplus\sl(2,\R) \\[3pt] 
\oplus\;\u(1)
\end{array}
$ & 3 \\[10pt]
$5$ & \quad 
\begin{picture}(260,30) \thicklines
\multiput(15,0)(30,0){5}{\circle{6}}
\multiput(18,0)(30,0){4}{\line(1,0){24}} 
\put(9,-14){$\theta_{D_5}$}
\put(39,-12){$\a_3$} 
\put(69,-12){$\a_4$} 
\put(99,-12){$\a_5$}
\put(129,-12){$\a_8$}
 \put(165,-3){ $\times$ }
\put(195,-3){$i\widetilde{H}^{[2]}$} 
\end{picture}\hspace{1.5mm} & $
\sl(6,\R)\oplus\u(1)$ & 4 \\[15pt] 
$4$ & \quad 
\begin{picture}(260,40) \thicklines
\multiput(21,-18)(0,36){2}{\circle{6}}
 \put(23,-16){\line(4,3){19}}
\put(23,16){\line(4,-3){19}} 
\multiput(45,0)(30,0){4}{\circle{6}}
\multiput(48,0)(30,0){3}{\line(1,0){24}}
 \put(2,15){$\theta_{D_5}$}
\put(4,-21){$\a_2$} \put(39,-12){$\a_3$}
 \put(69,-12){$\a_4$}
\put(99,-12){$\a_5$} \put(129,-14){$\a_8$} 
\put(165,-3){ $\times$ }
\put(195,-3){$i\widetilde{H}^{[2]}$} 
\end{picture}\hspace{1.5mm} & $
\so(6,6)\oplus\u(1)$ & 5 \\[15pt] 
$3$ & \quad \begin{picture}(260,40)
\thicklines \put(75,20){\circle{6}}
\put(75,17){\line(0,-1){14}}
 \multiput(15,0)(30,0){6}{\circle{6}}
\multiput(18,0)(30,0){5}{\line(1,0){24}} 
\put(81,15){$\theta_{D_5}$}
\put(11,-12){$\a_1$} 
\put(41,-12){$\a_2$} 
\put(71,-12){$\a_3$}
\put(101,-12){$\a_4$} 
\put(131,-12){$\a_5$} 
\put(161,-12){$\a_8$}
\put(190,-3){ $\times$ } 
\put(220,-3){$i\widetilde{H}^{[2]}$}
\end{picture}
& $\e_{7|7}\oplus \mathfrak{u}(1)$ & 6 \\[30pt] 
\hhline{~:==:~}
$2$ & \quad \begin{picture}(260,40) \thicklines
\put(90,20){\circle{6}}
\put(90,17){\line(0,-1){14}} 
\multiput(0,0)(30,0){7}{\circle{6}}
\multiput(3,0)(30,0){6}{\line(1,0){24}}
 \put(96,15){$\theta_{D_5}$}
\put(-4,-12){$\a_{0}$} 
\put(26,-12){$\a_1$} 
\put(56,-12){$\a_2$}
\put(86,-12){$\a_3$} 
\put(116,-12){$\a_4$}
 \put(146,-12){$\a_5$}
\put(176,-12){$\a_8$} 
\put(190,-3){ $\times$ }
\put(210,-3){$\left\{i\widetilde{H}^{[2]}_n\right\}_{n\in\Z}$} 
\end{picture}
& $\hat{\e}_{7|9}\oplus \mathcal{L}(\u(1))_{|-1}$ & 8 \\[15pt] 
$1$ & \quad 
\begin{picture}(275,40) \thicklines
\put(0,20){\circle{6}}
\put(0,20){\circle{9}} 
\put(0,15){\line(0,-1){12}} 
\put(8,15){$\b_{I}$}
\put(120,20){\circle{6}} 
\put(120,17){\line(0,-1){14}}
\multiput(0,0)(30,0){8}{\circle{6}} 
\multiput(3,0)(30,0){7}{\line(1,0){24}}
\put(126,15){$\theta_{D_5}$} 
\put(-4,-12){$\a_{-1}$} 
\put(26,-12){$\a_0$}
\put(56,-12){$\a_1$} 
\put(86,-12){$\a_2$} 
\put(116,-12){$\a_3$}
\put(146,-12){$\a_4$} 
\put(176,-12){$\a_5$} 
\put(206,-12){$\a_8$}
\put(220,-3){ $\times$ } 
\put(240,-3){$i\widetilde{H}^{[2]}$} 
\end{picture}
& $^2{\cal B}_{10|11}\oplus \u(1) $ & 8 \\[20pt] 
\end{tabular}
\end{center}
\caption{The split subalgebras $\ginv$ for $T^{9-D}\times T^2/\Z_{n>2}$ compactifications}
\label{T2/ZnDynkin}
\end{table}

\subsection{A Borcherds symmetry of orbifolded M-theory in $D=1$}

In $D=1$, finally, plenty of new $\mathfrak{sl}(10,\R)$-tensors
appear as roots of $\eten$, so it is now far from obvious whether 
the invariant subalgebra constructed from 
$\hat{\e}_{7}\Join\hat{\mathfrak{u}}(1)$ by adding the node $\a _{-1}$
exhausts all invariant objects. Moreover, the structure of such an algebra,
as well as its Dynkin diagram is not a priori clear, since we know of no
standard way to reconnect the two factors of the central product through the
extended node $\a_{-1}$. As a matter of fact, mathematicians are aware that 
invariant subalgebras of KMA under finite-order automorphisms might not be KMA, 
but can be Borcherds algebras or EALA~\cite{AzamBerYous,AllBerPian,AllBerPian2}.
Despite these preliminary reservations, we will show that the real invariant
subalgebra $\ginv$ in $D=1$ can nevertheless be described in a closed form
by a Satake diagram and the Conjecture \ref{extendedT2} below,
while its root system and root multiplicities can in principle be determined up to
arbitrary height by a proper level decomposition.

\begin{conjecture}
\label{extendedT2} \textit{The invariant subalgebra of $\eten$
under the automorphism $\mathcal{U}_{2}^{\Z_{n}}$ is the direct
sum of a $\mathfrak{u}(1)$ factor and a Borcherds algebra with degenerate
Cartan matrix characterized by one isotropic imaginary simple root $\b _{I}$ of
multiplicity one and nine real simple roots, modded out by its centre and its derivation.}
\end{conjecture}

As already mentioned in Section \ref{affT2}, we choose to represent, in Table \ref{T2/ZnDynkin}, 
the real form $\ginv$ before quotientation, by the
Dynkin diagram of its defining Borcherds algebra $^2{\cal B}_{10}$. Both are related through 
\begin{equation}
\label{kotienhyp}
\ginv=\u(1)\oplus \,^2{\cal B}_{10|11}\,/\,\{\mathfrak{z},d_I\}\,,
\end{equation}
which is an extension of the affine central product (\ref{kotien}) encountered in $D=2$,
provided we now set $\mathfrak{z}=H_{\d}-H_{I}$. We also define $H_{I}\doteq H_{\b_I}$ and 
$d_{I}\doteq d_{\b_I}$ as the Cartan generator dual to $\b_I$ and the derivation counting levels in $\b_I$.

More precisely, $^2{\cal B}_{10}$ has the following $10\times 10$ degenerate
Cartan matrix, with rank $r=9$
\begin{equation*}
A=\left( 
\begin{array}{cccc}
0 & -1 & 0 &  \\ 
-1 & 2 & -1 & \raisebox{1.5ex}[0cm][0cm]{\Large{$\O$}} \\ \cline{3-4}
0 & -1 & \multicolumn{2}{|c}{\multirow{2}{10mm}{$A(\hat{\e}_7)$}} \\ 
\multicolumn{2}{c|}{\text{{\Large {${\mathbbm{O}}$}}}} &  & 
\end{array}%
\right) \,,
\end{equation*}
and it can be checked that its null vector is indeed the centre $\z$ of the 
Borcherds algebra mentioned above. As for affine KMA, the Cartan subalgebra of
Borcherds algebras with a non-maximal $n\times n$ Cartan matrix has to be
supplemented by $n-r$ new elements that allow to discriminate between roots
having equal weight under Ad($H_{i}$), $\forall i=1,..,n$. Here, the
Cartan subalgebra of $^2{\cal B}_{10}$ thus contains a
derivation $d_I$ counting the level in $\b _{I}$, allowing,
for example, to distinguish between $2\b _{I}$,\footnote{In contrast to real
simple roots, we expect for isotropic simple roots of a Borcherds algebra that
$n\b _{I}\in \Delta $, $\forall n\in \Z$.} 
$\b _{I}+\delta $ and $2\delta ,$ which all have weights $-2$ under $H_{-1}$ and 
$0$ under all other Cartan generators dual to simple roots. However, the operator 
$d_I$ is not in $\eten$ and consequently not in $\ginv$, either. 
Hence, the quotient by $\left\{\z,\bar{d}\right\} $ in Conjecture \ref{extendedT2}
amounts to identifying $H_{I}$ with $H_{\d}$. Furthermore, since the roots $\a_{-1}$ 
and $\b_I$ are connected on the Dynkin diagram,  $-H_{-1}$ plays, already in 
$^2\mathcal{B}_{10}$, the same r\^ole as $d_{I}$ with respect to $\b_I$. So the
elimination of $d_{I}$ by the quotient (\ref{kotienhyp}) is equivalent to 
identifying it with $-H_{-1}$, which parallels the treatment of $H_{I}$ with $H_{\d}$.

These two processes reconstruct in $\ginv$ the 8-dimensional root space $(\ginv)_{\d}=
(\g_{^2{\cal B}_{10}})_{\d}\oplus (\g_{^2{\cal B}_{10}})_{\b_I}$
inherited from $\eten$.
\begin{table}[!t]
\begin{center}
$
\begin{array}{|c|c|c|c|}
\hline
l & \mathcal{R}(\Lambda ) & \Lambda & \dim \mathcal{R}(\Lambda ) \\ 
\hline\hline
0 & K_{(ij)} & [200000000] & 55=45+10\text{ Cartan} \\ \hline
1 & Z_{[ijk]} & [001000000] & 120 \\ \hline
2 & \widetilde{Z}_{[i_{1}\cdots i_{6}]} & [000001000] & 210 \\ \hline
3 & \widetilde{K}_{(i)[j_{1}\cdots j_{8}]} & [100000010] & 440=360+8\times 10_{[0]} \\ \hline
4 & (\widetilde{K}_{(i)}\otimes Z)_{[j_{1}\cdots j_{8}][k_{1}k_{2}k_{3}]} & 
[001000001] & 1155=840+7\times 45_{[0]} \\ \cline{2-4}
& A_{(ij)} & [200000000] & 55=10+45_{[0]}\\ \hline
5 &(\widetilde{K}_{(i)}\otimes \widetilde{Z})_{[j_{1}\cdots j_{8}][k_{1}\cdots
k_{6}]} & [000001001] & 1848=840+4\times 252_{[0]} \\ \cline{2-4}
 & B_{(i)[j_{1}\cdots j_{4}]} & [100100000] & 1848=840+4\times 252_{[0]} \\ \hline
6 &(\widetilde{K}_{(i)}\otimes \widetilde{K}_{(j)})_{[k_1\cdots k_8][l_1\cdots
l_8]} & [100000011] & 3200=720+2\times 840+16\times 45+8\times 10_{[0]} \\ \cline{2-4}
& (A\otimes \widetilde{Z})_{(ij)[k_1\cdots k_6]} & [010001000] & 
8250=3150+5\times 840+20\times 45 \\ \cline{2-4}
& D_{(i)[j_1\cdots j_7]} & [100000100] & 1155=840+7\times 45 \\ \cline{2-4}
 & S_{[i_1\cdots i_8]} & [000000010] & 45 \\ \hline
\end{array}
$
\end{center}
\caption{Representations of $\sl(10,\R)$ in $\eten$ up to $l=6$}
\label{Representations}
\end{table}

Formally, one decomposes:
$$E_{\d}^{a}|_{\ginv}=E_{\d}^{a}|_{^2{\cal B}_{10}}\,, \forall a=1,\ldots,7\,,\quad 
\text{ and }\;E_{\d}^{8}|_{\ginv}=E_{\b_{I}}\doteq \frac{1}{\sqrt{2}}\,
(E_{\d+\a_{7}}-E_{\d-\a_{7}})\,.$$
and $F_{\b_{I}}=(E_{\b_{I}})^{\dagger}$ in $\eten$.
One should thus pay attention to the fact that although $\b_I\sim\delta$ in $\ginv$, 
their corresponding ladder operators remain distinct. 
\begin{table}[!t]
\begin{center}
{\footnotesize $
\begin{array}{|c|c|c|c|c|c|}
\hline
\text{Generator} & \a & \text{Physical basis} & O_{w}^{10} & m(\a) & |\a|^2 \\ 
\hline\hline
K_{9\,10} & 
\begin{array}{ccccccccc}
&  &  &  &  &  & 0 &  &  \\ 
0 & 0 & 0 & 0 & 0 & 0 & 0 & 0 & 1%
\end{array}
& (0,0,0,0,0,0,0,0,1,-1) & 45 & 1 & 2 \\ \hline
Z_{[89\,10]} & 
\begin{array}{ccccccccc}
&  &  &  &  &  & 1 &  &  \\ 
0 & 0 & 0 & 0 & 0 & 0 & 0 & 0 & 0%
\end{array}
& (0,0,0,0,0,0,0,1,1,1) & 120 & 1 & 2 \\ \hline
\widetilde{Z}_{[56789\,10]} & 
\begin{array}{ccccccccc}
&  &  &  &  &  & 2 &  &  \\ 
0 & 0 & 0 & 0 & 1 & 2 & 3 & 2 & 1%
\end{array}
& (0,0,0,0,1,1,1,1,1,1) & 210 & 1 & 2 \\ \hline
\widetilde{K}_{(10)[3\cdots 10]} & 
\begin{array}{ccccccccc}
&  &  &  &  &  & 3 &  &  \\ 
0 & 0 & 1 & 2 & 3 & 4 & 5 & 3 & 1%
\end{array}
& (0,0,1,1,1,1,1,1,1,2) & 360 & 1 & 2 \\ \hline
\widetilde{K}_{(2)[3\cdots 10]} & 
\begin{array}{ccccccccc}
&  &  &  &  &  & 3 &  &  \\ 
0 & 1 & 2 & 3 & 4 & 5 & 6 & 4 & 2%
\end{array}
& (0,1,1,1,1,1,1,1,1,1) & 10_{[0]} & 8 & 0 \\ \hline
(\widetilde{K}_{(2)}\otimes Z)_{[3\cdots 10][89\,10]} & 
\begin{array}{ccccccccc}
&  &  &  &  &  & 4 &  &  \\ 
0 & 1 & 2 & 3 & 4 & 5 & 6 & 4 & 2%
\end{array}
& (0,1,1,1,1,1,1,2,2,2) & 840 & 1 & 2 \\ \hline
A_{(10\,10)} & 
\begin{array}{ccccccccc}
&  &  &  &  &  & 4 &  &  \\ 
1 & 2 & 3 & 4 & 5 & 6 & 7 & 4 & 1%
\end{array}
& (1,1,1,1,1,1,1,1,1,3) & 10 & 1 & 2 \\ \hline
\begin{array}{c}
(\widetilde{K}_{(2)}\otimes Z)_{[3\cdots 10][19\,10]} \\ 
A_{(9\,10)}
\end{array}
& 
\begin{array}{ccccccccc}
&  &  &  &  &  & 4 &  &  \\ 
1 & 2 & 3 & 4 & 5 & 6 & 7 & 4 & 2%
\end{array}
& (1,1,1,1,1,1,1,1,2,2) & 45_{[0]} & 8 & 0 \\ \hline
(\widetilde{K}_{(2)}\otimes \widetilde{Z})_{[3\cdots 10][5\cdots \,10]} & 
\begin{array}{ccccccccc}
&  &  &  &  &  & 5 &  &  \\ 
0 & 1 & 2 & 3 & 5 & 7 & 9 & 6 & 3%
\end{array}
& (0,1,1,1,2,2,2,2,2,2) & 840 & 1 & 2 \\ \hline
B_{(10)[7\cdots \,10]} & 
\begin{array}{ccccccccc}
&  &  &  &  &  & 5 &  &  \\ 
1 & 2 & 3 & 4 & 5 & 6 & 8 & 5 & 2%
\end{array}
& (1,1,1,1,1,1,2,2,2,3) & 840 & 1 & 2 \\ \hline
\begin{array}{c}
(\widetilde{K}_{(2)}\otimes \widetilde{Z})_{[3\cdots 10][16\cdots \,10]} \\ 
B_{(6)[7\cdots \,10]}%
\end{array}
& 
\begin{array}{ccccccccc}
&  &  &  &  &  & 5 &  &  \\ 
1 & 2 & 3 & 4 & 5 & 7 & 9 & 6 & 3%
\end{array}
& (1,1,1,1,1,2,2,2,2,2) & 252_{[0]} & 8 & 0 \\ \hline
(\widetilde{K}_{(2)}\otimes \widetilde{K}_{(10)})_{[3\cdots 10][3\cdots
\,10]} & 
\begin{array}{ccccccccc}
&  &  &  &  &  & 6 &  &  \\ 
0 & 1 & 3 & 5 & 7 & 9 & 11 & 7 & 3%
\end{array}
& (0,1,2,2,2,2,2,2,2,3) & 720 & 1 & 2 \\ \hline
(A\otimes \widetilde{Z})_{(9\,10)[5\cdots \,10]} & 
\begin{array}{ccccccccc}
&  &  &  &  &  & 6 &  &  \\ 
1 & 2 & 3 & 4 & 6 & 8 & 10 & 6 & 3%
\end{array}
& (1,1,1,1,2,2,2,2,3,3) & 3150 & 1 & 2 \\ \hline
\begin{array}{c}
(\widetilde{K}_{(1)}\otimes \widetilde{K}_{(10)})_{[24\cdots 10][3\cdots
\,10]} \\[2pt] 
(A\otimes \widetilde{Z})_{[4\,10][5\cdots \,10]} \\ 
D_{(10)[4\cdots 10]}%
\end{array}
& 
\begin{array}{ccccccccc}
&  &  &  &  &  & 6 &  &  \\ 
1 & 2 & 3 & 5 & 7 & 9 & 11 & 7 & 3%
\end{array}
& (1,1,1,2,2,2,2,2,2,3) & 840 & 8 & 0 \\ \hline
(\widetilde{K}_{(2)}\otimes \widetilde{K}_{(2)})_{[3\cdots 10][3\cdots \,10]}
& 
\begin{array}{ccccccccc}
&  &  &  &  &  & 6 &  &  \\ 
0 & 2 & 4 & 6 & 8 & 10 & 12 & 8 & 4%
\end{array}
& (0,2,2,2,2,2,2,2,2,2) & 10_{[0]} & 8 & 0 \\ \hline
\begin{array}{c}
(\widetilde{K}_{(1)}\otimes \widetilde{K}_{(10)})_{[2\cdots 9][3\cdots \,10]}
\\[2pt] 
(A\otimes \widetilde{Z})_{(34)[5\cdots \,10]} \\ 
D_{(3)[4\cdots 10]} \\[2pt] 
(\widetilde{K}_{(1)}\otimes \widetilde{K}_{(2)})_{[3\cdots 10][3\cdots \,10]}
\\[2pt] 
S_{[3\cdots 10]}
\end{array}
& 
\begin{array}{ccccccccc}
&  &  &  &  &  & 6 &  &  \\ 
1 & 2 & 4 & 6 & 8 & 10 & 12 & 8 & 4%
\end{array}
& (1,1,2,2,2,2,2,2,2,2) & 45 & 44 & -2 \\ \hline
\end{array}
$}
\end{center}
\caption{Decomposition of root spaces of $\eten$ into $\sl(10,\R)$ representations}
\label{TablTensor}
\end{table}

We have chosen to depict the Borcherds algebra under scrutiny by the Dynkin
diagram displayed in Table \ref{T2/ZnDynkin}. However such a GKMA, let alone
its root multiplicities, is not known in the literature. So at this stage,
one must bear in mind that the Dynkin diagram we associate to 
$^2{\cal B}_{10}$ is only meant to determine the correct root lattice for 
$(\ginv)^{\C}$. The root multiplicities, on the other hand, have
to be computed separately by decomposing root-spaces of $\eten$
into root-spaces of $(\ginv)^{\C}$. So we need \textit{both} the
Dynkin diagram of $^2{\cal B}_{10}$ \textit{and} the root
multiplicities listed in Table \ref{Borcherds2} in order to determine 
$\ginv$ completely.

In order to support the conjecture \ref{extendedT2}, we start by performing
a careful level by level analysis. We proceed by decomposing $\eten$ with respect 
to the coefficient of $\a_{8}$ into tensorial irreducible representations of 
$\sl(10,\mathbbm{R})$. Such representations, together with the multiplicity of 
the weights labelling them, are summarized up to level $l=6$ in $\a_{8}$ in Tables
\ref{Representations} and \ref{TablTensor}. These tables have been deduced from
the low-level decomposition of roots of $\eten$ that can be found up to level 18
in \cite{NicFisch}. Since we are more interested in the roots themselves 
and their multiplicities than in the dimension of the corresponding 
$\sl(10,\mathbbm{R})$ representations, we added, in column $\dim {\cal R}(\La)$
of Table \ref{Representations}, the way the dimension of each representation decomposes 
in generators corresponding to different sets of roots, obtained by all reflections 
by the Weyl group of $\sl(10,\mathbbm{R})$ (i.e. permutations of indices in the physical 
basis) on the highest weight and possibly other roots. In the first column of
Table~\ref{TablTensor}, the tensor associated to the highest weight is defined, the highest 
weight being obtained by setting all indices to their maximal values.

Note that roots that are permutations of the highest weight of no representation,
or in other words, have null outer multiplicity, do not appear in Table~\ref{Representations},
contrary to what is done in $\cite{NicFisch}$. However, these can be found in 
Table~\ref{TablTensor}. The order of the orbits under the Weyl group of $\sl(10,\R)$
is given in column $O^{10}_{w}$, in which representations of null outer multiplicity are 
designated by a $[0]$ subscript. Besides, column $m$ contains the root multiplicities, 
while column $|\La|^2$ contains the squared length, which, in the particular case of $\eten$, 
provide equivalent characterizations.

For example, the representation with Dynkin labels $[000001000]$ at level 3 is composed 
of the Weyl orbit of its highest weight generator $\widetilde{K}_{(10)[3\cdots 10]}$ together
with 8 Weyl orbits of the (outer multiplicity 0) root $\widetilde{K}_{(2)[3\cdots 10]}$ for 
a total size $360+8\times 10$. Similarly, the representation $[001000001]$ at level 4 is
composed of the 840 components of the associated tensor, together with 7 copies
of the anti-symmetric part of $A_{ij}$ that corresponds to a root of multiplicity 8
and outer multiplicity 0, for a total dimension $840+7\times 45$. The remaining 
eighth copy combines with the (inner and outer) multiplicity 1 diagonal part $A_{(ii)}$
to form a symmetric tensor $[200000000]$. Note that the $A_{(ij)}$ representation differs 
from the $K_{(ij)}$ one, first because the diagonal elements of the latter are given 
by Cartan elements and not ladder operators as in $A_{(ij)}$, and second because 
these two representations obviously correspond to roots of totally different
level, height and threshold. Clearly, isomorphic irreducible representations of 
$\sl(10,\R)$ can appear several times in the decomposition of $\eten$.

Moreover, and more interestingly, weights with different physical interpretations 
may live in the same representation of $\eten$. In particular, the third weight 
$\widetilde{K}_{(10)[3\cdots 10]}$ in Table~\ref{TablTensor} is clearly related 
to the corresponding Euclidean Kaluza-Klein monopole (KK7M), while the fourth weight 
$\widetilde{K}_{(2)[3\cdots 10]}$, though belonging to the same $[100000010]$ 
representation, corresponds, according to the proposal of \cite{Gan1} (cf. Table~\ref{ListBranes}) 
to the Minkowskian Kaluza-Klein particle (KKp) $G_{01}$. Similarly, the seventh weight $A_{(10\,10)}$
is associated to the conjectured Euclidean KK9M-brane $W_{(10\,10)[1\cdots\,10]}$, while
$A_{(9\,10)}$ is interpreted as the Minkowskian M2-brane $C_{09\,10}$. To complete the list of 
Minkowskian objects, we have in addition the weights $B_{(6)[7\cdots \,10]}$ and $D_{(10)[4\cdots 10]}$ 
related respectively to the exceptional M5-brane $\widetilde{C}_{06\cdots 10}$ and the Kaluza-Klein 
monopole (KK7M) $\widetilde{G}_{(10)04\cdots 10}$.

\begin{table}[p]
\begin{center}
{\footnotesize $
\begin{array}{|c|c|c|c|c|c|}
\hline
l & \text{Invariant Tensor} & \text{Physical eigenbasis of }\eten & 
O_w^8 & \a\in \Delta_+(^2{\cal B}_{10}) & m(\a) \\ 
\hline\hline
0 & K_{[78]} & (0,0,0,0,0,0,1,-1,0,0)' & 28 & 
\begin{array}{cccccccc}
0 &  &  &  & 0 &  &  &  \\ 
0 & 0 & 0 & 0 & 0 & 0 & 1 & 0
\end{array}
& 1 \\ \hline
1 & Z_{[89\,10]} & (0,0,0,0,0,0,0,1,1,1)' & 8 & 
\begin{array}{cccccccc}
0 &  &  &  & 0 &  &  &  \\ 
0 & 0 & 0 & 0 & 0 & 0 & 0 & 1
\end{array}
& 1 \\ \cline{2-6}
 & Z_{[678]} & (0,0,0,0,0,1,1,1,0,0)' & 56 & 
\begin{array}{cccccccc}
0 &  &  &  & 1 &  &  &  \\ 
0 & 0 & 0 & 0 & 0 & 0 & 0 & 0
\end{array}
& 1 \\ \hline
2 & \widetilde{Z}_{[56789\,10]} & (0,0,0,0,1,1,1,1,1,1)' & 70 & 
\begin{array}{cccccccc}
0 &  &  &  & 1 &  &  &  \\ 
0 & 0 & 0 & 0 & 1 & 1 & 1 & 1
\end{array}
& 1 \\ \cline{2-6}
 & \widetilde{Z}_{[345678]} & (0,0,1,1,1,1,1,1,0,0)' & 28 & 
\begin{array}{cccccccc}
0 &  &  &  & 2 &  &  &  \\ 
0 & 0 & 1 & 2 & 3 & 2 & 1 & 0
\end{array}
& 1 \\ \hline
3 & \widetilde{K}_{(9)[2\cdots 89]}-\widetilde{K}_{(10)[2\cdots 810]} & 
(0,1,1,1,1,1,1,1,1,1)'+(0^8,-1,1)' & 8 & 
\begin{array}{cccccccc}
1 &  &  &  & 0 &  &  &  \\ 
0 & 0 & 0 & 0 & 0 & 0 & 0 & 0
\end{array}
& 1 \\ \cline{2-6}
 & \widetilde{K}_{(8)[3\cdots 10]} & (0,0,1,1,1,1,1,2,1,1)' & 168 & 
\begin{array}{cccccccc}
0 &  &  &  & 2 &  &  &  \\ 
0 & 0 & 1 & 2 & 3 & 2 & 1 & 1
\end{array}
& 1 \\ \cline{2-6}
 & \widetilde{K}_{(2)[3\cdots 10]} & (0,1,1,1,1,1,1,1,1,1)' & 8 & 
\begin{array}{cccccccc}
0 &  &  &  & 2 &  &  &  \\ 
0 & 1 & 2 & 3 & 4 & 3 & 2 & 1
\end{array}
& 7 \\ \cline{2-6}
 & \widetilde{K}_{(8)[1\cdots 8]} & (1,1,1,1,1,1,1,2,0,0)' & 8 & 
\begin{array}{cccccccc}
0 &  &  &  & 3 &  &  &  \\ 
1 & 2 & 3 & 4 & 5 & 3 & 1 & 0
\end{array}
& 1 \\ \hline
4 & \begin{array}{l}
\big((\widetilde{K}\otimes Z)_{[1\cdots 9][789]} \\[2pt] 
\;-(\widetilde{K}\otimes Z)_{[1\cdots 8\,10][78\,10]}\big)
\end{array}
& (1,1,1,1,1,1,2,2,1,1)'+(0^8,-1,1)' & 28 & 
\begin{array}{cccccccc}
1 &  &  &  & 1 &  &  &  \\ 
1 & 1 & 1 & 1 & 1 & 0 & 0 & 0
\end{array}
& 1 \\ \cline{2-6}
 & (\widetilde{K}_{(2)}\otimes Z)_{[3\cdots 10][89\,10]} & (0,1,1,1,1,1,1,2,2,2)'
& 56 & 
\begin{array}{cccccccc}
0 &  &  &  & 2 &  &  &  \\ 
0 & 1 & 2 & 3 & 4 & 3 & 2 & 2
\end{array}
& 1 \\ \cline{2-6}
 & (\widetilde{K}_{(2)}\otimes Z)_{[3\cdots 10][678]} & (0,1,1,1,1,2,2,2,1,1)' & 
280 & 
\begin{array}{cccccccc}
0 &  &  &  & 3 &  &  &  \\ 
0 & 1 & 2 & 3 & 4 & 3 & 2 & 1
\end{array}
& 1 \\ \cline{2-6}
 & \begin{array}{c}
(\widetilde{K}_{(2)}\otimes Z)_{[3\cdots 10][19\,10]} \\[3pt]
A_{(9\,10)}%
\end{array}
& (1,1,1,1,1,1,1,1,2,2)' & 1 & 
\begin{array}{cccccccc}
0 &  &  &  & 2 &  &  &  \\ 
1 & 2 & 3 & 4 & 5 & 4 & 3 & 2
\end{array}
& 7 \\ \cline{2-6}
& \begin{array}{c}
(\widetilde{K}_{(2)}\otimes Z)_{[3\cdots 10][178]} \\[3pt] 
A_{(78)}
\end{array}
& (1,1,1,1,1,1,2,2,1,1)' & 28 & 
\begin{array}{cccccccc}
0 &  &  &  & 3 &  &  &  \\ 
1 & 2 & 3 & 4 & 5 & 3 & 2 & 1
\end{array}
& 7 \\ \cline{2-6}
 & A_{(9\,9)}-A_{(10\,10)} & (1,1,1,1,1,1,1,1,2,2)'+(0^8,-1,1)' & 1 & 
\begin{array}{cccccccc}
1 &  &  &  & 0 &  &  &  \\ 
1 & 1 & 1 & 1 & 1 & 1 & 1 & 1
\end{array}
& 1 \\ \cline{2-6}
 & A_{(88)} & (1,1,1,1,1,1,1,3,1,1)' & 8 & 
\begin{array}{cccccccc}
0 &  &  &  & 3 &  &  &  \\ 
1 & 2 & 3 & 4 & 5 & 3 & 1 & 1
\end{array}
& 1 \\ \hline
5 & \begin{array}{l}
\big((\widetilde{K}\otimes \widetilde{Z})_{[1\cdots 9][4\cdots \,9]} \\[2pt] 
\;-(\widetilde{K}\otimes \widetilde{Z})_{[1\cdots 810][4\cdots 8\,10]}\big)
\end{array}
& (1,1,1,2,2,2,2,2,1,1)'+(0^8,-1,1)' & 56 & 
\begin{array}{cccccccc}
1 &  &  &  & 2 &  &  &  \\ 
1 & 1 & 1 & 2 & 3 & 2 & 1 & 0
\end{array}
& 1 \\ \cline{2-6}
 & (\widetilde{K}_{(2)}\otimes \widetilde{Z})_{[3\cdots 10][5\cdots \,10]} & 
(0,1,1,1,2,2,2,2,2,2)' & 280 & 
\begin{array}{cccccccc}
0 &  &  &  & 3 &  &  &  \\ 
0 & 1 & 2 & 3 & 5 & 4 & 3 & 2
\end{array}
& 1 \\ \cline{2-6}
& (\widetilde{K}_{(2)}\otimes \widetilde{Z})_{[3\cdots 10][3\cdots \,8]} & 
(0,1,2,2,2,2,2,2,1,1)' & 56 & 
\begin{array}{cccccccc}
0 &  &  &  & 4 &  &  &  \\ 
0 & 1 & 3 & 5 & 7 & 5 & 3 & 1
\end{array}
& 1 \\ \cline{2-6}
 & B_{(8)[7\cdots \,10]} & (1,1,1,1,1,1,2,3,2,2)' & 56 & 
\begin{array}{cccccccc}
0 &  &  &  & 3 &  &  &  \\ 
1 & 2 & 3 & 4 & 5 & 3 & 2 & 2
\end{array}
& 1 \\ \cline{2-6}
 & B_{(8)[5\cdots \,8]} & (1,1,1,1,2,2,2,3,1,1)' & 280 & 
\begin{array}{cccccccc}
0 &  &  &  & 4 &  &  &  \\ 
1 & 2 & 3 & 4 & 6 & 4 & 2 & 1
\end{array}
& 1 \\ \hline
\end{array}
$}
\end{center}
\end{table}
\begin{table}[p]
\begin{center}
{\footnotesize $
\begin{array}{|c|c|c|c|c|c|}
\hline
5 & B_{(9)[678\,9]}-B_{(10)[678\,10]} & (1,1,1,1,1,2,2,2,2,2)'+(0^8,-1,1)'
& 56 & 
\begin{array}{cccccccc}
1 &  &  &  & 1 &  &  &  \\ 
1 & 1 & 1 & 1 & 1 & 1 & 1 & 1
\end{array}
& 1 \\ \cline{2-6}
 & \begin{array}{c}
(\widetilde{K}_{(2)}\otimes \widetilde{Z})_{[3\cdots 10][16\cdots \,10]} \\[3pt] 
B_{(6)[7\cdots \,10]}
\end{array}
& (1,1,1,1,1,2,2,2,2,2)' & 56 & 
\begin{array}{cccccccc}
0 &  &  &  & 3 &  &  &  \\ 
1 & 2 & 3 & 4 & 5 & 4 & 3 & 2
\end{array}
& 7 \\ \cline{2-6}
& \begin{array}{c}
(\widetilde{K}_{(2)}\otimes \widetilde{Z})_{[3\cdots 10][14\cdots \,8]} \\ [3pt]
B_{(4)[5\cdots \,8]}
\end{array}
& (1,1,1,2,2,2,2,2,1,1)' & 56 & 
\begin{array}{cccccccc}
0 &  &  &  & 4 &  &  &  \\ 
1 & 2 & 3 & 5 & 7 & 5 & 3 & 1
\end{array}
& 7 \\ \hline
6 & \begin{array}{l}
\big((\widetilde{K}\otimes \widetilde{K}_{(9)})_{[2\cdots 10][2\cdots \,9]} \\
[2pt] 
\;-(\widetilde{K}\otimes \widetilde{K}_{(10)})_{[2\cdots 8\,10][2\cdots 810]}\big)
\end{array}
& (0,2,2,2,2,2,2,2,2,2)'+(0^8,-1,1)' & 8 & 
\begin{array}{cccccccc}
2 &  &  &  & 0 &  &  &  \\ 
0 & 0 & 0 & 0 & 0 & 0 & 0 & 0
\end{array}
& 1 \\ \cline{2-6}
 & (\widetilde{K}_{(2)}\otimes \widetilde{K}_{(8)})_{[3\cdots 10][3\cdots \,10]}
& (0,1,2,2,2,2,2,3,2,2)' & 336 & 
\begin{array}{cccccccc}
0 &  &  &  & 4 &  &  &  \\ 
0 & 1 & 3 & 5 & 7 & 5 & 3 & 2
\end{array}
& 1 \\ \cline{2-6}
 & \begin{array}{c}
 \begin{array}{l}
\big((\widetilde{K}\otimes \widetilde{K}_{(9)})_{[1\cdots 9][3\cdots \,10]} \\[2pt] 
\;-(\widetilde{K}\otimes \widetilde{K}_{(10)})_{[1\cdots 8\,10][3\cdots 10]}\big)
\end{array} 
\\[4pt] 
\begin{array}{l}
\big((A\otimes \widetilde{Z})_{[89][3\cdots \,79]} \\[2pt] 
\;-(A\otimes \widetilde{Z})_{[8\,10][3\cdots \,7\,10]} \big)
\end{array}
\\[4pt] 
D_{(9)[3\cdots 9]}-D_{(10)[3\cdots 8\,10]}%
\end{array}
& (1,1,2,2,2,2,2,2,2,2)'+(0^8,-1,1)' & 28 & 
\begin{array}{cccccccc}
2 &  &  &  & 0 &  &  &  \\ 
1 & 0 & 0 & 0 & 0 & 0 & 0 & 0
\end{array}
& 8 \\ \cline{2-6}
 & \begin{array}{c}
(\widetilde{K}_{(2)}\otimes \widetilde{K}_{(8)})_{[3\cdots 10][1\cdots \,8]}
\\[3pt] 
(A\otimes \widetilde{Z})_{[78][2\cdots \,68]} \\[3pt] 
D_{(8)[2\cdots 8]}
\end{array}
& (1,2,2,2,2,2,2,3,1,1)' & 56 & 
\begin{array}{cccccccc}
0 &  &  &  & 5 &  &  &  \\ 
1 & 3 & 5 & 7 & 9 & 6 & 3 & 1
\end{array}
& 1 \\ \cline{2-6}
 & (A\otimes \widetilde{Z})_{(9\,10)[5\cdots \,10]} & (1,1,1,1,2,2,2,2,3,3)' & 70
& 
\begin{array}{cccccccc}
0 &  &  &  & 3 &  &  &  \\ 
1 & 2 & 3 & 4 & 6 & 5 & 4 & 3
\end{array}
& 1 \\ \cline{2-6}
 & (A\otimes \widetilde{Z})_{(78)[5\cdots \,10]} & (1,1,1,1,2,2,3,3,2,2)' & 420
& 
\begin{array}{cccccccc}
0 &  &  &  & 4 &  &  &  \\ 
1 & 2 & 3 & 4 & 6 & 4 & 3 & 2
\end{array}
& 1 \\ \cline{2-6}
 & (A\otimes \widetilde{Z})_{(78)[3\cdots \,8]} & (1,1,2,2,2,2,3,3,1,1)' & 420 & 
\begin{array}{cccccccc}
0 &  &  &  & 5 &  &  &  \\ 
1 & 2 & 4 & 6 & 8 & 5 & 3 & 1
\end{array}
& 1 \\ \cline{2-6}
 & \begin{array}{l}
\big((A\otimes \widetilde{Z})_{(89)[4\cdots \,89]} \\[2pt] 
\;-(A\otimes \widetilde{Z})_{(8\,10)[4\cdots 8\,10]}\big)
\end{array}
& (1,1,1,2,2,2,2,3,2,2)'+(0^8,-1,1)' & 280 & 
\begin{array}{cccccccc}
1 &  &  &  & 2 &  &  &  \\
1 & 1 & 1 & 2 & 3 & 2 & 1 & 1
\end{array}
& 1 \\[3pt] \cline{2-6}
 & \begin{array}{l}
\big((\widetilde{K}\otimes \widetilde{K}_{(1)})_{[1\cdots 9][2\cdots \,9]} \\[2pt] 
\;-(\widetilde{K}\otimes \widetilde{K}_{(1)})_{[1\cdots 8\,10][2\cdots 8\,10]}\big)
\end{array}
& (2,2,2,2,2,2,2,2,1,1)'+(0^8,-1,1)' & 8 & 
\begin{array}{cccccccc}
2 &  &  &  & 1 &  &  &  \\ 
2 & 2 & 2 & 2 & 2 & 1 & 0 & 0
\end{array}
& 8 \\[3pt] \cline{2-6}
 & (\widetilde{K}_{(2)}\otimes \widetilde{K}_{(2)})_{[3\cdots 10][3\cdots \,10]}
& (0,2,2,2,2,2,2,2,2,2)' & 8 & 
\begin{array}{cccccccc}
0 &  &  &  & 4 &  &  &  \\ 
0 & 2 & 4 & 6 & 8 & 6 & 4 & 2
\end{array}
& 7 \\ \cline{2-6}
 & \begin{array}{c}
(\widetilde{K}\otimes \widetilde{K}_{(10)})_{[1\cdots 9][3\cdots \,10]} \\[3pt] 
(A\otimes \widetilde{Z})_{(34)[5\cdots \,10]} \\[3pt] 
D_{(3)[4\cdots 10]} \\[3pt]  
(\widetilde{K}\otimes \widetilde{K}_{(2)})_{[13\cdots 10][3\cdots \,10]} \\[3pt]  
S_{[3\cdots 10]}%
\end{array}
& (1,1,2,2,2,2,2,2,2,2)' & 28 & 
\begin{array}{cccccccc}
0 &  &  &  & 4 &  &  &  \\ 
1 & 2 & 4 & 6 & 8 & 6 & 4 & 2
\end{array}
& 36 \\ \cline{2-6}
 & \begin{array}{c}
(\widetilde{K}\otimes \widetilde{K}_{(8)})_{[1\cdots 79\,10][1\cdots \,8]}
\\[3pt]  
(A\otimes \widetilde{Z})_{(12)[3\cdots \,8]} \\[3pt]  
D_{(1)[2\cdots 8]} \\[3pt]  
(\widetilde{K}\otimes \widetilde{K}_{(1)})_{[1\cdots 9][2\cdots \,810]} \\[3pt]  
S_{[1\cdots 8]}
\end{array}
& (2,2,2,2,2,2,2,2,1,1)' & 1 & 
\begin{array}{cccccccc}
0 &  &  &  & 5 &  &  &  \\ 
2 & 4 & 6 & 8 & 10 & 7 & 4 & 1
\end{array}
& 36 \\ \hline
\end{array}
$}
\end{center}
\caption{Decomposition of root spaces of $^2{\cal B}_{10}$ in representations of $\sl(8,\R)$}
\label{Borcherds2}
\end{table}

After this short excursion into weights and representations of $\eten$, let us 
come back to the characterization of $^2{\cal B}_{10}$.
Observing that objects commuting with $i\widetilde H^{[2]}=-i{\cal K}_{z\bar z}$
have the form $X_{\cdots(99)}-X_{\cdots (10\,10)}$ or $X_{\cdots \lbrack 9\,10]}$, 
we are naturally looking for invariant combinations of generators of $\eten$ 
with such tensorial properties. 
The latter can then be decomposed into $\sl(8,\R)$
tensors with Weyl orbits of order $O_{w}^{8}$ and identified with a root of 
$^2{\cal B}_{10}$. We have carried out such a decomposition up to $l=6$ in $\a_8$ and listed 
the corresponding roots of $^2{\cal B}_{10}$, together with their multiplicities $m$,
in Table \ref{Borcherds2}. 

In order to make clear how to retrieve the root system of $\ginv$ from 
Table \ref{Borcherds2}, we give in the third column the expression
of a given root of $^2{\cal B}_{10}$ in a generalized notation for
the physical basis, denoted {\it physical eigenbasis of $\eten$}.
This eigenspace basis is defined by\footnote{This basis will be used again
for computing shift vectors in Section \ref{ShiftSec}.}:
\begin{equation}
E_i'=E_i\,, \; \forall i=-1,\ldots,5,8 \, ,\quad
E_6'=\frac{1}{\sqrt{2}}(E_6+iE_{67})\, ,\quad
E_7'=\frac{1}{2}(H_7-i(E_7+F_7)) \,,
\end{equation}
so that
\begin{equation}\label{Eprime}
[E_{\a'},E_7'/F_7']=\mp i(E_{\a+\a_7}-E_{\a-\a_7}) \, ,
\end{equation}
for all $\a'$'s satisfying $|\a'|^2\leq 0$ and $\a'=\a$, where $\a$ is a root of 
$\eten$ in the original basis. In fact, all invariant generators in $\eten$ 
either satisfy $E_{\a'}=E_{\a}$, or are of the form~(\ref{Eprime}). In 
Table~\ref{Borcherds2}, we characterize the former by their root $\a'=\a$ 
in the physical eigenbasis, and the latter as the sum of a root $\a'=\a$ and 
$-\a_7'$, to emphasize the fact that they build separate root spaces of $^2{\cal B}_{10}$
that will merge in $\ginv$. Indeed, modding out $^2{\cal B}_{10}$ by $\{\z;\bar d\}$ 
eliminates the Cartan elements measuring the level in $\b_I=\d-\a_7'$ in $\ginv$, 
thus identifying $\b_I$ with $\d$.

As an example, consider the fourth and sixth root at $l=4$ in Table \ref{Borcherds2}.
Both are identified in $(\ginv)^{\C}$:
$$
(1,1,1,1,1,1,1,1,2,2)'+(0^8,-1,1)'\sim(1,1,1,1,1,1,1,1,2,2)'\,
$$
so that their respective generators: $A_{99}-A_{10\,10}$ on the one hand, and 
$A_{9\,10}$ plus 6 combinations of operators of the form 
$\widetilde{K}_{[1\cdots\hat{i}\cdots 8 9\,10]}\otimes Z_{[i 9\, 10]}\,,\; i=1,\ldots,6$ 
on the other hand, are now collected in a common 8-dimensional root space $(\ginv)_{\a}$ ¨
for $\a=\d+\a_{-1}+\ldots+\a_5+\a_8$. As a result, the root multiplicity of 
$\d+\a_{-1}+\ldots+\a_5+\a_8$ is conserved when reducing $\eten$ to $(\ginv)^{\C}$, 
even though its corresponding root space is spanned by (partly) different generators 
in each case. We expect this mechanism to occur for all imaginary roots of $\ginv$.

On the other hand, the multiplicity of isotropic roots in $^2{\cal B}_{10}$ splits
according to $8\rightarrow 1+7$, in which the root space of multiplicity one is of 
the form~(\ref{Eprime}). Likewise,
imaginary roots of $\eten$ of length $-2$ split in 
$^2{\cal B}_{10}$ as $44\rightarrow 8+36$, and we expect, though we did not
push the analysis that far, that imaginary roots of length $-4$ will split as
$192\rightarrow 44+148$. Generally, we predict root multiplicities of 
$^2{\cal B}_{10}$ to be 1, 7, 36, 148, 535, 1745,$\ldots$ Although
not our initial purpose, the method can thus be exploited to predict root
multiplicities of certain Borcherds algebras constructed as fixed-point
algebras of KMAs under a finite-order automorphism of order bigger than 2.

Finally, a remark on the real $\ginv$. As anticipated in eqn.(\ref{kotienhyp}), 
the Borcherds algebra involved is actually the split form $^2{\cal B}_{10|11}$. 
Its reality properties can  be inferred from the affine case, which has been worked 
out in detail in Section~\ref{affT2}, and the behaviour of the generators
$E_{n\b_{I}}\doteq~(1/\sqrt{2})(E_{n\d+\a_{7}}-E_{n\d-\a _{7}})$ and  
$F_{n\b_{I}}=(E_{n\b_{I}})^{\dagger}$ under the restriction $\phi$. 

Since $\phi(E_{n\b_{I}})=~-F_{n\b_{I}}$ and $\phi(F_{n\b_{I}})=-E_{n\b_{I}}$, 
both sets of operators combine symmetrically in the usual compact and non-compact 
operators $E_{n\b _{I}}\mp F_{n\b _{I}}$. Moreover, the Cartan generator $H_I$ 
has to match the reality property of $H_{\d}$ to which it gets identified under 
relation~(\ref{kotien}), and must therefore be non-compact in $^2{\cal B}_{10}$, 
which agrees with the definition of the split form $^2{\cal B}_{10|11}$.

As in the affine case, the signature of $\ginv$ remains finite and is completely
determined by the reality properties of the Cartan subalgebra. Taking into account 
the quotient~(\ref{kotien}), the signature is $\sigma=8$.

\section{The orbifolds $T^{4}/\Z_{n>2}$}
\label{secT4}

In this section, we will treat the slightly more involved orbifold $%
T^{7-D}\times T^{4}/Z_{n}$ for $n\geqslant 3$. A new feature appears in this
case: the invariant subalgebras will now contain generators that are complex
combinations of the original $\eten$ generators.
If the orbifold is chosen to act on the coordinates 
$\{x^{7},x^{8},x^{9},x^{10}\}$, it will only affect roots containing 
$\a_{4}$, $\a_{5}$, $\a_{6}$, $\a_{7}$ or $\a_{8}$, and the corresponding 
generators. This orbifold should thus be studied first in $D=6$
where $\g^{U}=\so(5,5)$, with the following action on the complex coordinates: 
\begin{equation}\label{2Complexification}
\qquad (z_{1},\bar{z}_{1})\rightarrow (e^{2\pi i/n}z_{1},e^{-2\pi i/n}\bar{z}_{1})\,,
\text{ \ \ \ }\qquad (z_{2},\bar{z}_{2})\rightarrow (e^{-2\pi i/n}z_{2},e^{2\pi i/n}\bar{z}_{2})\,.
\end{equation}
In other words, we choose the prescription $Q_{1}=+1$ and $Q_{2}=-1$ to ensure 
$\sum_{i}Q_{i}=0$.

The rotation operator 
${\ds \mathcal{U}_{4}^{\Z_{n}}=\prod_{k=1}^2 e^{-\frac{2\pi i}{n}Q_k\mathcal{K}_{z_k\bar z_k}}}$
with the above charge prescription leaves invariant the following objects: 
\begin{equation}
\label{Charge0forT4/Zn}
\begin{array}{c|ccl}
\cline{2-4}
Q_{A}=0 & K_{66} & = & \frac{1}{4}(5H_{4}+6H_{5}+4H_{6}+2H_{7}+3H_{8})\,,
\\ \cline{1-1}
& K_{z_{1}\bar{z}_{1}} & = & \frac{1}{2}(K_{5}+K_{6})
=\frac{1}{4}(H_{4}+4H_{5}+4H_{6}+2H_{7}+3H_{8})\,, 
\\[2pt] 
& K_{z_{2}\bar{z}_{2}} & = & \frac{1}{2}(K_{7}+K_{8})=
\frac{1}{4}(H_{4}+2H_{5}+3H_{8})\,, \\[2pt] 
& \left\{ 
\begin{array}{c}
K_{\bar{z}_{1}\bar{z}_{2}}\,/\,\frac{1}{2}\mathcal{K}_{\bar{z}_{1}bar{z}_{2}} \\[2pt]
K_{z_{1}z_{2}}\,/\,{\frac12}\mathcal{K}_{z_{1}z_{2}}
\end{array}
\right\} & = & \frac{1}{4}\big(E_{56}-E_{67}(+/-)(F_{56}-F_{67}) \\ 
&  &  & \qquad \pm i\big(E_{6}+E_{567}(+/-)(F_{6}+F_{567})\big)\big)\,,\\[4pt] 
& Z_{6z_{1}\bar{z}_{1}}\,/\,\frac{1}{2}\mathcal{Z}_{6z_{1}\bar{z}_{1}} & = 
& \frac{i}{2}(E_{45^{2}6^{2}78}(+/-)F_{45^{2}6^{2}78}) \\[2pt] 
& Z_{6z_{2}\bar{z}_{2}}\,/\,\frac{1}{2}\mathcal{Z}_{6z_{2}\bar{z}_{2}} & = 
& \frac{i}{2}(E_{458}(+/-)F_{458})\,, \\[2pt] 
& \left\{ 
\begin{array}{c}
Z_{6\bar{z}_{1}\bar{z}_{2}}\,/\,{\frac12}\mathcal{Z}_{6\bar{z}_{1}\bar{z}_{2}} \\[2pt] 
Z_{6z_{1}z_{2}}\,/\,\frac{1}{2}\mathcal{Z}_{6z_{1}z_{2}}
\end{array}
\right\} & = & \frac{1}{4}\big(E_{45^{2}678}-E_{4568}(+/-)(F_{45^{2}678}-F_{4568}) \\ 
&  &  & \qquad \pm i\big(E_{45^{2}68}+F_{45^{2}68}(+/-)(E_{45678}+F_{45678})\big)\big)\,. \\[4pt] 
& \mathcal{K}_{z_{1}\bar{z}_{1}}=i(E_{5}-F_{5})\,, &  & \qquad \qquad 
\mathcal{K}_{z_{2}\bar{z}_{2}}=i(E_{7}-F_{7})\,.
\end{array}
\end{equation}
Thus $\ginv$ has as before (conserved) rank 5. Note that the
invariant diagonal metric elements are in fact linear combinations of
the three basic Cartan generators satisfying $\a_{5}(H)=\a_{7}(H)=0$, namely 
$\{2H_{4}+H_{5},H_{5}+2H_{8},H_{5}+2H_{6}+H_{7}\}$.
Furthermore, we have various charged combinations: 
\begin{equation}
\label{Charge1forT4/Zn}
\begin{array}{c|ccl}
\cline{2-4}
Q_{A}=+1 & K_{6\bar{z}_{1}}\,/\,\frac{1}{2}\mathcal{K}_{6\bar{z}_{1}} & = 
& \frac{1}{2\sqrt{2}}\big(E_{4}(+/-)F_{4}+i(E_{45}(+/-)F_{45})\big)\,, \\ \cline{1-1}
& K_{6z_{2}}\,/\,\frac{1}{2}\mathcal{K}_{6z_{2}} & = & \frac{1}{2\sqrt{2}}
\big(E_{456}(+/-)F_{456}-i(E_{4567}(+/-)F_{4567})\big)\,, \\[2pt] 
& Z_{z_{1}\bar{z}_{1}z_{2}}\,/\,\frac{1}{2}\mathcal{Z}_{z_{1}\bar{z}_{1}z_{2}} & = 
& \frac{1}{2\sqrt{2}}\big(E_{568}(+/-)F_{568}+i(E_{5678}(+/-)F_{5678})\big)\,, \\[2pt] 
& Z_{\bar{z}_{1}z_{2}\bar{z}_{2}}\,/\,\frac{1}{2}\mathcal{Z}_{\bar{z}_{1}z_{2}\bar{z}_{2}} & = 
& \frac{1}{2\sqrt{2}}\big(-(E_{8}(+/-)F_{8})+i(E_{58}(+/-)F_{58})\big)\,,
\end{array}
\end{equation}
and their complex conjugates with $Q_{A}=-1$, along with: 
\begin{equation}
\label{Charge2forT4/Zn}
\begin{array}{c|ccl}
\cline{2-4}
Q_{A}=+2 & K_{\bar{z}_{1}\bar{z}_{1}} & = 
& \frac{1}{2}\big(H_{5}+i(E_{5}+F_{5})\big)\,, \\ \cline{1-1}
& K_{z_{2}z_{2}} & = & \frac{1}{2}\big(H_{7}-i(E_{7}+F_{7})\big)\,, \\[2pt] 
& K_{\bar{z}_{1}z_{2}}\,/\,\frac{1}{2}\mathcal{K}_{\bar{z}_{1}z_{2}} & = 
& \frac{1}{4}\big(E_{56}+E_{67}(+/-)(F_{56}+F_{67}) \\ 
&  &  & \qquad -i\big(E_{567}-E_{6}(+/-)(F_{567}-F_{6})\big)\big)\,, \\[2pt] 
& Z_{6\bar{z}_{1}z_{2}}\,/\,\frac{1}{2}\mathcal{Z}_{6\bar{z}_{1}z_{2}} & = 
& \frac{1}{4}\big(E_{45^{2}678}+E_{4568}(+/-)(F_{45^{2}678}+F_{4568}) \\ 
&  &  & \qquad -i\big(E_{45^{2}68}-E_{45678}(+/-)(F_{45^{2}68}-F_{45678})\big)\big)\,,
\end{array}
\end{equation}
and complex conjugates ($Q_{A}=-2$). Note that these five sectors are all
different in $T^{4}/\Z_{n}$ for $n\geqslant 5$, while the two sectors with 
$Q_{A}=\pm 2$ will clearly have the same charge assignment in $T^{4}/\Z_{4}$. 
Finally, the orbifold $T^{4}/\Z_{3}$ merges, on the one hand, the two sectors
with $Q_{A}= 2,-1$ and, on the other hand, the two remaining ones with $Q_{A}= 1,-2$,
giving rise to three main sectors instead of five. In string theory, these
three cases will lead to different twisted sectors, however, the untwisted
sector and the residual U-duality algebra do not depend on $n$ for any $n\geqslant 3$. 
The $n=2$ case will again be treated separately.

For clarity, we will start by deriving the general structure of the
(complex) invariant subalgebra, leaving aside, for the moment being, the
analysis of its reality property. To do so, we perform a change of basis in
the $Q_A=0$ sector, separating raising from lowering operators. Let 
$X_{\a}=(1/2)(E_{\a}+F_{\a})$ be the generator of any field
element of $\g^U$, we will resort to the combinations 
$X_{\a}^+ \doteq X_{\a}+\frac{1}{2}\mathcal{X}_{\a}=E_{\a}$ and 
$X_{\a}^- \doteq X_{\a}-\frac{1}{2}\mathcal{X}_{\a}=F_{\a}$ to derive 
$(\ginv)^{\C}$. First, the following generators can be shown to form a basis 
of the non-abelian part of $(\ginv)^{\C}$: 
\begin{eqnarray}
E_{\tilde{\a}}\!\!&\!\!=\!\!&\!\!-iE_{458}=Z_{6\bar{z}_{2}z_{2}}^{+}\,, \qquad\quad
F_{\widetilde{\a}}\,=\,iF_{458}=Z_{6z_{2}\bar{z}_{2}}^{-}\,, \notag \\
E_{\a_{\pm}} \!\!&\!\! =\!\!&\!\!\frac{1}{2}\left(E_{56}-E_{67}\pm i(E_{567}+E_{6})\right)=
(K_{\bar{z}_{1}\bar{z}_{2}}/K_{z_{1}z_{2}})^{+}\,,\label{A3GeneratorsT4/Zn} \\
F_{{\a }_{\pm }}\!\!& \!\!=\!\!&\!\!\frac{1}{2}\left( F_{56}-F_{67})\pm
i(-F_{567}-F_{6})\right) =(K_{z_{1}z_{2}}/K_{\bar{z}_{1}\bar{z}_{2}})^{-}\,.
\notag
\end{eqnarray}
Computing their commutation relations determines the remaining generators of
the algebra (for economy, we have omitted the lowering operators, which can
be obtained quite straightforwardly by $F_{\a}=(E_{\a})^{\dagger}$): 
\begin{equation}
\label{A3CommRel}
\begin{array}{rl}
{\ds E_{\widetilde{\a}+{\a}_{\pm}}\!\!\!}&{\ds \doteq\; 
\pm [E_{\widetilde{\a}},E_{\a_{\pm}}] \;=
\;\frac{1}{2}\left(E_{45^{2}678}-E_{4568}\pm i(E_{45678}+E_{45^{2}68})\right)} \\[4pt]
&=(Z_{6\bar{z}_{1}\bar{z}_{2}}/Z_{6z_{1}z_{2}})^{+}\,, \\[4pt]
{\ds E_{{\a }_{-}+\widetilde{\a}+{\a }_{+}} \!\!\! }& {\ds \doteq 
\lbrack E_{\a_{-}},E_{\widetilde{\a}+\a_{+}}] \;=\;-iE_{45^{2}6^{2}78}=(Z_{6\bar{z}_{1}z_{1}})^{+}\,,}\\[4pt]
{\ds H_{\widetilde{\a}} \!\!\! }& {\ds\doteq [E_{\widetilde{\a}},F_{\widetilde{\a}}] \;=
\;(H_{4}+H_{5}+H_{8})\,, }  \\[4pt]
{\ds H_{\a_{\pm}} } & {\ds \doteq [E_{\a_{\pm }},F_{\a_{\pm}}]\;=\;
\frac{1}{2}\big( H_{5}+2H_{6}+H_{7}\pm i(F_{5}-E_{5}+F_{7}-E_{7})\big)\,, }  \\[4pt]
{\ds H_{\widetilde{\a}+\a_{\pm}}\!\!\!} & {\ds \doteq 
[E_{\widetilde{\a}+\a_{\pm}},F_{\widetilde{\a}+\a_{\pm}}]} \\[3pt]
& {\ds =\;\frac{1}{2}\big(2H_{4}+3H_{5}+2H_{6}+H_{7}+2H_{8}\mp i(E_{5}-F_{5}+E_{7}-F_{7})\big) \,,}\\[4pt]
{\ds H_{{\a }_{-}+\widetilde{\a}+\a_{+}}\!\!\!} & {\ds \doteq 
[E_{\a_{-}+\widetilde{\a}+\a_{+}},F_{\a_{-}+\widetilde{\a}+\a_{+}}]=
H_{4}+2H_{5}+2H_{6}+H_{7}+H_{8}\,. }
\end{array}
\end{equation}
which shows that the non-abelian part of the complexified invariant
subalgebra is of type $\an_{3}\simeq \dn_{3}$.

The rest of the $Q_A=0$ sectors combines into two abelian contributions, so that
the whole $D=6$ $(\ginv)^{\C}$ reads
\vskip10pt 
\begin{picture}(400,20)\thicklines
\put(20,7){$\dn_3\oplus \C^{\oplus^2}\;:$}
\multiput(100,10)(30,0){3}{\circle{6}}
\multiput(103,10)(30,0){2}{\line(1,0){24}}
\put(94,-4){$\a_{-}$}
\put(126,-4){$\widetilde{\a}$}
\put(156,-4){$\a_{+}$}
\put(170,7){ $\times$ }
\put(187,7){$\{H_8-H_4\}$}
\put(240,7){ $\times$ }
\put(260,7){$\{E_5-F_5-E_7+F_7\}$}
\end{picture}
\vskip10pt

Concentrating on the non-abelian $\dn_3$ part of the real form $\ginv$, we remark that
it can be chosen to have Cartan subalgebra spanned by the basis
$\big\{i(H_{\a_+}-H_{\a_-});\,H_{\widetilde{\a}};\,H_{\a_+}+H_{\a_-}\big\}$
compatible with the restriction Fix$_{\tau_0}(\ginv^{\C})$.
Since, in this basis, all ladder operators combine under $\phi$ into pairs of one compact 
and one non-compact operator, the signature of the real $\dn_3$ is again
completely determined by the difference between non-compact and compact Cartan generators:
since $i(H_{\a_+}-H_{\a_-})$ is compact while the two remaining generators are non-compact,
$\sigma(\dn_3)=1$, which determines the real form to be $\su(2,2)\simeq \so(4,2)$.
The reality property of the invariant subalgebra is encoded in the
Satake diagram of Table~\ref{T4/ZnOrbifolds}.

In addition, the two abelian factors appearing in the diagram
restrict, under Fix$_{\tau_0}$ to $H^{[4]}=H_{8}-H_{4}$ and 
$i\widetilde{H}^{[4]}=(E_{5}-F_{5}-E_{7}+F_{7})$ and generate $\so(1,1)\oplus\u(1)$, 
similarly to the $T^2/\Z_{n>2}$ case. Their contributions to the signature
cancel out, so $\sigma(\ginv)=1$

If we refer to the Satake diagram of Table~\ref{T4/ZnOrbifolds}, we note that in contrast 
to the split case, the arrows now joining the roots $\a_{+}$ and $\a_{-}$ indeed change
the compactness of the Cartan subalgebra without touching the "split" structure of the ladder
operators. Moreover, the combinations $i(H_{\a_{+}}-H_{\a _{-}})$ and
$H_{\a_{+}}+H_{\a_{-}}$ are now directly deducible from the action of
$\phi$ on the set of simple roots.

Finally, as will be confirmed with the $T^6/\Z_{n>2}$ orbifold, if
the chief inner automorphism $\mathcal{U}_q^{\Z_n}$ produces $k$ pairs of Cartan
generators in $(\ginv)^{\C}$ taking value in $\h(\e_{r|r}) \pm i \k(\e_{r|r})$,
there will be $k$ arrows joining the dual simple roots in the Satake diagram. 

Compactifying further to $D=5$, the additional node $\a_{3}$ connects
to ${\widetilde{\a}}$ forming a $\dn_{4}$ subalgebra. As in the $T^{2}/\Z_{n>2}$ 
case, this extra split $\an_1$ will increase the total signature by one, 
yielding the real form $\so(5,3)$. Since $\a_{3}(H_{8}-H_{4})\neq 0$, 
the non-compact Cartan generator $H^{[4]}$ commuting with $\so(5,3)\oplus\u(1)$ 
is now any multiple of $H^{[4]}=2H_{3}+4H_{4}+3H_{5}+2H_{6}+H_{7}$.

In $D=4$, a new invariant root $\c=\a_{2}+2\a_{3}+3\a_{4}+3\a_{5}+2\a_{6}+\a_{7}+\a_{8}
\in \Delta^{+}(\e_{7})$ appears which enhances the $\so(1,1)$ factor to $sl(2,\R)$. 
The reality property of the latter abelian factor can be checked by rewriting 
$\c=\a_{2}+2\a_{3}+2\a_{4}+\a_{5}+\a_{-}+\widetilde{\a}+\a_{+}$, which tells us 
that $\phi(\c)=-\c$. In $D=3$, the additional node $\a_{1}$ extending $\ginv$ 
reconnects $\c$ to the Dynkin diagram, resulting in $\so(8,6)\oplus \u(1)$.

\subsection{Equivalence classes of involutive automorphisms of Lie algebras}

Before treating the affine case, we shall introduce a procedure extensively
used by \cite{GTP2}, \S 14.4, to determine real forms by translating the
adjoint action of the involutive automorphism on the generators by an
exponential action on the root system directly.

Our concern in this paper will be only with real forms generated from chief
\textit{inner} involutive automorphisms, in other words involutions which
can be written as $\vartheta =\Ad (e^{\overline{H}})$ for
$\overline{H}\in \h^{\C}$. In this case, a matrix realization of the
defining chief inner automorphism, denoted 
$\boldsymbol{\vartheta}=\Ad(e^{\overline{\mathbf{H}}})$ will act on
the compact real form $\g_{c}$ as:
\begin{equation}
\boldsymbol{\vartheta}=\Id_{r}\oplus \sum_{\a \in\Delta _{+}}\left( 
\begin{array}{cc}
\cosh (\a (\overline{H})) & i\sinh (\a (\overline{H})) \\ 
-i\sinh (\a (\overline{H}) & \cosh (\a (\overline{H}))
\end{array}
\right) \text{ },\text{ \ \ \ \ }\forall \a \in \Delta _{+}\text{ .}
\label{automorf}
\end{equation}
Being involutive $\boldsymbol{\vartheta}^{2}=\Id$
implies $\cosh (2\a (\overline{H}))=1$, leading to
\begin{equation}
e^{\a (\overline{H})}=\pm 1,\text{ \ \ }\forall \a \in \Delta _{+}
\text{ .}  \label{condexp}
\end{equation}
In particular, this should hold for the simple roots: 
$e^{\a _{i}(\overline{H})}=\pm 1,$ $\forall \a _{i}\in \Pi $. Then, how
one assigns the $\pm $-signs to the simple roots completely determines the action 
of $\vartheta $ on the whole root lattice (\ref{condexp}). For a $r$-rank
algebra, there are then $2^{r}$ inner involutive automorphisms, but in
general far less non-isomorphic real forms of $\g$.

We are now ready to implement the procedure (\ref{sqr2}), first by splitting
the positive root system $\Delta_{+}$ into two subsets%
\begin{equation}
\Delta_{(\pm 1)}\doteq \left\{ \a \in \Delta _{+}\text{ }|
\text{ }e^{\a (\overline{H})}=\pm 1\right\}  \label{SpacePlusMin}
\end{equation}
and then by acting with the linear operator (\ref{sqr}) in its matrix
realisation (\ref{automorf}) on the base of $\g_{c}$. 
Then, the eigenspaces with eigenvalue $(\pm 1)$ can be shown to be spanned by 
\begin{equation}
\k=\text{Span}\left\{iH_{\a _{j}}\text{, }\forall \a_{j}\in \Pi \text{; }
(E_{\a}-F_{\a})\text{ and }i(E_{\a}+F_{\a }),\text{ }\forall \a \in \Delta _{(+1)}\right\}
\label{kset}
\end{equation}
and
\begin{equation}
\p=\text{Span}\left\{ i(E_{\a }-F_{\a })\text{, }
(E_{\a}+F_{\a}),\text{ }\forall \a \in \Delta _{(-1)}\right\} 
\text{ .}  \label{pset}
\end{equation}
In this approach, the signature determining all equivalence classes of
involutive automorphisms (\ref{condexp}) takes the handy form
\begin{equation}\label{signatForm}
-\sigma =\text{Tr}\boldsymbol{\vartheta}=\big(r+2\sum_{\a\in\Delta^{+}}
\cosh (\a(\overline{H}))\big)=r+2(\dim \Delta_{(+1)}-\dim \Delta_{(-1)})\text{ .}
\end{equation}

\subsection{A matrix formulation of involutive automorphisms of affine KMAs}

\label{Matrx}

This analysis can be extended to real forms of affine extension of Lie
algebras. The general method based on a matrix reformulation of the
involutive automorphism has been developed in \cite{Corn92a} and
successfully applied to the $\widehat{A}_{r}$, $\widehat{B}_{r}$, 
$\widehat{C}_{r}$ and $\widehat{D}_{r}$ cases 
in~\cite{Corn92b,Corn92c,ClarkCorn1,ClarkCorn2,ClarkCorn3}. Here, we will
only present the very basics of the method, and refer the reader to these
articles for more details.

There are two ways of handling involutive automorphisms of untwisted affine
Lie algebras. The first (classical) one is based on the study of 
Cartan-preserving automorphisms. Since every conjugacy class of the automorphism
group contains at least one such automorphism, one can by this means arrive
at a first classification of the involutive automorphisms of a given affine
KMA. This procedure would be enough for determining all real forms of a
finite Lie algebra, but would usually overcount them for affine KMA, because
in this case some Cartan-preserving automorphisms can be conjugate via
non-Cartan-preserving ones within Aut$(\widehat{\g})$. This
will obviously reduce the number of conjugacy classes and by the same token
the number of real forms of an untwisted affine KMA. A matrix formulation of 
automorphisms has been proposed in~\cite{Corn92a} precisely to treat these cases.

The first method takes advantage of the fact that Cartan-preserving
automorphisms can be translated into automorphisms of the root system that leave
the root structure of $\widehat{\g}$ invariant. Let us call $\phi$
such an automorphism acting on $\Delta(\widehat\g)$. It can be
constructed from an automorphism $\phi_{0}$ acting on the basis of simple
roots $\Pi(\g)$, for rk$\g=r$, as
$\phi_{0}(\a_{i})=\sum_{j=1}^{r}(\phi _{0})_{i}^{\phantom{i}j}\a _{j}$ for $i=1,..,r$.
\footnote{Not to confuse with the Cartan involution acting on the root system, as
given from the Satake diagram. In the finite case, if $\phi_{0}$ is
non-trivial, it typically corresponds to outer automorphisms of the algebra.}
Define the linear functional $\Omega\in P(\widehat{\g})$ such that
\begin{equation}
\phi(\delta)=\mu \delta \,,\qquad \phi (\a_{i})=\phi _{0}(\a_{i})-(\phi _{0}(\a _{i})|
\Omega)\mu\delta \,,\text{ }\forall i=1,..,r\,.  \label{aut1}
\end{equation}
with $(\a_{i}|\Omega)=n_{i}\in \Z$. This automorphism will be root-preserving if 
\begin{equation*}
\mu =\pm 1\,,\quad \text{and}\quad (\phi_{0})_{i}^{\phantom{i}j}\in \Z\,.
\end{equation*}
All root-preserving automorphisms can thus be characterized by the triple 
$\mathcal{D}_{\phi}=\{\phi_{0},\Omega,\mu\}$, with the composition law:
\begin{equation}
\mathcal{D}_{\phi_1}\mathcal{D}_{\phi_2}=\{(\phi_1)_{0}\cdot(\phi_2)_{0},
\mu_2\Omega_1+(\phi_1)_0(\Omega_2),\mu_1\mu_2\}
\end{equation}

The action of $\phi$ lifts to an algebra automorphism $\vartheta_{\phi }$.
The first relation in expression (\ref{aut1}) implies $\vartheta_{\phi}(c)=\mu c$,
while we have:
\begin{equation}
\begin{array}{l}
\vartheta_{\phi}(z^{n}\otimes E_{\a})=C_{\a+n\d}\,
z^{\mu(n-(\phi_{0}(\a_{i})|\Omega))}\otimes E_{\phi_{0}(\a)}\,, \\[3pt]
\vartheta_{\phi}(z^{n}\otimes H_{\a})=C_{n\d}\,z^{\mu n}\otimes
H_{\phi_{0}(\a)}\,,\qquad \vartheta_{\phi}(d)=\mu d+H_{\Omega}-\frac{1}{2}|\Omega |^{2}\mu c\,.
\end{array}
\label{varauto}
\end{equation}
on the rest of the algebra.
By demanding that $\vartheta_{\phi }$ preserves the affine algebra (\ref{affAlg}),
we can derive the relations, for $\a$ and $\b\in \Delta(\g)$: $C_{n\d}C_{m\d}=C_{(m+n)\d}$, 
$C_{\a+n\d}=C_{n\d}C_{\a}$, and 
$\mathcal{N}_{\a,\b}C_{\a+\b}=\mathcal{N}_{\phi_0 (\a),\phi_0 (\b)}C_{\a}C_{\b}$
with $C_0=1$ and $C_{-\a}=C_{\a}^{-1}$.
The condition for $\vartheta_{\phi}$ to be involutive is analogous to the
requirement (\ref{condexp}), namely
\begin{equation*}
e^{\a_{i}(\overline{H})}=\pm 1\,,\qquad \forall i=0,1,..,r\,,
\end{equation*}
where $i=0$ is this time included, and $\overline{H}=\sum c_{i}H_{i}+c_{d}d$, 
with $c_{i},c_{d}\in {\C}$.

In particular, for a Cartan-preserving chief inner automorphism of type 
$\vartheta =e^{\ad(\overline{H})}$, we have:
\begin{equation}\label{Cartpres}
\phi_{0}=\Id\,,\qquad \Omega=0\,,\qquad \mu=1\,,\qquad
C_{\a}=e^{\a(\overline{H})}\,,\text{ }\forall \a \in \Delta (\g)\,.  
\end{equation}
Possible real forms of an untwisted affine KMA are then determined by
studying conjugacy classes of triples $\mathcal{D}_{\phi}$, for various
involutive automorphisms $\phi$. However, from the general structure~(\ref{Cartpres}),
we see that a chief inner automorphism cannot be conjugate through a 
Cartan-preserving automorphism to an automorphism associated with
a Weyl reflection, for instance. They could, however, be conjugate under
some more general automorphism (note that this could not happen in the 
finite context). The above method might thus lead to overcounting the 
number of equivalence classes of automorphisms, and consequently, of real 
forms of an affine Lie algebra.

This problem has been solved by a newer approach due to Cornwell, which is
based on a matrix reformulation of the set of automorphisms for a given
affine KMA. Choosing a faithfull $d_{\Gamma}$-dimensional representation of 
$\g$ denoted by $\Gamma$, we can represent any element of $\mathcal{L}(\g)$ 
by $A(z)=\sum_{b=1}^{r}\sum_{n=-\infty}^{\infty}
a_{n}^{\phantom{n}b}\,z^{n}\otimes \Gamma (X_{b})$, for $X_{b}\in \mathfrak{g}$. 
Then any element of $\widehat\g$ may be written as: 
\begin{equation*}
\widehat{A}(z)=A(z)+\mu _{c}c+\mu _{d}d
\end{equation*}
where the $+$ are clearly not to be taken as matrix additions.

It has been pointed out in~\cite{Corn92a} that all automorphisms of complex 
untwisted KMAs are classified in this matrix formulation according to four
types, christened: type 1a, type 1b, type 2a and type 2b.

A type 1a automorphism will act on $A(z)$ through an invertible 
$d_{\Gamma}\times d_{\Gamma}$ matrix $U(z)$ with components given by Laurent
polynomials in $z$:
\begin{equation}
\varphi (A(z))=U(z)A(uz)U(z)^{-1}+\frac{1}{\c_{\Gamma}}\oint \frac{dz}{2\pi iz}
\Tr\left[\left(\frac{d}{dz}\ln U(z)\right) A(uz)\right] c\,,
\label{InvolU}
\end{equation}
where $\c_{\Gamma}$ is the Dynkin index of the representation, and 
$u\in \C^{\ast}$ (this parameter corresponds, in the preceding
formulation, to a Cartan preserving automorphism of type $\vartheta =e^{\ad(d)}$).
The remaining three automorphisms are defined as above, by replacing 
$A(uz)\rightarrow \{-\widetilde{A}(uz);A(uz^{-1});-\widetilde{A}(uz^{-1})\}$ 
on the RHS of expression~(\ref{InvolU}) for, respectively, type
\{1b;2a;2b\} automorphisms.

Here the tilde denotes the contragredient representation $-\widetilde{\Gamma}$. 
The action on $c$ and $d$ is the same for all four automorphisms, namely:
\begin{eqnarray*}
\varphi (c) &=&\mu c\,, \\
\varphi (d) &=&\mu \Phi (U(z))+\lambda c+\mu d\,,
\end{eqnarray*}
with $\mu =1$ for type 1a and 1b, and $\mu =-1$ for type 2a and 2b, and the
matrix: 
\begin{equation}\label{PhiU}
\Phi (U(z))=-z\frac{d}{dz}\ln U(z)+\frac{1}{d_{\Gamma}}\Tr\left( 
z\frac{d}{dz}\ln U(z)\right) \,\Id\,.
\end{equation}
An automorphism $\varphi$ can then be encoded in the triple:
$\mathcal{D}_{\varphi}=\{U(z),u,\lambda \}$, and, as before, conjugation classes of
automorphisms can be determined by studying equivalence classes of triples 
$\mathcal{D}_{\varphi}$. In this case, the more general structure of the
matrix $U(z)$ as compared to $\phi_{0}$, which acts directly on the
generators of $\widehat{\g}$ in a given representation, allows conjugation 
of two Cartan-preserving automorphisms via both Cartan-preserving and 
non-Cartan-preserving ones.

Finally, the conditions for $\varphi$ to be involutive are, for type 1a:
\begin{equation}
u^{2}=1\,,  \label{invcond1}
\end{equation}
and 
\begin{equation}  \label{invcond2}
\begin{array}{rcl}
{\displaystyle U(z)U(uz)} \!\! & \!\!=\!\! & \!\!{\displaystyle \zeta z^{k}
\Id\,,\quad\text{ with }k\in \N\text{ and }\zeta \in {\C} }\,, \\[3pt] 
{\displaystyle \lambda}\!\! & \!\!=\!\! & \!\!
{\displaystyle -\frac{1}{2\c_{\Gamma}}\oint \frac{dz}{2\pi iz}
\Tr\left[ \left( \frac{d}{dz}\ln U(z)\right) \Phi (U(uz))\right] \,. }
\end{array}
\end{equation}
For a type 1b automorphism, the first condition~(\ref{invcond1}) remains the
same, while we have to replace $U(uz)\rightarrow \widetilde{U}(uz)^{-1}$ and 
$\Phi(U(uz))\rightarrow-\widetilde{\Phi}(U(uz))$ in the two last
conditions~(\ref{invcond2}).

Involutive automorphisms of type 2a and 2b are qualitatively different since
they are already involutive for any value of $u$ (so  that condition~(\ref{invcond1}) 
can be dropped), provided the last two conditions~(\ref{invcond2}) are met, 
with the substitutions $U(uz)\rightarrow U(uz^{-1})$ in the first and 
$\Phi (U(uz))\rightarrow \Phi (U(uz^{-1}))$ in the second one for type
2a, and $U(uz)\rightarrow \widetilde{U}(uz^{-1})^{-1}$ in the first and 
$\Phi (U(uz))\rightarrow -\Phi (\widetilde{U}(uz^{-1}))$ in the second one
for type 2b. In both cases, we are free to set $u=1$.

When studying one particular class of involutive automorphisms, one will
usually combine both the method based on root-preserving automorphisms and
the one using the more elaborate matrix formulation to get a clearer picture
of the resulting real form.

\subsection{The non-split real invariant subalgebra in $D=2$}
\label{AffineT4}

The affine extension in $D=2$ yields a real form of $\hat{\dn}_{7}\Join \hat{\u}(1)$. 
We will show that this real form, obtained from projecting from $\e_{9|10}$ 
all charged states, builds a $\widehat{\so}(8,6)\Join \hat u_{|1} (1)$, where, 
by $\widehat{\so}(8,6)$, we mean the affine real form described by the $D=2$ 
Satake diagram of Table~\ref{T4/ZnOrbifolds} as determined in \cite{TripPati}. 
The proof requires working in a basis of $\ginv$ in which the Cartan subalgebra 
is chosen compact. It will be shown that such a basis can indeed be constructed 
from the restriction $(\ginv)^{\C}\cap \e_{9|10}$. Then, by determining the action 
of $\phi$ on the latter, we will establish that, following \cite{Corn92a},
the vertex operator ({\it or} Sugawara) construction of $\ginv$  reproduces
exactly the Cartan decomposition of $\widehat{\so}(8,6)$ expected
from \cite{ClarkCorn2}. Finally, we will show how the reality properties of 
$\hat{\dn}_{7}$, entail, through the affine central product, those of the 
$\hat{\u}(1)$ factor. 

Concentrating first on $\hat{\dn}_{7}$, we follow for a start the matrix method
outlined in the preceding Section \ref{Matrx}. In this case, the
automorphism~(\ref{InvolU}) restricted to the transformation 
$A(z)\rightarrow U(z)A(z)U(z)^{-1}$ has to preserve the defining condition:
$$
A(z)^{\intercal}G+GA(z)=0\,,
$$
where $G$ is the metric kept invariant by $SO(14)$ matrices in the rep $\Gamma$. 

We start by choosing, for $\dn_7\subset \hat{\dn}_7$, the $14$-dimensional
representation given in Appendix~\ref{AppendixC} with Dynkin index $\c_{\Gamma}=1/\sqrt{42}$, 
whose generators will be denoted $\Gamma(E_{\a})$ and $\Gamma(H_{i})$. The affine 
extension of these operators is obtained as usual by the Sugawara construction, 
and the involutive automorphism $U(z)$ will be represented by a $14\times 14$ 
matrix. This representation  $\Gamma$ is in fact equivalent to its contragredient one 
$-\widetilde{\Gamma}$ in the sense that one can find a $14\times 14$ non-singular 
matrix $C$ such that:
$$
\Gamma(X)=-C \,\widetilde{\Gamma}(X)\, C^{-1}\,, \;\; \forall X\in \dn_7\,.
$$
One readily sees from eqn.~(\ref{InvolU}) and subsequent arguments that, in this case,
type 1b and 2b automorphisms coincide respectively with type 1a and 2a, which leaves us, 
for $\hat{\dn}_{7}$, with just two classes of involutive automorphisms,
characterizing, roughly, real forms where the central charge $c$ and the
scaling operator $d$ are both compact or both non-compact.

Since we do not expect the restriction $\phi$ of the Chevalley involution to $\ginv$ 
to mix levels in $\d$ in this case, this in principle rules out all involutive automorphisms 
of type 2a, which explicitly depend on $z$. In turn, it tells us that the central charge 
and the scaling operator are now both compact in $\ginv$, contrary to, for instance, 
the $T^{7}\times T^{2}/\Z_{n>2}$ case analyzed in Section~\ref{SecT2}, and will be written 
$ic'\doteq iH_{\d'_{D_{7}}}=ic$ and $id'$. Neither is the involution $\phi$ likely to involve 
different compactness properties for even and odd levels in $\d$. These considerations lead us
to select $u=+1$. The $z$-independent automorphism of type 1a with $u=+1$ which seems
to be a good candidate, in the sense that it reduces to $\so(8,6)$ when we
restrict to the finite Lie algebra ${\dn}_{7}\subset \hat{\dn}_{7}$, is
\begin{equation}
U(z)=\Id_{4}\oplus (-\Id_{6})\oplus \Id_{4}\,,  \label{U}
\end{equation}
so that eq.(\ref{InvolU}) reduces to $\varphi(A(z))= U(z)A(z)U(z)^{-1}$.

Obviously, we have $\Phi(U(z))=0$ from expression~(\ref{PhiU}) and the condition~(\ref{invcond2}) 
for the automorphism to be involutive determines $\lambda=0$. Now, since both central charge 
and scaling operator are compact in the new primed basis, we have $\mu=1$. All these considerations 
put together lead to:
\begin{equation}
\label{phiCD}
\varphi(ic')=ic'\,, \qquad \varphi(id')=id'\,,
\end{equation}
from which we can determine the two triples:
$$
\mathcal{D}_{\varphi}=\left\{\Id_{4}\oplus (-\Id_{6})\oplus \Id_{4};+1;0\right\}\, \leftrightarrow \,
\mathcal{D}_{\phi}=\left\{\Id_{7};0;+1\right\}\,,
$$
the structure of $\mathcal{D}_{\phi}$ clearly showing that we are dealing with a chief inner involutive 
automorphism. A natural choice for the primed basis of the Cartan subalgebra of $\dn_7$ is to pick
it compact, so that its affine extension 
$\tilde{\h}=\{iH'_{1},\ldots ,iH'_{7},ic',id'\}$ is compact, as well.

We will check that the real form of $\hat{\dn}_{7}$ generated by
the automorphism (\ref{U}) and the one determined by the Cartan involution
$\phi$ are conjugate, and thus
lead to isomorphic real forms. Let us, for a start, redefine the basis of
simple roots of $\hat{\dn}_{7}\subset \ginv$ appearing in Table \ref{T4/ZnOrbifolds}:
\begin{equation}
\b_{1}\equiv \a_{-}\,,\quad \b_{2}\equiv \tilde{\a}\,,\quad
\b_{3}\equiv \a_{+}\,,\quad \b_{4}\equiv \a_{3}\,,\quad
\b_{5}\equiv \a_{2}\,,\quad \b_{6}\equiv \a_{1}\,,\quad
\b_{7}\equiv \c \quad
\b_{0}\equiv \a_0\,.  \label{convRoots}
\end{equation}
The lexicographical order we have chosen ensures that the convention for the
structure constants is a natural extension of the $D=6$ case. As for
$\e_{9}$, we introduce an abbreviated notation 
$E_{\b_{6}+2\b_{5}+2\b_{4}+\b_{1}+2\b_{2}+\b_{3}}\doteq E_{\underline{65^{2}4^{2}12^{2}3}}$. 
Conventions and a method for computing relevant structure constants are given in 
Appendix~\ref{AppendixC}. 

We can now construct the compact Cartan subalgebra $\tilde{\h}$ by selecting
combinations of elements of ${\dn}_{7}\subset \hat{\dn}_{7}\subset \ginv$ 
which commute and are themselves combinations of compact generators of $\e_{9|10}$. 
The Cartan generators in this new basis are listed below, both in terms of 
$\dn_{7}\subset (\ginv)^{\C}$ and $\e_{8}\subset \g^{U}$ generators: 
\begin{eqnarray}
iH'_{\underline{1}} &=&i(E_{\underline{2}}+F_{\underline{2}})\equiv
E_{458}-F_{458}\,,  \notag \\
iH'_{\underline{2}}&=&
\frac{i}{2}\big(\eta_{1}(H_{\underline{3}}-H_{\underline{1}})-E_{\underline{2}}-
F_{\underline{2}}-E_{\underline{123}}-F_{\underline{123}}\big)  \notag \\
&\equiv &\frac{1}{2}\big(\eta_{1}(E_{5}-F_{5}+E_{7}-F_{7})-E_{458}+F_{458}-
E_{45^{2}6^{2}78}+F_{45^{2}6^{2}78}\big)\,,  \notag \\
iH'_{\underline{3}}&=&i(E_{\underline{123}}+F_{\underline{123}})\equiv
E_{45^{2}6^{2}78}-F_{45^{2}6^{2}78}\,,  \notag \\
iH'_{\underline{4}}&=&
\frac{1}{2}\big(\eta_{2}(E_{\underline{54^{2}12^{2}3}}-F_{\underline{54^{2}12^{2}3}})-
i\eta_{1}(H_{\underline{3}}-H_{\underline{1}})-E_{\underline{5}}+F_{\underline{5}}\big)\notag \\
&\equiv &-\frac{1}{2}\big(\eta_{2}(E_{234^{2}5^{3}6^{2}78^{2}}-F_{234^{2}5^{3}6^{2}78^{2}})+
\eta_{1}(E_{5}-F_{5}+E_{7}-F_{7})+E_{2}-F_{2}\big)\,,  \label{CartD} \\
iH'_{\underline{5}}&=&E_{\underline{5}}-F_{\underline{5}}\equiv E_{2}-F_{2}\,,
\notag \\
iH'_{\underline{6}}&=&
\frac{1}{2}\big(\eta_{3}(E_{\underline{76^{2}5^{2}4^{2}12^{2}3}}-
F_{\underline{76^{2}5^{2}4^{2}12^{2}3}})-E_{\underline{7}}+F_{\underline{7}}-E_{\underline{5}}+
F_{\underline{5}}-\eta_{2}(E_{\underline{54^{2}12^{2}3}}-F_{\underline{54^{2}12^{2}3}})\big) 
\notag \\
&\equiv &-\frac{1}{2}
\big(\eta _{3}(E_{\theta_{E_{8}}}-F_{\theta_{E_{8}}})+E_{\c }-F_{\c }+E_{2}-F_{2}-
\eta_{2}(E_{234^{2}5^{3}6^{2}78^{2}}-F_{234^{2}5^{3}6^{2}78^{2}})\big)\,,  \notag
\\
iH'_{\underline{7}}&=&E_{\underline{7}}-F_{\underline{7}}\equiv E_{\c}-F_{\c}\,,  \notag
\end{eqnarray}
where the factors $\eta_{i}=\pm 1$, $\forall i=1,2,3$, determine equivalent
solutions. 

The Cartan generator attached to the affine root $\b_0'$ is constructed from
the above (\ref{CartD}) in the usual way:
\begin{eqnarray*}
iH'_{\underline{0}}&=&iH_{\delta'_{D_{7}}}-\eta _{3}(E_{\underline{76^{2}5^{2}4^{2}12^{2}3}}-
F_{\underline{76^{2}5^{2}4^{2}12^{2}3}}) \\
&=&ic' +\eta_{3}(E_{\theta_{E_{8}}}-F_{\theta_{E_{8}}})\,,
\end{eqnarray*}
which commutes with $\tilde{\h}$ (\ref{CartD}) and is indeed
compact, as expected from expression (\ref{phiCD}).

We find the associated ladder operators by solving the set of equations
$[H'_{\underline{j}},E'_{\underline{i}}]=A_{\underline{ij}}
E'_{\underline{i}}$, $[E'_{\underline{i}},F'_{\underline{j}}]=\d_{\underline{ij}}H'_{\underline{i}}$ 
and $[E'_{\underline{i}},E'_{\underline{j}}]=\mathcal{N}_{\underline{i},\underline{j}}E'_{\underline{ij}}$ 
(the corresponding commutation relations for the lowering operators are then automatically satisfied). 
Here we write $E'_{\underline{i}}\equiv E_{\b'_{i}}$ and 
$F'_{\underline{i}}\equiv E_{-\b'_{i}}$ for short, for the set 
$\Pi'=\{\b'_{0},\ldots,\b'_{7}\}$ of simple roots dual to the Cartan basis~(\ref{CartD}). Thus: 
\begin{equation}  
\label{base1D7}
\begin{array}{rcl}
{\ds E'_{\underline{1}}/F'_{\underline{1}}}&=& {\ds H_{\underline{2}}\mp 
(E_{\underline{2}}-F_{\underline{2}})\equiv H_{458}\pm i(E_{458}+F_{458})\,,}  \\[4pt]
{\ds E'_{\underline{2}}}&=&{\ds E_{\underline{1}}-F_{\underline{3}}-
\eta_{1}(F_{\underline{12}}-F_{\underline{23}})\equiv E_{\a_{-}}-F_{\a_{+}}-
\eta_{1}(E_{\a_{-}+\widetilde{\a}}-F_{\widetilde{\a}+\a_{+}})\,, }  \\[4pt]
{\ds E'_{\underline{3}}/F'_{\underline{3}}} &=&{\ds H_{\underline{123}}\mp 
(E_{\underline{123}}-F_{\underline{123}})\equiv H_{45^{2}6^{2}78}\pm
i(E_{45^{2}6^{2}78}+F_{45^{2}6^{2}78})\,, }\\[4pt]
{\ds E'_{\underline{4}}} &=& {\ds E_{\underline{4(23\leftrightarrow 12)}}+
iE_{\underline{54(23\leftrightarrow 12)}}-\eta_{2}(F_{\underline{4(12\leftrightarrow 23)}}
+iF_{\underline{54(12\leftrightarrow 23)}})\,,\quad\text{for }\eta_{1}=
\pm 1 \,, } \\[4pt]
{\ds E'_{\underline{5}}/F'_{\underline{5}}}&=&{\ds H_{\underline{5}}\pm 
i(E_{\underline{5}}+F_{\underline{5}})\equiv H_{2}\pm i(E_{2}+F_{2})\,, } 
\end{array}
\end{equation}
together with $F'_{i}=(E'_{i})^{\dag}$, for $i=2,4$. In the expression in~(\ref{base1D7}) for
$E'_{\underline{4}}$ , the $\leftrightarrow$ gives the two possible values of the last two indices
depending on the choice of $\eta_1=\pm 1$. It can be checked that 
$[E'_{\underline{2}},E'_{\underline{4}}]=0$.

The raising operator $E'_6$ is independent of $\eta_{1}$ and takes the form: 
\begin{eqnarray}
E'_{\underline{6}}&=&
(E_{\underline{6}}+iE_{\underline{65}}-iE_{\underline{76}}+E_{\underline{765}})+
\eta_{2}\eta_{3}(F_{\underline{6}}+iF_{\underline{65}}-iF_{\underline{76}}+F_{\underline{765}})\notag \\
&&-\eta_{2}(iE_{\underline{654^{2}12^{2}3}}+E_{\underline{65^{2}4^{2}12^{2}3}}+
E_{\underline{7654^{2}12^{2}3}}-iE_{\underline{765^{2}4^{2}12^{2}3}})\label{base2D7} \\
&&-\eta _{3}(iF_{\underline{654^{2}12^{2}3}}+F_{\underline{65^{2}4^{2}12^{2}3}}+
F_{\underline{7654^{2}12^{2}3}}-iF_{\underline{765^{2}4^{2}12^{2}3}})\,, 
\notag
\end{eqnarray}
the corresponding lowering operator is obtained from the above by hermitian
conjugation. Moreover, it can be verified after some tedious algebra
that indeed $[E'_{\underline{4}},E'_{\underline{6}}]=0$. Note that we have
translated the primed generators into $\e_{8}$ ones only when the
expression is not too lengthy. Hereafter, such substitutions will be made
only when necessary.

The two remaining pairs of ladder operator enhancing $\dn_6$ to $\hat{\dn}_7$ are:
\begin{equation}  
\label{base1D7fin}
\begin{array}{rcl}
{\ds E'_{\underline{7}}/F'_{\underline{7}}} 
 &=&{\ds H_{\underline{7}}\pm i(E_{\underline{7}}+F_{\underline{7}})\equiv 
 H_{\c}\pm i(E_{\c}+F_{\c})\,, } \\[4pt]
{\ds E'_{\underline{0}}/F'_{\underline{0}}} &=&
{\ds -\eta_3\,t^{\pm 1}\otimes \big( H_{\underline{\theta}_{D_7}}\mp i
(E_{\underline{\theta}_{D_7}}+F_{\underline{\theta}_{D_7}})\big)
}\\[4pt]
&\equiv& {\ds \eta_3\,t^{\pm 1}\otimes \big( H_{\theta_{E_8}}\mp i
(E_{\theta_{E_8}}+F_{\theta_{E_8}})\big)\,,
}
\end{array}
\end{equation}
where $\underline{\theta}_{D_7}\doteq \b_7+2(\b_6+\b_5+\b_4+\b_2)+\b_1+\b_3$.

At this stage it is worth pointing out that the affine real form $\ginv$ is realized as
usual as a central extension of the loop algebra of the finite $(\dn_7)_{0}$
which may or may not descend to a real form of $\dn_7$ (in our case, it does since 
it will be shown that $(\dn_7)_{0}=\dn_{7|5}$)
\begin{equation} \notag
\ginv / \mathcal{L}(\mathfrak{u}(1))=\R[t,t^{-1}]\otimes (\dn_7)_{0}\oplus \R ic' \oplus \R id'\,.
\end{equation}
The difference is that we are now tensoring with an algebra of Laurent polynomials 
$\mathcal{L}=\R[t,t^{-1}]$ in the (indeterminate) variable $t$ defined as follows
\begin{eqnarray} \label{def-t}
t=\frac{1}{2}\big((1-i)+(1+i)\vartheta_C\big)z\equiv 
\frac{1}{1+i}\big(1+i\vartheta_C\big)z\,.
\end{eqnarray}
The second term of the equality~(\ref{def-t}) is clearly reminescent from
the operator $\sqrt{\vartheta}$~(\ref{sqr}). The inverse transformation yields:
$$
z\,=\,\frac{(1+i)t+\sqrt{2i(t^2-2)}}{2}\,, \qquad 
z^{-1}\equiv\bar z\,=\,\frac{(1+i)t-\sqrt{2i(t^2-2)}}{2i}\,.
$$ 
On can check that under the Chevalley involution: $\vartheta_C(t)=t$ and 
$\vartheta_C(t^{-1})=t^{-1}$.
Moreover, using 
$$
t^n=\frac{1}{(1-i)^n z^n} \sum_{k=0}^n \binom{n}{k}(-iz^2)^{k}
$$
one can check that $\vartheta_C(t^n)=t^n$ $\forall n\in\Z^*$,
as required by the affine extension of the basis (\ref{CartD}-\ref{base1D7fin}),
which will become clearer when we give the complete realization
of the real $\ginv/\mathcal{L}(u(1))$ (\ref{kD7+}-\ref{pD7+}).

Finally, we may now give the expression of the compact scaling operator in 
the primed basis:
$$
id'=\frac{(1 +i\vartheta_C)z}{(1-i\vartheta_C)z}\,id\,, 
$$
which can be shown to be Hermitian.

Now that we know the structure of the generators $E'_{\underline{i}}$ and 
$F'_{\underline{i}}$, $i=0,\ldots,7$, we are in the position of determining 
Fix$_{\tau_0}(\hat{\dn}_{7})$ and, by acting with $\phi$ on the latter, 
are able to reconstruct the eigenvalues of the representation 
$\phi=\Ad(e^{\overline{H}})$ of the Cartan involution on the basis
(\ref{kset})-(\ref{pset}), namely 
$\phi\cdot(\hat{\dn}_7)_{\b'}=e^{\b'(\overline{H})}\,(\hat{\dn}_7)_{\b'}$, with
$e^{\b'_i(\overline{H})}=\pm 1$, $\forall \b_i'\in\Pi'$. 
We will then show that the four automorphisms determined by this method
corresponding to all possible values of $\eta_{i}$, $i=1,2,3$, are conjugate 
to the action of the $U(z)$ given in~(\ref{U}) on the representation $\Gamma$ 
of $\dn_7$.

Reexpressing, for instance, the second line of the list~(\ref{base1D7}) in
terms of the original basis~(\ref{A3GeneratorsT4/Zn}) and~(\ref{A3CommRel}), 
and taking Fix$_{\tau_0}(\hat{\dn}_{7})$ yields the two following
generators of $\ginv$:
\begin{equation}  \label{compact}
\begin{array}{l}
{\ds \frac{1}{2}(E'_{\underline{2}}-F'_{\underline{2}})=
\frac{1}{2}\Big(E_{56}-F_{56}-E_{67}+F_{67}-
\eta_{1}\big(E_{45^{2}6^{2}78}-F_{45^{2}6^{2}78}-E_{4568}+F_{4568}\big)\Big)\,,  }
\\[3pt]
{\ds \frac{i}{2}(E'_{\underline{2}}+F'_{\underline{2}})=
\frac{1}{2}\Big(E_{567}-F_{567}+E_{6}-F_{6}-
\eta_{1}\big(E_{45678}-F_{45678}+E_{45^{2}678}-F_{45^{2}68}\big)\Big)\,.} 
\end{array}
\end{equation}
Both are obviously invariant under $\phi$, since they are linear
combinations of compact generators. According to Section~\ref{condexp}, we have 
$e^{\b'_{2}(\overline{H})}=+1$. The same reasoning applies to the pairs 
$E'_{\underline{4}}/F'_{\underline{4}}$ and $E'_{\underline{6}}/F'_{\underline{6}}$.
In contrast to the $E'_{\underline{2}}/F'_{\underline{2}}$ case,
these two couples of generators will be alternatively compact or non-compact depending on the sign 
of $\eta_{2}$ and $\eta_{3}$. In particular, since $E'_{\underline{4}}$
has basic structure $[E_{\a},E_{\a_{\pm}}]-\eta_2[F_{\a_{\mp}},F_{\a}]$, 
the choice $\eta_{2}=+1$ will produce the two compact combinations
$2^{-1}(E'_{\underline{4}}-F'_{\underline{4}})$ and $2^{-1}i(E'_{\underline{4}}+F'_{\underline{4}})$,
while the opposite choice selects the two non-compact ones, by flipping the reciprocal sign between 
$E$ and $F$. From expression (\ref{base2D7}), we see that the $E'_{\underline{6}}/F'_{\underline{6}}$ 
case is even more straightforward, compactness and non-compactness being selected by 
$\eta_{2}\eta_{3}=\pm 1$ respectively. At this stage, our analysis thus leads to the four possibilities: 
$e^{\b'_{4}(\overline{H})}=\pm 1$ and $e^{\b'_{6}(\overline{H})}=\pm 1$.

Finally, the remaining ladder operators $E'_{\underline i}$ and $F'_{\underline i}$ for $i=0,1,3,5,7$
 combine in purely non-compact expressions, for instance
\begin{equation}  \label{ncompact}
\frac{1}{2}(E'_{\underline{1}}+F'_{\underline{1}})=H_{458}\,,\qquad \frac{i}{2}%
(E'_{\underline{1}}-F'_{\underline{1}})=-(E_{458}+F_{458})\,,
\end{equation}
The $E'_{\underline{0}}/F'_{\underline{0}}$ case is a bit more subtle because of the
presence of the $(t,t^{-1})$ loop factors, and requires adding 
$E_{\theta_{D_7}+\d}/F_{\theta_{D_7}+\d}=\eta_3\,t^{\pm 1}\otimes \big(H_{\theta_{E_8}}
\pm i (E_{\theta_{E_8}}+F_{\theta_{E_8}})\big)$ into the game. Computing Fix$_{\tau_0}$ for 
all of these four operators results in four non-compact combinations. This is in accordance with 
$\theta'_{D_7}$ which we now know to statisfy $e^{\theta'_{D_7}(\overline{H})}=-1$ for all four 
involutive automorphisms, and tells us in addition that: $e^{\b'_{0}(\overline{H})}=-1$.

Collecting all previous results, the eigenvalues of the four involutive automorphisms \\$\phi_{(\eta_2,\eta_3)}=~\text{Ad}(e^{\overline{H}_{(\eta_2,\eta_3)}})$ are summarized 
in the table (\ref{tablAuto}) below.
\begin{equation}
\begin{array}{|c|c||c|c|c|c|}
\hline
\eta _{2} & \eta _{3} & e^{\b'_{2}(\overline{H})} & e^{\b'_{4}(\overline{H})} & 
e^{\b'_{6}(\overline{H})} & 
e^{\b'_{i\neq 2,4,6}(\overline{H})} \\ \hline
+1 & -1 & +1 & +1 & +1 & -1 \\ 
+1 & +1 & +1 & +1 & -1 & -1 \\ 
-1 & +1 & +1 & -1 & +1 & -1 \\ 
-1 & -1 & +1 & -1 & -1 & -1 \\ \hline
\end{array}
\label{tablAuto}
\end{equation}
The Cartan element $\overline{H}$ defining the involution $\phi$ can be read off
table (\ref{tablAuto}).  The most general solution is given by 
$\overline{H}=\pi i\sum_{i=0}^7 c_i H'_{\underline i}+ \pi i c_{d'} d'$ with
\begin{eqnarray*}
c_1=c_3= \frac{\kappa+1}{2}+\Z\,, \quad c_4=\kappa-1\,, \quad c_5=\kappa-\frac{\eta_2-1}{2}+\Z\,, \quad c_6=\kappa+\eta_2-1\,, \\
c_7=\frac{\kappa+\eta_2}{2}+\Z\,, \quad c_0=\frac{\kappa+\eta_2-\eta_3+1}{2}+\Z\,, \quad c_d=\eta_3-1
\hspace{2cm}
\end{eqnarray*}
where $c_2\doteq\kappa\in\C$ is a free parameter.

Restricted to $\dn_{7}$, the four inner automorphisms defined in the table (\ref{tablAuto}) 
are all in the same class of equivalence, and thus determine the same real form, namely $\so(8,6)$ 
as expected from $\ginv$ in $D=3$. In Appendix~\ref{AppendixC}, we have computed the two sets of roots 
$\Delta_{(+1)}$ and $\Delta_{(-1)}$~(\ref{SpacePlusMin}) generating the Cartan decomposition
(\ref{kset})-(\ref{pset}) of the real form. It can be checked that, in these four cases, the signature 
$\sigma |_{\dn_{7}}=~-(7+2(\dim \Delta_{(+1)}-\dim \Delta_{(+1)}))=5$, in accordance with $\so(8,6)$.

The involutive automorphism (\ref{U}) in turn can be shown to split the
root system of $\dn_{7}$ according to
\begin{equation}
\b'=\varepsilon _{i}\pm \varepsilon _{j}\rightarrow \left\{ 
\begin{array}{cc}
\b'\in \Delta_{(+1)} & 1\leqslant i<j\leqslant 4\text{ and }
4<i<j\leqslant 7 \\ 
\b'\in \Delta_{(-1)} & 1\leqslant i\leqslant 4<j\leqslant 7
\end{array}
\right. \,,  \label{splitting}
\end{equation}
which can be verified by computing $U(z)\c (E_{\a })U(z)^{-1}$ for
the representation $\c $ (\ref{Ed7}). We can check that
we have again: $\sigma |_{{\dn}_{7}}=5$, for the splitting 
(\ref{splitting}), since the automorphism~(\ref{U}) corresponds, in our
previous formalism to the involution
$e^{\b'_{4}(\overline{H})}=-1$ and 
$e^{\b'_{i}(\overline{H})}=+1$, $\forall i\neq 4$.

Since they are conjugate at the $\dn_7$ level and all of them preserve 
the central charge and scaling element, the four automorphisms~(\ref{tablAuto})
lift to conjugate automorphisms of $\hat{\dn}_7$. All four of them
are again clearly conjugate to $U(z)$ defined by properties~(\ref{U}),~(\ref{phiCD}) 
and~(\ref{splitting}). These five Cartan preserving inner involutive
automorphisms lead to equivalent Cartan decomposition $\k\oplus ^{\bot}\p$~(\ref{chief})
given by generalizing  the basis~(\ref{compact}) and~(\ref{ncompact})
found previously to the affine case:
\begin{eqnarray}
\k &:&\bullet \;iH'_{\underline{k}}\;(\forall k=1,..,7);\;ic';\;id';\; \label{kD7+}\\
&&\bullet \;\frac{1}{2}(t^{n}-t^{-n})\otimes H'_{\underline{k}}\quad \text{and}
\quad \frac{i}{2}(t^{n}+t^{-n})\otimes H'_{\underline{k}}\qquad
(\forall k=1,..,7;\,n\in \mathbbm{N}^{\ast})\,; \notag\\
&&\bullet \;\frac{1}{2}(t^{n}\otimes E_{\b'}-t^{-n}\otimes
F_{\b'})\quad \text{and}\quad \frac{i}{2}(t^{n}\otimes E_{\b'}+t^{-n}\otimes F_{\b'})\,,
\;n\in \Z\, \notag\\
&&\qquad \qquad (\forall \b'\in \Delta_{(+1)} \text{ defined by }
(\ref{splitting}),\,
(\ref{Delta1}),\, (\ref{Delta2}),\, (\ref{Delta3}) \text{ and } (\ref{Delta4}))
\notag\\
\p &:&\bullet \;\frac{i}{2}(t^{n}\otimes E_{\b'}-t^{-n}\otimes F_{\b'})\quad \text{and}\quad 
\frac{1}{2}(t^{n}\otimes E_{\b'}+t^{-n}\otimes F_{\b'}),\, \;n\in \Z\label{pD7+}\\
&&\qquad \qquad (\b'\in \Delta_{(-1)}=\Delta_+(D_7) \backslash\Delta_{(+1)})\,. \notag
\end{eqnarray}
These decompositions define isomorphic real forms,
which we denote by $\widehat{\so}(8,6)$, encoded in the affine Satake diagram of
Table \ref{T4/ZnOrbifolds} (see for instance \cite{TripPati} for
a classification of untwisted and twisted affine real forms). 

We have checked before the behaviour of the ladder operators of the finite $\dn_7$
subalgebra of $\ginv$. The verification can be performed in a similar manner for
the level $n\geqslant 1$ roots $\b'+n\d'_{D_7}$. Applying for example Fix$_{\tau_0}$ to
the four generators $2^{-1}(t^{\pm n}\otimes E_{\underline{4}}-t^{\mp n}\otimes F_{\underline{4}})$ and
$2^{-1}i(t^{\pm n}\otimes~E_{\underline{4}}+~t^{\mp n}\otimes F_{\underline{4}})$
for example, one obtains the following combinations
$$
\begin{array}{l}
\frac{1}{2}(t^{n}+t^{-n})\otimes \big( E_{345^2 678}-\eta_2 F_{345^2 678}-E_{34568}+\eta_2 F_{34568}\\[3pt]
\hspace{3cm}-E_{2345678}+\eta_2 F_{2345678}-E_{2345^2 68}+\eta_2 F_{2345^2 68}\big)\,, \\[3pt]
\frac{1}{2}(t^{n}-t^{-n})\otimes \big( E_{345678}-\eta_2 F_{345678}+E_{345^2 68}-\eta_2 F_{345^2 68}
\\[3pt]
\hspace{3cm}+E_{2345^2 678}-\eta_2 F_{2345^2 678}-E_{234568}+\eta_2 F_{234568}\big)\,,
\end{array}
$$
which, since now $\vartheta_C(t^n\pm t^{-n})=t^n\pm t^{-n}$, are all either non-compact if $\eta_2=+1$,
or compact otherwise, by virtue of table~(\ref{tablAuto}). This is in accordance with the Cartan 
decomposition~(\ref{kD7+}-\ref{pD7+}). The compactness of the remaining $n\geqslant 1$ ladder operators 
can be checked in similar and straightforward fashion by referring once again to the table~(\ref{tablAuto}).

In contrast to the split case, a naive extension of the signature, which we denote by $\hat{\sigma}$, 
is not well defined since it yields in this case an infinite result:
\begin{equation}
\hat{\sigma}=3+2\times 5\times \infty \,. \label{signD7+}
\end{equation}
In the first finite contribution, we recognize the signature of $\so(8,6)$ together with 
the central charge and scaling element, while the infinite towers of vertex operator 
contribute the second part. As mentioned before in the $D=4$ case, the signature for the finite
$\dn_{7}$ amounts to the difference between compact and non-compact Cartan generators, for the 
following alternative choice of basis for the Cartan algebra 
$\{H_{\c};H_{1};H_{2};H_{3};H_{\tilde{\a}};H_{\a _{+}}+H_{\a_{-}};i(H_{\a _{+}}-H_{\a _{-}})\}$. 
This carries over to the infinite contribution in expression~(\ref{signD7+}), 
where it counts the number of overall compact towers, with an additional factor of 2 coming from 
the presence of both raising and lowering operators.

Care must be taken when defining the real affine central product $\ginv$. The
real Heisenberg algebra
\begin{eqnarray*}
\hat{\mathfrak{u}}(1)_{|1}=\sum_{n=0}^{\infty}\R
(z^{n}+z^{-n})\otimes i\widetilde{H}^{[4]}\,+ \,\sum_{n=1}^{\infty}\R(z^{n}-z^{-n})\otimes
\widetilde{H}^{[4]} + \R c + \R d\,
\end{eqnarray*}
is in this case isomorphic to the one appearing in the $T^2/\Z_{n>2}$ orbifold (\ref{invol}).
Clearly both scaling operators and central charge are, in contrast to $\widehat{\so}(8,6)$
non-compact. The identification required by the affine central product formally
takes place before changing basis in $\widehat{\so}(8,6)$ to the \textit{primed} operators.
The central charge and scaling operators acting on both subspaces of 
$\ginv=\widehat{\so}(8,6)\Join \hat{\u}_{|1}(1)$ are then redefined as $d\oplus d\rightarrow id'\oplus d$ 
and $c\oplus c\rightarrow ic'\oplus c\equiv ic\oplus c$. Then we can write 
$\ginv=\widehat{\so}(8,6)\oplus \hat{\u}_{|1}(1)/\{\z,\bar{d}\}$, with $\z=c-c'$
and $\bar{d}=d-\frac{\sqrt{2i(t^2-2)}}{(1+i)t}d'$.  The signature $\hat{\sigma}$ of $\ginv$ is undefined.
\begin{table}[t!]
\begin{center}
\begin{tabular}{c|c|c|c}
$D$ & $(\Pi_0,\phi)$ & $\ginv$ & $\sigma(\mathfrak{g}_{\text{inv}})$ \\ \hline\hline
$6$ &  
\begin{picture}(260,40) 
\thicklines
\multiput(30,0)(30,0){3}{\circle{6}}
\multiput(33,0)(30,0){2}{\line(1,0){24}} 
\put(60,-7){\psline{<->}(-1,0)(1,0)} 
\put(24,10){$\a_-$}
\put(56,10){$\widetilde{\a}$} 
\put(84,10){$\a_+$} 
\put(110,-3){ $\times$ }
\put(140,-3){$H^{[4]}$} 
\put(160,-3){ $\times$ }
\put(188,-3){$i\widetilde{H}^{[4]}$} 
\end{picture}\hspace{1.5mm} & $
\begin{array}{c}
\so(4,2)\oplus\so(1,1) \\[8pt] 
\oplus \;\u(1)
\end{array}
$ & 1 \\[15pt]
$5$ &  
\begin{picture}(260,40) 
\thicklines
\multiput(30,0)(30,0){2}{\circle{6}}
\put(33,0){\line(1,0){24}}
\put(62,2){\line(4,3){19}} 
\put(62,-2){\line(4,-3){19}}
\multiput(84,-18)(0,36){2}{\circle{6}}
\put(85,0){\psline{<->}(0,.47)(0,-.47)}
\put(26,-13){$\a_3$}
\put(54,-13){$\widetilde{\a}$}
\put(92,15){$\a_{+}$} 
\put(92,-21){$\a_{-}$}
\put(110,-3){ $\times$ } 
\put(140,-3){$H^{[4]}$} 
\put(160,-3){ $\times$ }
\put(188,-3){$i\widetilde{H}^{[4]}$} 
\end{picture}\hspace{1.5mm} & $
\begin{array}{c}
\so(5,3)\oplus\so(1,1) \\[8pt] 
\oplus\;\u(1)
\end{array}
$ & 2 \\[15pt] 
$4$ &  
\begin{picture}(260,40) 
\thicklines
\multiput(15,0)(30,0){3}{\circle{6}} 
\multiput(18,0)(30,0){2}{\line(1,0){24}}
\multiput(99,-18)(0,36){2}{\circle{6}}
 \put(77,2){\line(4,3){19}}
\put(77,-2){\line(4,-3){19}} 
\put(100,0){\psline{<->}(0,.47)(0,-.47)}
\put(150,0){\circle{6}}
\put(11,-13){$\a_2$} 
\put(41,-13){$\a_3$}
\put(69,-13){$\widetilde{\a}$} 
\put(107,15){$\a_{+}$} 
\put(107,-21){$\a_{-}$}
\put(144,-13){$\c$} 
\put(177,-3){ $\times$ }
\put(210,-3){$i\widetilde{H}^{[4]}$}
 \end{picture}\hspace{1.5mm} & $
\begin{array}{c}
\so(6,4)\oplus\sl(2,\R) \\[8pt] 
\oplus\;\u(1)
\end{array}
$ & 3 \\[15pt] 
$3$ &  
\begin{picture}(260,40) 
\thicklines
\multiput(15,0)(30,0){5}{\circle{6}}
\multiput(18,0)(30,0){4}{\line(1,0){24}}
\multiput(159,-18)(0,36){2}{\circle{6}} 
\put(137,2){\line(4,3){19}}
\put(137,-2){\line(4,-3){19}} 
\put(160,0){\psline{<->}(0,.47)(0,-.47)}
\put(11,-13){$\c$} 
\put(41,-13){$\a_1$}
\put(71,-13){$\a_2$}
\put(101,-13){$\a_3$} 
\put(129,-13){$\widetilde{\a}$} 
\put(167,15){$\a_{+}$}
\put(167,-21){$\a_{-}$}
\put(192,-3){ $\times$ }
\put(225,-3){$i\widetilde{H}^{[4]}$} 
\end{picture} 
& $\so(8,6)\oplus\u(1) $ & 4 \\[35pt] 
\hhline{~:==:~}
$2$ & 
\begin{picture}(260,40) 
\thicklines
\put(21,0){\psline{<->}(0,.47)(0,-.47)}
\multiput(21,-18)(0,36){2}{\circle{6}} 
\put(23.5,-16.5){\line(4,3){19}}
\put(23.5,16.5){\line(4,-3){19}} 
\multiput(45,0)(30,0){4}{\circle{6}}
\multiput(48,0)(30,0){3}{\line(1,0){24}}
\put(137,2){\line(4,3){19}} 
\put(137,-2){\line(4,-3){19}}
\multiput(159,-18)(0,36){2}{\circle{6}}
\put(160,0){\psline{<->}(0,.47)(0,-.47)}
\put(41,-13){$\a_1$} 
\put(71,-13){$\a_2$} 
\put(101,-13){$\a_3$}
\put(129,-13){$\widetilde{\a}$} 
\put(167,15){$\a_{+}$} 
\put(167,-21){$\a_{-}$}
\put(4,15){$\c$} 
\put(4,-21){$\a_0$} 
\put(182,-3){ $\times$ }
\put(205,-3){$\left\{i\widetilde{H}^{[4]}_n\right\}_{n\in\Z}$} 
\end{picture}
& $\widehat{\so}(8,6)\oplus \mathcal{L}(\u(1))_{|-1}$ & - \\[25pt] 
$1$ &
\begin{picture}(260,40) 
\thicklines \put(15,20){\circle{9}}
\put(15,20){\circle{6}} 
\put(15,15.5){\line(0,-1){12}}
\put(75,20){\circle{6}} 
\put(75,17){\line(0,-1){14}} 
\put(81,17){$\c$}
\multiput(15,0)(30,0){6}{\circle{6}}
\multiput(18,0)(30,0){5}{\line(1,0){24}}
\multiput(189,-18)(0,36){2}{\circle{6}} 
\put(167.5,2){\line(4,3){18.5}}
\put(167.5,-2){\line(4,-3){18.5}} 
\put(190,0){\psline{<->}(0,.47)(0,-.47)}
\put(22,17){$\xi_{I}$} 
\put(11,-13){$\a_{-1}$} 
\put(41,-13){$\a_0$}
\put(71,-13){$\a_1$} 
\put(101,-13){$\a_2$} 
\put(131,-13){$\a_3$}
\put(159,-13){$\widetilde{\a}$} 
\put(197,15){$\a_{+}$} 
\put(197,-21){$\a_{-}$}
\put(210,-3){ $\times$ } 
\put(235,-3){$i\widetilde{H}^{[4]}$} 
\end{picture}
& $^4 \mathcal{B}_{10(Ib)}\oplus \u(1) $ & - \\[25pt] 
\end{tabular}
\end{center}
\caption{The real subalgebras $\ginv$ for $T^{7-D}\times T^{4}/\Z_{n>2}$ compactifications}
\label{T4/ZnOrbifolds}
\end{table}

\subsection{The non-split real Borcherds symmetry in $D=1$}
\label{HypT4}

The analysis of the $D=1$ invariant subalgebra closely resembles the $T^{8}\times T^{2}/\Z_{n>2}$ 
case. The central product of Section~\ref{AffineT4} is extended to a direct sum of a $\u(1)$ 
factor with the quotient of a Borcherds algebra by an equivalence relation similar to the one 
stated in Conjecture~\ref{extendedT2}. The Borcherds algebra $^4\mathcal{B}_{10}$ found here is
defined by a $10\times 10$ degenerate Cartan matrix of rank $r=9$. Its unique isotropic imaginary simple
root
(of multiplicity one) $\xi_I$ is now attached to the raising operator
$E_{\xi_{I}}=~(1/2)(E_{\d+\a_{5}}-E_{\d-\a_{5}}-E_{\d+\a_{7}}+E_{\d-\a_{7}})$, so that 
the equivalence relation defining $(\ginv)^{\C}$ from $^4\mathcal{B}_{10}\oplus\u(1)$
identifies the Cartan generator $H_I\doteq~H_{\xi_{I}}$ with $H_{\d}$ and removes 
the derivation operator $d_I\doteq d_{\xi_{I}}$.

Moreover, the splitting of multiplicities should occur as in the $T^{8}\times T^{2}/\Z_{n>2}$ 
example, since $\dim\,(\hat{\dn}_{7})_{\d_{D_7}}=\dim\,(\hat{\e}_{7})_{\d}$.
It might a priori seem otherwise from the observation that both 
$\widetilde{K}_{(7)[2\cdots 679\,10]}-\widetilde{K}_{(8)[2\cdots 689\,10]}$ and 
$\widetilde{K}_{(9)[2\cdots 89]}-\widetilde{K}_{(10)[2\cdots 8\,10]}$ are separately invariant. 
However, the combination:
\begin{equation*}
\widetilde{K}_{(7)[2\cdots 679\,10]}-\widetilde{K}_{(8)[2\cdots 689\,10]}+
\widetilde{K}_{(9)[2\cdots 89]}-\widetilde{K}_{(10)[2\cdots 8\,10]}=
\widetilde{K}_{[2\cdots 10]}\otimes i(H_{\a_{+}}-H_{\a_{-}})\in {\dn}_{7}^{\wedge\wedge}
\end{equation*}
contributes to the multiplicity of $\d$, while we may rewrite
\begin{equation*}
\frac{1}{2}(\widetilde{K}_{(7)[2\cdots 679\,10]}-\widetilde{K}_{(8)[2\cdots 689\,10]}-
\widetilde{K}_{(9)[2\cdots 89]}+\widetilde{K}_{(10)[2\cdots 8\,10]})=E_{\xi_{I}}\in \,^4\mathcal{B}_{10}\,,
\end{equation*}
which is the unique raising operator spanning $(^4\mathcal{B}_{10})_{\xi_{I}}$. Thus, though root 
multiplicities remain unchanged, we have to group invariant objects in representations of $\sl(6,\R)$, 
which are naturally shorter than in the $T^{2}/\Z_{n}$ case. We will not detail all such representations 
here, since they can in principle be reconstructed by further decomposition and/or regrouping of the 
results of Table~\ref{Borcherds2}. 

The real invariant subalgebra can again be formally realized as
$$
\ginv=\u(1)\,\oplus\, ^4\mathcal{B}_{10(\small{Ib})}/\{\z,d_I\}
$$
where $\z=H_{\d}-H_I$. We denote by $\mathcal{B}_{10(\small{Ib})}$ the real Borcherds algebra
obtained from Fix$_{\tau_0}(^4\mathcal{B}_{10})$ and represented in Table~\ref{T4/ZnOrbifolds}, 
choosing $Ia$ to refer to the split form. The disappearance of the diagram automorphism which,
in the $D=2$ case, exchanged the affine root $\a_0$ with $\c$ leads to non-compact $H_{-1}$ and $H_{\d}$,
in contrast to what happened with the $\widehat{\so}(8,6)\subset\ginv$ factor in $D=2$. This is reflected
by $\xi_I$ being a white node with no arrow attached to it. Note that a black  isotropic imaginary
simple root connected to a white real simple root would, in any case, be forbidden,
since such a diagram is not given by an involution on the root system. Moreover, an imaginary simple 
root can only be identified by an arrow to another imaginary simple root (and similarly for real 
simple roots).

\section{The orbifolds $T^{6}/\Z_{n > 2}$}
\label{secT6}

The orbifold compactification $T^{5-D}\times T^{6}/\Z_{n}$ for $n\geqslant 3$ can
be carried out similarly. We fix the orbifold action in the directions 
$\{x^{5},x^{6},x^{7},x^{8},x^{9},x^{10}\}$, so that it will only be felt by the set
of simple roots $\{\a_{2},..,\a_{8}\}$ defining the $\e_{7|7}$ subalgebra of 
$\g^{U}=\text{Split}(\e_{11-D})$ from $D=4$ downward. Thus, we may start again by 
constructing the appropriate charged combinations of generators for $\g^{U}=\e_{7|7}$, 
and then extend the result for $D\leq 3$ in a straightforward fashion. Since $\e_{7}$ 
has 63 positive roots, we will restrict ourselves to the invariant subalgebra, and list
only a few noteworthy charged combinations of generators. In this case, a new
feature appears: the invariant algebra is not independent of $n$, $\forall n\geqslant 3$,
as before. Instead, the particular cases $T^{6}/\Z_{3}$ and $T^{6}/\Z_{4}$ are non-generic 
and yield invariant subalgebras larger than the $n\geqslant 5$ one.

More precisely, we start by fixing the orbifold action to be 
\begin{equation}
\begin{array}{l}
\qquad (z_{i},\bar{z}_{i})\rightarrow (e^{2\pi i/n}z_{i},e^{-2\pi i/n}\bar{z}_{i}\,)
\text{ \ \ for }i=1,2\,,\qquad (z_{3},\bar{z}_{3})\rightarrow 
(e^{-4\pi i/n}z_{3},e^{4\pi i/n}\bar{z}_{3})\,,
\end{array}
\label{3Complexification}
\end{equation}
in other words, we choose $Q_{1}=+1$, $Q_{2}=+1$ and $Q_{3}=-2$ to ensure $\sum_{i}Q_{i}=0$. 
Note that for values of $n$ that are larger than four, there are other possible choices, 
like $Q_{1}=1$, $Q_{2}=2$ and $Q_{3}=-3$ for $T^{6}/\Z_{6}$ or $Q_{1}=1$, $Q_{2}=3$ and 
$Q_{3}=-4$ for $T^{6}/\Z_{8}$ and so on. Indeed, the richness of $T^{6}/Z_{n}$ orbifolds 
compared to $T^{4}/Z_{n}$ ones stems from these many possibilities. Though interesting 
in their own right, we only treat the first of the above cases in detail, though any choice 
of charges can in principle be worked out with our general method. One has to keep in mind,
however, that any other choice than the one we made in expression~(\ref{3Complexification})
may lead to different non-generic values of $n$.

\subsection{The generic $n\geqslant 5$ case}

Concentrating on the invariant subalgebra $(\ginv)^{\C}$ for $n\geqslant 5$, it turns out 
that the adjoint action of the rotation operator defining the orbifold charges 
${\ds \mathcal{U}_{6}^{\Z_{n}}=\prod_{k=1}^3 e^{-\frac{2\pi i}{n}Q_k\mathcal{K}_{z_k\bar z_k}}}$
leaves invariant the following diagonal components of the metric: 
\begin{equation}
\label{Charge0forT6/Zn}
\begin{array}{rcl}
{\ds K_{44}} \!&\!\!=
\!\!&\!{\ds \frac{1}{2}(3H_{2}+4H_{3}+5H_{4}+6H_{5}+4H_{6}+2H_{7}+3H_{8})\,, }
\\[4pt]
{\ds K_{z_{1}\bar{z}_{1}}}\!&\!\!=\!\!&\!
{\ds \frac{1}{2}(H_{2}+3H_{3}+5H_{4}+6H_{5}+4H_{6}+2H_{7}+3H_{8})\,, }\\[4pt]
{\ds K_{z_{2}\bar{z}_{2}}}\! &\!\!=\!\!&\!
{\ds \frac{1}{2}(H_{2}+2H_{3}+3H_{4}+5H_{5}+4H_{6}+2H_{7}+3H_{8})\,, }  \\[4pt]
{\ds K_{z_{3}\bar{z}_{3}}} \!&\!\!=\!\!&\!
{\ds \frac{1}{2}(H_{2}+2H_{3}+3H_{4}+4H_{5}+2H_{6}+H_{7}+3H_{8})\,,} 
\end{array}
\end{equation}
as well as various fields corresponding to non-zero roots: 
\begin{eqnarray}
\label{ChargeNon0forT6/Zn}
K_{z_{1}\bar{z}_{2}}/K_{\bar{z}_{1}z_{2}}\!\! &\!\!=\!\!&\!\!
\frac{1}{4}\left(E_{34}+F_{34}+E_{45}+F_{45}\pm i(E_{345}+F_{345}-(E_{4}+F_{4}))\right) \,, \notag \\
Z_{4z_{1}\bar{z}_{1}} \!\! &\!\!=\!\!&\!\!
\frac{i}{2}(E_{23^{2}4^{3}5^{3}6^{2}78}+F_{23^{2}4^{3}5^{3}6^{2}78})\,, \notag\\
\quad Z_{4z_{2}\bar{z}_{2}}\!\!&\!\!=\!\!&\!\!\frac{i}{2}(E_{2345^{2}6^{2}78}+F_{2345^{2}6^{2}78})\,,\notag\\
Z_{4z_{3}\bar{z}_{3}} \!\! &\!\!=\!\!&\!\!\frac{i}{2}(E_{23458}+F_{23458})\,, \notag \\
\widetilde{Z}_{z_{1}\bar{z}_{1}z_{2}\bar{z}_{2}z_{3}\bar{z}_{3}} \!\! &\!\!=\!\!&\!\!
-\frac{i}{2}(E_{34^{2}5^{3}6^{2}78^{2}}+F_{34^{2}5^{3}6^{2}78^{2}})\,, \\
Z_{4z_{1}\bar{z}_{2}}/Z_{4\bar{z}_{1}z_{2}} \!\! &\!\!=\!\!&\!\!
\frac{1}{4}\Big(E_{23^{2}4^{2}5^{3}6^{2}78}+F_{23^{2}4^{2}5^{3}6^{2}78}+E_{234^{2}5^{2}6^{2}78}+
F_{234^{2}5^{2}6^{2}78}\notag\\
&&\pm i\big(E_{23^{2}4^{2}5^{2}6^{2}78}+F_{23^{2}4^{2}5^{2}6^{2}78}-(E_{234^{2}5^{3}6^{2}78}+
F_{234^{2}5^{3}6^{2}78})\big)\Big)\,,\notag \\
Z_{z_{1}z_{2}z_{3}}/Z_{\bar{z}_{1}\bar{z}_{2}\bar{z}_{3}} \!\! &\!\!=\!\!&\!\!
\frac{1}{4\sqrt{2}}\Big(E_{345^{2}678}+F_{345^{2}678}-E_{34568}-F_{34568}-E_{45^{2}68}\notag \\
&&-F_{45^{2}68}-E_{45678}-F_{45678}
\pm i\big(-E_{345^{2}68}-F_{345^{2}68}-E_{345678}\notag\\
&&-F_{345678}-E_{45^{2}678}-F_{45^{2}678}+E_{4568}+F_{4568}\big)\Big)\,, \notag
\end{eqnarray}
together with their compact counterparts, supplemented by the
generators of the orbifold action $\mathcal{K}_{z_{1}\bar{z}_{1}}=-i(E_{3}-F_{3})$, 
$\mathcal{K}_{z_{2}\bar{z}_{2}}=-i(E_{5}-F_{5})$ and $\mathcal{K}_{z_{3}\bar{z}_{3}}=-i(E_{7}-F_{7})$, 
which bring the Cartan subalgebra to rank 7, ensuring rank conservation again.

Note that the 4 invariant combinations in the list~(\ref{ChargeNon0forT6/Zn}) are in fact spanned
by the elementary set of linearly independent Cartan elements satisfying $[H,E_{\a}]=0$ for 
$\a\in \{\a_{3},\a_{5},\a_{7}\}$, namely: 
$\{2H_{2}+H_{3};H_{3}+2H_{4}+H_{5};H_{5}+2H_{8};H_{5}+2H_{6}+H_{7}\}$.

Furthermore, let us recall that the objects listed in expression~(\ref{Charge0forT6/Zn}) 
form the minimal set of invariant ladder operators for $n\geqslant 5$ . In the non-generic cases 
$n=3,4$, this set is enhanced, and so is the size of $(\ginv)^{\C}$. We will treat these cases later
on, and, for the moment being, focus on the generic invariant subalgebra for $T^{6}/\Z_{n\geqslant 5}$ 
only.

As before, we extract the generators corresponding to simple roots of the
invariant subalgebra (the negative-root generators are omitted, since they
can be obtained in a straighforward manner as $F_{\a}=(E_{\a})^{\dagger}$): 
\begin{equation}
\label{SimpRootsT6/Zn}
\begin{array}{rcl}
{\ds E_{\tilde{\b}}} \!&\!\!=\!\!&\!{\ds-iE_{2345^{2}6^{2}78}=
\left( Z_{4\bar{z}_{2}z_{2}}\right) ^{+}\,,  } \\[4pt]
{\ds E_{\b_{\pm}}} \!&\!\!=\!\!&\!
{\ds \frac{1}{2}\left( E_{34}+E_{45}\pm i(E_{345}-E_{4})\right) =
(K_{z_{1}\bar{z}_{2}}/K_{\bar{z}_{1}z_{2}})^{+}\,,}
 \\[4pt]
{\ds E_{\c_{\pm}} }\!&\!\!=\!\!&\!
{\ds \frac{1}{2\sqrt{2}}\Big(E_{345^{2}678}-E_{34568}-E_{45^{2}68}-E_{45678}}\\[4pt]
\!&\!\!\!\!&\!{\ds \quad\mp i\big(E_{345^{2}68}+E_{345678}+E_{45^{2}678}-E_{4568}\big)\Big)
=(Z_{z_{1}z_{2}z_{3}}/Z_{\bar{z}_{1}\bar{z}_{2}\bar{z}_{3}})^{+}\,,}\\[4pt]
{\ds E_{\epsilon }} \!&\!\!=\!\!&\!{\ds-iE_{23458}=\left( Z_{4\bar{z}_{3}z_{3}}\right)^{+}\,, }
\end{array}
\end{equation}

These generators define a complex invariant subalgebra $(\ginv)^{{\C}}$ of type 
$\dn_{3}\oplus \an_{2}\oplus \an_{1}\oplus \C$ with the following root
labeling 
\vskip10pt 
\begin{picture}(300,25)\thicklines
\multiput(30,13)(30,0){3}{\circle{6}}
\multiput(33,13)(30,0){2}{\line(1,0){24}}
\put(26,-3){$\b_{-}$}
\put(56,-3){$\tilde{\b}$}
\put(86,-3){$\b_{+}$}
\multiput(130,13)(30,0){2}{\circle{6}}
\multiput(133,13)(30,0){1}{\line(1,0){24}}
\put(126,-1){$\c_{-}$}
\put(156,-1){$\c_{+}$}
\put(200,13){\circle{6}}
\put(196,-1){$\epsilon$}
\put(215,10){ $\times$ }
\put(240,10){$\{E_3-F_3+E_5-F_5-2(E_7-F_7)\}$
}
\end{picture}\vskip10pt 
The detailed structure of the ${\dn}_{3}\subset (\ginv)^{\C}$ 
is encoded in the following commutation relations: 
\begin{eqnarray}
E_{\tilde{\b}+\b_{\pm}} \!\!& \!\!\doteq\!\! &\!\!\pm[E_{\tilde{\b}},E_{\b_{\pm}}]=
\frac{1}{2}\left(E_{23^{2}4^{2}5^{3}6^{2}78}+E_{234^{2}5^{2}6^{2}78}\pm
i(E_{23^{2}4^{2}5^{2}6^{2}78}-E_{234^{2}5^{3}6^{2}78})\right)  \notag \\
\!\!&\!\!&\!\!=(Z_{4z_{1}\bar{z}_{2}}/Z_{4\bar{z}_{1}z_{2}})^{+}\,,  \notag \\[1pt]
E_{\b_{-}+\tilde{\b}+\b_{+}} \!\!&\!\!\doteq \!\!&\!\!
[E_{\b_{-}},E_{\tilde{\b}+\b_{+}}]=iE_{23^{2}4^{3}5^{3}6^{2}78}=(Z_{4z_{1}\bar{z}_{1}})^{+}\,,\label{D3CommRel}\\[2pt]
H_{\tilde{\b}} \!\!&\!\!\doteq\!\! &\!\![E_{\tilde{\b}},F_{\tilde{\b}}]=(H_{2}+H_{3}+H_{4}+2H_{5}+2H_{6}+H_{7}+H_{8})\,,  \notag \\
H_{\b_{\pm}} \!\!&\!\!\doteq \!\!&\!\![E_{\b_{\pm}},F_{\b_{\pm}}]=
\frac{1}{2}\left( H_{3}+2H_{4}+H_{5}\pm i(-E_{3}+F_{3}+E_{5}-F_{5})\right) \,. 
\notag
\end{eqnarray}
The ${\an}_{2}\subset (\ginv)^{{\C}}$ factor is characterized as follows:
\begin{equation}
\label{A2CommRel} 
\begin{array}{rcl}
{\ds E_{\c_{-}+\c_{+}} }  \!&\!\!\doteq \!\!&\! {\ds[E_{\c_{-}},E_{\c_{+}}]=iE_{34^{2}5^{3}6^{2}78^{2}}=
(\widetilde{Z}_{z_{1}\bar{z}_{1}z_{2}\bar{z}_{2}z_{3}\bar{z}_{3}})^{+}\,, } \\[4pt]
{\ds F_{\c_{-}+\c_{+}}}\!&\!\!\doteq \!\!&\! {\ds[F_{\c_+},F_{\c_-}]=-iF_{34^{2}5^{3}6^{2}78^{2}}=
(\widetilde{Z}_{z_{1}\bar{z}_{1}z_{2}\bar{z}_{2}z_{3}\bar{z}_{3}})^{-}\,,} \\[4pt]
{\ds H_{\c_{\pm}}}\!&\!\!\doteq \!\!&\!{\ds [E_{\c_{\pm}},F_{\c_{\pm}}]=
\frac{1}{2}\left( H_{3}+2H_{4}+3H_{5}+2H_{6}+H_{7}+2H_{8}\right) }\\[2pt]
\!&\!\!\!\!&\!{\ds\pm \frac{i}{2}(E_{3}-F_{3}+E_{5}-F_{5}-E_{7}+F_{7})\,. } 
\end{array}
\end{equation}
Finally, the Cartan generator of the remaining $\an_{1}\subset (\ginv)^{{\C}}$ is given by 
$[E_{\epsilon},F_{\epsilon}]=H_{2}+H_{3}+H_{4}+H_{5}+H_{8}$. One can verify that all
three simple subalgebras of $(\ginv)^{\C}$ indeed commute and that the compact abelian factor
$i\widetilde{H}^{[6]}=(E_{3}-F_{3}+E_{5}-F_{5}-2(E_{7}-F_{7}))$ is the centre of $\ginv$. 
The structure of $i\widetilde{H}^{[6]}$ can be retrieved from rewriting the orbifold automorphism 
as $\mathcal{U}_{6}^{\Z_{n}}=~\exp \big((2\pi /n)\,(E_{3}-F_{3}+E_{5}-F_{5}-2(E_{7}-F_{7}))\big)$,
and noting that it preserves $\ginv$. 

\begin{table}[!t]
\begin{center}
\begin{tabular}{c|c|c|c}
$D$ & $(\Pi_0,\phi)$ & $\ginv$ & $\sigma(\ginv)$ \\ \hline\hline
$4$ & \quad 
\begin{picture}(260,40) \thicklines
\put(45,-7){\psline{<->}(1,0)(-1,0)} 
\multiput(15,0)(30,0){3}{\circle{6}}
\multiput(18,0)(30,0){2}{\line(1,0){24}} 
\put(11,10){$\b_{-}$}
\put(41,10){$\tilde\b$} 
\put(71,10){$\b_{+}$}
\put(120,-7){\psline{<->}(.5,0)(-.5,0)}
\multiput(105,0)(30,0){2}{\circle{6}}
\multiput(108,0)(30,0){1}{\line(1,0){24}} 
\put(101,11){$\c_{-}$}
\put(131,11){$\c_{+}$} 
\put(165,0){\circle{6}} 
\put(161,-17){$\epsilon$}
\put(183,-3){ $\times$ } 
\put(210,-3){$i\widetilde{H}^{[6]}$} 
\end{picture}
\hspace{1.5mm} & $
\begin{array}{c}
\su(2,2)\oplus \su(2,1) \\[8pt] 
\oplus\; \sl(2,\R)\oplus\u(1)
\end{array}%
$ & 1 \\[10pt]
$3$ & \quad 
\begin{picture}(260,40) \thicklines
\multiput(15,0)(30,0){3}{\circle{6}}
\multiput(18,0)(30,0){2}{\line(1,0){24}}
\multiput(99,-18)(0,36){2}{\circle{6}} 
\put(77,2){\line(4,3){19}}
\put(77,-2){\line(4,-3){19}} 
\put(100,0){\psline{<->}(0,.47)(0,-.47)}
\put(11,-13){$\epsilon$} 
\put(41,-13){$\a_1$} 
\put(67,-15){$\tilde\b$}
\put(107,15){$\b_{+}$}
\put(107,-21){$\b_{-}$}
\put(153,-7){\psline{<->}(.5,0)(-.5,0)}
\multiput(138,0)(30,0){2}{\circle{6}}
\multiput(141,0)(30,0){1}{\line(1,0){24}} 
\put(134,11){$\c_{-}$}
\put(164,11){$\c_{+}$} 
\put(186,-3){ $\times$ }
\put(212,-3){$i\widetilde{H}^{[6]}$} 
\end{picture} & $%
\begin{array}{c}
\so(6,4)\oplus \su(2,1) \\[8pt] 
\oplus \;\u(1)
\end{array}
$ & 2 \\[25pt] 
\hhline{~:==:~}
$2$ & \quad 
\begin{picture}(270,40) \thicklines
\put(22,0){\psline{<->}(0,.47)(0,-.47)}
\multiput(45,0)(30,0){2}{\circle{6}}
\multiput(48,0)(30,0){1}{\line(1,0){24}}
\multiput(22,-18)(0,36){2}{\circle{6}} 
\put(24,-16){\line(4,3){19}}
\put(24,16){\line(4,-3){19}} 
\multiput(98,-18)(0,36){2}{\circle{6}}
\put(77,2){\line(4,3){18.5}} 
\put(77,-2){\line(4,-3){18.5}}
\put(99,0){\psline{<->}(0,.47)(0,-.47)}
\put(41,-13){$\a_1$}
\put(67,-15){$\tilde\b$} 
\put(107,15){$\b_{+}$} 
\put(107,-21){$\b_{-}$}
\put(8,15){$\epsilon$}
\put(5,-21){$\a_0$}
\multiput(147,-12)(36,0){2}{\circle{6}} 
\put(150,-12){\line(1,0){30}}
\put(164,13){\circle{6}} 
\put(165,-19){\psline{<->}(.5,0)(-.5,0)}
\put(161,12){\line(-2,-3){14}} 
\put(167,12){\line(2,-3){14}}
\put(143,-29){$\c_{-}$}
\put(180,-29){$\c_{+}$} \
\put(161,20){$\a'_0$}
\put(196,-3){ $\times$}
\put(220,-3){$\left\{i\widetilde{H}^{[6]}_n\right\}_{n\in\Z}$}
 \end{picture}
& 
$
\begin{array}{c}
\widehat{\so}(6,4)\oplus \widehat{\su}(2,1) \\[8pt] 
\oplus \;\mathcal{L}(\u(1))_{|-1}
\end{array}
$ & - \\[25pt] 
$1$ & \quad 
\begin{picture}(280,40) \thicklines 
\put(17,15){\line(0,-1){30}}
\multiput(40,0)(30,0){5}{\circle{6}}
\multiput(43,0)(30,0){4}{\line(1,0){24}}
\multiput(183,-18)(0,36){2}{\circle{6}} 
\put(162,3){\line(4,3){18}}
\put(162,-3){\line(4,-3){18}}
 \put(184,0){\psline{<->}(0,.47)(0,-.47)}
\put(70,20){\circle{6}} 
\put(70,20){\circle{9}} 
\put(70,15){\line(0,-1){12}}
\put(78,15){$\zeta_{I}$} 
\put(130,20){\circle{6}}
\put(130,17){\line(0,-1){14.5}} 
\put(138,18){$\epsilon$}
\put(10,0){\psline{<->}(0,.47)(0,-.47)}
\multiput(17,-18)(0,36){2}{\circle{6}} 
\put(20,-17){\line(4,3){19}}
\put(20,17){\line(4,-3){19}}
 \put(37,-13){$\a'_{0}$} 
 \put(66,-13){$\a_{-1}$}
\put(96,-13){$\a_0$} 
\put(126,-13){$\a_1$} 
\put(152,-15){$\tilde\b$}
\put(192,15){$\b_{+}$} 
\put(192,-21){$\b_{-}$} 
\put(-6,17){$\c_-$}
\put(-6,-23){$\c_+$}
 \put(215,-3){ $\times$ }
\put(244,-3){$i\widetilde{H}^{[6]}$}
\end{picture} & 
$^6 \mathcal{B}_{11(I\!I)}\oplus\u(1) $ & - \\[30pt] 
\end{tabular}
\end{center}
\caption{The real subalgebras $\ginv$ for $T^{5-D}\times T^{6}/\Z_{n\geqslant 5}$ 
compactifications}
\label{T6/ZnOrbifolds}
\end{table}

We determine the real form $\ginv$ by a manner similar to the $T^4/\Z_{\n\geqslant 3}$ case. 
Applying procedure (\ref{chainFix}), we find that, in a given basis, the Cartan combinations 
$i(H_{\b_+}-H_{\b_-})$ and $i(H_{\c_+}-H_{\c_-})$ are compact, resulting, for both $\an_{2}$
and the $\dn_{3}$ subalgebras, in maximal tori $(S^{1})\oplus \R^{\oplus^{r-1}}$, for $r=2,3$,
respectively. Taking into account the remaining $\u(1)$ factor, it is easy to see that
$\ginv=\su(2,2)\oplus \su(2,1)\oplus \sl(2,\R)\oplus \u(1)$, with overall signature 
$\sigma(\ginv)=1$.

This and further compactifications of the theory are listed in Table~\ref{T6/ZnOrbifolds}.
In $D=3$ the roots $\epsilon $ and $\tilde{\b}$ listed in expression~(\ref{SimpRootsT6/Zn}) connect through 
$\a_{1}$, producing the invariant real form $\ginv=\so(6,4)\oplus \su(2,1)\oplus \u(1)$.

In $D=2$, the invariant subalgebra is now a triple affine central product 
$\hat{\dn}_{5}\Join\hat{\an}_{2}\Join\hat{\u}(1)\equiv \hat{\dn}_{5}\Join(\hat{\an}_{2}\Join\hat{\u}(1))$,
associatively. For convenience, we have once again depicted in Table~\ref{T6/ZnOrbifolds} the direct 
sum before identification of centres and scaling operators. When carrying out the identification, 
the affine root $\a'_{0}$ has thus to be understood as a non-simple root in $\Delta _{+}(E_{9})$,
enforcing: $\a'_{0}=\d-(\c_{+}+\c_{-})$. Moreover, it can be checked that 
$\d_{D_{5}}\doteq \a _{0}+\epsilon+2(\a_{1}+\tilde\b)+\b_{+}+\b_{-}=\d$, resulting in 
$H_{\d_{D_{5}}}=H_{\d_{A_{2}}}=c_{\hat{\u}(1)}$, for $\d_{A_{2}}=\a'_{0}+\c _{+}+\c_{-}$.
This identification carries over to the three corresponding scaling operators. 

The reality properties of $\ginv$ can be inferred from the finite case,
by extending the analysis of the $T^5\times T^4/\Z_{n\geqslant n}$ orbifold 
in Section \ref{AffineT4} separately to the $\hat{\dn}_5$ and $\hat{\an}_2$ factors. 
Since both subalgebras are "next to split", the $\hat{\dn}_5$ case is directly retrievable 
from the construction exposed in Section~\ref{AffineT4} by reducing the rank by two.
The real $\widehat{\su}(2,1)$ factor is also characterized by an automorphism
$U(z)$ of type 1a, with $u=1$, which can be found, along with further
specifications, in \cite{Corn92c}. The rest of the analysis is similar to the
discussion for the $T^5\times T^4/\Z_{n\geqslant n}$ case. The signatures $\hat{\sigma}$
of both non-abelian factors are infinite again, and can be decomposed as
in expression (\ref{signD7+}).

In $D=1$, the Borcherds algebra $^6\mathcal{B}_{11}$ resulting from reconnecting
the three affine KMAs appearing in $D=2$ through the extended root $\a_{-1}$ is defined 
this time by a Cartan matrix of corank 2, with simple imaginary root $\zeta_{I}$ attached 
to the raising operator 
$E_{\zeta_{I}}=(1/2)(E_{\d+\a_{3}}-E_{\d-\a_{3}}+E_{\d+\a_{5}}-E_{\d-\a_{5}}-2E_{\d+\a_{7}}+2E_{\d-\a _{5}})$. 
Since the triple extension is not successive, the ensuing algebra is more involved than an EALA. 
Writing for short $\d_2\doteq \d_{A_2}= \a_{0}^{\prime }+\c _{+}+\c_{-}$, it possesses two centres, 
namely $\{\z_{1}=H_{\d}-H_I,\z_{2}=H_{\d_2}-H_{\d}\}$ 
and two scaling elements $\{d_I,d_2\}$ counting the levels in $\zeta_{I}$ and $\d_2$. 
The signature $\hat\sigma$ of $\ginv$ is again undefined.

Denoting by $^6\mathcal{B}_{11(I\!I)}$ the real Borcherds algebra represented in 
Table~\ref{T6/ZnOrbifolds}, the $I\!I$ referring to the
two arrows connecting respectively $\c_{\pm}$ and $\b_{\pm}$ in the Satake diagram,
the real form $\ginv$ is given by
\begin{equation*}
\ginv=\mathfrak{u}(1)\oplus\, 
^6\mathcal{B}_{11(I\!I)}/\{\z_{1};\z_{2}; d_I; d_2\}\,.
\end{equation*}
By construction, $-H_{-1}$ will replace $d_I$ and $d_2$ after the quotient is performed.

\subsection{The non-generic $n=4$ case}

As we mentioned at the beginning of this section, there is a large number of consistent 
choices for the charges of the $T^{6}/\Z_{n}$ orbifold. Moreover, non-generic invariant 
subalgebras appear for particular periodicities $n$. For our choice of orbifold charges,
the non-generic cases appear in $n=3,4$, and are singled out from the generic one by the 
absence of a $\u(1)$ factor in the invariant subalgebra. In $D=1$, this entails the appearance 
of simple invariant Kac-Moody subalgebras in place of the simple Borcherds type ones encountered 
up to now. These KMA will be denoted by $\mathcal{KM}$.

The novelty peculiar to the $T^{6}/\Z_{4}$ orbifold lies in the invariance of the root $\a_{7}$, 
which is untouched by the mirror symmetry $(z_3,\bar z_3)\rightarrow (-z_3,-\bar z_3)$, so that 
the generators $E_{7}$, $F_{7}$ and $H_{7}$ are now conserved separately. Furthermore, several 
new invariant generators appear related to $Z_{z_{1}z_{2}\bar{z}_{3}}$ and 
$Z_{\bar{z}_{1}\bar{z}_{2}z_{3}}$: 
\begin{equation}
\begin{array}{l}
Z_{z_{1}z_{2}\bar{z}_{3}}/Z_{\bar{z}_{1}\bar{z}_{2}z_{3}} =\\[5pt] 
\qquad{\displaystyle\frac{1}{4\sqrt{2}}\big(E_{345^{2}678}+F_{345^{2}678}+E_{34568}+F_{34568}+
E_{45^{2}68}+F_{45^{2}68}-E_{45678}-F_{45678}} \\[5pt]
\qquad \pm
i(E_{345^{2}68}+F_{345^{2}68}-E_{345678}-F_{345678}-
E_{45^{2}678}-F_{45^{2}678}-E_{4568}-F_{4568})
\big)\,, \notag
\end{array}
\end{equation}
together with the corresponding compact generators.

\begin{table}[!t]
\begin{center}
\begin{tabular}{c|c|c|c}
$D$ & $(\Pi_0,\phi)$ & $\ginv$ & $\sigma(\ginv)$ \\
\hline\hline[3pt]
$4$ &
\begin{picture}(240,40)
\thicklines
\put(60,-7){\psline{<->}(1,0)(-1,0)}
\multiput(30,0)(30,0){3}{\circle{6}}
\multiput(33,0)(30,0){2}{\line(1,0){24}}
\put(26,10){$\b_{-}$}
\put(56,10){$\tilde\b$}
\put(86,10){$\b_{+}$}
\put(150,-7){\psline{<->}(1,0)(-1,0)}
\multiput(120,0)(30,0){3}{\circle{6}}
\multiput(123,0)(30,0){2}{\line(1,0){24}}
\put(116,11){$\lambda_{-}$}
\put(146,11){$\a_7$}
\put(176,11){$\lambda_{+}$}
\put(210,0){\circle{6}}
\put(206,10){$\epsilon$}
\end{picture}\hspace{1.5mm}
&
$\su(2,2)^{\oplus^2}\oplus \sl(2,\R)$ & 3
\\[20pt] 
$3$ & 
\begin{picture}(240,40)
\thicklines
\multiput(15,0)(30,0){3}{\circle{6}}
\multiput(18,0)(30,0){2}{\line(1,0){24}}
\multiput(99,-18)(0,36){2}{\circle{6}}
\put(77,2){\line(4,3){19}}
\put(77,-2){\line(4,-3){19}}
\put(100,0){\psline{<->}(0,.47)(0,-.47)}
\put(11,-13){$\epsilon$}
\put(41,-13){$\a_1$}
\put(67,-15){$\tilde\b$}
\put(107,15){$\b_{+}$}
\put(107,-21){$\b_{-}$}
\put(183,-7){\psline{<->}(1,0)(-1,0)}
\multiput(153,0)(30,0){3}{\circle{6}}
\multiput(156,0)(30,0){2}{\line(1,0){24}}
\put(149,11){$\lambda_{-}$}
\put(179,11){$\a_7$}
\put(209,11){$\lambda_{+}$}
\end{picture}
& $
\so(6,4)\oplus \su(2,2) 
$ & 4\\[35pt] 
\hhline{~:==:~}
$2$ & 
\begin{picture}(240,40)
\thicklines
\put(22,0){\psline{<->}(0,.47)(0,-.47)}
\multiput(45,0)(30,0){2}{\circle{6}}
\multiput(48,0)(30,0){1}{\line(1,0){24}}
\multiput(22,-18)(0,36){2}{\circle{6}}
\put(24,-16){\line(4,3){19}}
\put(24,16){\line(4,-3){19}}
\multiput(98,-18)(0,36){2}{\circle{6}}
\put(77,2){\line(4,3){18.5}}
\put(77,-2){\line(4,-3){18.5}}
\put(99,0){\psline{<->}(0,.47)(0,-.47)}
\put(41,-13){$\a_1$}
\put(67,-15){$\tilde\b$}
\put(107,15){$\b_{+}$}
\put(107,-21){$\b_{-}$}
\put(8,15){$\epsilon$}
\put(5,-21){$\a_0$}
\multiput(153,-12)(33,0){3}{\circle{6}}
\put(186,-19){\psline{<->}(1,0)(-1,0)}
\multiput(156,-12)(33,0){2}{\line(1,0){27}}
\put(186,13){\circle{6}}
\put(183,12){\line(-4,-3){28}}
\put(189,12){\line(4,-3){28}}
\put(149,-30){$\lambda_{-}$}
\put(182,-30){$\a_7$}
\put(215,-30){$\lambda_{+}$}
\put(181,20){$\a_0''$}
\end{picture}
& 
$
\widehat{\so}(6,4)\oplus \widehat{\su}(2,2) 
$ & -
\\[30pt] 
$1$ & 
{\hspace{-8mm}
\begin{picture}(240,40)
\thicklines
\put(54,0){\psline{<->}(0,.52)(0,-.52)}
\put(34,0){\circle{6}}
\put(15,-3){$\a_7$}
\put(36,3){\line(1,1){15}}
\put(54,20){\circle{6}}
\put(72,3){\line(-1,1){15}}
\put(62,20){$\lambda_{+}$}
\put(36,-3){\line(1,-1){15}}
\put(54,-20){\circle{6}}
\put(72,-3){\line(-1,-1){15}}
\put(62,-24){$\lambda_{-}$}
\put(75,-13){$\a_0''$}
\put(164,3){\line(0,1){15}}
\put(164,21){\circle{6}}
\multiput(74,0)(30,0){5}{\circle{6}}
\multiput(77,0)(30,0){4}{\line(1,0){24}}
\multiput(218,-18)(0,36){2}{\circle{6}}
\put(219,0){\psline{<->}(0,.47)(0,-.47)}
\put(196,2){\line(4,3){19}}
\put(196,-2){\line(4,-3){19}}
\put(100,-13){$\a_{-1}$}
\put(130,-13){$\a_0$}
\put(170,18){$\epsilon$}
\put(160,-13){$\a_1$}
\put(185,-14){$\tilde\b$}
\put(224,17){$\b_{+}$}
\put(224,-23){$\b_{-}$}
\end{picture}}
& $^6\mathcal{KM}_{11(I\!I)}$ & -\\[30pt]
\end{tabular}
\end{center}
\caption{The real subalgebras $\ginv$ for $T^{5-D}\times T^6/\Z_{4}$ compactifications}
\label{T6/Z4Orbifolds}
\end{table}

The invariant subalgebra is now more readily derived by splitting
the $Z$ generators into combinations containing or not an overall
Ad$E_{7}$ factor (in other words, we "decomplexify" $z_{3}$ into 
$x^{9}$ and $x^{10}$): 
\begin{equation}
\label{A3TerGeneratorsT6/Zn}
\begin{array}{rcl}
{\ds E_{\lambda_{\pm}}} \!&\!\!=\!\!&\!
{\ds -\frac{i}{2}\big(E_{34568}+E_{45^{2}68}\pm i(E_{345^{2}68}-E_{4568})\big)=
(Z_{z_{1}z_{2}10}/-Z_{\bar{z}_{1}\bar{z}_{2}10})^{+}\,,  } \\[4pt]
{\ds E_{\a_{7}+\lambda_{\pm}}}\!&\!\!=\!\!&\!
{\ds \frac{1}{2}\big(E_{345^{2}678}-E_{45678}\pm i(E_{345678}+E_{45^{2}678})\big)=
(Z_{\bar{z}_{1}\bar{z}_{2}9}/Z_{z_{1}z_{2}9})^{+}\, }
\end{array}
\end{equation}
which verify the following algebra:
\begin{eqnarray}
E_{\a_{7}+\lambda_{\pm}} \!\!&\!\!\doteq \!\!&\pm [E_{\a_{7}},E_{\lambda_{\pm}}]\,,\notag \\
E_{\lambda_{-}+\a_{7}+\lambda_{+}} \!\!&\!\!\doteq \!\!&\!\![E_{\lambda_{-}},E_{\a_{7}+\lambda_{+}}]=
-E_{34^{2}5^{3}6^{2}78^{2}}=-i(\widetilde{K}_{z_{1}\bar{z}_{1}z_{2}\bar{z}_{2}z_{3}\bar{z}_{3}})^{+},
\label{A3CommRelTer} \\
H_{\lambda_{\pm}} \!\!&\!\!\doteq \!\!&\!\![E_{\lambda_{\pm}},F_{\lambda_{\pm}}]=
\frac{1}{2}\left(H_{3}+2H_{4}+3H_{5}+2H_{6}+2H_{8}\pm i(E_{3}-F_{3}+E_{5}-F_{5})\right) \,,  \notag
\end{eqnarray}
so that the former $\an_{2}$ factor for $n$ generic is now enhanced to $\an_{3}$.
One compact combination $i(H_{\lambda_+}-H_{\lambda_-})$ results from
the action of Fix$_{\tau_0}$ on the algebra formed by the generators in~(\ref{A3CommRelTer}), which determines the corresponding real form to be $\su(2,2)$.
The chain of invariant subalgebras resulting from further compactifications
follows as summarized in Table~\ref{T6/Z4Orbifolds}. 

In $D=2$, we have the identification $\a''_{0}=\d-(\lambda_{+}+\a_{7}+\lambda_{-})$
leading to the now customary affine central product $\widehat{\so}(6,4)\Join\widehat{\su}(2,2)$, 
represented for commodity as a direct sum in Table~\ref{T6/Z4Orbifolds}. The corresponding Satake 
diagram defines the real form $\ginv$. Its signature $\hat\sigma$ is infinite with the correspondence $\hat\sigma(\ginv|_{D=2})= 2+2\times 4\times \infty=
\sigma(\ginv|_{D=3})-2+2\times \sigma(\ginv|_{D=3})\times \infty$.

In $D=1,$ $\ginv$ is defined by the quotient of a simple KMA: $^6\mathcal{KM}_{11}$, by its centre 
$\z=H_{\d}-H_{\d_3}$, where $\d_3=\a''_0+\lambda_{+}+\a_{7}+\lambda_{-}$, and by the derivation $d_3$.
As for affine KMAs, $^6\mathcal{KM}_{11}$ is characterized by a degenerate Cartan matrix with 
rank $r=2\times 11-\dim\h=10$. However, the principal minors of its Cartan matrix are not all strictly 
positive, so that $^6\mathcal{KM}_{11}$ does not result from the standard affine extension of any finite 
Lie algebra. The real form $\ginv$ is determined from the Satake diagram in Table~\ref{T6/Z4Orbifolds}
and the relation:
\begin{equation}
\label{kotienT6}
\ginv= \,^6\mathcal{KM}_{11(I\!I)}/\{\z; d_3\}\,.
\end{equation}
Our convention denotes by $I\!I$ the class of real
forms of $^6\mathcal{KM}_{11}$ for which the Cartan
involution exhibits both possible diagram 
symmetries, exchanging $\phi(\lambda_{\pm})=\mp \lambda_{\pm}$ 
and $\phi(\b_{\pm})=\mp \b_{\pm}$.

\subsection{The standard $n=3$ case}

\begin{table}[!t]
\begin{center}
\begin{tabular}{c|c|c|c}
$D$ & $(\Pi_0,\phi)$ & $\ginv$ & $\sigma(\ginv)$ \\
\hline\hline[3pt]
$4$ & \quad
\begin{picture}(235,40)
\thicklines
\multiput(15,0)(30,0){5}{\circle{6}}
\multiput(18,0)(30,0){4}{\line(1,0){24}}
\put(75,-6){\psline{<->}(-1,0)(1,0)}
\put(75,-11){\psline{<->}(-2,0)(2,0)}
\put(9,10){$\b_-$}
\put(41,10){$\a_-$}
\put(71,10){$\epsilon$}
\put(101,10){$\a_+$}
\put(133,10){$\b_+$}
\put(195,-7){\psline{<->}(.5,0)(-.5,0)}
\multiput(180,0)(30,0){2}{\circle{6}}
\multiput(183,0)(30,0){1}{\line(1,0){24}}
\put(176,11){$\c_{-}$}
\put(206,11){$\c_{+}$}
\end{picture}\hspace{1.5mm}
&
$\su(3,3)\oplus \su(2,1)$ & 1
\\[20pt]
$3$ & \quad
\begin{picture}(235,40)
\thicklines
\put(75,20){\circle{6}}
\put(75,17){\line(0,-1){14}}
\multiput(15,0)(30,0){5}{\circle{6}}
\multiput(18,0)(30,0){4}{\line(1,0){24}}
\put(75,-6){\psline{<->}(-1,0)(1,0)}
\put(75,-11){\psline{<->}(-2,0)(2,0)}
\put(79,25){$\a_1$}
\put(9,10){$\b_-$}
\put(41,10){$\a_-$}
\put(64,6){$\epsilon$}
\put(101,10){$\a_+$}
\put(133,10){$\b_+$}
\put(195,-7){\psline{<->}(.5,0)(-.5,0)}
\multiput(180,0)(30,0){2}{\circle{6}}
\multiput(183,0)(30,0){1}{\line(1,0){24}}
\put(176,11){$\c_{-}$}
\put(206,11){$\c_{+}$}
\end{picture}
& $
\e_{6|2}\oplus \su(2,1) 
$ &2\\[35pt] 
\hhline{~:==:~}
$2$ & \quad
\begin{picture}(235,60)
\thicklines
\put(75,20){\circle{6}}
\put(75,40){\circle{6}}
\put(75,17){\line(0,-1){14}}
\put(75,37){\line(0,-1){14}}
\multiput(15,0)(30,0){5}{\circle{6}}
\multiput(18,0)(30,0){4}{\line(1,0){24}}
\put(75,-6){\psline{<->}(-1,0)(1,0)}
\put(75,-11){\psline{<->}(-2,0)(2,0)}
\put(81,37){$\a_0$}
\put(81,17){$\a_1$}
\put(9,10){$\b_-$}
\put(41,10){$\a_-$}
\put(64,6){$\epsilon$}
\put(101,10){$\a_+$}
\put(133,10){$\b_+$}
\multiput(180,-12)(36,0){2}{\circle{6}}
\put(183,-12){\line(1,0){30}}
\put(197,13){\circle{6}}
\put(198,-19){\psline{<->}(.5,0)(-.5,0)}
\put(194,12){\line(-2,-3){14}}
\put(200,12){\line(2,-3){14}}
\put(176,-29){$\c_{-}$}
\put(213,-29){$\c_{+}$}
\put(194,20){$\a'_0$}
\end{picture}
& 
$\hat{\e}_{6|2}\oplus \widehat{\su}(2,1) $ & -
\\[45pt] 
$1$ & \quad
\begin{picture}(235,40)
\thicklines
\put(17,15){\line(0,-1){30}}
\multiput(40,0)(30,0){5}{\circle{6}}
\multiput(43,0)(30,0){4}{\line(1,0){24}}
\put(10,0){\psline{<->}(0,.47)(0,-.47)}
\multiput(17,-18)(0,36){2}{\circle{6}}
\put(20,-17){\line(4,3){19}}
\put(20,17){\line(4,-3){19}}
\multiput(183,-18)(0,36){2}{\circle{6}}
\put(162,3){\line(4,3){18}}
\put(162,-3){\line(4,-3){18}}
\put(183,0){\psline{<->}(0,.47)(0,-.47)}
\multiput(206,-36)(0,72){2}{\circle{6}}
\put(185,21){\line(4,3){18}}
\put(185,-21){\line(4,-3){18}}
\put(206,0){\psline{<->}(0,1.08)(0,-1.08)}
\put(37,-13){$\a'_{0}$}
\put(66,-13){$\a_{-1}$}
\put(96,-13){$\a_0$}
\put(126,-13){$\a_1$}
\put(153,-13){$\epsilon$}
\put(168,25){$\a_{+}$}
\put(172,-29){$\a_{-}$}
\put(212,33){$\b_{+}$}
\put(212,-40){$\b_{-}$}
\put(-6,17){$\c_-$}
\put(-6,-23){$\c_+$}
\end{picture}
& $^{6'}\mathcal{KM}_{11(I\!I\!I)}$ & -\\[45pt]
\end{tabular}
\end{center}
\caption{The real subalgebras $\ginv$ for $T^{5-D}\times T^6/\Z_{3}$ compactifications}
\label{T6/Z3Orbifolds}
\end{table}

Starting in $D=4$, the $\Z_{3}$-invariant subalgebra builds up the semi-simple 
$(\ginv)^{\C}=\an_{5}\oplus\an_{2}$. The $\an_{5}$ part follows from enhancing the 
$\an_{3}\oplus\an_{1}$ semi-simple factor of the generic invariant 
subalgebra~(\ref{SimpRootsT6/Zn}) by the following additional invariant generators: 
\begin{equation}
\label{Charge0forT6/Z3}
\begin{array}{rcl}
{\ds K_{z_{1}\bar{z}_{3}}/K_{\bar{z}_{1}z_{3}}} \!\!&\!=\!&\!\!
{\ds \frac{1}{4}\big(E_{3456}+F_{3456}+E_{4567}+F_{4567}}\\[4pt]
\!\!&\!\!&\!\!\quad{\ds \pm i(E_{34567}+F_{34567}-E_{456}-F_{456})\big)\,, }  \\[4pt]
{\ds K_{z_{2}\bar{z}_{3}}/K_{\bar{z}_{2}z_{3}}} \!\!&\!=\!&\!\! 
{\ds \frac{1}{4}\big(E_{56}+F_{56}+E_{67}+F_{67}\pm i(E_{567}+F_{567}-E_{6}-F_{6})\big)\,,} 
\\[4pt]
{\ds Z_{4z_{1}\bar{z}_{3}}/Z_{4\bar{z}_{1}z_{3}}}\!\!&\!=\!&\!\! 
{\ds \frac{1}{4}\big(E_{23^{2}4^{2}5^{2}678}+F_{23^{2}4^{2}5^{2}678}+E_{234^{2}5^{2}68}+F_{234^{2}5^{2}68}}\\[4pt]
\!\!&\!\!&\!\!\quad {\ds \pm i(E_{23^{2}4^{2}5^{2}68}+F_{23^{2}4^{2}5^{2}68}-
E_{234^{2}5^{2}678}-F_{234^{2}5^{2}678})\big)\,, }\\[4pt]
{\ds Z_{4z_{2}\bar{z}_{3}}/Z_{4\bar{z}_{2}z_{3}}}\!\!&\!=\!&\!\! 
{\ds \frac{1}{4}\big(E_{2345^{2}678}+F_{2345^{2}678}+E_{234568}+F_{234568}} \\[4pt]
\!\!&\!\!&\!\!\quad{\ds \pm i(E_{2345^{2}68}+F_{2345^{2}68}-E_{2345678}-F_{2345678})\big)\,,} 
\end{array}
\end{equation}
(together with their corresponding compact generators). It becomes clear that one is dealing with an
$\an_{5}$-type algebra, when recasting the whole system in the basis: 
\begin{equation}
\label{A5GeneratorsT6/Z3}
\begin{array}{rcl}
{\ds E_{\epsilon}} \!\!&\!\doteq \!&\!\! {[E_2,[E_3,E_{\tilde{\a}}]]=\ds -iE_{23458}
=(Z_{4z_3\bar z_3})^+\,, }\\[4pt]
{\ds E_{\a_{\pm}}} \!\!&\!\doteq \!&\!\! {\ds \frac{1}{2}\left(E_{56}+E_{67}\pm
i(E_{567}-E_{6})\right) =(K_{z_{2}\bar{z}_{3}}/K_{\bar{z}_{2}z_{3}})^{+}\,,}\\[4pt]
{\ds E_{\b_{\pm}}} \!\!&\!\doteq \!&\!\! {\ds \frac{1}{2}\left( E_{34}+E_{45}\pm
i(E_{345}-E_{4})\right) =(K_{z_{1}\bar{z}_{2}}/K_{\bar{z}_{1}z_{2}})^{+}\,,}\\[4pt]
{\ds E_{\a_{\pm}+\b_{\pm}} } \!\!&\!\doteq \!&\!\! {\ds \frac{1}{2}\left( E_{3456}+E_{4567}\pm
i(E_{34567}-E_{456})\right) =(K_{z_{2}\bar{z}_{3}}/K_{\bar{z}_{2}z_{3}})^{+}\,, } \\[4pt]
{\ds E_{\epsilon+\a_{\pm}} }\!\!&\!\doteq \!&\!\!{\ds\frac{1}{2}\left(E_{2345^{2}678}+E_{234568}\pm
i(E_{2345^{2}68}-E_{2345678})\right)}\\[4pt]
\!\!&\! \!&\!\!\quad{=\ds(Z_{4z_{2}\bar{z}_{3}}/Z_{4\bar{z}_{2}z_{3}})^{+}\,,} \\[4pt]
{\ds E_{\epsilon+\a_{\pm}+\b_{\pm}} } \!\!&\!\doteq \!&\!\!
{\ds\frac{1}{2}\left(E_{23^{2}4^{2}5^{2}678}+E_{234^{2}5^{2}68}\pm
i(E_{23^{2}4^{2}5^{2}68}-E_{234^{2}5^{2}678})\right)}\\[4pt]
\!\!&\! \!&\!\! {=\ds(Z_{4z_{1}\bar{z}_{3}}/Z_{4\bar{z}_{1}z_{3}})^{+}\,, }\\[4pt]
{\ds E_{\a_{-}+\a_{23458}+\a_{+}}}\!\!&\!\doteq \!&\!\! 
{\ds iE_{2345^{2}6^{2}78}=(Z_{4z_{2}\bar{z}_{2}})^{+}\,, }\\[4pt]
{\ds E_{\a_{-}+\epsilon+\a_{+}+\b_{\pm}} } \!\!&\!\doteq \!&\!\! 
{\ds\frac{1}{2}\left( E_{23^{2}4^{2}5^{3}6^{2}78}+E_{234^{2}5^{2}6^{2}78}\pm
i(E_{23^{2}4^{2}5^{2}6^{2}78}-E_{234^{2}5^{3}6^{2}78})\right)} \\[4pt]
\!\!&\! \!&\!\! {=\ds(Z_{4z_{1}\bar{z}_{2}}/Z_{4\bar{z}_{1}z_{2}})^{+}} \\[4pt]
{\ds E_{\b_{-}+\a_{-}+\epsilon+\a_{+}+\b_{+}}}
\!\!&\!\doteq \!&\!\! {\ds iE_{23^{2}4^{3}5^{3}6^{2}78}=(Z_{4z_{1}\bar{z}_{1}})^{+}\,, }
\end{array}
\end{equation}
Combining expressions~(\ref{D3CommRel}) and~(\ref{A5GeneratorsT6/Z3}) reproduces the
commutation relations of an $\an_{5}$-type algebra. The remaining factor $\an_{3}\subset(\ginv)^{\C}$ 
is kept untouched from the generic case. Choosing an appropriate basis for the Cartan subalgebra of 
$\ginv$ produces three compact combinations $i(H_{\c_+}-H_{\c_-})$, $i(H_{\a_+}-H_{\a_-})$
and $i(H_{\b_+}-H_{\b_-})$, leaving four non-compact ones, which determines 
$\ginv=\mathfrak{su}(3,3)\oplus \mathfrak{su}(2,1)$.

The $T^{5-D}\times T^6/\Z_3$ chain of real invariant subalgebras follows as depicted in 
Table~\ref{T6/Z3Orbifolds}, for $D=4,\ldots,1$. In $D=2$, as for the $n=4$ case, we can associate
a formal signature $\hat\sigma$ to the real form $\ginv=\hat{\e}_{6|2}\Join \widehat{\su}(2,1)$, 
which is infinite but keeps a trace of the $D=3$ finite case for which $\sigma(\ginv)=2$, as 
$\hat\sigma(\ginv|_{D=2})=\sigma(\ginv|_{D=3})-2+2\times \sigma(\ginv|_{D=3})\times \infty=0+2\times 2\times\infty$.
In $D=1$, the $I\!I\!I$ subscript labeling the real form appearing in Table~\ref{T6/Z3Orbifolds} refers, 
as before, to the number of arrows in its defining Satake diagram, and the real invariant subalgebra
is retrieved from modding out $^{6'}\mathcal{KM}_{11(I\!I\!I)}$ in a way similar, modulo the required 
changes, to expression~(\ref{kotienT6}).

\section{Non-linear realization of the $\Z_{n}$-invariant sector of M-theory}

In this section, we want to address one last issue concerning the invariant (untwisted)
sector of these orbifolds, namely how the residual symmetry $\ginv$ can be made manifest 
in the equations of motion of the orbifolded supergravity in the finite-dimensional case, 
and in the effective $D=1$ $\sigma$-model description of M-theory near a space-time singularity
in the infinite-dimensional case. The procedure follows the theory of non-linear 
$\sigma$-models realization of physical theories from coset spaces, more particularly 
from the conjectured effective Hamiltonian on $E_{10|10}/K(E_{10|10})$ presented in 
Sections~\ref{NonLinSugra} and~\ref{sec:sing}. 

It is customary in this context to choose the Borel gauge to fix the class representatives
in the coset space. To do this in our orbifolded case, we need the Iwasawa decomposition 
of the real residual U-duality algebra $\ginv$, which can be deduced from its restricted-root 
space decomposition (see Section~\ref{secSatak}). 
For this purpose, we build the set of restricted roots $\Sigma_{0}$~(\ref{restSigm})
for $\ginv$ and partition it into a set of positive and a set of negative restricted roots 
$\Sigma_{0}=\Sigma_{0}^{+}\cup \Sigma_{0}^{-}$, based on some lexicographical ordering. 
Then, following definition~(\ref{restric}), we build:
\begin{equation} \label{Nspace}
\mathfrak{n}_{\text{inv}}^{(\pm)}=\bigoplus_{\bar{\a}\in \Sigma_{0}^{\pm}}(\ginv)_{\bar{\a}}\,,
\text{ \ \ \ with }\phi (\mathfrak{n}_{\text{inv}}^{(\pm)})=\mathfrak{n}_{\text{inv}}^{(\mp)} \, .
\end{equation}
Identifying the nilpotent algebra as $\mathfrak{n}_{\text{inv}}\equiv \mathfrak{n}_{\text{inv}}^{(+)}$, 
the Iwasawa decomposition of $\ginv$ is given by 
$\ginv=\k_{\text{inv}}\oplus\an_{\text{inv}}\oplus\mathfrak{n}_{\text{inv}}$, and the coset parametrization 
in the Borel gauge by $\ginv/\k_{\text{inv}}=\an_{\text{inv}}\oplus \mathfrak{n}_{\text{inv}}$.

The cases where $\ginv$ is a finite real Lie algebra are easily handled. 
For $8\leqslant D\leqslant 3$, the Satake diagrams listed in Tables~\ref{T2/ZnDynkin},
~\ref{T4/ZnOrbifolds},~\ref{T6/ZnOrbifolds},~\ref{T6/Z4Orbifolds} and~\ref{T6/Z3Orbifolds} 
describing the residual U-duality algebras under $T^{q}/\Z_{n}$ orbifolds
for $q=2,4,6$ and $n\geqslant 3$ are well known, and the corresponding
Dynkin diagrams for the basis $\overline{\Pi}_{0}$ of restricted roots 
$\Sigma_{0}$ can be found, together with the associated multiplicities $m_{r}(\bar{\a}_{i})$, 
in \cite{Helga}. We will not dwell on the $D=2$ case, which only serves as a stepping stone 
to the understanding of the $D=1$ case. Moreover, all the arguments we present here 
regarding $D=1$ also apply, with suitable restrictions, to the $D=2$ case.

In the $D=1$ case, since the restricted root space is best determined from the Satake 
diagram of the real form, we will replace $\ginv$ in eqn.~(\ref{Nspace}) and all
ensuing formulae by the real Borcherds and indefinite KM algebras described by the 
corresponding Satake diagrams in Tables~\ref{T2/ZnDynkin} and~\ref{T4/ZnOrbifolds}-\ref{T6/Z3Orbifolds}.
This procedure leads to the Dynkin diagrams and root multiplicities represented in Table~\ref{RestrDyn}
for the bases of restricted roots $\overline{\Pi}_{0}$, given for all but the split case of Table~\ref{T2/ZnDynkin}
(for which normal and restricted roots coincide). The multiplicities appear in bold beside the corresponding 
restricted root, while we denote the multiplicity of $2\bar\a$ with a 2 subscript whenever it is also a 
root of $\Sigma_0$.
\hspace{-2cm}
\begin{table}[!t]
\begin{center}
\begin{tabular}{c|c}
$$ & $\overline{\Pi}_0$ \\
\hline\hline[3pt]
$
\begin{array}{c}
T^6\times T^4/\Z_{n>2}:
 \\ \\
^4\mathcal{B}_{10(Ib)}
\end{array}
$ & 
{\hspace{-10mm}
\begin{picture}(270,40)
\thicklines
\put(74,20){\circle{6}}
\put(74,20){\circle{9}} 
\put(74,16){\line(0,-1){12}} 
\put(82,15){$\bar\b_{I}$} \put(95,15){{\bf 1}}
\put(77,0){\line(1,0){23.5}}
\multiput(74,0)(30,0){7}{\circle{6.5}}
\multiput(107,0)(30,0){4}{\line(1,0){23.5}}
\put(164,3.5){\line(0,1){14}}
\put(164,21){\circle{6.5}}
\put(227,2){\line(1,0){23.5}}
\put(227,-2){\line(1,0){23.5}}
\put(238,-2.5){\footnotesize{$\rangle$}}
\put(70,-15){$\bar\a_{-1}$} \put(70,-30){{\bf 1}}
\put(100,-15){$\bar\a_0$} \put(100,-30){{\bf 1}}
\put(130,-15){$\bar\a_1$} \put(130,-30){{\bf 1}}
\put(170,18){$\bar\c$}     \put(180,18){{\bf 1}}
\put(160,-15){$\bar\a_2$}  \put(160,-30){{\bf 1}}
\put(190,-15){$\bar\a_3$}  \put(190,-30){{\bf 1}}
\put(220,-15){$\bar{\tilde{\a}}$}  \put(220,-30){{\bf 1}}
\put(250,-15){$\bar\a_+$}  \put(250,-30){{\bf 2}}
\end{picture}}
\\[25pt]  \hline
$
\begin{array}{c}
T^4\times T^6/\Z_{n\geq 5}:
 \\ \\
^6\mathcal{B}_{11(I\!I)}
\end{array}
$
& 
{\hspace{-10mm}
\begin{picture}(270,40)
\thicklines
\put(134,20){\circle{6}}
\put(134,20){\circle{9}} 
\put(134,16){\line(0,-1){12}} 
\put(142,15){$\bar\xi_{I}$} \put(153,15){{\bf 1}}
\put(77,3){\line(1,0){24}}
\put(77,1){\line(1,0){23.5}}
\put(77,-1){\line(1,0){23.5}}
\put(77,-3){\line(1,0){24}}
\put(86.5,-2.5){\small{$\langle$}}
\multiput(107,0)(30,0){2}{\line(1,0){23.5}}
\multiput(74,0)(30,0){2}{\circle{6.5}}
\multiput(134,0)(30,0){5}{\circle{6.5}}
\multiput(167,0)(30,0){2}{\line(1,0){23.5}}
\put(194,3.5){\line(0,1){14}}
\put(194,21){\circle{6.5}}
\put(227,2){\line(1,0){23.5}}
\put(227,-2){\line(1,0){23.5}}
\put(238,-2.5){\footnotesize{$\rangle$}}
\put(70,-16){$\bar\c_+$}     \put(62,-31){({\bf 2},{\bf 1}$_{2}$)}
\put(100,-16){$\bar\a'_0$}   \put(100,-31){{\bf 1}}
\put(130,-16){$\bar\a_{-1}$}  \put(130,-31){{\bf 1}}
\put(200,18){$\bar\epsilon$} \put(210,18){{\bf 1}}
\put(160,-16){$\bar\a_0$}     \put(160,-31){{\bf 1}}
\put(190,-16){$\bar\a_1$}     \put(190,-31){{\bf 1}}
\put(220,-18){$\bar{\tilde\b}$}  \put(220,-31){{\bf 1}}
\put(250,-18){$\bar{\b}_+$}     \put(250,-31){{\bf 2}}
\end{picture}}
\\[25pt]  \hline
 $
\begin{array}{c}
T^4\times T^6/\Z_{4}:
 \\ \\
^6\mathcal{KM}_{11(I\!I)}
\end{array}
$
 & 
{\hspace{-10mm}
\begin{picture}(270,43)
\thicklines
\multiput(47,2)(30,0){2}{\line(1,0){23.5}}
\multiput(47,-2)(30,0){2}{\line(1,0){23.5}}
\put(58,-2.5){\footnotesize{$\rangle$}}
\put(86.5,-2.5){\footnotesize{$\langle$}}
\multiput(44,0)(30,0){8}{\circle{6.5}}
\multiput(107,0)(30,0){4}{\line(1,0){23.5}}
\put(194,3.5){\line(0,1){14}}
\put(194,21){\circle{6.5}}
\put(227,2){\line(1,0){23.5}}
\put(227,-2){\line(1,0){23.5}}
\put(238,-2.5){\footnotesize{$\rangle$}}
\put(40,-16){$\bar\a_7$}        \put(40,-31){{\bf 1}}
\put(70,-16){$\bar\lambda_+$}    \put(70,-31){{\bf 2}}
\put(100,-16){$\bar\a''_0$}       \put(100,-31){{\bf 1}}
\put(130,-16){$\bar\a_{-1}$}     \put(130,-31){{\bf 1}}
\put(200,18){$\bar\epsilon$}   \put(207,18){{\bf 1}}
\put(160,-16){$\bar\a_0$}     \put(160,-31){{\bf 1}}   
\put(190,-16){$\bar\a_1$}    \put(190,-31){{\bf 1}}
\put(220,-18){$\bar{\tilde\b}$}  \put(220,-31){{\bf 1}}
\put(250,-18){$\bar{\b}_+$}    \put(250,-31){{\bf 2}}
\end{picture}}
\\[25pt]  \hline
$
\begin{array}{c}
T^4\times T^6/\Z_3:
 \\ \\
^{6'}\mathcal{KM}_{11(I\!I\!I)}
\end{array}
$
 & 
{\hspace{-10mm}
\begin{picture}(270,40)
\thicklines
\put(47,3){\line(1,0){24}}
\put(47,1){\line(1,0){23.5}}
\put(47,-1){\line(1,0){23.5}}
\put(47,-3){\line(1,0){24}}
\put(56.5,-2.5){\small{$\langle$}}
\multiput(44,0)(30,0){2}{\circle{6.5}}
\multiput(104,0)(30,0){6}{\circle{6.5}}
\multiput(77,0)(30,0){4}{\line(1,0){23.5}}
\put(227,0){\line(1,0){23.5}}
\put(197,2){\line(1,0){23.5}}
\put(197,-2){\line(1,0){23.5}}
\put(208,-2.5){\footnotesize{$\rangle$}}
\put(40,-16){$\bar\c_+$}     \put(32,-31){({\bf 2},{\bf 1}$_2$)}
\put(70,-16){$\bar\a'_0$}       \put(70,-31){{\bf 1}}
\put(100,-16){$\bar\a_{-1}$}   \put(100,-31){{\bf 1}}
\put(130,-16){$\bar\a_0$}      \put(130,-31){{\bf 1}}
\put(160,-16){$\bar\a_1$}      \put(160,-31){{\bf 1}}
\put(190,-16){$\bar\epsilon$}   \put(190,-31){{\bf 1}}
\put(220,-16){$\bar{\a}_+$}     \put(220,-31){{\bf 2}}
\put(250,-17){$\bar{\b}_+$}      \put(250,-31){{\bf 2}}
\end{picture}}
 \\[25pt]  \hline
\end{tabular}
\end{center}
\caption{Restricted Dynkin diagrams for very-extended $\ginv$ subalgebras}
\label{RestrDyn}
\end{table}

We may now give a general prescription to compute the algebraic field
strength $\mathcal{G}=g^{-1}dg$ of the orbifolded theory, which also applies to the 
infinite-dimensional case, where $\mathcal{G}$ is the formal coset element~(\ref{coselmn}).
There are two possible equivalent approaches, depending on whether we set to zero the dual field 
associated to the possible centres and derivations of the Borcherds/KMA algebras leading to $\ginv$ 
before or after the computation of $\mathcal{G}$. 

Let us consider an algebra with $A$ pairs of non-compact centres and derivations: 
$\{\z_{a},d_a\}_{a=1,\ldots,A}$ and a maximal non-compact abelian subalgebra $\an$ 
of dimension $n_s$. We introduce a vector of $n_s-2A$ scale factors $\bar{\varphi}$ and a vector of
auxiliary fields $\psi$ and develop them on the basis of $\an(\g)$ as:
\begin{equation}
\bar{\varphi}=\sum_{\bar{\textit{\i}}=1}^{n_s-2A} \bar{\varphi}^{i} H_{\bar{\textit{\i}}}\, , \qquad 
\psi=\sum_{a=1}^{A} \left( \psi^{a}\z_a + \psi^{A+a}d_a \right)\,.
\end{equation}
For example, in the case of the $T^6\times T^4/\Z_n$ orbifold, $\an_{\text{inv}}$ is given by:
\begin{equation*}
\begin{array}{rcl}
\an(^4 \mathcal{B}_{10(Ib)})&=&\Span_{\R}\{H_{\bar{\a}_{-1}};\ldots;H_{\bar{\a}_{3}};
H_{\bar{\tilde{\a}}}=H_4+H_5+H_8;H_{\bar{\a}_{+}}=H_5+2H_6+H_7 ;\\[2pt]
&& H_{\bar{\c}}=H_2+2H_3+3H_4+3H_5+2H_6+H_7+H_8;\z;d_I\} \,.
\end{array}
\end{equation*}
in which the centre is $\z=H_{\bar{\b}_I}-(H_{\bar{\a}_{-1}}+\ldots+H_{\bar{\a}_{3}}+H_{\bar{\tilde{\a}}}+
H_{\bar{\a}_{+}}+H_{\bar{\c}})=H_{\b_I}-H_{\d}$. The generator $H_{\bar{\textit{\i}}}$ will be understood to represent
the $i$-th element of the above list, for $\bar{\textit{\i}}=1,\ldots,8$.

In general, a central element obviously does not contribute much to $\mathcal{G}$, except for a term 
$\propto d\psi^{a}/dt$, so that it does not matter whether we impose 
the physical constraint $\psi^a=0$, $\forall a=1,\ldots,A$ before or after the computation
of $\mathcal{G}$. The derivations $d_a$ also create terms $\propto d\psi^{A+a}/dt$, but $\psi^{A+a}$ 
appears in exponentials in front of generators for roots containing an imaginary/affine ($\neq \a_0$) 
simple root, as well. However, there is no difference between setting the auxiliary field $\psi$ to 
zero directly in $g$ or, later, in the exponentials in $\mathcal{G}$. Indeed, the counting of levels in these
imaginary/affine simple roots is taken care of by $H_{\bar{\a}_{-1}}$ anyway. Finally, a term proportional to
$\z_a$ other than $\z_a\, d\psi^{a}/dt$ can not be produced either, since we work in the Borel gauge, so that
terms containing commutators like $[E_{\b_I},F_{\b_I}]$ are absent. It is thus not necessary to impose $\z_a=0$ $\forall a$
again in the end. 

More generally, let us now set $n_{\text{inv}}=\bigoplus_{\bar{\a}\in \Sigma_{0}^{+}}(\ginv)_{\bar{\a}}$ 
with $\dim(\ginv)_{\bar{\a}}=m(\bar{\a})\cdot m_{r}(\bar{\a})$, the (formal) group element\footnote{Note
that we adopt here a perspective that is different from~\cite{HenryJulPaul1}
when associating restricted roots to the metric and $p$-form potential of
orbifolded 11$D$ supergravity / M-theory. In particular the authors of~\cite{HenryJulPaul2}
were concerned with super-Borcherds symmetries of supergravity with non-split U-duality groups,
which form the so-called real magic triangle, i.e. which are oxidations of
pure supergravity in 4 dimensions with $\mathcal{N}=0,..,7$ supersymmetries.
When doubling the fields of these theories by systematically introducing
Hodge duals for all original $p$-form fields (but not for the metric), the
duality symmetry of this enlarged model can be embedded in a larger 
Borcherds superalgebra. The self-duality equations for all $p$-forms of 
these supergravities can be recovered by a certain choice of truncation in
the Grassmanian coefficients of the superalgebra.
In contrast to our approach however, one positive restricted root was related 
to one $p$-form potential term in~\cite{HenryJulPaul2}, whereas, 
we associate a restricted root generator to one component of the
potential. This is the reason why these authors drop the sum over $m_{r}(\bar{\a})$ 
in expression (\ref{FieldST}) (not mentioning the sum over $m(\bar{\a})$ in the $D=1$ 
case, which we keep since we do not want to discard any higher order contributions 
to classical 11D supergravity).}
\begin{equation}\label{FieldST}
g =e^{\bar{\varphi}+\psi}\cdot \prod_{\bar{\a}\in\Sigma_{0}^{+}}\prod_{s=1}^{m_{r}(\bar{\a})}
\prod_{a=1}^{m(\bar{\a})}e^{C_{\bar{\a},(s,a)}E_{\bar{\a}}^{(k,a)}} \,, 
\end{equation}
can be used to compute the Maurer-Cartan one-form:
\begin{equation}
\mathcal{G} =\big[g^{-1}dg\big]_{\psi^a=0,a=1,\ldots,2A}
\notag
\end{equation}
in which the coefficients $\{\bar{\varphi}^{i};C_{\bar{\a}}\}$ correspond, at levels $l=0,1$,
to the invariant dilatons and potentials of orbifolded classical 11D supergravity, and, 
at higher levels, participate to (invariant) contributions from M-theory.
Their exact expressions can be reconstructed from the material of Sections~\ref{SecT2}-\ref{secT6} 
and the Satake diagrams of Tables~\ref{T2/ZnDynkin} and~\ref{T4/ZnOrbifolds}-\ref{T6/Z3Orbifolds}.

The Maurer-Cartan equation $d\mathcal{G}=\mathcal{G}\wedge \mathcal{G}$ will then reproduce the 
equations of motion of the untwisted sector of the reduced supergravity theory in the finite case,
of M-theory on $M=\R_{+}\times T^{11-D-2p}\times T^{2p}/\Z_{n}$ in $D=1$, which will make manifest
the residual symmetry $\ginv$. Finally, one can write down an effective invariant Hamiltonian as 
in expression~(\ref{ham2}), by performing a Legendre transform of 
$\mathcal{L}=\frac{1}{4n}\big[\Tr(\partial\mathcal{M}^{-1}\partial\mathcal{M})\big]_{\psi^a=0,a=1,\ldots,2A}$,
the orbifolded version of expression~(\ref{LagNL}).

\section{Shift vectors and chief inner automorphisms}
\label{ShiftSec}

We have dedicated the first few chapters of this article to explaining the
characterization of fixed-point subalgebras under finite-order automorphisms
of U-duality algebras. In physical words, we have computed the residual 
U-duality symmetry of maximally supersymmetric supergravities compactified on 
certain toric orbifolds. In $D\geq 3$, the quotient of this residual algebra by
its maximal compact subalgebra is in one-to-one correspondence with the physical 
spectrum of 11$D$~supergravity surviving the orbifold projection. 
In string theory language, this corresponds to the untwisted sector of 
the orbifolded theory. Extrapolating this picture to $D=1$, the orbifold
spectrum gets enhanced by a whole tower of massive string states and/or 
non-perturbative states. 

Although the interpretation of most of these higher level $\eten$ 
roots is still in its infancy, an interesting proposal
was made in~\cite{Gan2} for a restricted number of them, namely for those 
appearing as shift vectors describing $\Z_{2}$ orbifold actions. 
They were interpreted as the extended objects needed for local
anomaly and charge cancellations in brane models of certain M-theory orbifolds
and type IIA orientifolds. 

In this section, a general method to compute the shift vectors of any 
$T^{p}\times T^{q}/\Z_{n}$ orbifolds will be given, as well as
explicit results for $q=2,4,6$. Then, an empirical technique to obtain 
$\eten$ roots that are physically interpretable will be presented,
exploiting the freedom in choosing a shift vector from its equivalence class. 
Our results in particular reproduce the one given in~\cite{Gan2} for 
$T^{6}\times T^{4}/\Z_{2}$. Note that the method will allow to
differentiate, for example, between $T^{4}/\Z_{3}$ and 
$T^{4}/{\mathbbm{Z}}_{4}$ despite the fact that they lead to the same invariant
subalgebra, which gives a clue on the r\^{o}le of the $n$-dependent part of
the shift vectors. Finally, we will see how to extract the roots describing
a $\Z_{n}$ orbifolds from all level $3n$ roots of $\eten$.

We first remark that the complex combinations of generators corresponding to
the complexified physical fields are the eigenvectors of the automorphisms 
$\mathcal{U}_{q}^{\Z_{n}}$ with eigenvalues $\exp (i\frac{2\pi }{n}Q_{A})$, 
for $Q_{A}\in\Z_{n}$. Having a basis of eigenvectors suggests that there is 
a conjugate Cartan subalgebra $\h'$ inside of the $Q_{A}=0$ eigenspace $\g^{(0)}$ 
for which the automorphism is diagonal. We can then reexpress the orbifold action 
as an automorphism that leaves this new Cartan subalgebra invariant, in other
words as a chief inner automorphism of the form $\Ad(\exp (i\frac{2\pi}{n}H'))$ 
for some $H'\in \h'_{\Q}$. As already noticed in the case of $\Z_{2}$
orbifolds, such a chief inner automorphism simply acts as 
$\exp (i\frac{2\pi}{n}\a'(H'))$ on every root subspace $\g_{\a'}$ where $\a'$ is
a root defined with respect to $\h'$. In particular, we can find a (non-unique) 
weight vector $\La'$ corresponding to $H'$ so that 
\begin{equation*}
\Ad(e^{i\frac{2\pi}{n}H'})\g_{\a'}
=e^{i\frac{2\pi}{n}(\La'|\a')}\g_{\a'}\,.
\end{equation*}
Such a weight vector is commonly called shift vector. It turns out
that all automorphisms of a given simple Lie algebra can be classified by
all weights $\La'=\sum_{i=1}^{r}l_{i}\La^{'i} $, $l_{i}\in\Z_{n}$ without 
common prime factor, so that, according to~\cite{KacPet}, 
\begin{equation}\label{KacPetGolden}
(\La|\theta_{G})\leq n 
\end{equation}
(here, the fundamental weights $\La^{'i}$ are defined to be dual to the new 
simple roots $\a'_{i}$, i.e. $(\La^{'i}|\a'_{j})=\d_{j}^{i}$ 
$\forall i,j=9-r,\ldots,8$). Furthermore, there is a simple way, see~\cite{Choi}, 
to deduce the invariant subalgebra from $\La$ by guessing its action on the 
extended Dynkin diagram, if $\La$ satisfies the above condition. Here, we will 
first show how to obtain the shift vectors in the cases we are interested in 
and then describe the above-mentioned diagrammatic method with the help of 
these examples.

\subsection{A class of shift vectors for $T^2/{\mathbbm{Z}}_n$ orbifolds}
\label{Shift2}

Let us start by the particularly simple case of a $T^{2}/\Z_{n}$
orbifold in $T^{3}$. We can directly use the decomposition in eigensubspaces
obtained in equation~(\ref{AlgebraicT2/ZnCharges}). The first task is to
choose a new Cartan subalgebra, or equivalently a convenient Cartan-Weyl
basis. In other words, we are looking for a new set of simple roots for 
$\an_{2}\oplus \an_{1}$, so that all Cartan generators are in the $Q_{A}=0$ 
eigensubspace $\g^{(0)}$. Since the $\an_{1}$ does not feel the orbifold action, 
we can simply take $H_{8}'=H_{8}$. On the other hand, we should take for 
$H'_{6}$ and $H'_{7}$ some combinations of $2H_{6}+H_{7}$ and $E_{7}-F_{7} $. 
A particularly convenient choice is given by the following Cartan-Weyl basis: 
\begin{equation*}
\begin{array}{lll}
E'_{6}=\frac{1}{\sqrt{2}}(E_{6}+iE_{67})\,, & 
F'_{6}=\frac{1}{\sqrt{2}}(F_{6}-iF_{67})\,, & 
H'_{6}=\frac{1}{2}(2H_{6}+H_{7}-i(E_{7}-F_{7}))\,, \\[2pt] 
E'_{7}=\frac{1}{2}(H_{7}-i(E_{7}+F_{7}))\,, & 
F'_{7}=\frac{1}{2}(H_{7}+i(E_{7}+F_{7}))\,, & 
H'_{7}=i(E_{7}-F_{7})\,, \\[2pt] 
E'_{67}=\frac{1}{\sqrt{2}}(E_{6}-iE_{67})\,, & 
F'_{67}=\frac{1}{\sqrt{2}}(F_{6}+iF_{67})\,, & 
H'_{67}=\frac{1}{2}(2H_{6}+H_{7}+i(E_{7}-F_{7}))\,,
\end{array}
\end{equation*}
This gives the following simple decomposition in eigensubspaces: 
\begin{equation*}
\begin{array}{rlrl}
\g^{(0)}= & \Span\{H'_{6};H'_{7};E'_{8};F'_{8};H'_{8}\}\,,\qquad & 
\g^{(n-1)}= & \Span\{E'_{67};F'_{6}\}\,, \\[2pt] 
\g^{(1)}= & \Span\{E'_{6};F'_{67}\}\,, & 
\g^{(n-2)}= & \Span\{E'_{7}\}\,, \\[2pt] 
\g^{(2)}= & \Span\{F'_{7}\}\,. &  & 
\end{array}
\end{equation*}
Notice that $\g^{(n-i)}$ is obtained from $\g^{(i)}$ by the substitution 
$E\leftrightarrow F$, so that we will only give the latter explicitly in the following 
examples. Furthermore, since $\mathcal{U}_{2}^{\Z_{n}}$ actually defines a gradation 
$\g=\bigoplus_{i=0}^{n-1}\g^{(i)}$, we have the property 
$[\g^{(i)},\g^{(j)}]\subseteq \g^{(i+j)}$.
This implies that if we can find a weight $\La^{'\{2\}}$ that acts as 
$\exp (i\frac{2\pi}{n}(\La^{'\{2\}}|\a'_{i}))$ on $\g_{\a'_{i}}$ for all new
simple roots $\a'_{i}$, $i\in I$, it will induce the correct charges 
for all new generators. Here, we should choose $\La^{'\{2\}}$ so that it has scalar 
product 0 with $\a'_{8}$, 1 with $\a'_{6}$ and $n-2$ with $\a'_{7}$, which suggests
to take: 
\begin{equation}\label{T2Shift}
\La^{'\{2\}}=\La^{'6}+(n-2)\La^{'7}\,.
\end{equation}
Note first that the same set of charges can be obtained with all choices of
the form: 
$\La^{'\{2\}}=~(c_{1}n+1)\La^{'6}+(c_{2}n-2)\La^{'7}+c_{3}n\La^{'8} $ for 
any set of $\{c_{i}\}_{i=1}^{3}\in \Z^{3}$. In particular, there exists one 
weight vector that is valid for automorphisms of all finite orders, here 
$\La^{'\{2\}}=\La^{'6}-2\La^{'7}$. However, in equation~\ref{T2Shift},
we took all coefficients in $\Z_{n}$ as is required for the Kac-Peterson
method to work. Since the $\an_{1}$ is obviously invariant, we can restrict 
our attention to the $\an_{2}$ part. One can verify that $\La^{'\{2\}}$ 
satisfies the condition~(\ref{KacPetGolden}) since $\theta_{A_2}=\a'_{6}+\a'_{7}$
implies $(\La^{'\{2\}}|\theta_{A_{2}})=n-1$. 

In general, for a U-duality group $G$, we can define an $(r+1)$-th component 
of $\La'$ as $l_{9}^{G}=n-(\La',\theta_{G})$ ($l_{9}^{A_{1}\times A_{2}}=1$ in 
the above case). On the basis of this extended vector, one can apply the following 
diagrammatic method to obtain the invariant subalgebra in the finite-dimensional 
case (a simple justification of this method can be found in~\cite{Choi}):

\begin{itemize}
\item Replace the Dynkin diagram of $\g$ by its extended Dynkin
diagram, adding an extra node corresponding to the (non-linearly
independent) root $\a'_9=-\theta_G$. We denote the extended diagram 
by $\g^+$ to distinguish it from the affine $\hat{\g}$ in which
the extra node $\a_0=\d-\theta_G$ is linearly independent.

\item Discard all nodes corresponding to roots $\a'_i$ such that 
$l_i\neq 0$ and keep all those such that $l_i=0$, $i\in \{9-r;\ldots;9\}$.

\item Let $p$ be the number of discarded nodes, the (usually reductive)
subalgebra left invariant by the automorphism $\mathcal{U}_{2}^{\Z_{n}}$ 
is given by the (possibly disconnected) remaining diagram times $p-1$
abelian subalgebras.
\end{itemize}

In particular, for $T^{2}/\Z_{n}$ in $T^{3}$ for $n\geq 3$, we
see that $l_{6}$,$l_{7}$ and $l_{9}$ are non-zero, leaving invariant only 
$\a'_{8}=\a_{8}$ which builds an $\an_{1}$ diagram. Since $p=3$, we should 
add two abelian factors for a total (complexified) invariant subalgebra 
$\an_{1}\oplus \C^{\oplus^{2}}$, the same conclusion we arrived at 
from Table~\ref{AlgebraicT2/ZnCharges}. On the other hand, for $T^{2}/\Z_{2}$
in $T^{3}$, $l_{7}=n-2=0$ and we have one more conserved node, leaving a 
total (complexified) invariant subalgebra $\an_{1}^{\oplus^{2}}\oplus \C$.

\begin{figure}[!t]
\begin{picture}(350,50)\put(0,22){$T^3$:}
\thicklines
\put(20,22){$\an^+_2\oplus\an_1\rightarrow \an_1\oplus\C^2$:}
\put(150,25){\circle{6}}
\put(146,13){$\a'_8$}
\multiput(190,15)(36,0){2}{\circle{6}}
\put(193,15){\line(1,0){30}}
\put(208,40){\circle{6}}
\put(205,39){\line(-2,-3){14}}
\put(211,39){\line(2,-3){14}}
\put(204,47){$\a'_9$}
\put(186,3){$\a'_6$}
\put(222,3){$\a'_7$}
\put(250,22){$\rightarrow$}
\put(290,25){\circle{6}}
\put(286,13){$\a'_8$}
\put(305,22){ $\times$ }
\put(330,22){$\C^2$}
\end{picture}
\vskip 15pt 
\begin{picture}(350,50)\put(0,22){$T^4$:}
\thicklines
\put(20,22){$\an^+_4\rightarrow \an_2\oplus\C^2$:}
\multiput(135,15)(30,0){4}{\circle{6}}
\multiput(138,15)(30,0){3}{\line(1,0){24}}
\put(180,38){\circle{6}}
\put(177,37){\line(-2,-1){40}}
\put(183,37){\line(2,-1){40}}
\put(177,45){$\a'_9$}
\put(131,3){$\a'_8$}
\put(161,3){$\a'_5$}
\put(191,3){$\a'_6$}
\put(221,3){$\a'_7$}
\put(240,22){$\rightarrow$}
\multiput(270,25)(30,0){2}{\circle{6}}
\put(273,25){\line(1,0){24}}
\put(266,13){$\a'_8$}
\put(296,13){$\a'_5$}
\put(313,22){ $\times$ }
\put(335,22){$\C^2$}
\end{picture}
\vskip 15pt 
\begin{picture}(350,50)\thicklines
\put(0,27){$T^5$:}
\put(20,27){$\dn^+_5\rightarrow \an_3\oplus\an_1\oplus\C$:}
\multiput(146,12)(0,36){2}{\circle{6}}
\put(148,14){\line(4,3){19}}
\put(148,46){\line(4,-3){19}}
\multiput(169,30)(30,0){2}{\circle{6}}
\put(172,30){\line(1,0){24}}
\multiput(223,12)(0,36){2}{\circle{6}}
\put(201,32){\line(4,3){19}}
\put(201,28){\line(4,-3){19}}
\put(130,10){$\a'_4$}
\put(130,45){$\a'_8$}
\put(165,18){$\a'_5$}
\put(191,18){$\a'_6$}
\put(231,45){$\a'_{9}$}
\put(231,9){$\a'_{7}$}
\put(250,27){$\rightarrow$}
\multiput(280,30)(30,0){3}{\circle{6}}
\multiput(283,30)(30,0){2}{\line(1,0){24}}
\put(277,18){$\a'_4$}
\put(307,18){$\a'_5$}
\put(337,18){$\a'_8$}
\put(377,30){\circle{6}}
\put(374,18){$\a'_9$}
\put(389,27){ $\times$ }
\put(412,27){$\C$}
\end{picture}
\vskip 15pt 
\begin{picture}(350,60)\thicklines
\put(0,22){$T^6$:}
\put(20,22){$\e^+_6\rightarrow \an_5\oplus\C$:}
\multiput(155,28)(0,15){3}{\circle{6}}
\multiput(155,31)(0,15){2}{\line(0,1){9}}
\multiput(132,17)(-23,-12){2}{\circle{6}}
\multiput(152,27)(-23,-12){2}{\line(-2,-1){17}}
\multiput(178,17)(23,-12){2}{\circle{6}}
\multiput(158,27)(23,-12){2}{\line(2,-1){17}}
\put(116,1){$\a'_3$}
\put(138,13){$\a'_4$}
\put(162,56){$\a'_9$}
\put(162,41){$\a'_8$}
\put(162,26){$\a'_5$}
\put(180,20){$\a'_6$}
\put(203,9){$\a'_7$}
\put(210,25){$\rightarrow$}
\multiput(240,28)(30,0){5}{\circle{6}}
\multiput(243,28)(30,0){4}{\line(1,0){24}}
\put(236,16){$\a'_3$}
\put(266,16){$\a'_4$}
\put(296,16){$\a'_5$}
\put(326,16){$\a'_8$}
\put(356,16){$\a'_9$}
\put(370,25){ $\times$ }
\put(391,25){$\C$}
\end{picture}
\vskip 15pt 
\begin{picture}(340,50)\thicklines
\put(0,22){$T^7$:}
\put(20,22){$\e^{+}_7\rightarrow \dn_6\oplus \C$:}
\put(185,45){\circle{6}}
\put(185,42){\line(0,-1){14}}
\multiput(110,25)(25,0){7}{\circle{6}}
\multiput(113,25)(25,0){6}{\line(1,0){19}}
\put(191,42){$\a_8'$}
\put(106,13){$\a_2'$}
\put(131,13){$\a_3'$}
\put(156,13){$\a_4'$}
\put(181,13){$\a_5'$}
\put(206,13){$\a_6'$}
\put(231,13){$\a_7'$}
\put(256,13){$\a_9'$}
\put(270,25){ $\rightarrow$ }
\multiput(294,25)(25,0){4}{\circle{6}}
\multiput(297,25)(25,0){3}{\line(1,0){19}}
\multiput(393,7)(0,36){2}{\circle{6}}
\put(371,27){\line(4,3){19}}
\put(371,23){\line(4,-3){19}}
\put(290,13){$\a_2'$}
\put(315,13){$\a_3'$}
\put(340,13){$\a_4'$}
\put(376,22){$\a_5'$}
\put(398,40){$\a_8'$}
\put(398,4){$\a_6'$}
\put(398,22){ $\times$ }
\put(417,22){$\C$}
\end{picture}
\vskip 15pt 
\begin{picture}(340,40)\thicklines
\put(0,12){$T^8$:}
\put(20,12){$\e^{+}_8\rightarrow \e_7\oplus \C$:}
\put(205,35){\circle{6}}
\put(205,32){\line(0,-1){14}}
\multiput(105,15)(20,0){8}{\circle{6}}
\multiput(108,15)(20,0){7}{\line(1,0){14}}
\put(211,32){$\a_8'$}
\put(101,3){$\a_9'$}
\put(121,3){$\a_1'$}
\put(141,3){$\a_2'$}
\put(161,3){$\a_3'$}
\put(181,3){$\a_4'$}
\put(201,3){$\a_5'$}
\put(221,3){$\a_6'$}
\put(241,3){$\a_7'$}
\put(255,15){ $\rightarrow$ }
\multiput(280,15)(20,0){6}{\circle{6}}
\multiput(283,15)(20,0){5}{\line(1,0){14}}
\put(340,35){\circle{6}}
\put(340,32){\line(0,-1){14}}
\put(276,3){$\a_2'$}
\put(296,3){$\a_3'$}
\put(316,3){$\a_4'$}
\put(336,3){$\a_5'$}
\put(346,32){$\a_8'$}
\put(356,3){$\a_6'$}
\put(376,3){$\a_7'$}
\put(388,12){ $\times$ }
\put(405,12){$\C$}
\end{picture}
\vskip 7pt
\caption{Diagrammatic method for $T^{2}/\Z_{n>2}$ orbifolds of
M-theory}
\label{KacPetDiag}
\end{figure}
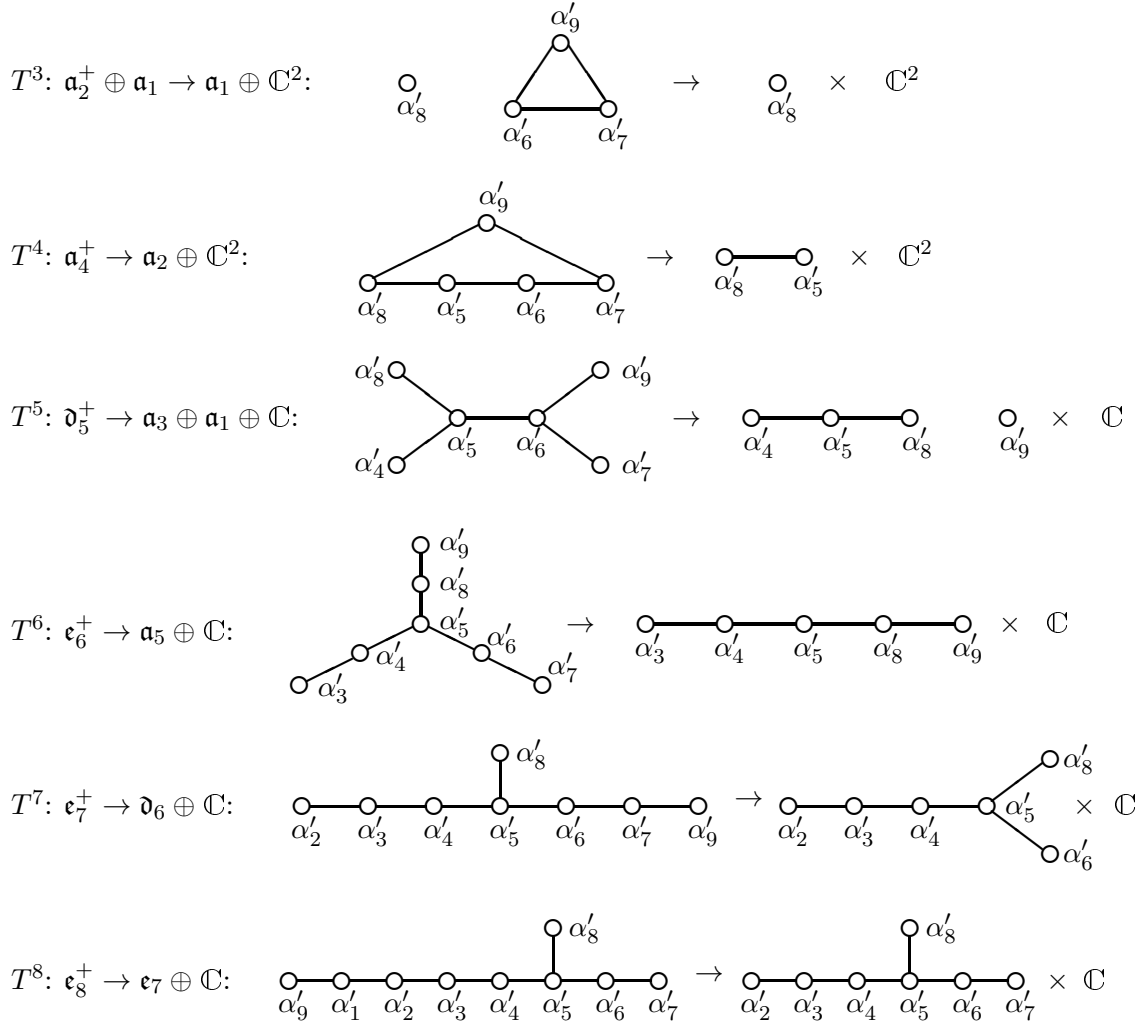

Since the orbifold is not acting on other space-time directions, it seems
logical to extend this construction by taking $\a'_{i}=\a_{i}$ and 
$l_{i}=0$ $\forall i<6$ for all $T^{p}\times T^{2}/\Z_{n}$ orbifolds. 
Indeed, for $p\leq 4$, we obtain 
$(\La'^{\{2\}}|\theta_{A_{4}})=n-1\rightarrow l_{9}^{A_{4}}=1$, 
$(\La'^{\{2\}}|\theta_{D_{5}})=n\rightarrow l_{9}^{D_{5}}=0$ and 
$(\La'^{\{2\}}|\theta_{E_{6}})=n\rightarrow l_{9}^{E_{6}}=0$, giving 
the results in Figure~\ref{KacPetDiag}. Comparing Figure~\ref{KacPetDiag} 
with Table~\ref{T2/ZnDynkin} in $D=6$, we can identify 
$\a'_{9}=-\theta_{D_5}$.

However, looking at the respective invariant subdiagrams in $D=5$, it is
clear that one should not choose $\a'_{i}=\a_{i}$ $\forall i<6$. Looking at 
Table~\ref{T2/ZnDynkin}, one guesses that $\a'_{3}=-\theta_{D_{5}}$, 
$\a'_{4}=-\a_{3}$ and $\a'_{5}=-\a_{4}$. Since there is only one element 
in the eigensubspace $\g^{(n-2)}$, we also have to take 
$E'_{7}=\frac{1}{2}(H_{7}-i(E_{7}+F_{7}))$, as before. On the other hand, 
there are now plenty of objects in $\g^{(1)}$, all of them not commuting 
with $E'_{7}$. One should find one that commutes with $F_{\theta_{D_5}}$ 
and $F_{\a_3}$ but not with $F_{\a_4}$. This suggests to set 
$E'_{6}=\frac{1}{\sqrt{2}}(E_{34^{2}5^{2}68}-iE_{34^{2}5^{2}678})$. 
Finally, we also take $E'_{8}=F_{5}$. Computing the expression of the generator 
corresponding to the highest root in this new basis gives 
$E'_{9}=F_{\theta_{E_6}}=iF_{8}$, as one would expect. Since the shift vector 
simply takes the same form $\La^{'\{2\}}=\La^{'6}+(n-2)\La^{'7}$ on a new basis 
of fundamental weights, the naive guess above was correct.

From $D=4$ downwards, this ceases to be true, since naively
$(\La^{'\{2\}}|\theta_{E_7})=2n-1>n$. In fact, we should again change basis 
in $\e_{7}$. Comparing again Figure~\ref{KacPetDiag} with Table~\ref{T2/ZnDynkin}, 
we see that there are 2 different equivalent ways to choose the 2 roots to be discarded, 
on the left ($\a'_{2}$ and $\a'_{3}$) or on the right ($\a'_{7}$ and $\a'_{9}$). 
We choose the latter, since it will be easier to generalize to $\e_{8}$. 
Indeed, in the extended diagram of $\e_{8}$, the Coxeter label of $\a'_{9}$ will be 
the only one to be 1, making $l_{9}^{E_{8}}=n-2$ the only possible choice. 
Further inspection of Figure~\ref{KacPetDiag} and Table~\ref{T2/ZnDynkin} suggests 
to take the new basis as follows: $\a'_{2}=-\a_{8}$, $\a'_{3}=-\a_{5}$, $\a'_{4}=-\a_{4}$, 
$\a'_{5}=-\a_{3}$, $\a'_{6}=-\theta_{D_{5}}$, $\a'_{8}=-\a_{2}$ and finally 
\begin{equation*}
E'_{7}=\frac{1}{\sqrt{2}}(E_{23^{2}4^{3}5^{4}6^{3}78^{2}}-iE_{23^{2}4^{3}5^{4}6^{3}7^{2}8^{2}})
\in \g^{(1)}\,.
\end{equation*}
A lengthy computation allows to show that this choice leads to \\
$E'_{9}=F_{\theta'_{E_7}}=(i/2)(H_{7}-i(E_{7}+F_{7}))\in \g^{(n-2)}$ 
as it should, giving the shift vector: 
$\La=\La^{'7}$ with $l_{9}^{E_{7}}=n-2$.

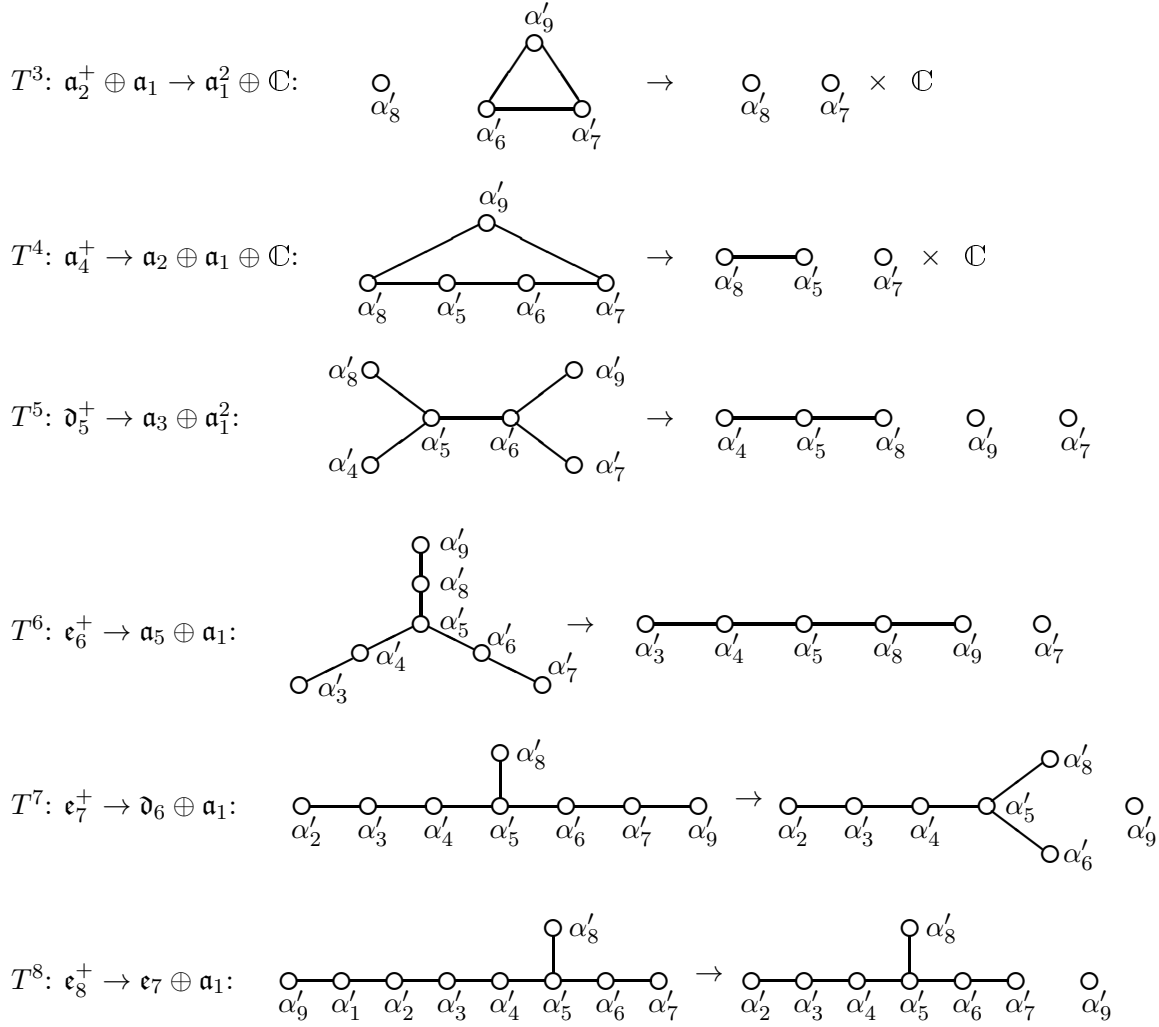
\begin{figure}[!t]
\begin{picture}(350,50)\thicklines
\put(0,22){$T^3$:}
\put(20,22){$\an^+_2\oplus\an_1\rightarrow \an_1^2\oplus\C$:}
\put(140,25){\circle{6}}
\put(136,13){$\a'_8$}
\multiput(180,15)(36,0){2}{\circle{6}}
\put(183,15){\line(1,0){30}}
\put(198,40){\circle{6}}
\put(195,39){\line(-2,-3){14}}
\put(201,39){\line(2,-3){14}}
\put(194,47){$\a'_9$}
\put(176,3){$\a'_6$}
\put(212,3){$\a'_7$}
\put(240,22){$\rightarrow$}
\multiput(280,25)(30,0){2}{\circle{6}}
\put(276,13){$\a'_8$}
\put(306,13){$\a'_7$}
\put(320,22){ $\times$ }
\put(340,22){$\C$}
\end{picture}
\vskip 15pt 
\begin{picture}(350,50)\thicklines
\put(0,22){$T^4$:}
\put(20,22){$\an^+_4\rightarrow \an_2\oplus\an_1\oplus\C$:}
\multiput(135,15)(30,0){4}{\circle{6}}
\multiput(138,15)(30,0){3}{\line(1,0){24}}
\put(180,38){\circle{6}}
\put(177,37){\line(-2,-1){40}}
\put(183,37){\line(2,-1){40}}
\put(177,45){$\a'_9$}
\put(131,3){$\a'_8$}
\put(161,3){$\a'_5$}
\put(191,3){$\a'_6$}
\put(221,3){$\a'_7$}
\put(240,22){$\rightarrow$}
\multiput(270,25)(30,0){3}{\circle{6}}
\put(273,25){\line(1,0){24}}
\put(266,13){$\a'_8$}
\put(296,13){$\a'_5$}
\put(326,13){$\a'_7$}
\put(340,22){ $\times$ }
\put(360,22){$\C$}
\end{picture}
\vskip 15pt 
\begin{picture}(350,50)\thicklines
\put(0,27){$T^5$:}
\put(20,27){$\dn^+_5\rightarrow \an_3\oplus\an_1^2$:}
\multiput(136,12)(0,36){2}{\circle{6}}
\put(138,14){\line(4,3){19}}
\put(138,46){\line(4,-3){19}}
\multiput(159,30)(30,0){2}{\circle{6}}
\put(162,30){\line(1,0){24}}
\multiput(213,12)(0,36){2}{\circle{6}}
\put(191,32){\line(4,3){19}}
\put(191,28){\line(4,-3){19}}
\put(120,10){$\a'_4$}
\put(120,45){$\a'_8$}
\put(155,18){$\a'_5$}
\put(181,18){$\a'_6$}
\put(221,45){$\a'_{9}$}
\put(221,9){$\a'_{7}$}
\put(240,27){$\rightarrow$}
\multiput(270,30)(30,0){3}{\circle{6}}
\multiput(273,30)(30,0){2}{\line(1,0){24}}
\put(267,18){$\a'_4$}
\put(297,18){$\a'_5$}
\put(327,18){$\a'_8$}
\put(365,30){\circle{6}}
\put(362,18){$\a'_9$}
\put(400,30){\circle{6}}
\put(397,18){$\a'_7$}
\end{picture}
\vskip 15pt 
\begin{picture}(350,60)\thicklines
\put(0,22){$T^6$:}
\put(20,22){$\e^+_6\rightarrow \an_5\oplus\an_1$:}
\multiput(155,28)(0,15){3}{\circle{6}}
\multiput(155,31)(0,15){2}{\line(0,1){9}}
\multiput(132,17)(-23,-12){2}{\circle{6}}
\multiput(152,27)(-23,-12){2}{\line(-2,-1){17}}
\multiput(178,17)(23,-12){2}{\circle{6}}
\multiput(158,27)(23,-12){2}{\line(2,-1){17}}
\put(116,1){$\a'_3$}
\put(138,13){$\a'_4$}
\put(162,56){$\a'_9$}
\put(162,41){$\a'_8$}
\put(162,26){$\a'_5$}
\put(180,20){$\a'_6$}
\put(203,9){$\a'_7$}
\put(210,25){$\rightarrow$}
\multiput(240,28)(30,0){5}{\circle{6}}
\multiput(243,28)(30,0){4}{\line(1,0){24}}
\put(236,16){$\a'_3$}
\put(266,16){$\a'_4$}
\put(296,16){$\a'_5$}
\put(326,16){$\a'_8$}
\put(356,16){$\a'_9$}
\put(390,28){\circle{6}}
\put(387,16){$\a'_7$}
\end{picture}
\vskip 15pt 
\begin{picture}(340,50)\thicklines
\put(0,22){$T^7$:}
\put(20,22){$\e^{+}_7\rightarrow \dn_6\oplus \an_1$:}
\put(185,45){\circle{6}}
\put(185,42){\line(0,-1){14}}
\multiput(110,25)(25,0){7}{\circle{6}}
\multiput(113,25)(25,0){6}{\line(1,0){19}}
\put(191,42){$\a_8'$}
\put(106,13){$\a_2'$}
\put(131,13){$\a_3'$}
\put(156,13){$\a_4'$}
\put(181,13){$\a_5'$}
\put(206,13){$\a_6'$}
\put(231,13){$\a_7'$}
\put(256,13){$\a_9'$}
\put(270,25){ $\rightarrow$ }
\multiput(294,25)(25,0){4}{\circle{6}}
\multiput(297,25)(25,0){3}{\line(1,0){19}}
\multiput(393,7)(0,36){2}{\circle{6}}
\put(371,27){\line(4,3){19}}
\put(371,23){\line(4,-3){19}}
\put(290,13){$\a_2'$}
\put(315,13){$\a_3'$}
\put(340,13){$\a_4'$}
\put(376,22){$\a_5'$}
\put(398,40){$\a_8'$}
\put(398,4){$\a_6'$}
\put(425,25){\circle{6}}
\put(422,13){$\a'_9$}
\end{picture}
\vskip 15pt 
\begin{picture}(340,40)\thicklines
\put(0,12){$T^8$:}
\put(20,12){$\e^{+}_8\rightarrow \e_7\oplus \an_1$:}
\put(205,35){\circle{6}}
\put(205,32){\line(0,-1){14}}
\multiput(105,15)(20,0){8}{\circle{6}}
\multiput(108,15)(20,0){7}{\line(1,0){14}}
\put(211,32){$\a_8'$}
\put(101,3){$\a_9'$}
\put(121,3){$\a_1'$}
\put(141,3){$\a_2'$}
\put(161,3){$\a_3'$}
\put(181,3){$\a_4'$}
\put(201,3){$\a_5'$}
\put(221,3){$\a_6'$}
\put(241,3){$\a_7'$}
\put(255,15){ $\rightarrow$ }
\multiput(280,15)(20,0){6}{\circle{6}}
\multiput(283,15)(20,0){5}{\line(1,0){14}}
\put(340,35){\circle{6}}
\put(340,32){\line(0,-1){14}}
\put(276,3){$\a_2'$}
\put(296,3){$\a_3'$}
\put(316,3){$\a_4'$}
\put(336,3){$\a_5'$}
\put(346,32){$\a_8'$}
\put(356,3){$\a_6'$}
\put(376,3){$\a_7'$}
\put(408,15){\circle{6}}
\put(405,3){$\a'_9$}
\end{picture}
\vskip 7pt
\caption{Diagrammatic method for $T^{2}/\Z_{2}$ orbifolds of M-theory}
\label{KacPetDiag2}
\end{figure}

In $\e_{8}$, a similar game leads to 
$\a'_{2}=-\a_{8}$, $\a'_{3}=-\a_{5}$, $\a'_{4}=-\a_{4}$, 
$\a'_{5}=-\a_{3}$, $\a'_{6}=-\a_{2}$, $\a'_{7}=-\a_{1}$, 
$\a'_{8}=-\theta_{D_{5}}$ and finally 
\begin{equation*}
E'_{1}=\frac{1}{\sqrt{2}}(E_{12^{2}3^{3}4^{4}5^{5}6^{3}78^{3}}-iE_{12^{2}3^{3}4^{4}5^{5}6^{3}7^{2}8^{3}})\in 
\g^{(1)}\,,
\end{equation*}
leading to $E'_{9}=F_{\theta'_{E_8}}=\frac{i}{2}(H_{7}-i(E_{7}+F_{7}))\in \g^{(n-2)}$ 
with shift vector $\La'=\La^{'1}$, while $l_{9}^{E_{8}}=n-2$. \vskip5pt 
\vskip5pt The results for $n=2$ can be obtained directly by putting $l_{7}=0$, 
adding one more node to the diagrams instead of the abelian $\u(1)$ factor. 
The results are summarized in Figure~\ref{KacPetDiag2}. 
\vskip 5pt 
It was instructive to compare our method based on automorphisms induced by algebraic
rotations and the standard classification of Lie algebra automorphisms based
on shift vectors defining chief inner automorphisms. However, the mapping from one language 
to the other can be fairly obscure, in particular for orbifolds more complicated than 
the $T^{2}/\Z_{n}$ case treated above. In fact, in the $\an_{r}$ serie of Lie algebras, 
for which the Coxeter labels are all equal, the necessary change of basis can be computed 
only once and trivially extended to larger algebras in the serie. In general, and in 
particular for exceptional algebras, one has to perform a different change of Cartan-Weyl 
basis whenever we consider the \textit{same} orbifold in a larger U-duality symmetry algebra 
(or, geometrically speaking, when we compactify one more dimension).

Our method based on non-Cartan preserving automorphisms is thus more
appropriate to treat a few particular orbifolds in a serie of algebras that
are successively included one into the other, as is the case for the
U-duality algebras of compactified supergravity theories. On the other hand,
the method based on chief inner automorphisms is more amenable to classify all
orbifolds of a unique algebra, for example all possible breakings of a given
gauge group under an orbifold action. For instance, the breakings of the 
$E_{8}\times E_{8}$ gauge group of heterotic string theory have been
treated this way by~\cite{Koba,Choi}. It is also easier to generalize
the method based on algebraic rotations to the infinite-dimensional case, 
since we can use the decomposition of $\eten$ in tensorial representations of 
$\sl(10)$ and our intuition on the behaviour of tensorial indices under 
a physical rotation to identify non-trivial invariant objects.

We can draw a related conclusion from the explicit forms of the above 
basis transformation: when the orbifold is expressed in terms of the standard
shift vector satisfying $(\La|\theta)\leq n$, the geometric interpretation of 
the orbifold action gets blurred. More precisely, the directions in which 
the rotation is performed is determined above by the roots $\a'_i$ with 
coefficients $l_{i}=n-2$. For example, in $\e_{7}$, our original Lorentz 
rotation by ${\cal K}_{9\, 10}$ represented by $\a_7$ appears in the standard 
basis as a gauge transformation generated by $\widetilde{\mathcal{Z}}_{456789}$. 
Similarly, in $\e_{8}$, it seems that we are rotating in a direction
corresponding to $(\widetilde{\mathcal{K}}_{3})_{3456789\,10}$. Of course,
mathematically, all conjugate Cartan-Weyl basis in a Lie algebra give rise
to an isomorphic gradation of $\g$, but the physical interpretation based 
on the decomposition of $\e_{r}$ in tensorial representations of $\sl(r)$ is
obscured by the conjugation.

Indeed, our $T^{q}/\Z_{2}$ and $T^{q}/\Z_{n}$ orbifolds for $q=2,4$ all appear 
in the classification of $T^{6}/\Z_{n}$ orbifolds given in~\cite{Koba}, where 
they are interpreted as $T^{6}/\Z_{n}$ orbifolds with particularly small breakings of
the gauge group and degenerate shift vectors (in the sense of having lots of 0). 
It is however clear in our formalism that this degeneracy should actually be 
seen as having considered a rotation of null angle in certain directions.

\subsection{Classes of shift vectors for $T^{q}/\Z_{n}$ orbifolds, for $q=4,6$}\label{Shift46}

In the more complicated cases of $T^{4}/\Z_{n}$ and $T^{6}/\Z_{n}$ orbifolds, 
we will not give in detail the basis transformations necessary to obtain the 
standard shift vectors satisfying $(\La'|\theta )\leq n$ for the whole serie of U-duality
algebras. Rather, we will give the shift vectors in their universal form, which
is valid for the whole serie of U-duality algebras. In particular, for $T^{4}/\Z_{n}$, 
the gradation of $\dn_{5}$ by eigensubspaces of $\mathcal{U}_{4}^{\Z_{n}}$ has been given in 
expressions~(\ref{Charge0forT4/Zn}),~(\ref{Charge1forT4/Zn}) and~(\ref{Charge2forT4/Zn}). 
A particularly natural choice of diagonal Cartan-Weyl basis for this
decomposition is obtained by taking: 
\begin{equation}
\begin{array}{ll}
E'_{4}=\frac{1}{\sqrt{2}}(E_{4}+iE_{45})\,, & 
H'_{4}=\frac{1}{2}(2H_{4}+H_{5}-i(E_{5}-F_{5}))\,, \\[2pt] 
E'_{5}=\frac{1}{2}(H_{5}-i(E_{5}+F_{5}))\,, & 
H'_{5}=i(E_{5}-F_{5})\,, \\[2pt] 
E'_{6}=\frac{1}{\sqrt{2}}(E_{56}-E_{67}+i(E_{567}+E_{6}))\,, & 
H'_{6}=\frac{1}{2}(H_{5}+2H_{6}+H_{7}-i(E_{5}-F_{5}+E_{7}-F_{7}))\,,\\[2pt] 
E'_{7}=\frac{1}{\sqrt{2}}(H_{7}-i(E_{7}+F_{7}))\,, & 
H'_{7}=i(E_{7}-F_{7})\,, \\[2pt] 
E'_{8}=-\frac{1}{\sqrt{2}}(E_{8}-iE_{58})\,, & 
H'_{8}=\frac{1}{2}(2H_{8}+H_{5}-i(E_{5}-F_{5}))\,,
\end{array}
\label{DiagBasisT4/Zn}
\end{equation}
while $F'_{i}$ is obtained from $E'_{i}$ as above by conjugation and exchange 
of $E$ and $F$. This leads to the following eigensubspace decomposition of $\dn_{5}$: 
\begin{equation}
\begin{array}{rl}
\g^{(0)}= & \Span\{H'_{4};H'_{5};H'_{6};H'_{7};H'_{8};E'_{6};E'_{567};E'_{458};E'_{4568};
E'_{45^{2}678};E'_{45^{2}6^{2}78};E'\leftrightarrow F'\}\,,\\[2pt] 
\g^{(1)}= & \Span\{E'_{4};E'_{8};E'_{4567};E'_{5678};F'_{45};F'_{58};F'_{456};F'_{568}\}\,, \\[2pt] 
\g^{(2)}= & \Span\{E'_{7};E'_{67};E'_{45678};F'_{5};F'_{56};F'_{45^{2}68}\}\,.
\end{array}
\label{GraduationT4/Z2}
\end{equation}
The shift vector corresponding to this basis is given by: 
\begin{equation*}
\La^{'\{4\}}=\La^{'4}+(n-2)\La^{'5}+2\La^{'7}+\La^{'8}\,,
\end{equation*}
which clearly reduces to $\La^{'\{4\}}=\La^{'4}+\La^{'8}$ in the case of $T^{4}/\Z_{2}$.
By simply taking $E'_{i}=E_{i}$ for any additional roots that are unaffected by the orbifold 
action, this shift vector is valid in $\e_r$, for $r=6,..,10$, as well.

For the case of $T^{6}/\Z_{n}$, we take the following Cartan-Weyl basis: 
\begin{equation}
\begin{array}{ll}
E'_{2}=\frac{1}{\sqrt{2}}(E_{2}+iE_{23})\,, & 
H'_{2}=\frac{1}{2}(2H_{2}+H_{3}-i(E_{3}-F_{3}))\,, \\[2pt] 
E'_{3}=\frac{1}{2}(H_{3}-i(E_{3}+F_{3}))\,, & 
H'_{3}=i(E_{3}-F_{3})\,, \\[2pt] 
E'_{4}=\frac{1}{\sqrt{2}}(E_{34}+E_{45}-i(E_{345}-E_{4}))\,, & 
H'_{4}=\frac{1}{2}(H_{3}+2H_{4}+H_{5}+i(-E_{3}+F_{3}+E_{5}-F_{5}))\,,\\[2pt] 
E'_{5}=\frac{1}{2}(H_{5}+i(E_{5}+F_{5}))\,, & 
H'_{5}=-i(E_{5}-F_{5})\,, \\[2pt] 
E'_{6}=\frac{1}{\sqrt{2}}(E_{56}-E_{67}-i(E_{567}+E_{6}))\,, & 
H'_{6}=\frac{1}{2}(H_{5}+2H_{6}+H_{7}+i(E_{5}-F_{5}+E_{7}-F_{7}))\,,\\[2pt] 
E'_{7}=\frac{1}{\sqrt{2}}(H_{7}+i(E_{7}+F_{7}))\,, & 
H'_{7}=-i(E_{7}-F_{7})\,, \\[2pt] 
E'_{8}=-\frac{1}{\sqrt{2}}(E_{8}+iE_{58})\,, & 
H'_{8}=\frac{1}{2}(2H_{8}+H_{5}+i(E_{5}-F_{5}))\,,
\end{array}
\label{DiagBasisT6/Zn}
\end{equation}
that leads to the universal shift vector: 
\begin{equation*}
\La^{'\{6\}}=\La^{'2}+(n-2)\La^{'3}+2\La^{'5}+\La^{'6}+(n-4)\La^{'7}+(n-1)\La^{'8}\,,
\end{equation*}
that is valid in $\e_{8}$, $\e_{9}$ and $\e_{10}$, as well. It is obvious in this 
form that the degeneration of the coefficient $l^7$ when $n=4$ leads to a larger invariant 
subalgebra with fewer abelian factors. On the other hand, as the invariant subalgebras for 
$T^{6}/\Z_{4}$ and $T^{6}/\Z_{3}$ both have no abelian factors, the coefficients of 
$\La^{'\{6\}}$ does not allow to discriminate between them. Another fact worth noting
is that setting $n=2$ leads to $\La^{'\{6\}}=\La^{'2}+\La^{'6}+\La^{'8}$, corresponding to a
$T^{4}/\Z_{2}$ orbifolds with respect to the nodes $\a_{3}$ and $\a_{5}$ and not to a 
$T^{6}/\Z_{2}$ orbifolds. This is natural since we chose the charge in the $(x^9,x^{10})$-plane
to be $Q_{3}=-2$, so that it reduces to the identity rotation for $n=2$.

\subsection{Roots of $\eten$ as physical class representatives}
\label{Classrep}

The universal shift vectors are mathematically interesting, but the original
motivation to compute them was actually to give a physical interpretation of
certain roots of $\eten$. Typically, our universal shift vectors 
$\La'$ are not roots, but we can use the self-duality of $Q(\eten)$ and 
the periodicity modulo $n$ of the orbifold action to replace $\La'$ by a root
$\xi$ generating the same orbifold action.

Self-duality of $Q(\eten)$ relates the weight $\La'$ to a vector in the root lattice 
satisfying $(\La'|\a')=(\tilde{\xi}|\a')$ $\forall \a' \in\Delta(\g,\h')$. However, 
every root lattice vector is not a root. One should thus use the equivalence modulo $n$: 
$\La'\equiv \La'+n\sum_{i=-1}^8 c_i \La'^i=\overline{\La}'$, for any 10-dimensional vector 
$\vec{c}\in\Z^{10}$, to find a weight $\overline{\La}'=\sum_{i=-1}^{8}l_{i}\La^{'i}$ such that: 
\begin{equation*}
\xi=\sum_{i,j=-1}^{8}(A^{-1})^{ij}l_{j}\a'_{i}
\end{equation*}
is a root of $\eten$. In fact, such a condition does not fix $\xi$ uniquely either. 
However, it seems that there is a unique way to choose $\vec{c_{q}}$ so that $\xi^{[q,n]}$ is 
a root describing the orbifold $T^{10-q}\times T^{q}/\Z_{n}$ for all values of $n$.

From that point of view, we can see the shift vector as containing two parts: 
the \textit{universal part}, that reflects the choices of orbifold directions and charges, 
and the \textit{$n$-dependent part}, that defines the orbifold periodicity.

Concretely, it seems that $\vec{c_{q}}$ can always be chosen to be dual to a
Weyl reflection of $\d$ (at least for even orbifolds). In the case of $T^{2}/\Z_{2}$, 
for example, we had the universal part $\La^{'\{2\}}=\La^{'6}-2\La^{'7}$, which is dual 
to $-\a'_{7}$. Adding $n(\La^{'7}-\La^{'8})$, i.e. the root 
$n\tilde{\d}^{[2]}=n(\d'+\sum_{i=-1}^7\a'_i)$, we obtain the desired form of shift vector 
in the physical basis: 
\begin{equation*}
\xi^{[2,n]}=(n,n,n,n,n,n,n,n,n-1,1)\,.
\end{equation*}
From the tables of~\cite{NicFisch} it is easy to verify that this is a root of $\eten$ with 
$l=3n$ for all values of $n\leq 6$, and it is very likely to be a root for any integer value 
of $n$. Note also that translating the results back in the original basis gives: 
\begin{equation*}
e^{i\frac{2\pi}{n}(\xi^{[2,n]}|\a')}{\mathfrak{g}}_{\a'}=
e^{i2\pi (\a'|\tilde{\d}^{[2]})}\,\Ad e^{\frac{2\pi}{n}(E_{7}-F_{7})}\g_{\a'}\,,
\end{equation*}
where the first factor does not contribute to the charge, so that the
equivalence between the two descriptions, one in terms of shift vectors and the 
other of in terms of non-Cartan preserving inner automorphisms, is obvious.

For $T^{4}/\Z_{n}$, we similarly take 
$\overline{\La}^{'\{4\}}=\La^{'4}-2\La^{'5}+2\La^{'7}+\La^{'8}+n(\La^{'5}-\La^{'6}-\La^{'8})$,
which corresponds to
\begin{eqnarray*}
\xi^{[4,n]} &=&-\a'_{5}+\a'_{7}+n(\d'+\a'_{-1}+\a'_{0}+\a'_{1}+\a'_{2}+\a'_{3}+\a'_{4}+\a'_{5})\\
&=&(n,n,n,n,n,n,n-1,1,n+1,n-1)\,.
\end{eqnarray*}%
Again, this is indeed a root $\forall n\leq 6$ and it can be checked to
reduce to one of the 4 possible permutations proposed in~\cite{Gan2} for $n=2$. 
Furthermore, $H_{\xi^{[4,n]}}=nH_{\tilde{\d}^{[4]}}-i(E_{5}-F_{5}-E_{7}+F_{7})$ 
as one would expect.

Finally, for $T^{6}/\Z_{n}$, one can check that
$\overline{\La}^{'\{6\}}=\La^{'2}-2\La^{'3}+2\La^{'5}+\La^{'6}-4\La^{'7}-\La^{'8}
+n(\La^{'7}-\La^{'8})$ has all desired properties.
It is dual to 
\begin{eqnarray*}
\xi^{[6,n]} &=&-\a'_{3}+\a'_{5}-2\a'_{7}+n(\d'+\a'_{-1}+\a'_{0}+\a'_{1}+\a'_{2}+
\a'_{3}+\a'_{4}+\a'_{5}+\a'_{6}+\a'_{7})\\
&=&(n,n,n,n,n-1,n+1,n+1,n-1,n-2,2)\,,
\end{eqnarray*}
where the factor of -2 in front of $\a'_{7}$ reminds us of the charge assignment $Q_{3}=-2$.
On the other hand, the - sign in front of $\a'_{3}$ does not contradict our choice of $Q_{1}=+1$,
but is rather due to our Cartan-Weyl basis~(\ref{DiagBasisT6/Zn}), in which $H'_{3}$ has a different
conventional sign compared to $H'_{5}$ and $H'_{7}$. Accordingly, one obtains:
$H_{\xi^{[6,n]}}=nH_{\tilde{\d}^{[6]}}-i(E_{3}-F_{3}+E_{5}-F_{5}-2(E_{7}+F_{7}))$ as it should.

It is now easy to guess the general form of the shift vector for all 
$T^{10-q}\times T^{q}/\Z_{n}$ orbifolds, in which the orbifold projections 
are taken independently on each of the $(q/2)$ $T^{2}$ subtori (in other words, 
we exclude for example a $\Z_{3}\times \Z_{3}$ orbifold of $T^{6}$ for which one 
$\Z_{3}$ acts on the planes $\{x^5,x^6\}$ and $\{x^7,x^8\}$ and the other on the 
planes $\{x^7,x^8\}$ and $\{x^9,x^{10}\}$, since it contains two independent 
projections on the same $T^{2}$ subtorus).

By translating the tables of $\eten$ roots established by~\cite{NicFisch} in the 
physical basis, we can identify the roots which constitute class representatives 
of shift vectors (satisfying the conditions mentioned above) for orbifolds 
with various charge assignments, and build the classification represented 
in Tables~\ref{Rep1} and~\ref{Rep2}. These listings deserve a few comments.

First of all, what we are really classifying are inner automorphisms of the
type~(\ref{OrbifoldAction}) with all different charges assignments (up to
permutations of the shift vectors). Though some of these automorphisms allow
to take a geometrical orbifold projection and descend to well-defined type
IIA orbifolds, like the $T^{4}/\Z_{n}$ and the $T^{6}/\Z_{n}$ 
cases\footnote{This kind of $T^{6}/\Z_{n}$ orbifold with charge 
assignment $(1,1,-2)$ is denoted $T^{4}/\Z_{n}\times T^{2}/\Z_{n/2}$ 
in Tables~\ref{Rep1} and~\ref{Rep2} to distinguish it from the one
with charge assignment $(1,-1,1)$.} we studied in Sections~\ref{secT4} 
and~\ref{secT6} for $n=2,3,4,6$, the Lefschetz fixed point
formula would give a non-integer number of fixed points for most of the
others. Clearly, such cases do not correspond to compactifications on
geometrical orbifolds that can be made sense of in superstring theory (let
alone preserve some supersymmetry). However, whether compactifications on
such peculiar spaces makes sense in M-theory is, on the other hand, an open
question. The invariant subalgebras and "untwisted" sectors can in any case be
defined properly.

Second, we chose not to consider as different two shift vectors differing
only by a permutation of $\tilde{\d},$ but exhibiting the same universal part, 
for example $(3,3,3,3,3,3,3,3,2,1)$ and $(3,3,3,3,3,3,3,0,2,4)$.

Finally, looking at the Tables~\ref{Rep1} and~\ref{Rep2} in an horizontal way, 
one can identify series of shift vectors defining orbifold charges which appear 
as "subcharges" one of the others, when some $Q_{i}$'s are set to zero. 
For example, starting from $T^{8}/\Z_{6}\times T^{2}/\Z_{3}$ for $q=10$, 
one obtains successively $T^{6}/\Z_{6}\times T^{2}/\Z_{3}$, 
$T^{4}/\Z_{6}\times T^{2}/\Z_{3}$, $T^{2}/\Z_{6}\times T^{2}/\Z_{3}$
and $T^{2}/\Z_{3}$ for $q=8,6,4,2$, with shift vectors of
monotonally decreasing squared lengths -8, -10, -12, -14 and -16. 

Though, for evident typographical reasons, we were not able to accomodate 
all shift vectors related in this way on the same line, we have done so 
whenever possible to highlight the appearance of such families of class 
representatives. This explains the blank lines, whenever there was no such 
correspondence. An attentive study of Tables~\ref{Rep1} and~\ref{Rep2} shows 
that those families end up when a root of the serie reaches squared length 2. 
For example, going backwards and starting instead from $T^{6}/\Z'_{3}$ with
a shift vector of null squared length for $q=6$, one finds 
$T^{2}/\Z_{6}\times T^{6}/\Z_{3}$ with a vector of squared length 2 for $q=8$, 
but there is no $T^{4}/\Z_{6}\times T^{6}/\Z_{3}$ for $q=10$, since it would 
have to be generated by a vector of length 4, which is of course not a root.

To extend this classification to orbifolds that are not induced by an
automorphism of type (\ref{OrbifoldAction}), further computations are
nevertheless necessary (to obtain the correct form of the universal parts).
However, exactly the same methods can in principle be applied and we leave
this matter for further research. When tables of roots of ${\mathfrak{e}_{10}%
}$ will be available up to higher levels in $\alpha _{8}$, one could also
study orbifolds for higher values of $n.$ Of physical interest are perhaps
values of $n$ up to 12, which would in principle require knowledge of roots
of levels up to 36.

A more speculative question is whether these orbifold-generating roots all
have another physical interpretation, for example as solitonic M-theory
objects with or without non-trivial fluxes, just as in~\cite{Gan2}. A first look at
the general shape of these roots in the physical basis seems to confirm this
view, since the first $(10-q)$ $n$'s remind of a $(10-q)$-brane transverse
to the orbifolded torus, while the other components might be given an
interpretation as fluxes through the orbifold. Indeed, both are expected to
contribute to local anomaly cancellation at the orbifold fixed points.
We do not have a general realization of this idea, yet, but we will describe a number 
of more concrete constructions in the following and discuss in particular all of
the  $\Z_{2}$ cases in detail, hinting at a possible interpretation 
of the general $\Z_n$ ones.

\renewcommand\arraystretch{1.2}
\begin{sidewaystable}[p]
\begin{center}
\begin{tabular}{l}
{\large {\bf q=2}} \\[5pt]
$
\begin{array}{|c|c|c|c|c||c|c|c|c|c|}\hline
n & \text{Orbifold} & \text{shift vector} & Q_{1} & |\xi |^{2}&  n & \text{Orbifold}& 
\text{shift vector}  &  Q_{1} & |\xi|^{2} \\ \hline\hline
3 & T^{2}/\Z_{3} & (3,3,3,3,3,3,3,3,2,1) & 1 & -4  &5 & T^{2}/\Z_{5} & 
(5,5,5,5,5,5,5,5,4,1) & 1 & -8 \\
&  &  &  &  &  &  T^{2}/\Z'_{5}& (5,5,5,5,5,5,5,5,3,2) & 2 & -12 \\  \hline
4 & T^{2}/\Z_{4} & (4,4,4,4,4,4,4,4,3,1) & 1 & -6 & 6 & T^{2}/\Z_{6} & 
(6,6,6,6,6,6,6,6,5,1) & 1 & -10 \\ 
& T^{2}/\Z'_{2} & (4,4,4,4,4,4,4,4,2,2) & 2 & -8 &  & T^{2}/\Z'_{3} & (6,6,6,6,6,6,6,6,4,2) & 2 & -16 \\ 
&  &  &  &  & &  T^{2}/\Z''_{2} & (6,6,6,6,6,6,6,6,3,3) & 3 & 
-18\\ \hline
\end{array}
$
\\[5pt] \\
{\large {\bf q=4}} \hspace{9.7cm}{\large {\bf q=6}}\\[5pt]
$
\begin{array}{|c|c|c|c|c|l|c|c|c|c|c|} \cline{1-5}\cline{7-11}
n & \text{Orbifold} & \text{shift vector} & (Q_{1},Q_{2}) & |\xi |^{2} & &n & \text{Orbifold} & 
\text{shift vector} & (Q_{1},Q_{2},Q_{3}) & |\xi |^{2} \\ \hhline{:=====:~:=====:}
3 & T^{4}/\Z_{3} & (3,3,3,3,3,3,4,2,2,1) & (1,-1) & -2 & &3 
& T^{6}/\Z_{3} & (3,3,3,3,2,4,4,2,2,1) & (-1,1,-1) & 0 \\ 
\cline{1-5}\cline{7-11} 
4& T^{4}/\Z_{4} & (4,4,4,4,4,4,5,3,3,1) & (1,-1) & -4 & &4 
& T^{6}/\Z_{4} & (4,4,4,4,3,5,5,3,3,1) & (-1,1,-1) & -4 \\ 
& T^{2}/\Z_{4}\times T^{2}/\Z_{2} & (4,4,4,4,4,4,5,3,2,2) & (1,-2) & -6 & & 
& T^{4}/\Z_{4}\times T^{2}/\Z_{2} & (4,4,4,4,3,5,5,3,2,2) & (1,1,-2) & -4 \\ 
& T^{4}/\Z_{2} & (4,4,4,4,4,4,6,2,2,2) & (2,-2) & 0 & & & 
T^{2}/\Z_{4}\times T^{4}/\Z_{2} & (4,4,4,4,5,3,6,2,2,2) & (1,-2,2) & 2 \\ 
& T^{2}/\Z_{2}\times T^{2}/\Z_{4} & (4,4,4,4,4,4,6,2,3,1) & (2,-1) & 2 & & &&&&\\  
\cline{1-5}\cline{7-11}
5 & T^{4}/\Z_{5} & (5,5,5,5,5,5,6,4,4,1) & (1,-1) & -6 & & 
5 & T^{6}/\Z_{5} & (5,5,5,5,4,6,6,4,4,1) & (-1,1,-1) & -4 \\ 
& T^{4}/\Z'_{5} & (5,5,5,5,5,5,6,4,3,2) & (1,-2) & -10 & & 
& T^{6}/\Z'_{5} & (5,5,5,5,4,6,6,4,3,2) & (-1,1,-2) & -8 \\ 
& T^{4}/\Z''_{5} & (5,5,5,5,5,5,7,3,4,1) & (2,-1) & 0 &  & 
& T^{6}/\Z''_{5} & (5,5,5,5,4,6,7,3,4,1) & (-1,2,-1) & 2 \\ 
& T^{4}/\Z'''_{5} & (5,5,5,5,5,5,7,3,3,2) & (2,-2) & -2&
& & T^{6}/\Z'''_{5} & (5,5,5,5,4,6,7,3,3,2) & (-1,2,-2) & -2\\ 
\cline{1-5}\cline{7-11}
6 & T^{4}/\Z_{6} & (6,6,6,6,6,6,7,5,5,1) & (1,-1) & -8 & & 
6 & T^{6}/\Z_{6} & (6,6,6,6,5,7,7,5,5,1) & (-1,1,-1) & -6 \\ 
& T^{2}/\Z_{3}\times T^{2}/\Z_{6} & (6,6,6,6,6,6,8,4,5,1) & (2,-1) & 2 &  
& & T^{2}/\Z_{3}\times T^{4}/\Z_{6} & (6,6,6,6,4,8,7,5,5,1) & (-2,1,-1) & 2 \\ 
& T^{2}/\Z_{6}\times T^{2}/\Z_{3} & (6,6,6,6,6,6,7,5,4,2) & (1,-2) & -14 & & 
& T^{4}/\Z_{6}\times T^{2}/\Z_{3} & (6,6,6,6,5,7,7,5,4,2) & (-1,1,-2) & -12 \\ 
& T^{4}/\Z_{3}^{\prime } & (6,6,6,6,6,6,8,4,4,2) & (2,-2) & -8 & & 
& T^{2}/\Z_{6}\times T^{4}/\Z_{3} & (6,6,6,6,5,7,8,4,4,2) & (-1,2,-2) & -6 \\ 
& & & & & & & T^{6}/\Z'_{3} & (6,6,6,6,4,8,8,4,4,2) & (-2,2,-2) & 0 \\ 
& T^{2}/\Z_{2}\times T^{2}/\Z_{3} & (6,6,6,6,6,6,9,3,4,2) & (3,-2) & 2 & & & &  &  &  \\ 
& T^{2}/\Z_{6}\times T^{2}/\Z_{2} & (6,6,6,6,6,6,7,5,3,3) & (1,-3) & -16 & &
& T^{4}/\Z_{6}\times T^{2}/\Z_{2} & (6,6,6,6,5,7,7,5,3,3) & (-1,1,-3) & -14 \\ 
& T^{4}/\Z'_{2} & (6,6,6,6,6,6,9,3,3,3) & (3,-3) & 0 & & 
& T^{2}/\Z_{6}\times T^{4}/\Z_{2} & (6,6,6,6,5,7,9,3,3,3) & (-1,3,-3) & 2\\ 
\cline{1-5}\cline{7-11}
\end{array}
$
\end{tabular}
\end{center}
\caption{$\eten$ roots as class representatives of shift vectors for $\Z_n$ orbifolds}
\label{Rep1}
\end{sidewaystable}

 
\renewcommand\arraystretch{1.2}

\begin{sidewaystable}[p]
\begin{center}
\begin{tabular}{l}
\hspace{-1cm}
{\large {\bf q=8}} \hspace{12cm}{\large {\bf q=10}}\\[5pt]
\hspace{-1cm}
$
\begin{array}{|c|c|c|c|c|c|c|c|c|c|c|}\cline{1-5}\cline{7-11}
n & \text{Orbifold} & \text{shift vector} & (Q_{1},\ldots,Q_{4}) & |\xi |^{2} & & n & \text{Orbifold} 
& \text{shift vector} & (Q_{1},\ldots,Q_{5}) & |\xi |^{2} \\ \hhline{:=====:~:=====:}
3 & T^{8}/\Z_{3} & (3,3,2,4,2,4,4,2,2,1) & (1,-1,1,-1) & 2 & & &  &  &  & \\ 
\cline{1-5}
4 & T^{8}/\Z_{4} & (4,4,5,3,3,5,5,3,3,1) & (1,-1,1,-1) & -2 & &4 
& T^{10}/\Z_{4} & (3,5,5,3,3,5,5,3,3,1) & (-1,1,-1,1,-1) & 2 \\ 
& T^{6}/\Z_{4}\times T^{2}/\Z_{2} & (4,4,5,3,3,5,5,3,2,2) & (1,-1,1,-2) & -2&
&  & T^{8}/\Z_{4}\times T^{2}/\Z_{2} & (3,5,5,3,3,5,5,3,2,2) & (-1,1,-1,1,-2) & 0 \\ 
\cline{1-5}\cline{7-11}
5& T^{8}/\Z_{5} & (5,5,6,4,4,6,6,4,4,1) & (1,-1,1,-1) & -2 & 
& 5 & T^{10}/\Z_{5} & (4,6,6,4,4,6,6,4,4,1) & (-1,1,-1,1,-1) & 0  \\ 
& T^{8}/\Z'_{5} & (5,5,6,4,4,6,6,4,3,2) & (1,-1,1,-2) & -6 && 
5 & T^{10}/\Z_{5} & (4,6,6,4,4,6,6,4,3,2) & (-1,1,-1,1,-2) & -4 \\ 
& T^{8}/\Z'''_{5} & (5,5,6,4,4,6,7,3,3,2) & (1,-1,2,-2) & 0 & &
& T^{10}/\Z'''_{5} & (4,6,6,4,4,6,7,3,3,2) & (-1,1,-1,2,-2) & 2 \\ 
\cline{1-5}\cline{7-11}
6& T^{8}/\Z_{6} & (6,6,7,5,5,7,7,5,5,1) & (1,-1,1,-1) & -4 & & 
6 & T^{10}/\Z_{6} & (5,7,7,5,5,7,7,5,5,1) & (-1,1,-1,1,-1) & -2 \\ 
& T^{2}/\Z_{3}\times T^{6}/\Z_{6} & (6,6,8,4,5,7,7,5,5,1) & (2,-1,1,-1) &  2 &
&  &  &  &  &  \\ 
& T^{6}/\Z_{6}\times T^{2}/\Z_{3} & (6,6,7,5,5,7,7,5,4,2) & (1,-1,1,-2) & -10 &
&  & T^{8}/\Z_{6}\times T^{2}/\Z_{3} & (5,7,7,5,5,7,7,5,4,2) & (-1,1,-1,1,-2) & -8 \\ 
& T^{4}/\Z_{6}\times T^{4}/\Z_{3} & (6,6,7,5,5,7,8,4,4,2) & (1,-1,2,-2) & -4 &
&  & T^{6}/\Z_{6}\times T^{4}/\Z_{3} & (5,7,7,5,5,7,8,4,4,2) & (-1,1,-1,2,-2) & -2 \\ 
& T^{2}/\Z_{6}\times T^{6}/\Z_{3} & (6,6,7,5,4,8,8,4,4,2) & (1,-2,2,-2) & 2& & & & & & \\ 
& T^{6}/\Z_{6}\times T^{2}/\Z_{2} & (6,6,7,5,5,7,7,5,3,3) & (1,-1,1,-3) & -12& &  
& T^{8}/\Z_{6}\times T^{2}/\Z_{2} & (5,7,7,5,5,7,7,5,3,3) & (-1,1,-1,1,-3) & -10 \\ 
& T^{4}/\Z_{6}\times T^{2}/\Z_{3}\times T^{2}/\Z_{2} & (6,6,7,5,5,7,8,4,3,3) & (1,-1,2,-3) & -6 & & 
& T^{6}/\Z_{6}\times T^{2}/\Z_{3}\times T^{2}/\Z_{2} & (5,7,7,5,5,7,8,4,3,3) & (-1,1,-1,2,-3) & -4 \\ 
& T^{2}/\Z_{6}\times T^{4}/\Z_{3}\times T^{2}/\Z_{2} & (6,6,7,5,4,8,8,4,3,3) & (1,-2,2,-3) & 0 & & 
& T^{4}/\Z_{6}\times T^{4}/\Z_{3}\times T^{2}/\Z_{2} & (5,7,7,5,4,8,8,4,3,3) & (-1,1,-2,2,-3) & 2 \\ 
\cline{1-5}\cline{7-11}
\end{array}
$
\end{tabular}
\end{center}
\caption{$\eten$ roots as class representatives of shift vectors for $\Z_n$ orbifolds}
\label{Rep2}
\end{sidewaystable}

\newpage
\renewcommand\arraystretch{1}


\section{$\Z_{2}$ orbifolds}

The case of $\Z_{2}$-orbifolds is slightly degenerated and must
be considered separately. In \cite{Gan1}, the orbifold $T^{4m}/\Z_{2}$, 
$m=1,2$ and $T^{4m^{\prime }+1}/\Z_{2}$, $m^{\prime }=0,1,2$ 
have been worked out, and the orbifold charges have been shown to be generated 
by a generic Minkowskian brane required \cite{DasMuk1,Witten2} for
anomaly cancellation, living in the transverse space.

In this section, we will show how to treat all $T^{q}/\Z_{2}$ orbifolds
for $q\in\{1,..,9\}$. In Section~\ref{SecOrbi1}, it will be shown how the algebraic results 
for invariant subalgebras in~\cite{Gan2}, which we henceforth refer to as 
$\Z_2$ orbifolds of the {\it first kind}, are recovered as a particular case in the
more general framework of Section~\ref{ShiftSec}. 

In Section~\ref{SecOrbi2}, we investigate in detail the $q=2,3,6,7$ cases, 
or $\Z_2$ orbifolds of {\it the second kind}, which have not been considered
in \cite{Gan2}. Let us stress that by \textit{orbifolds of the second kind} 
we mean the purely algebraic implementation of the $\Z_2$ projection in the
U-duality algebra. Then, we will extract from the construction of Section~\ref{Classrep} 
the roots of $\eten$ defining the representatives of classes of shift vectors 
for these orbifolds of M-theory and give a tentative physical interpretation.

Concretely, let $i,j,k$ be transverse spacelike coordinates and $A,B,C$ coordinates on
the orbifold, under a $\Z_{2}$-transformation, $11D$-supergravity
and fields have charge assignment
\begin{eqnarray*}
\text{(all)} &\text{:}&g_{ij}\rightarrow +g_{ij}\,,\qquad g_{iA}\rightarrow
-g_{iA}\,,\qquad g_{AB}\rightarrow +g_{AB}\,, \\
\binom{\text{odd}}{\text{even}} &:&C_{ijk}\rightarrow \mp C_{ijk}\,,\quad
C_{ijA}\rightarrow \pm C_{ijA}\,,\quad C_{iAB}\rightarrow \mp
C_{iAB}\,,\quad C_{ABC}\rightarrow \pm C_{ABC}\,,
\end{eqnarray*}
\textit{odd} and \textit{even} referring to the dimension of the orbifold
torus.

In contrast to the $\Z_{n>2}$ case, where the inner automorphisms
generating the orbifold charges were pure $SO(r)$ rotations, the action of a 
$\Z_{2}$-orbifold can be regarded as an element of the larger 
$O(r)=\Z_{2}\times SO(r)$. This distinctive feature of $\Z_{2}$-orbifold 
can be ascribed to the fact that while even orbifolds act as central symmetries 
and may be viewed as $\pi$-rotations, hence falling in $O^{+}(r)$ (positive 
determinant elements connected to the identity), odd orbifolds behave as 
mirror symmetries, and thus fall in $O^{-}(r)$. Concretely, negative determinant 
orthogonal transformations will contain, in the $\eten$ language, a rotation in 
the $\a_{8}$ direction, namely Ad$(e^{\pi \,(E_{8}-F_{8})})$ or Ad$(e^{i\pi \,H_{8}})$,
which, in this framework, behaves as a mirror symmetry.

The even case can be dealt with in a general fashion by applying the
following theorem:

\begin{theorem}
\label{evenZ2 copy(1)}{\it Let $T^{q}/\Z_{2}$ be a $\Z_{2}$
toroidal orbifold of $\e_{r}$, for $q\in\{1,\ldots,9\}$, $r\in\{q+1,\ldots,10\}$.
Let $q$ be either $2m$ or $2m+1$. Given a set of (possibly non-simple)
roots $\Delta_{\Z_{2}}=\{\b_{(p)}\}_{p=1,..,m}$ satisfying 
$(\b_{(p)}|\b_{(l)})=c_{p}\delta _{p,l}$, with $c_{p}\leqslant 2$ and
provided the orbifold acts on the U-duality algebra $\g^{U}$
with the operator $\mathcal{U}_{2m}^{\Z_{2}}\in G^{U}$ defined,
according to expression~(\ref{OrbifoldAction}), by
\begin{equation}
\mathcal{U}_{q}^{\Z_{2}}=\prod_{p=1}^{m}\Ad(e^{\pi
\,(E_{\b _{(p)}}-F_{\b _{(p)}})})\,,  \label{thmZ2op}
\end{equation}
then, the orbifold action decomposes on the root-subspace $\g_{\a}^{U}\subset \g^{U}$ as
\begin{equation}
\mathcal{U}_{q}^{\Z_{2}}\cdot \g_{\a}^{U}\equiv
\prod_{p=1}^{m}\Ad(e^{i\pi \,H_{\b_{(p)}}})\cdot \g_{\a}^{U}=
(-1)^{\sum_{p=1}^{m}(\b_{(p)}|\a)}\,\g_{\a}^{U}\,,\quad \,\forall {\a}\in \Delta (\g^{U})\,.
\label{thmZ2charge}
\end{equation}}
\end{theorem}

If the $\Z_{2}$-orbifold is restricted to extend in successive directions, starting 
from $x^{10}$ downwards, it can be shown that for any root 
$\a=\sum_{j=-1}^{8}k^{j}\a_{j}\in \Delta (\g^{U})$, expression~(\ref{thmZ2charge}) assumes 
the simple form
\begin{equation}\label{autoZ2even}
\mathcal{U}_{2m}^{\Z_{2}}\cdot \g_{\a}^{U}\equiv 
\Ad (e^{i\pi \,\sum_{i=1}^{m}H_{9-2i}})\cdot \g_{\a}^{U}=\left\{
\begin{array}{c}
(-1)^{k^{6}}\,\g_{\a}^{U}\,,\,\text{\ for }m=1 \\ 
(-1)^{k^{8-2m}+k^{8}}\,\g_{\a}^{U}\,,\,\text{\ for }
2\leqslant m \leqslant 4
\end{array}
\right. \text{ .}
\end{equation}
for even orbifold. For odd ones, we note the appearance of the mirror
operator we mentioned above
\begin{equation}
\mathcal{U}_{2m+1}^{\Z_{2}}\cdot \g_{\a}^{U}\equiv 
\Ad\left( e^{i\pi \,\left(H_{8}+H_{6}+\sum_{i=1}^{m}H_{8-2i}\right) }\right) 
\cdot \g_{\a}^{U}=(-1)^{k^{7-2m}}\,\g_{\a}^{U}\,,\,\text{\ for }m\geqslant 0\text{ .}  \label{autoZ2odd}
\end{equation}

Following Section \ref{Shift2}, we are free to recast the orbifold charges
resulting from expressions~(\ref{autoZ2even}) and~(\ref{autoZ2odd}), by
resorting to a shift vector $\xi^{[q,2]}$ such that:
\begin{equation*}
\mathcal{U}_{q}^{\Z_{2}}\cdot \g_{\a}^{U}=(-1)^{(\xi^{[q,2]}|\a)}\,\g_{\a}^{U}\,,
\,\text{\ }\forall q=1,\ldots,9\text{ .}
\end{equation*}
The subalgebra invariant under $T^{q}/\Z_{2}$ is now reformulated
as a KMA with root system
\begin{equation}
\Delta_{\text{inv}}=\left\{ {\a}\in \Delta (\g^{U})\,|
(\xi^{[q,2]}|\a)=0\,\mbox{mod}\,2\right\} \,. \label{deltainv}
\end{equation}
This is definition of the $\Z_{2}$-charge used in~\cite{Gan1,Gan2}.

\subsection{$\Z_{2}$ orbifolds of M-theory of the first kind}

\label{SecOrbi1}

The orbifolds of M-theory with $q=1,4,5,8,9$ have already been studied in
\cite{Gan2}, and a possible choice of shift vectors has
been shown to be, in these cases, dual to prime isotropic roots, identified
in \cite{Gan1} as Minkowskian branes. As such, they were
interpreted as representatives of the 16 transverse M-branes stacked at the
$2^{q}$ orbifold fixed points and required for anomaly cancellation in the corresponding M-theory orbifolds~\cite{Witten2,DasMuk1}.

In this section about $\Z_{2}$ orbifolds of the first kind, we will show
how to rederive the results of~\cite{Gan2} about shift vectors and invariant subalgebra, 
from the more general perspective we have developed in Section~\ref{ShiftSec} by resorting 
to the Kac-Peterson formalism. After this cross check, we will generalize this construction 
to the $q=2,3,6,7$ cases, which have not been considered so far, and show how
the roots $\xi_{2}^{[q,2]}$ are related to D-branes and involved in the cancellation 
of tadpoles due to O-planes of certain type 0B' orientifolds.
For this purpose, we start by summarizing in Table~\ref{twisted1} the shift
vectors for the $q=1,4,5,8,9$ cases found in~\cite{Gan2}, specifying in addition 
the $SL(10,\R)$-representation they belong to. 
\begin{table}[h]
\begin{center}
$
\begin{array}{|c|c|c|c|}
\hline
q & \xi^{[q,2]} & \text{physical basis} & \text{Dynkin labels}
\\ \hline
1 & \a _{(-1)^{2}0^{4}1^{6}2^{8}3^{10}4^{12}5^{14}6^{9}7^{6}8^{7}} & 
(2,2,2,2,2,2,2,2,4,1) & [200000001] \\ 
4 & \a _{(-1)^{2}0^{4}1^{6}2^{8}3^{10}4^{12}5^{13}6^{8}7^{5}8^{6}} & 
(2,2,2,2,2,2,1,1,3,1) & [100000100] \\ 
5 & \a _{(-1)^{2}0^{4}1^{6}2^{8}3^{10}4^{11}5^{12}6^{8}7^{4}8^{5}} & 
(2,2,2,2,2,1,1,1,1,1) & [000010000] \\ 
8 & \a _{(-1)^{2}0^{4}1^{5}2^{6}3^{7}4^{8}5^{9}6^{6}7^{3}8^{4}} & 
(2,2,1,1,1,1,1,1,1,1) & [010000000] \\ 
9 & \delta & (0,1,1,1,1,1,1,1,1,1) & [000000001] \\ \hline
\end{array}
$
\end{center}
\caption{Physical class representatives for $T^{10-q}\times T^q/\Z_2$ orbifolds of M-theory of the first kind}
\label{twisted1}
\end{table}
Next we will show how the results of Table~\ref{twisted1} for even orbifolds
can be retrieved as special cases of the general solutions computed in 
Section~\ref{Shift2}. 

The root of $E_{10}$ relevant to the $q=4$ orbifold can be determined as a special 
case of $T^{4}/\Z_{n}$ shift vectors, namely:
\begin{equation*}
\xi^{[4,2]}=2(\La^{5}-\La^{6}-\La^{8})+\La^{\{4\}}=2\tilde{\d}^{[4]}-\a _{5}+\a _{7}\,,
\end{equation*}
which coincides with the results of Table~\ref{twisted1}. This choice of
weight is far from unique, but is the lowest height one corresponding to a root
of $E_{10}$ (given that $\La^{\{4\}}$ is not a root). Likewise $\xi^{[8,2]}$ can in 
principle be deduced from the generic weight $\La^{\{8\}}$ determining the 
$T^{8}/\Z_{n}$ charges.

Shift vectors for odd orbifolds of Table \ref{twisted1} can also be recast
in a similar form, even though they do not generalize to $n>2$. We can
indeed rewrite:
\begin{equation*}
\xi^{[1,2]}=2(2\La^{7}-\La^{6})-\La^{7},\text{ \ \ \ }
\xi^{[5,2]}=2(-\La^{8})+\La^{3},\text{ \ \ \ }
\xi^{[9,2]}=2(-\La^{-1})+\La^{-1}.
\end{equation*}

The last four $E_{10}$ roots listed in Table \ref{twisted1} were identified in \cite{Gan1} 
as, respectively, Minkowskian Kaluza-Klein monopole (KK7M), M5-brane, M2-brane and 
Kaluza-Klein particle (KKp), with spatial extension in the transverse directions and
have been presented in Table~\ref{ListBranes}. The first root is the mysterious M-theory 
lift of the type IIA D8-brane, denoted as KK9M in this paper. In the language of 
Table~\ref{TablTensor}, these roots correspond to the representation weights 
$(A\otimes \widetilde{K})_{(99)[1\cdots 9]}$, $D_{(10)[1\cdots 6\,10]}$, $B_{(1)[2\cdots \,5]}$, 
$A_{(12)}$, $\widetilde{K}_{(2)[3\cdots 10]}$.

Furthermore, it has been shown in~\cite{Witten2,DasMuk1} that the consistency of
$\Z_2$ orbifolds of M-theory of the first kind requires the presence at the fixed points
of appropriate solitonic configurations. For $T^{q=5,8}/\Z_{2}$, one needs
respectively 16 M5-branes/M2-branes to ensure anomaly cancellation.
In the case of $T^{9}/\Z_{2}$, 16 units of Kaluza-Klein momentum are needed, 
while Kaluza-Klein monopoles with a total Chern class of the KK gauge bundle 
amounting to 16 is required in the case of $T^4/\Z_2$.

For $q=4,5,8,9$, the transverse Minkowskian objects of Table~\ref{twisted1} having 
all required properties were interpreted as generic representatives of these 
non-perturbative objects. However, their total multiplicity/charge cannot be inferred from
the shift vectors. It was proposed in~\cite{Gan2} that these numbers could be deduced from
an algebraic point of view from the embedding of $\ginv$ into a real form of the conjectured 
heterotic U-duality symmetry $\mathfrak{de}_{18}$. However, this idea seems to be difficult to 
generalize to the new examples treated in the present paper and we will not discuss 
it further. 

For $q=1$, the analysis is a bit more subtle, and needs to be carried out 
in type IIA language. To understand the significance of the shift vector in this case, 
it is convenient to reduce from M-theory on 
$T^{8}\times S^{1}/\Z_{2}\times S^{1}$ to type
IIA theory on $T^{8}\times S^{1}/\Omega I_1$, where $I_1$ 
is the space parity-operator acting on the $S^{1}$ as the original $\Z_{2}$ inversion, 
while $\Omega$ is the world-sheet parity operator. In this setup, the appropriate shift
vector is $\xi_{\sigma}^{[1,2]}=(2,2,2,2,2,2,2,2,1,4)$, which can be interpreted as
a KK9M of M-theory with mass:
\begin{equation*}
M_{\text{KK9M}}=M_{P}^{-9}V^{-1}e^{\left\langle \xi|H_m\right\rangle}
=M_{P}^{12}R_{1}\cdots R_{8}R_{10}^{3}\,.
\end{equation*}
Upon reduction to type IIA theory, we reexpress it in string units by setting 
\begin{equation}
R_{10}=g_{A}M_{s}^{-1}\,,\text{ \ \ and \ \ }M_{P}=g_{A}^{-1/3}M_{s}\,,
\label{reducR10}
\end{equation}
and take the limit $M_{P}R_{10}\rightarrow 0$:
\begin{equation*}
M_{\text{KK9M}}\rightarrow M_{\text{D}8}=\frac{M_{s}^{9}}{g_{A}}R_{1}\cdots R_{8}\,.
\end{equation*}
The resulting mass is that of a D8-brane of type IIA theory. The appearance
of this object reflects the need to align 8 D8-branes on each of the two O8
planes at both ends of the orbifold interval to cancel locally the $-8$ units
of D8-brane charge carried by each O8 planes, a setup known as type I' theory.

The chain of invariant subalgebras $\ginv$ in Table 
\ref{S1/Z2Orbifolds} is obtained by keeping only those root spaces of 
$\g^{U}$ which have eigenvalue $+1$ under the action of 
$\mathcal{U}_{2m}^{\Z_{2}}$ (\ref{autoZ2even}) or 
$\mathcal{U}_{2m+1}^{\Z_{2}}$ (\ref{autoZ2odd}). We can use the 
set of invariant roots (\ref{deltainv}) to build the Dynkin diagram of 
$\ginv$, but this is not enough to determine root 
multiplicities in $D=1$. In the hyperbolic case indeed, one will need to 
know the dimension of the root spaces $\g_{\a }^{U}\subset 
\g^{U}=\e_{10|10}$ which are invariant under the actions 
(\ref{autoZ2even}) or (\ref{autoZ2odd}) to determine the size of 
$\ginv$. We will come back to this issue at the end 
of this section.

\begin{table}[!t]
\begin{center}
\begin{tabular}{c|c|c|c}
$D$ & $(\Pi_0,\phi)$ & $\ginv$ & $\sigma(\ginv)$ \\
\hline\hline[3pt]
$8$ & 
\begin{picture}(218,30)
\thicklines
\multiput(75,0)(30,0){2}{\circle{6}}
\put(126,-3){ $\times$ }
\put(158,-3){$H$}
\end{picture}
& 
$
\mathfrak{so}(2,2)\oplus\so(1,1)
$ & 3\\[15pt]
$7$ & 
\begin{picture}(215,30)
\thicklines
\multiput(60,0)(30,0){3}{\circle{6}}
\multiput(63,0)(30,0){2}{\line(1,0){24}}
\put(140,-3){ $\times$ }
\put(168,-3){$H$}
\end{picture}
& $\mathfrak{so}(3,3)\oplus\so(1,1)$& 4\\[15pt]
$6$ & 
\begin{picture}(215,40)
\thicklines
\multiput(60,0)(30,0){2}{\circle{6}}
\put(63,0){\line(1,0){24}}
\multiput(114,-18)(0,36){2}{\circle{6}}
\put(92,2){\line(4,3){19}}
\put(92,-2){\line(4,-3){19}}
\put(140,-3){ $\times$ }
\put(173,-3){$H$}
\end{picture}
& $\mathfrak{so}(4,4)\oplus\so(1,1)$ & 5\\[15pt]
$5$ & 
\begin{picture}(215,40)
\thicklines
\multiput(45,0)(30,0){3}{\circle{6}}
\multiput(48,0)(30,0){2}{\line(1,0){24}}
\multiput(129,-18)(0,36){2}{\circle{6}}
\put(107,2){\line(4,3){19}}
\put(107,-2){\line(4,-3){19}}
\put(147,-3){ $\times$ }
\put(178,-3){$H$}
\end{picture}
& $\mathfrak{so}(5,5)\oplus\so(1,1)$ & 6\\[15pt]
$4$ & 
\begin{picture}(215,40)
\thicklines
\multiput(30,0)(30,0){4}{\circle{6}}
\multiput(33,0)(30,0){3}{\line(1,0){24}}
\multiput(144,-18)(0,36){2}{\circle{6}}
\put(122,2){\line(4,3){19}}
\put(122,-2){\line(4,-3){19}}
\put(180,-3){\circle{6}}
\end{picture}
& $\mathfrak{so}(6,6)\oplus \mathfrak{sl}(2,\R) $ & 7 \\[15pt]
$3$ & 
\begin{picture}(215,40)
\thicklines
\multiput(15,0)(30,0){6}{\circle{6}}
\multiput(18,0)(30,0){5}{\line(1,0){24}}
\multiput(189,-18)(0,36){2}{\circle{6}}
\put(167,2){\line(4,3){19}}
\put(167,-2){\line(4,-3){19}}
\end{picture}
& $\mathfrak{so}(8,8)$ & 8
\\[20pt]
$2$ & 
\begin{picture}(215,40)
\thicklines
\multiput(45,0)(30,0){5}{\circle{6}}
\multiput(48,0)(30,0){4}{\line(1,0){24}}
\multiput(21,-18)(0,36){2}{\circle{6}}
\put(23.5,-16.5){\line(4,3){19}}
\put(23.5,16.5){\line(4,-3){19}}
\multiput(189,-18)(0,36){2}{\circle{6}}
\put(167,2){\line(4,3){19}}
\put(167,-2){\line(4,-3){19}}
\end{picture}
& $\hat{\dn}_{8|10}$ & 10\\[35pt] 
\hhline{~:==:~}
$1$ & 
{\hspace{2mm}
\begin{picture}(215,40)
\thicklines
\put(60,20){\circle{6}}
\put(60,17){\line(0,-1){14}}
\multiput(0,0)(30,0){7}{\circle{6}}
\multiput(3,0)(30,0){6}{\line(1,0){24}}
\multiput(204,-18)(0,36){2}{\circle{6}}
\put(182,2){\line(4,3){19}}
\put(182,-2){\line(4,-3){19}}
\end{picture}}
& $\mathfrak{de}_{10|10}$ & 10 \\[25pt]
\end{tabular}
\end{center}
\caption{The split subalgebras $\ginv$ for $\Z_2$ orbifolds of the first kind}
\label{S1/Z2Orbifolds}
\end{table}

This construction of the root system leads, $\forall q=1,4,5,8,9$, to a
unique chain of invariant subalgebras, depicted in Table~\ref{S1/Z2Orbifolds}. 
Thus, we verify that the statement made in \cite{Gan2} for the hyperbolic case 
is true for all compactifications of type $T^{11-(D+q)}\times T^{q}/\Z_{2}$ 
with $q=1,4,5,8,9$. There, this isomorphism was
ascribed to the fact that, in $D=1$, the shift vectors~(\ref{twisted1}) are
all prime isotropic and thus Weyl-equivalent to one another. The
mathematical origin of this fact lies in the general method developed by
Kac-Peterson explained in Section~\ref{Shift2}, which states that
equivalence classes of shift vectors related by Weyl transformation and/or
translation by $n$ times any weight lattice vector lead to isomorphic
fixed-point subalgebras. Since a Weyl reflection in $\g^{U}$
generates a U-duality transformation in M-theory and its low-energy supergravity, 
this isomorphism seems to indicate that all such
orbifolds are dual in M-theory, as pointed out in~\cite{DasMuk1,Sen1}.
In fact, if one had chosen to reduce $\Z_2$ orbifolds of the first kind
on a toroidal direction for $q$ odd, and on an orbifolded direction for
$q$ even, one would have realized that they are all part of the serie of
mutually T-dual orientifolds (a T-duality on $x^{i}$ is denoted by $\mathcal{T}_{i}$):
\begin{equation}\label{T-DualOrient}
\begin{array}{c}
{\ds \text{type IIB on } T^{9}/\Omega \overset{\mathcal{T}_{9}}{\longrightarrow} 
\mbox{type IIA on } T^{9}/\Omega I_1
\overset{\mathcal{T}_{8}}{\longrightarrow} \mbox{ type IIB on } T^{9}/(-1)^{F_L}\Omega I_2 
\overset{\mathcal{T}_{7}}{\longrightarrow}}\\[3pt] 
{\ds \rightarrow\mbox{ type IIA on } T^{9}/(-1)^{F_L}\Omega I_3
\overset{\mathcal{T}_{6}}{\longrightarrow} \mbox{ type IIB on } T^{9}/\Omega I_4 \overset{\mathcal{T}_{8}}{\longrightarrow} \ldots}
\end{array}
\end{equation}
where $I_{r}$ denotes the inversion of the last $r$ space-time coordinates, 
while $\Omega$ is as usual the world-sheet parity. The {\it space-time} left-moving 
fermions number $(-1)^{F_L}$ appear modulo 4 in these dualities.

The reality properties of $\ginv$ are easy to
determine. Since the original balance between Weyl and Borel generators is
preserved by the orbifold projection, the non-abelian part of 
$\ginv$ remains split. In $D=8,\ldots ,5$, the abelian 
$\mathfrak{so}(1,1)$ factor in $\ginv$ is generated by the
non-compact element $H^{[q]}$ which also appears in $T^{q}/\Z_{n>2}$
orbifolds. For $q=4$, it is, for instance, given by $H^{[4]}=H_{8}-H_{4}$ in $D=6$
and $2H_{3}+4H_{4}+3H_{5}+2H_{6}+H_{7}$ in $D=5 $, as detailed in Section~\ref{secT4}, 
and is enhanced, in $D=4$, to the $\mathfrak{sl}(2,\R)$ factor appearing 
in Table \ref{S1/Z2Orbifolds} when $H^{[4]}$ becomes the root
$\c =\a_{23^{2}4^{3}5^{3}6^{2}78}\in \Delta_{+}(\e_{7})$. 
The procedure is similar for $q=1,5,8,9$, for different combinations $H^{[q]}$ and
positive roots $\c$.

The root multiplicities in $\ginv$ is only relevant
to the two cases $D=2,1$, for which the root multiplicities are
inherited from $\e_{9}$ and $\eten$. For 
$\g^{U}=\e_{9}$ we have $\ginv=\hat{\dn}_{8}$, since $\d_{\ginv}=\d$, 
and since $\d$ and $\d_{\hat{\dn}_{8}}$ both have multiplicity 8.

In $D=1$, the story is different. In \cite{Gan2}, it has been shown that 
$\ginv$ contains a subalgebra of type $\mathfrak{de}_{10}$. 
The authors have performed a low-level decomposition of both 
$\ginv$ and $\mathfrak{de}_{10}$.
For a generic over-extended algebra $\g^{\wedge\wedge}$, such a
decomposition with respect to its null root $\delta _{G^{\wedge\wedge}}$ is given 
by $(\g^{\wedge\wedge})_{[k]}\doteq \bigoplus\limits_{(\a,\d_{G^{\wedge\wedge}})=
k}\g_{\a}^{\wedge\wedge}$. 
In particular, they define: 
$(\ginv)_{[k]}\doteq \ginv\cap (\eten)_{[k]}$ and show that:
\begin{equation*}
(\ginv)_{[1]}\simeq (\mathfrak{de}_{10})_{[1]}\,,\qquad (\ginv)_{[2]}\supset (\mathfrak{de}_{10})_{[2]}\,.
\end{equation*}
The first equality is a reformulation of 
$\ginv=\hat{\dn}_{8}$ for $\g^{U}=\e_{9}$. The second result comes from the fact 
that the orbifold projection selects certain preserved root subspaces without
affecting their dimension. This feature is similar to what we have observed in the case 
of $\Z_n$ orbifolds, where the original root multiplicities are restored after modding out 
the Borcherds or KM algebras appearing in $D=1$ by their centres and derivations.

\subsection{${\mathbbm{Z}}_{2}$ orbifolds of M-theory of the second kind and
orientifolds with magnetic fluxes}
\label{SecOrbi2}

Let us first recall that by \textit{orbifolds of the second kind} we mean the purely algebraic
implementation of $T^{10}/I_q$, $q=2,3,6,7$, in the U-duality algebra. In this case,
the connection to orbifolds of M-theory will be shown to be more subtle than in Section~\ref{SecOrbi1}.
Indeed, since the algebraic orbifolding procedure does not discriminate between two theories 
with the same bosonic untwisted sectors and different fermionic degrees of freedom,
there are in principle several candidate orbifolds on the M-theory side to which these
orbifolds of the second kind could be related.

The first (naive) candidate one can consider is to take M-theory directly on 
$T^{10}/I_q$, $q=2,3,6,7$. Then following the analysis of 
Section~\ref{SecOrbi1}, a reduction of such orbifolds to type II string theory 
would result in a chain of dualities similar to expression~(\ref{T-DualOrient}), 
with the important difference that the $(-1)^{F_L}$ operator now appears in the 
opposite places. It is well known that the spectrum of such theories cannot be 
supersymmetric. Referring to the chain of dualities~(\ref{T-DualOrient})
with the required extra factor of the left-moving fermionic number operator, 
one observes that such theories do not come from a consistent truncation of type IIB
string theory, since $(-1)^{F_L}\Omega$ is not a symmetry thereof, and
all of them are therefore unstable.

A more promising candidate is M-theory on $T^q/\{(-1)^{F}S,\Z_2\}$, where
$(-1)^{F}$ is now the total {\it space-time} fermion number and $S$ represents a $\pi$ shift in 
the M-theory direction. In contrast to the preceding case, these orbifolds
are expected to be dual to orientifolds of type 0 theory, which are non-supersymmetric
but are nonetheless believed to be stable, so that tadpole cancellation makes sense in such setups.

Let us work out, in these type 0 cases, a chain of dualities similar to expression~(\ref{T-DualOrient}).
To start with, we review the argument stating that M-theory on $S^1/(-1)^{F}S$  
is equivalent to the non-supersymmetric type 0A string theory in the small radius limit~\cite{BergGab}.

Considering the reduction of M-theory on $S^1\times S^1/(-1)^{F}S$ to type IIA string theory 
on $S^1/(-1)^{F}S$, one can determine the twisted sector of this orbifold (with no fixed point)
and perform a flip on $\{x^9,x^{10}\}$ to obtain the spectrum of M-theory on $S^1/(-1)^{F}S$.
At the level of massless string states, all fermions are projected out from the untwisted
sector and there appears a twisted sector that doubles the RR sector and adds a NSNS tachyon,
leading to type 0A string theory in 10 dimensions. Interestingly, type 0 string theories
have more types of $\Z_2$ symmetries and thus more consistent truncations. In $D=10$,
type 0A theory is symmetric under the action of $\Omega$, while type 0B theory is symmetric
under $\Omega$, $\Omega (-1)^{f_{L}}$ and $\Omega (-1)^{F_{L}}$, where $f_{L}$ and $F_{L}$
are respectively the {\it world-sheet} and {\it space-time} left-moving fermion numbers.
Furthermore, their compactified versions on $T^{10}$ each belong to a serie of
orientifold theories similar to~(\ref{T-DualOrient}). 
Among these four chains of theories, one turns out to be tachyon-free, the chain descending
from type 0B string theory on $\Omega (-1)^{f_{L}}$.
Let us concentrate on this family of orientifolds and show that the M-theory orbifolds of 
the second kind can all be seen to reduce to an orientifold from this serie in the small radius
limit.

In order to see this, one can mimick the procedure used for~(\ref{T-DualOrient}) and reduce 
on a toroidal direction for $q$ odd, and on an orbifolded direction for $q$ even. 
The untwisted sectors of our orbifolds then turns out to correspond to those of an orientifold
projection by $\Omega(-1)^{f_{L}}I_{(q)}$, resp. $\Omega(-1)^{f_{R}}I_{(q-1)}$, on type 
0A string theory. A projection by $\Omega(-1)^{f_{L/R}}I_{(q)}$ has the following effects 
in type 0A string theory: it eliminates the tachyon and half of the doubled RR sector, 
the remaining half being distributed over the untwisted and twisted sectors of M-theory 
on $S^1/(-1)^{F}S$. Consequently, one expects to obtain theories that belong to the following
chain of dual non-supersymmetric orientifolds:
\begin{equation}\label{T-DualOrient0}
\begin{array}{c}
{\ds \text{type 0B on } T^{9}/(-1)^{f_L}\Omega \overset{\mathcal{T}_{9}}{\longrightarrow} 
\mbox{type 0A on } T^{9}/(-1)^{f_R}\Omega I_1
\overset{\mathcal{T}_{8}}{\longrightarrow} }
 \mbox{ type 0B on } T^{9}/(-1)^{f_R}\Omega I_2 
 \\[3pt]
{\ds \overset{\mathcal{T}_{7}}{\longrightarrow}
\mbox{ type 0A on } T^{9}/(-1)^{f_L}\Omega I_3
\overset{\mathcal{T}_{6}}{\longrightarrow} \mbox{ type 0B on } T^{9}/(-1)^{f_L}\Omega I_4 \overset{\mathcal{T}_{8}}{\longrightarrow} \ldots}
\end{array}
\end{equation}
where $(-1)^{f_L}$ and $(-1)^{f_R}$ again appear modulo 4 in these dualities\footnote{From 
the point of view of M-theory, these dualities sometimes exchange the untwisted and twisted 
sectors under $(-1)^{F}S$}. Complications might however arise at the twisted sector level
when reducing to type 0A string theory on an orbifolded direction, since one should take
into account a possible non-commutativity between the small radius limit and the orbifold limit.
We will come back to this point later.

Instead, we first want to remind the reader that, as was shown in~\cite{BlumFontLust}, 
type 0B string on $(-1)^{f_L}\Omega$ can be made into a consistent non-supersymmetric
string theory by cancelling the tadpoles from the two RR 10-forms through the addition
of 32 pairs of D9- and D9'-branes for a total $U(32)$ gauge symmetry. This setup is usually called 
type 0'. There is also a NSNS dilaton tadpole, but this does not necessarily render
the theory inconsistent. Rather, it leads to a non-trivial cosmological constant through 
the Fischler-Susskind mechanism~\cite{FischSuss1,FischSuss2}. It was also shown
in~\cite{BlumFontLust} that there is no force between the D9- and D9'-branes and that 
twisted sector open strings stretched between them lead to twisted massless fermions 
in the ${\bf 496}\oplus{\bf \overline{496}}$ representation of $U(32)$. 
Even though the latter are chiral Majorana-Weyl fermions, it was shown
in~\cite{Sagnotti1,Sagnotti2} that a generalized Green-Schwarz mechanism ensures 
anomaly cancellation.

To characterize the twisted sectors of such orientifolds of type 0' string theory algebraicly, 
we will again use the equivalence classes of shift vectors that generate the orbifolds on the 
U-duality group. The simplest elements of these classes which are also roots give the
set of real roots in Table~\ref{twisted2}. 
\begin{table}[!h]
\begin{center}
$%
\begin{array}{|c|c|c|c|}
\hline
q & \tilde{\xi}^{[q,2]} & \text{physical basis} & \text{generator} \\ \hline
2 & \alpha _{7} & (0,0,0,0,0,0,0,0,1,-1) & \mathcal{K}_{[9\,10]} \\ 
3 & \alpha _{8} & (0,0,0,0,0,0,0,1,1,1) & \mathcal{Z}_{[89\,10]} \\ 
6 & \alpha _{34^{2}5^{3}6^{2}78^{2}} & (0,0,0,0,1,1,1,1,1,1) & \widetilde{%
\mathcal{Z}}_{[5\cdots 10]} \\ 
7 & \alpha _{1^{2}2^{3}3^{4}4^{5}5^{6}6^{4}7^{2}8^{3}} & 
(0,0,2,1,1,1,1,1,1,1) & \widetilde{\mathcal{K}}_{(3)[3\cdots 10]} \\ \hline
\end{array}
$%
\end{center}
\caption{Universal shift vectors for $\Z_2$ orbifolds of the second kind}
\label{twisted2}
\end{table}

As is obvious from the second and third column, all such roots are in 
$\Delta _{+}(\mathfrak{e}_{8})$ and correspond to instantons completely wrapping
the orbifolded torus. Since they are purely $\mathfrak{e}_{8}$
roots, we do not expect them to convey information on the string theory twisted sectors. 
As such, this set of shift vectors does not lend itself to an interesting physical
interpretation, but gives however certain algebraic informations. All these roots being 
in the same orbit of the Weyl subgroup of $E_{8}$, the resulting invariant subalgebras 
are again isomorphic (when existing) $\forall q=2,3,6,7$. We list the invariant
subalgebras for M-theory on $T^2/\{(-1)^F S,\Z_2\}$ together with their Dynkin
diagrams in Table \ref{T2/Z2Orbifolds}. The same invariant subalgebras appear
for all values of $q$, but of course start to make sense only in lower dimensions. 

The invariant subalgebras are not simple for $D\geq 3$ and all of them contain at least one $\mathfrak{sl}(2,{\mathbbm{R}})$ factor with simple root $\tilde{\xi}^{[q,2]}$. 
When an abelian factor is present, it coincides with the non-compact
Cartan element $H^{[q]}$ encountered in $T^{q}/{\mathbbm{Z}}_{n>2}$
orbifolds. Furthermore, in contrast to the connected $\hat{{\mathfrak{d}}}_{8}$ 
diagram obtained for $q=1,4,5,8,9$, the invariant subalgebra for $q=2,3,6,7$ 
is given in $D=2$ by an affine central product, as in the $T^{2,4,6}/{\mathbbm{Z}}_{n>2}$ 
cases treated before. In $D=1$, the invariant subalgebra is the following quotient of
the KMA, whose Dynkin diagram is drawn in Table~\ref{T2/Z2Orbifolds}:
\begin{equation*}
\mathfrak{g}_{\text{inv}}=\,^{2}\mathcal{KM}_{11|12}/\{\mathfrak{z},d_{1}\}\,,
\end{equation*}
where ${\mathfrak{z}}=c_{\e_{7}}-c_{\hat{{\mathfrak{a}}}_{1}}$. As in
the $T^{6}/{{\mathbbm{Z}}}_{3,4}$ cases, $^{2}\mathcal{KM}_{11} $ has a singular 
Cartan matrix with similar properties.

\begin{table}[!t]
\begin{center}
\begin{tabular}{c|c|c|c}
$D$ & $(\Pi_0,\phi)$ & $\mathfrak{g}_{\text{inv}}$ & 
$\sigma(\mathfrak{g}_{\text{inv}})$ \\
\hline\hline[3pt]
$8$ & \quad
\begin{picture}(240,30)
\thicklines
\multiput(75,0)(30,0){2}{\circle{6}}
\put(126,-3){ $\times$ }
\put(158,-3){$H$}
\end{picture}
& 
$
\mathfrak{sl}(2,\R)^{\oplus^2}\oplus\so(1,1)$ & 3
\\[15pt]
$7$ & \quad
\begin{picture}(240,30)
\thicklines
\multiput(60,0)(30,0){3}{\circle{6}}
\multiput(63,0)(30,0){1}{\line(1,0){24}}
\put(140,-3){ $\times$ }
\put(168,-3){$H$}
\end{picture}
& $
\begin{array}{c}
\mathfrak{sl}(3,\R)\oplus \mathfrak{sl}(2,\R) \\[5pt]
\oplus\;\so(1,1) 
\end{array}$ & 4\\
$6$ & \quad
\begin{picture}(240,40)
\thicklines
\multiput(45,0)(30,0){5}{\circle{6}}
\multiput(48,0)(30,0){2}{\line(1,0){24}}
\end{picture}
& $\mathfrak{sl}(4,\R)\oplus\sl(2,\R)^{\oplus^2}$ & 5\\[15pt]
$5$ & \quad
\begin{picture}(240,40)
\thicklines
\multiput(30,0)(30,0){6}{\circle{6}}
\multiput(33,0)(30,0){4}{\line(1,0){24}}
\end{picture}
& $\mathfrak{sl}(6,\R)\oplus\sl(2,\R)$ & 6\\[15pt]
$4$ & \quad
\begin{picture}(240,40)
\thicklines
\multiput(60,0)(30,0){5}{\circle{6}}
\multiput(63,0)(30,0){3}{\line(1,0){24}}
\multiput(36,-18)(0,36){2}{\circle{6}}
\put(38.5,-16.5){\line(4,3){19}}
\put(38.5,16.5){\line(4,-3){19}}
\end{picture}
& $\mathfrak{so}(6,6)\oplus \mathfrak{sl}(2,\R) $ & 7 \\[15pt]
$3$ & \quad
\begin{picture}(240,40)
\thicklines
\put(75,20){\circle{6}}
\put(75,17){\line(0,-1){14}}
\multiput(15,0)(30,0){7}{\circle{6}}
\multiput(18,0)(30,0){5}{\line(1,0){24}}
\end{picture}
& $\mathfrak{e}_{7|7}\oplus \sl(2,\R)$ & 8
\\[35pt] \hhline{~:==:~}
$2$ & \hspace{10mm}
\begin{picture}(240,40)
\thicklines
\put(75,20){\circle{6}}
\put(75,17){\line(0,-1){14}}
\multiput(-15,0)(30,0){9}{\circle{6}}
\multiput(-12,0)(30,0){6}{\line(1,0){24}}
\put(198,2){\line(1,0){24}}
\put(198,-2){\line(1,0){24}}
\put(200,-2.5){\footnotesize{$\langle$}}
\put(215.5,-2.5){\footnotesize{$\rangle$}}
\end{picture}
& $\hat{\mathfrak{e}}_{7|9}\oplus \hat{\mathfrak{a}}_{1|3}$ & 10 \\[20pt]
$1$ & \hspace{10mm}
\begin{picture}(240,40)
\thicklines
\put(105,20){\circle{6}}
\put(-15,20){\circle{6}}
\put(105,17){\line(0,-1){14}}
\multiput(-15,0)(30,0){9}{\circle{6}}
\multiput(18,0)(30,0){7}{\line(1,0){24}}
\put(-12,0){\line(1,0){24}}
\put(-16.5,17){\line(0,-1){14}}
\put(-13.5,17){\line(0,-1){14}}
\put(-19.3,7){$\big{\Updownarrow}$}
\end{picture}
& $^2\mathcal{KM}_{11|12} $ & 10 \\[20pt]
\end{tabular}
\end{center}
\caption{The split subalgebras $\mathfrak{g}_{\text{inv}}$ for 
$\Z_2$ orbifolds of the second kind}
\label{T2/Z2Orbifolds}
\end{table}

We will now show that certain equivalent shift vectors
can be interpreted as configurations of D9 and D9'-branes 
cancelling R-R tadpoles in a type 0' string theory orientifold.
This can be achieved by adding an appropriate
weight lattice vector $\La^{[q]}$ to $\tilde{\xi}^{[q,2]}$ that do
not change the scalar products modulo 2. It should be chosen so that
$\La^{[q]}+\tilde{\xi}^{[q,2]}$ is a root, and gives insight on the possible M-theory 
lift of such constructions. More precisely, we want to convince the reader that 
certain choices of shift vectors generating M-theory orbifolds of the second kind can be seen 
as representing either magnetized D9-branes or their image D9'-branes in some type 0' theory with
orientifold planes. Such branes carry fluxes that contribute to the overall D$(9-q)$-brane charge 
for even $q$ and D$(10-q)$-brane charge for odd $q$, but not to the higher ones.

Let us first study the example of a $T^{3}/\Z_{2}\times S^1/(-1)^{F}S$ orbifold of M-theory. 
Following the above construction, it should reduce in the limit $M_{P} R_{10}\rightarrow 0$
to type 0A string theory on $T^6\times T^{3}/(-1)^{f_L}\Omega I_3$, which is T-dual to 
type 0B string theory on $T^7\times T^{2}/(-1)^{f_R}\Omega I_2$. We summarize these dualities
in the diagram below: 
\begin{equation*}
\begin{CD} \mbox{M-theory on } T^{6}\times T^{3}/\Z_2\times
S^{1}/(-1)^{F}S\\ @VV{M_{P} R_{10}\rightarrow 0}V \\ \mbox{type 0A on }T^{6}\times
(S^{1}\times T^{2})/(-1)^{f_L}\Omega I_3 @>{\mathcal{T}_7}>> \mbox{type 0B on }
T^{6}\times S^{1}\times T^{2}/(-1)^{f_R}\Omega I_2 \end{CD}
\end{equation*}

In this last type 0B orientifold, there will be one orientifold plane carrying 
-4 units of D7- and D7'-brane charge at each of the four orbifold fixed points. 
Suppose that we consider $N$ pairs of magnetized D9- and D9'-branes carrying fluxes 
in the orbifolded plane $(x'^{8},x^{9})$. This system induces two Chern-Simons
couplings on the world-volume of the space-time filling branes: 
\begin{equation*}
\frac{M_{s}^{10}}{2(2\pi )^{9}}\int_{\R\times T^{9}}C_{8}\wedge
2\pi\a'\mathrm{Tr}(F_{2})=\frac{M_{s}^{8}}{(2\pi)^7}\int_{\R\times T^{7}}
C_{8}\cdot \frac{1}{2\pi }\int_{T^{2}/\Z_{2}}\mathrm{Tr}(F_{2})\,,
\end{equation*}
and a similar expression involving $C'_{8}$. The quantized fluxes can
then be chosen in such a way that the resulting total positive D7- and D7'-brane
charges cancel the negative charges from the orientifold planes and ensure
tadpole cancellation. Note that these charges are determined by the first Chern
class $c_{1}$ of the $U(N)$ gauge bundle.

We can use an analogy with the supersymmetric case, where the system of O7-planes 
and magnetized D9-branes in a $T^8\times T^2/(-1)^{F_L}\Omega I_2$ type IIB
orientifold has a well-known T-dual equivalent~\cite{BlumGoerKorLust,Marchesano} 
built from D8-branes at angle with O8-planes in a $T^9\times S^1/(-1)^{F_L}\Omega I_1$
type IIA orientifold, in which the flux is replaced by an angle $\chi$ in the following way: 
\begin{equation}\label{AngleDef}
2\pi\a'F_{89}=\frac{c_1}{N}\frac{\Id_{N}}{M_s^2 R'_8 R_9} 
\overset{\mathcal{T}_{8}}{\longrightarrow}\cot(\chi)=\frac{c_1}{N} \frac{R_8}{R_{9}} \,
\end{equation}
and where the type IIB radius\footnote{Note that 
both the signs of the orientifold plane charges and this angle $\chi$
would also be sensitive to the presence of a quantized Kalb-Ramond
background flux $\int B_{89}\, dx'^{8}dx^9$, but we neglect this
possibility here, since it would be the sign of a tilted geometry in the 
$(x^8;x^9)$ plane on the type A side (non-trivial complex structure of the
torus).} is $R'_8=1/M_s^2 R_8$. In fact, the appearance of D7-brane charges 
in the absence of D9-brane ones on the type IIB side can be understood, 
in the dual setup, as the addition (resp. cancellation) of the
charges due to the tilted D8-branes to those of their image branes with 
respect to the orientifold O8-plane. The image D8-brane indeed exhibits 
a different angle, being characterized by wrapping numbers $(c_1,-N)$ 
instead of $(c_1,N)$ around the directions along the orbifold. In our non-supersymmetric
case, however, one should keep in mind that the image brane of a D9-brane 
under $\Omega(-1)^{f_R}$ is a D9'-brane.

We will now show that the appropriate shift vector encodes not only explicit
information on the presence of type 0' pairs of D9- and D9'-branes, but also 
on the tilting of their dual type 0A D8- and D8'-branes with respect to the 
O8-planes. One can then deduce the presence of fluxes from the angle $\chi$.

To understand how this comes about, we note that both string theory D9- and
D8-branes descend from the (conjectured) KK9M soliton of M-theory described by the $\eten$ 
roots that are permutations of $\xi=~((2)^{6},1,2,2,4)$ in the following way: 
\begin{equation*}
M_{\text{KK9M}}=M_{P}^{-9}V^{-1}e^{\left\langle \xi |H_{R}\right\rangle
}=M_{P}^{12}R_{1}\cdots R_{6}R_{8}R_{9}R_{10}^{3}\,.
\end{equation*}
Following the chain of dualities~(\ref{Dualities}), we successively obtain
the D8- and D9-brane mass formulae:
\begin{equation}
\begin{CD} {\displaystyle M_{P}^{12}R_{1}\cdots R_{6}R_{8}R_9 R_{10}^{3} }
@>{M_{P}R_{10}\rightarrow 0}>> {\displaystyle
M_{\text{D8}}=\frac{M_{s}^{9}}{g_A}R_{1}\cdots R_{6}R_{8} R_{9}
\overset{\mathcal{T}_{7}}{\longrightarrow}
M_{\text{D9}}=\frac{M_{s}^{10}}{g_B}R_{1}\cdots R_{6} R'_7 R_{8}R_{9}\,} \end{CD}
\label{MassFormulae}
\end{equation}

Now, we select one definite shift vector $\tilde{\xi}^{[3,2]}$ from all equivalent ones,
which has the particularity to correspond like $\xi$ to a root of level 4. It is obtained from a permutation of the root $\tilde{\xi}^{[3,2]}$ that describes an orbifold 
in the directions $(x^{7};x^{8};x^{9})$, namely
$\tilde{\xi}_{\sigma}^{[3,2]}=((0)^{6},(1)^{3},0)$, as:
\begin{equation}
\xi^{[3,2]}=2(\La^{6}-\La^{7}-\La^{8})-
\tilde{\xi}_{\sigma}^{[3,2]}=((2)^{6},3,3,1,2)\,.
\end{equation}
Let us first blindly compute the ensuing mass formula, reduce it on $x^{10}$
and T-dualize it on $x^{7}$: 
\begin{equation}
\begin{CD} {\displaystyle M_{P}^{12}R_{1}\cdots R_{6}(R_7 R_{8})^{2} R_{10}
} @> {M_{P}R_{10}\rightarrow 0}>> {\displaystyle \frac{M_{s}^{11}}{g_A^3}
R_{1}\cdots R_6 (R_7 R_{8})^{2} \overset{\mathcal{T}_{7}}{\longrightarrow}
M_{\text{D9}}=\frac{M_{s}^{10}}{g_B^3}R_{1}\cdots R'_7 R^2_{8}\,.} \end{CD}
\label{RootReduction}
\end{equation}
On the type B side, $R_{8}$ and $R_{9}$ form the pair of
orbifolded directions. Comparing with~(\ref{MassFormulae}), we immediately
see that we will have a D9-brane if: $R_{9}\propto R_{8}/g_{B}^{2}$. As hinted above,
we need to find an angle in the dual type 0A setup to identify the flux. In this perspective, 
we perform a further T-duality along $x^{8}$ that brings us to a 
$S^{1}/(-1)^{f_R}\Omega I_{1}$ orientifold of type 0A string theory in which the 
type 0B flux is mapped to an angle between the O8-plane, and the D8-brane obtained
from~(\ref{RootReduction}) as
\begin{equation}
\begin{CD} 
M_{\text{D9}}={\displaystyle \frac{M_{s}^{10}}{g_B^3}R_{1}\cdots R'_7 R^2_{8}}
\overset{\mathcal{T}_{8}}{\longrightarrow} {\displaystyle \frac{M_{s}^{9}}{g_A^3}
R_{1}\cdots R_6 R'_7 R'_{8}
\,.} \end{CD}
\label{FurtherDuality}
\end{equation}
This implies that there is a dual relation to $R_{9}\propto R_{8}/g_{B}^{2}$ on the 0A 
side that has the same form: 
\begin{equation*}
R_{9}\propto R_{8}/g_{B}^{2}\;
\overset{\mathcal{T}_{8}}{\longrightarrow }\;R_{9}\propto R'_{8}/g_{A}^{2}\,.
\end{equation*}
Indeed, plugging back this dual relation in~(\ref{FurtherDuality}) clearly
identifies the corresponding object with a D8-brane of type A string
theory. Interestingly, (\ref{AngleDef}) implies that there can be a
non-right angle between the D8-brane and the O8-plane with 
$\cot(\chi)=\frac{c_{1}}{N}\frac{R_{8}}{R_{9}}\propto \frac{c_{1}}{N}g_{A}^{2}$.
Unfortunately, our purely algebraic formalism does not allow us to see the
individual values of $c_{1}$, $N$ and the proportionality constant, but they
must be physically chosen so that: 
$\frac{1}{2\pi}\int_{T^{2}/\Z_{2}}\Tr(F_{89})dx^{8}dx^{9}=c_{1}=16$.
This is similar to the case of~\cite{Gan2}, where the type of brane
necessary for anomaly cancellation was obtained from the shift vector,
but not their number.

Let us then study the case of a $T^{2}/\{(-1)^{F}S,\Z_2\}$ orbifold of
M-theory (where the shift operator $S$ only acts on $x^{10}$). We want to show that it gives an alternative M-theory
lift of the same type 0' $T^{2}/(-1)^{f_R}\Omega I_{2}$ orientifold that 
we have just studied. Before we discuss the brane configuration, it is necessary to discuss
the case of M-theory on $S^{1}/\{(-1)^{F}S,\Z_2\}$ to understand the effect of taking both
orbifold projections on the same circle. We first remark that the orientifold group has
four elements: $\{\Id,(-1)^{F}S, I_1,(-1)^{F}I'_1\}$, where $I'_1=S I_1$. While $I_1$ is 
a reflexion of the coordinate $x^{10}$ with respect to $x^{10}=0$, $I_1'$ is a reflection
of $x^{10}$ with respect to $x^{10}=\pi/2$. In particular, $I_1$ has two fixed points at
$x^{10}=0$ and $\pi$, while $I'_1$ has two fixed points at
$x^{10}=\pi/2$ and $3\pi/2$, and $S$ has no fixed point. Consequently, the fundamental
domain is an interval $[0,\pi/2]$ and there are three types of twisted sectors, the usual
bosonic closed string twisted sector of $(-1)^{F}S$ that leads to a type 0 spectrum and 
two open strings twisted sectors sitting at each pair of fixed points. What is not known,
however, is the precise resulting gauge symmetry and twisted spectrum.
There is a dual picture of the same model, where one first uses the $S$ symmetry to reduce
the circle by half, and then considers the projection by $I_1$ which replaces the circle by
the interval. This second picture resembles the non-supersymmetric heterotic orbifold
of M-theory discussed in~\cite{FabHora}, except that these authors did not include a closed
string twisted sector, which hopefully helps stabilizing the non-supersymmetric theory.
We now conjecture that M-theory on $T^{2}/\{(-1)^{F}S,\Z_2\}$ is the strong coupling limit
of the $S^{1}/(-1)^{f_R}\Omega I_{1}$ orientifold of type 0A string theory and is thus
T-dual to the $T^{3}/\Z_{2}\times S^1/(-1)^{F}S$ orbifold of M-theory through a double
T-duality, modulo the appropriate breaking of gauge groups by Wilson lines.

Let us be more concrete. We need to reduce to type 0A string theory on an orbifolded direction,
then T-dualize to type 0B on a normal toroidal direction to reach a type 0'
$T^{2}/(-1)^{f_R}\Omega I_{2}$ orientifold, as in the following mapping: 
\begin{equation}
\begin{CD}
\mbox{M-theory on }T^{8}\times T^2/\{(-1)^{F}S,\Z_2\}  \\
@VV{M_{P} R_{10}\rightarrow 0}V \\
\mbox{type 0A on } T^8\times S^1/(-1)^{f_R}\Omega I_1
@>{\mathcal{T}_8}>> \mbox{type 0B on } T^7\times T^2/(-1)^{f_R}\Omega I_2 
\end{CD}
\label{Dualities}
\end{equation}

First, we have to select one definite shift vector from all equivalent ones.
We take the one that has the particularity to descend from the more general $T^{2}/\Z_{n}$
serie of shift vectors of the form $n\tilde{\delta}^{[2]}-\a_{7}$, namely: 
\begin{equation}
\xi^{[2,2]}=2(\La^{7}-\La^{8})+\La^{\{2\}}=2\tilde{\d}^{[2]}
-\tilde{\xi}^{[2,2]}=((2)^{8},1,1)\,, \label{xi2}
\end{equation}
which lead to the mass formulae: 
\begin{equation}
\begin{CD} {\displaystyle M_{P}^{-9}V^{-1}e^{\left\langle \xi^{[2,2]},
H_R\right\rangle}= M_{P}^{9}R_{1}\cdots R_{8}}@>M_{P}R_{10}\rightarrow 0>>
{\displaystyle \frac{M_{s}^{9}}{g_A^3}R_{1}\cdots R_8
\overset{\mathcal{T}_{8}}{\longrightarrow} \frac{M_{s}^{10}}{g_B^3}
R_{1}\cdots R_{7}R_8^{'2}\,.} \end{CD}
\label{RootReductionT2}
\end{equation}
We immediately see that we end up with the same objects as in~(\ref{RootReduction}) 
and~(\ref{FurtherDuality}) and the analysis of fluxes and angles is completely parallel.
In a sense, the presence of these two different M-theory lifts of the same string orientifold 
reflects the equivalence between T-dualizing $S^{1}/\Omega\Z_{2}$ in the transverse
space and T-dualizing $T^{3}/\Omega\Z_{2}$ along an orbifold direction.
We will use a similar property later to relate $\xi^{[6,2]}$ and $\xi^{[7,2]}$.

We can now turn to the $T^{6}/\{(-1)^{F}S,\Z_{2}\}$ orbifold of M-theory. 
We will study this case along the same line as 
$T^{2}/{\mathbbm{Z}}_{2}$, first reducing on an orbifolded direction 
to a type 0A theory orientifold, then T-dualizing along a transverse direction
to a type 0B orientifold: 
\begin{equation}\label{Dualities2}
\begin{CD} 
\mbox{M-theory on }T^{4}\times T^6/\{(-1)^{F}S,\Z_2\} \\
@VV{M_{P}R_{10}\rightarrow 0}V \\ \mbox{type 0A on }T^{4}\times T^{5}/(-1)^{f_{R}}\Omega I_{5}
@>{\mathcal{T}_{4}}>> \mbox{type 0B on } T^{3}\times T^{6}/(-1)^{f_{R}}\Omega I_{6}\, 
\end{CD}  
\end{equation}
We will again have a system of $N$ pairs of magnetized D9- and D9'-branes, 
now contributing to cancel the -1/4 units of negative D3-brane charge carried 
by each of the 64 O3-planes. This can be achieved by the Chern-Simons coupling: 
\begin{equation*}
\frac{M_{s}^{4}}{(2\pi)^{3}}\int_{{\mathbbm{R}}\times T^{3}}C_{4}\cdot 
\frac{1}{(2\pi)^{3}}\int_{T^{6}/{\mathbbm{Z}}_{2}}\mathrm{Tr}(F_{2}\wedge
F_{2}\wedge F_{2})\,.
\end{equation*}
In the case of a factorizable metric, we can separate $T^{6}/\Z_2$ into 3 
$T^{2}/{\mathbbm{Z}}_{2}$ sub-orbifolds, and only $F_{56}$, $F_{78}$ and 
$F_{49}$ yield non-trivial fluxes. Instead of $c_{1}$ and $N$, we now 
introduce for each pair of coordinates $(x^{i};x^{j})$ of the $T^{2}$'s
pairs of quantized numbers denoted by $(m_{ij}^{a},n_{ij}^{a})$~\cite{Marchesano}. 
The index $a$ here numbers various stacks of $N_{a}$ pairs of branes, 
with different fluxes. In the dual 0A picture, the $m_{ij}^{a}$ and $n_{ij}^{a}$'s 
give wrapping numbers around the directions parallel, respectively perpendicular, 
to the O6-planes and $a$ distinguishes between wrappings of branes around different 
homology cycles. With an appropriate normalization of cohomology bases on the homology
cycles, one obtains: 
\begin{equation*}
\frac{1}{(2\pi)^{3}}\mathrm{Tr}\left( \int_{T^{2}/\Z_2}F_{56}\,dx^{5}dx^{6}
\int_{T^{2}/{\mathbbm{Z}}_{2}}F_{78}\,dx^{7}dx^{8}
\int_{T^{2}/{\mathbbm{Z}}_{2}}F_{49}\,dx^{^{\prime }4}dx^{9}\right)
=\sum_{a}N_{a}m_{56}^{a}m_{78}^{a}m_{49}^{a}=16
\end{equation*}
On the other hand, Chern-Simons couplings to higher forms such as $C_{5}$, $C_{7}$
and $C_{9}$ are determined by expressions which also include $n_{ij}^{a}$ factors. 
For example, the D9-charge is related to $\sum_{a}N_{a}n_{56}^{a}n_{78}^{a}n_{49}^{a}$. 
The wrapping numbers should then be chosen in a way that all those other total charges cancel. 
There are in principle several ways to achieve this, but it is not our main focus, so we
will not give a specific example here (see~\cite{Maillard} for concrete
realizations in the supersymmetric case). Rather, following the $T^{6}/\Z_{2}$ 
case above, one wishes to study the magnetized D9-brane action given by our algebraic
method, deduce from it that certain pairs of radii are related and then
perform a triple T-duality along $(x^{4};x^{6};x^{7})$ to exchange the
fluxes against tilting angles between O6-planes and pairs of D6-branes
and their image D6'-branes.

Keeping this framework in mind, we first recall the choice of shift vector
that comes from the general $T^{6}/\Z_n$ orbifold serie. It is given by: 
\begin{equation}
\xi^{[6,2]}=2(\La^{7}-\La^{8})+\tilde{\xi}^{[6,2]}=
2\tilde{\d}^{[6]}-\a_{3}+\a_{5}-\a_{7}=((2)^{4},1,3,3,1,1,1)\,.
\label{xi6}
\end{equation}
where $\tilde{\xi}^{[6,2]}$ differs from its expression $\xi^{[6,n]}$ for $n=2$ given in Section~\ref{Classrep} because the charge $Q_{3}$ is now $-1$ instead of $-2$. Let us again follow the dualities~(\ref{Dualities2}) to see
how the D9-brane is expressed in this formalism:
\begin{equation*}
\begin{CD} 
{\displaystyle \frac{e^{\left\langle \xi^{[6,2]}|H_R\right\rangle}}{M_{P}V}=
M_{P}^{9}R_{1}\cdot\cdot R_{4} (R_6 R_7)^2} @>{M_{P}R_{10}\rightarrow 0}>>
{\displaystyle \frac{M_{s}^{9}}{g_A^3}R_{1}\cdot\cdot R_{4} (R_6 R_7)^2
\overset{\mathcal{T}_{4}}{\longrightarrow}
\frac{M_{s}^{10}}{g_B^3}R_{1}\cdot\cdot R_{3}(R^{\prime}_{4} R_6 R_7)^2 \,.}
\end{CD}
\end{equation*}
This can match the action of a D9-brane if $R_{5}\propto R_{6}$, $R_{7}\propto R_{8}$ 
and $R_{9}\propto R'_{4}/g_{B}^{2}$. On the type A side, this again means that 
$R_{9}\propto R_{4}/g_{A}^{2}$, and one verifies easily that $\mathcal{T}_{4}$ 
indeed maps the D9-brane to a D8-brane extended along all directions except $x^{4}$. 
This D8-brane is tilted with respect to the O4-plane in the $(x^4;x^9)$-plane by an angle
$\cot(\chi _{49}^{a})=\frac{m_{49}^{a}R_{4}}{n_{49}^{a}R_{9}}$ and still carries
magnetic fluxes $F_{56}$ and $F_{78}$. T-dualizing further along $x^{6}$ and $x^{7}$ 
leads to a D6-brane extended in the hypersurface along 
$(x^{0};x^{1};x^{2};x^{3};x^{5};x^{8};x^{9})$ with mass: 
\begin{equation*}
\frac{M_{s}^{7}}{g_{A}}R_{1}\cdots R_{4}R_{6}R_{7}\sim 
\frac{M_{s}^{7}}{g_{A}^{3}}R_{1}\cdots R_{3}R_{5}R_{8}R_{9}\,.
\end{equation*}
Then, we can interpret this brane as one of the $N_{a}$ D6-branes exhibiting
two additional non-right angles with respect to the orientifold O6-plane, given by 
$\cot(\chi_{56}^a)=\frac{m_{56}^{a}R_{6}}{n_{56}^{a}R_{5}}$ and 
$\cot(\chi_{78}^{a})=\frac{m_{78}^{a}R_{7}}{n_{78}^{a}R_{8}}$. It is of course understood
in this discussion that the appropriate image D$p'$-branes are always present.

Finally, we still wish to study $T^{7}/\{(-1)^{F}S,\Z_{2}\}$ orbifolds of
M-theory. For this purpose, we use the permutation of $\tilde{\xi}^{[7,2]}$
describing an orbifold in $(x^{3};\ldots;x^{9})$ given by 
$\tilde{\xi}_{\sigma}^{[7,2]}=(0,0,(1)^{7},2)$ in the following fashion: 
\begin{equation*}
\xi^{[7,2]}=2(\La^{2}-\La^{3}+\La^{5}-2\La^{8})+
\tilde{\xi}_{\sigma}^{[7,2]}=(2,2,3,3,1,3,3,1,1,2).
\end{equation*}
This time, we follow the successive mappings
\begin{equation*}
\begin{CD}
\mbox{M-theory on } T^{2}\times T^{8}/(-1)^{F}S \\
@VV{M_{P} R_{10}\rightarrow 0}V \\ \mbox{type 0A on }\times T^{2}\times
T^{7}/(-1)^{f_L}\Omega I_7 @>{\mathcal{T}_3}>> \mbox{type 0B on }
T^{3}\times T^{6}/(-1)^{f_R}\Omega I_6
\end{CD}
\end{equation*}
leading to the mass formulae: 
\begin{equation*}
\begin{CD} {\displaystyle M_{P}^{12}R_{1} R_2 (R_3 R_{4} R_6 R_7)^2 R_{10} }
@>{M_{P}R_{10}}>>{\displaystyle \frac{M_{s}^{11}}{g_A^3} R_{1} R_2 (R_3
R_{4} R_6 R_7)^2 \overset{\mathcal{T}_{3}}{\longrightarrow}
\frac{M_{s}^{10}}{g_B^3}R_{1}\cdots R^{\prime}_{3}(R_{4} R_6 R_7)^2 \,.}
\end{CD}
\end{equation*}
and we obtain again the same type 0B $T^{6}/(-1)^{f_R}\Omega I_6$ orientifold
as above, while tilting angles in the dual type IIA picture can again be
obtained by ${\mathcal{T}_{467}}$.

\begin{table}[!h]
\begin{center}
$
\begin{array}{|c|c|c|c|c|}
\hline
q & \xi^{[q,2]} & \text{physical basis} & \text{Dynkin label} & \left\vert
\Lambda \right\vert ^{2} \\ \hline
2 & \alpha _{(-1)^{2}0^{4}1^{6}2^{8}3^{10}4^{12}5^{14}6^{10}7^{5}8^{6}} & 
(2,2,2,2,2,2,2,2,1,1) & [000000010] & -2 \\ 
3 & \alpha _{(-1)^{2}0^{4}1^{6}2^{8}3^{10}4^{12}5^{15}6^{11}7^{5}8^{7}} & 
(2,2,2,2,2,2,3,3,1,2) & [010000001] & -2 \\ 
6 & \alpha _{(-1)^{2}0^{4}1^{6}2^{8}3^{9}4^{12}5^{15}6^{10}7^{5}8^{6}} & 
(2,2,2,2,1,3,3,1,1,1) & [010001000] & 2 \\ 
7 & \alpha _{(-1)^{2}0^{4}1^{7}2^{10}3^{11}4^{14}5^{17}6^{11}7^{5}8^{7}} & 
(2,2,3,3,1,3,3,1,1,2) & [000100100] & 2 \\ \hline
\end{array}
$
\end{center}
\caption{Physical class representatives for $T^{10-q}\times T^q/\Z_2$ orbifolds of M-theory of the second kind}
\label{twisted3}
\end{table}

Overall, we have a fairly homogeneous approach to these four different
orbifolds of M-theory and it should not be too surprising that their
untwisted sectors build the same algebra. We finally summarize the shift vectors we
used for physical interpretation in Table~\ref{twisted3}. 
It is remarkable that these roots are found at level 6 and 7 in $\a_{8}$, 
showing again that a knowledge of the $\eten$ root space at high levels is essential 
for the algebraic study of M-theory orbifolds.

Another fact worth mentioning is that our $\Z_{2}$ shift vectors
either have norm 2 or -2, in contrast to the null shift vectors of
Section~\ref{SecOrbi1}. This lightlike characteristic has been proposed 
in~\cite{Gan1,Gan2} to be a general algebraic property characterizing Minkowskian
branes in M-theory. Similarly, these authors associated instantons with real
roots of $\eten$, viewed as extensions of roots of ${\mathfrak{e}}_{8}$,
that all have norm 2. However, we have just shown that Minkowskian objects
can just as well have norm 2, or -2, and perhaps almost any. We suggest that
the deciding factor is the threshold rather than the norm (at least for
objects coupling to forms, forgetting for a while the exceptional case of
Kaluza-Klein particles that have negative threshold, when they are instantonic 
and null threshold, when they are Minkowskian). Indeed, instantonic objects have 
threshold 0, while Minkowskian ones have threshold 1. This approach is compatible with
the point of view of~\cite{DamNic1}, as explained in Section~\ref{Thresh},
as well as with the results of this subsection. Some higher threshold roots
also appear in Table~\ref{Rep1} and~\ref{Rep2}, however, but we leave their
interpretation for further investigation.

\section{Shift vectors for $\Z_n$ orbifolds: an interpretative prospect}
\label{ShiftN}

Now that we have an apparently coherent framework to treat $\Z_2$ M-theory orbifolds, it is tempting to try to generalize it to all $\Z_n$ orbifolds.
To understand how this could be done, it is instructive to look at Tables~\ref{Rep1} and \ref{Rep2}. 
As mentioned at the end of Section~\ref{Classrep}, one notices that shift vectors for $T^q/\Z_n$ orbifolds can typically be grouped in series, for successive values of $q$ and $n$. As an illustration, we  
give one such serie (i.e. relating orbifolds with all charges $\pm 1$) in the following table:
\begin{equation*}\hspace{-0.7cm}
\begin{array}{|c|c|c|c|c|c|}
\hline
n \backslash q & 2 & 4 & 6 & 8 & 10 \\ \hline
2 & ((2)^8,1,1) & ((2)^6,3,1,1,1) & ((2)^4,1,3,3,1,1,1)
& (2,2,3,1,1,3,3,1,1,1) & / \\ 
3 & ((3)^8,2,1) & ((3)^6,4,2,2,1) & ((3)^4,2,4,4,2,2,1) & 
(3,3,4,2,2,4,4,2,2,1) & (2,4,4,2,2,4,4,2,2,1) \\ 
4 & ((4)^8,3,1) & ((4)^6,5,3,3,1) & ((4)^4,3,5,5,3,3,1) & 
(4,4,5,3,3,5,5,3,3,1) & (3,5,5,3,3,5,5,3,3,1) \\ 
5 & ((5)^8,4,1) & ((5)^6,6,4,4,1) & ((5)^4,4,6,6,4,4,1) & 
(5,5,6,4,4,6,6,4,4,1) & (4,6,6,4,4,6,6,4,4,1) \\ 
6 & ((6)^8,5,1) & ((6)^6,7,5,5,1) & ((6)^4,5,7,7,5,5,1) & 
(6,6,7,5,5,7,7,5,5,1) & (5,7,7,5,5,7,7,5,5,1) \\ \hline
\end{array}
\end{equation*}

From this table, it should be immediately apparent that typical shift vectors for 
$T^q/\Z_n$ orbifolds, with $q\in2\N$ are given by (some permutation of):
\begin{eqnarray*}
\xi &=& n\tilde{\delta}+\sum_{i=1}^{q/2}(-1)^{q/2-i}\,p_{i}\,\a_{7-q+2i}\\
&=&((n)^{10-q},n+p_1,n-p_1,n-p_2,n+p_2,\ldots,n-p_{q/2},p_{q/2})	\,
\end{eqnarray*}
and have a threshold bigger or equal to 1 since 
$1\leq q_i\leq n-1$, $\forall i=1,\ldots,q/2$. In analogy with the $\Z_2$ orientifold cases, 
it is tempting to think of the "average" value $((n)^{9},0)$ as spacetime-filling
branes, and of the deviations $q_{i}\a_{7-q+2i}$ as fluxes in successive pairs 
of (orbifolded) dimensions. Of course, the fluxes are only directly interpretable as such 
after the reduction to string theory. In the $\Z_2$ examples, they appeared because an M-theory 
orbifold turns into a string theory orientifold with open strings twisted sectors exhibiting
non-abelian Chan-Paton factors. This allowed us to invoke Chern-Simons couplings of the form:
\begin{equation}
\int C_{10-q}\cdot \int \Tr(F^{q/2})
\end{equation}
on the world-volume of the space-filling branes that participate to tadpoles cancellation at 
the orbifold fixed points. Geometrically speaking, the more orbifolded directions, the more non-trivial 
fluxes can be switched on, producing higher non-zero Chern numbers that reflect the increasingly 
complex topology in the presence of several conifold singularities at each fixed point.
A further research direction is to determine which kind of flux could appear in which $\Z_n$ orbifolds.

In any case, one should not forget that the orbifolded directions in the string theory 
limit are not exactly the same as in the original M-theory orbifold, so that a bit of 
caution is required when trying to interpret the shift vector directly, without going 
through a chain of dualities leading to a better-known string theory soliton.

Our proposal is to regard the mass formulae associated to these shift vectors as
M-theory lifts of the resulting string theory brane configurations, that are somehow necessary
for the M-theory orbifolds to be well-defined, in a sense which remains to be understood.

It also remains unclear how the change of average value of the components of the shift
vector from 2 to $n$ determines the fact that we have a higher order orbifold. Intuitively,
it should reflect the presence of more twisted sectors, but is a priori not related to the
different number of fixed points.

All these questions are of course of primary interest to obtain non-trivial physical
information from our algebraic toolkit and we will pursue them in forthcoming research
projects. They will be addressed in future publications.

\section{Conclusion}
In this paper, we have aimed at developing a rigorous and general algebraic procedure 
to study orbifolds of supergravity theories using their U-duality symmetry. We were
particularly interested in the $\e_{11-D|11-D}$ serie of real split U-duality algebras
for $D=1,\ldots,8$. Essentially, the procedure can be decomposed in the following 
successive steps. First, one constructs a finite order non-Cartan preserving 
inner automorphism describing the orbifold action in the complexified algebra $\e_{11-D}$. 
This $n$th-order rotation automorphism reproduces the correct $\Z_n$-charges of the physical states
of the theory, when using the "duality" mapping relating supergravity fields
and directions in the coset $\e_{11-D|11-D}/\mathfrak{k}(\e_{11-D|11-D})$
(in the symmetric gauge). Next, one derives the complexified invariant subalgebra 
satisfied by the null charge sector and fixes its real properties by taking its 
fixed point subalgebra under the restricted conjugation. One then moves to 
an eigenbasis, on which the orbifold action takes the form of a Cartan-preserving 
(or chief) inner automorphism, and computes, in terms of weights, the classes of shift vectors 
reproducing the expected orbifold charges for all root spaces of $\e_{11-D}$. 
In $D=1$, one uses the invariance modulo $n$ to show that every such class contains 
a root of $\eten$, which can be used as the class representative. In a number of cases, 
these roots can be identified with Minkowskian objects of M-theory or of the lower-dimensional 
string theories, and interpreted as brane configurations necessary for anomaly 
cancellation in the corresponding orbifold/orientifold setups.

In fact, for a given $T^q/\Z_n$ orbifold, the first two steps only have to be carried 
out explicitly once in $\e_{q+1}$ for the compactification space $S^1\times T^q/\Z_n$, 
and need not be repeated for all $T^p\times T^q/\Z_n$. Rather, one can deduce in which way 
the Dynkin diagram of the invariant subalgebra will get extended upon further compactifications.
This is relatively straightforward until $D=3$, but requires some more care in $D=2,1$, when
the U-duality algebra becomes infinite-dimensional. In $\eten$, in particular, a complete 
determination of the root system of the invariant subalgebra requires in principle to look for
all invariant generators. This could in theory be done, provided we know the full decomposition of
$\eten$ in representations of $\sl(10,\R)$. However, one of the conclusions of our analysis is that
once we understand the structure of $\ginv$ at low-level, its complete root system can be inferred
from the general structure of Borcherds algebras.

By doing so, however, one realizes that there are three qualitatively different possible
situations from which all cases can be inferred. The determining factor is the invariant 
subalgebra in $D=3$. If this subalgebra of $\e_{8|8}$ is simple, its extension in $\e_{10|10}$ 
is hyperbolic and non-degenerate. This happens for $T^q/\Z_2$ for $q=1,4,5,8,9$, as already
shown in~\cite{Gan2} by alternative methods. 
If it is on the other hand semi-simple, we obtain, in $D=2$, what we called an affine
central product. It denotes a product of the affinization of all simple factors present in $D=3$,
in which the respective centres and derivations of all factors are identified. 
Descending to $D=1$, all affine factors reconnect through $\a_{-1}$ in a simple Dynkin diagram,
leading to a degenerate hyperbolic Kac-Moody algebra, but without its natural centre(s) 
and derivation(s). This is the case for all remaining $\Z_2$ orbifolds, as well as for 
$T^6/\Z_n$ orbifolds with $n=3,4$. Finally, if an abelian factor is present in $\e_8$, its 
affinization in $\e_9$, $\hat{\u}(1)$, turns into all multiples of an imaginary root in 
$\eten$, which also connects through $\a_{-1}$ to the main diagram, thus leading to a 
Borcherds algebra with one isotropic simple root. 
Although it was conceptually clear to mathematicians that Borcherds algebras can emerge 
as fixed-point subalgebras of Kac-Moody algebras under automorphisms, we found here 
several explicit constructions, demonstrating how this comes about in examples of a kind
that does not seem to appear in the mathematical literature.

In the first case, the multiplicity of invariant roots is inherited from $\eten$, 
in the other two cases, however, great care should be taken in understanding how the 
original multiplicities split between different root spaces. In fact, the Borcherds/indefinite 
KM algebras appearing in these cases provide first examples of a splitting of multiplicities 
of the original KMA into multiplicities of several roots of its fixed point subalgebra. This 
is strictly speaking the case only for the algebras as specified by their Dynkin diagram, but 
one should keep in mind that the quotient by its possible derivations suppresses the operators 
that could differentiate between these roots, and recombines them into root spaces of 
the original dimension, albeit with a certain redistribution of the generators. In fact, 
it is likely that a computation of the root multiplicities by an appropriate Kac-Weyl formula 
for GKMA based on the root system of the Dynkin diagram would predict slightly smaller 
root spaces than those of the fixed-point subalgebra that are obtained from our method. 
However, it is not absolutely clear what is the right procedure to compute root multiplicities 
in GKMA. This is a still largely open question in pure mathematics, on which our method will 
hopefully shed some light.

Along the way, we also explicitly showed, in the $T^4/\Z_n$ case, how to go
from our completely real basis for $\ginv$, described by a fixed-point 
subalgebra under the restricted conjugation, to
the standard basis of its real form, obtained from the Cartan decomposition. This is especially
interesting in the affine case, where we obtained the relation between the two affine
parameters and their associate derivations.

Even though the present paper was focused on the breakings of U-duality symmetries, 
it is clear that, in another perspective, the same method can in principle be applied 
to obtain the known classification of (symmetric) breaking patterns of 
the $E_8\times E_8$ gauge symmetry of heterotic string theory (or any other gauge 
symmetry) by orbifold projections. Indeed, our result in $D=3$ for breakings of $\e_8$ 
can be found in the tables of~\cite{Koba,Koba2}, where they are derived from the 
Kac-Peterson method using chief inner automorphisms. Reciprocally,  one might wonder 
why we did not use the Kac-Peterson method to study U-duality symmetry, too. It is certainly 
a beautiful and simple technique, very well suited to classify all possible non-isomorphic 
symmetry breakings of one group by various orbifold actions. However, calculating with 
$\Z_n$-rotation automorphisms instead of Cartan-preserving ones has a number of advantages
when dealing with U-duality symmetries. In the Kac-Peterson method, one first fixes $n$,
then lists all shift vectors satisfying the condition $(\La,\theta_{G})\leq n$ of 
Section~\ref{ShiftSec}, which allows to obtain all non-isomorphic breakings. In the end, 
however, one has sometimes to resort to different techniques to associate these breakings 
with a certain orbifold with determined dimension and charges. 

Here, we adopt a quite opposite philosophy, by resorting to
non-Cartan preserving inner automorphisms with  a clear geometrical interpretation. 
In this perspective, one starts by fixing the dimension and charges of the orbifold and 
then computes the corresponding symmetry breaking, which allows to discriminate easily between 
a degenerate finite order rotation and an effective one. Only then do we reexpress this 
automorphism in an eigenbasis of the orbifold action, in which it takes the form of a chief 
inner automorphism, and compute the class of associated shift vectors. 
Doing so, we can unambiguously assign a particular class of shift vectors to a definite orbifold 
projection in space-time. Note that such shift vectors will typically not satisfy 
$(\La,\theta_{G})\leq n$, so that a further change of basis is required to relate them to 
their conjugate shift vector in the Kac-Peterson formalism (we have shown in 
Section~\ref{ShiftSec} how to perform this change of basis explicitly). However, this process 
may obscure the number of orbifolded dimensions and the charge assignment on the 
Kac-Peterson side.

Furthermore, another reason for not resorting to the Kac-Peterson method
is that we are not so much interested in all possible breakings of one particular group, 
say $E_8$, as in determining the fixed-point subalgebras for the whole $E_r$ serie.
Consequently, we can concentrate on the $T^q/\Z_n$ orbifold action in $E_{q+1}$ and 
then extend the result to the whole serie without too many additional computations, 
since the orbifold rotation acts trivially on the additional compactified dimension 
and the natural geometrical interpretation of the $SL(r,\R)\in E_{r}$ generators 
has been preserved. On the other hand, the change of basis necessary to obtain a 
shift vector satisfying the Kac-Peterson condition can be completely different in $E_r$ 
compared to $E_{r-1}$. Accordingly, starting from such a shift vector for $E_{r-1}$, 
there is no obvious way to obtain its extension describing the same orbifold in $E_{r}$. 
Finally, and much more important to us is the fact that there is no known way 
to extend the Kac-Peterson method to the infinite-dimensional case.

The above discussion has concentrated on the part of this work where
the invariant subalgebras of $\mathfrak{e}_{11-D|11-D}$ under an $n$th-order
inner automorphism were derived. In $D\geq 3$, these describe the residual
 U-duality symmetry and bosonic spectrum 
of supergravity theories compactified on orbifolds and map to the massless bosonic spectrum 
of the untwisted sector of orbifolded string theories. In such cases, these results have been
known for a long time. They are however new in $D=2,1$, which was the main focus of this research
project. In particular, the $D=1$ case is very interesting, since the hyperbolic U-duality
symmetry encountered there is expected to contain non-perturbative information, as well.
Indeed, the specific class representatives of shift vectors we find correspond to higher 
level roots of $\eten$ which have no direct interpretation as supergravity fields. It is thus 
tempting to try to relate them to non-classical effects in M-theory which might give us
information on the twisted sector of orbifolds/orientifolds of the descendant string theories.

Let us now discuss this more physical interpretative aspect inspired from the work 
of~\cite{Gan2}, where the shift vectors for a restricted class of $\Z_2$ orbifolds 
of M-theory were shown to reproduce the mass formulae of Minkowskian branes, 
which turned out to be the correct objects to be placed at each orbifold 
fixed point to ensure anomaly/tadpole cancellation. We have extended this analysis
to incorporate other $\Z_2$ orbifolds of M-theory, which are non-supersymmetric and 
should be considered in bosonic M-theory. They have the particularity to break 
the infinite U-duality algebra to indefinite KMAs. These orbifolds reduce to
$T^2/(-1)^{f_R}\Omega\Z_2$ and $T^6/(-1)^{f_R}\Omega\Z_2$ orientifolds of the type 0B 
string theory in which pairs of magnetized D9- and D9'-branes are used to cancel the 
O7- (resp. O3-)plane charges. They are part of a chain of dual orientifolds starting from
type 0B string theory on $(-1)^{f_L}\Omega$, a tachyon-free theory believed to be 
well-defined, usually referred to as type 0' string theory.
We have then shown that the $\eten$ roots playing the r\^ole of class 
representatives of shift vectors in these cases can be interpreted as 
such space-time filling D9-branes carrying the appropriate configuration of magnetic fluxes. 
This identification could in turn serve as a proposal for M-theory lifts of such type 
0B orientifolds, as generated by certain exotic objects corresponding to $\eten$ roots
that are not in $\e_9$. Finally, these type IIB setups have an alternative reading
in the T-dual type IIA pictures where the magnetic fluxes appear as tilting angles 
between O8- (resp. O6-)planes and D8- (resp. D6-)branes and their image branes,
our analysis providing an algebraic characterization of this tilting angle.

As for $\Z_{n\geq 3}$ case, even though we have treated only a few examples explicitly, 
we have noticed that their associated shift vectors fall into series of roots of $\eten$,
for successive values of $q$ and $n$, with remarkable regularity. This has provided
us with a facilitated procedure for constructing shift vectors for any $T^q/\Z_n$ 
orbifold which acts separately on each of the $(q/2)$ $T^2$ subtori. These roots of level 
$3n$ are classified in Tables~\ref{Rep1} and~\ref{Rep2}. Despite the remarkably regular 
structure of such roots, it is not completely clear how to extract information on
the correct anomaly/tadpole-cancelling brane configurations of the corresponding orbifolds.
In particular, the components of the shift vectors transverse to the orbifold increase 
monotonously with $n$, so that their interpretation requires novel ideas.
However, it is clearly of interest to generalize 
the identification of such brane constructions for $\Z_n$-shift vectors with $n>2$, 
and to understand their possible relation to twisted sectors and/or fluxes present 
in the related string orbifolds. Hopefully, this can be done in a systematic manner, 
reproducing what is known about string theory orbifolds/orientifolds and leading to 
predictive results about less-known types of M-theory constructions.

Another future direction of research would consist in investigating more complicated 
orbifold setups in our algebraic framework, in which, for instance, several projections 
of various orders are acting on the same directions. This could possibly lead to new 
interesting classes of GKMAs. In general, however, not only one, but two or more shift 
vectors will be necessary to generate such orbifolds and should from a physical perspective
be interpreted separately. This will hopefully open the door to working 
out the physical identification of yet a larger part of the $\eten$ root system, 
and constitute another step in the understanding of the precise relation between 
M-theory and $\eten$.
\vspace{0.7cm}
\newline
{\bf {\Large Acknowledgements}}
\vspace{0.5cm}

During the preparation of this work, we profited from numerous discussions with our colleagues
on specific parts of this project. We wish to thank Tatsuo Kobayashi, Tristan Maillard, Claudio Scrucca,
Bernhard Kroetz, Shinya Mizoguchi, Hermann Nicolai, Jean-Pierre Derendinger, Ori Ganor, Arjan Keurentjes, 
Axel Kleinschmidt, Marios Petropoulos and Nikolaos Prezas for their remarks and ideas. We are particularly
indebted to Matthias Gaberdiel for correspondence and discussion regarding non-supersymmetric strings 
and to Jun Morita for his help pertaining to infinite-dimensional Lie algebras.
M.B. also wants to acknowledge the financial support from the Swiss National Science Foundation
(SNSF) under the grant PBNE2-102986 and from the Japanese Society for the Promotion 
of Science (JSPS) under grant number P0477, as well as the friendly support from Hikaru Kawai 
and his particle theory group at Ky\={o}to University. L.C. acknowledges the financial support 
from the Swiss National Science Foundation (SNSF) and by the Commission of the European Communities 
under contract MRTN-CT-2004-005104.

\appendix
\section{Highest roots, weights and the Matrix $R$}
\label{AppendixA}
i) \underline{The matrix $R$:} herebelow, we give the expression of the matrix $R$ used in Section~\ref{ExE} to define the root lattice metric $g_{\varepsilon}$
(\ref{metrepsi}) in the physical basis:
\[
R=\left( 
\begin{array}{cccccccccc}
1 & 1 & 1 & 1 & 1 & 1 & 1 & 2/3 & 1/3 & 1/3 \\ 
0 & 1 & 1 & 1 & 1 & 1 & 1 & 2/3 & 1/3 & 1/3 \\ 
0 & 0 & 1 & 1 & 1 & 1 & 1 & 2/3 & 1/3 & 1/3 \\ 
0 & 0 & 0 & 1 & 1 & 1 & 1 & 2/3 & 1/3 & 1/3 \\ 
0 & 0 & 0 & 0 & 1 & 1 & 1 & 2/3 & 1/3 & 1/3 \\ 
0 & 0 & 0 & 0 & 0 & 1 & 1 & 2/3 & 1/3 & 1/3 \\ 
0 & 0 & 0 & 0 & 0 & 0 & 1 & 2/3 & 1/3 & 1/3 \\ 
0 & 0 & 0 & 0 & 0 & 0 & 0 & 2/3 & 1/3 & 1/3 \\ 
0 & 0 & 0 & 0 & 0 & 0 & 0 & -1/3 & 1/3 & 1/3 \\ 
0 & 0 & 0 & 0 & 0 & 0 & 0 & -1/3 & -2/3 & 1/3
\end{array}
\right) \,.
\]
\newline
ii) \underline{Highest roots of the exceptional $E_r$ chain:}
We list the highest roots of the finite Lie algebras of the chain $\an_{1}\subset \an_{2}\subset \ldots
\subset \an_{4}\subset \dn_{5}\subset \e_{6}\subset \e_{7}\subset \e_{8}$, appearing throughout this
article:
\begin{eqnarray*}
\theta _{A_{1}} &=&\a _{8}\,, \\
\theta _{A_{i}} &=&\a _{8-i}+..+\a _{7}\,,\text{ \ \ }i=2,3 \\
\theta _{A_{4}} &=&\a _{5}+\a _{6}+\a _{7}+\a _{8}\,, \\
\theta _{D_{5}} &=&\a _{4}+2\a _{5}+2\a _{6}+\a _{7}+\a
_{8}\,, \\
\theta _{E_{6}} &=&\a _{3}+2\a _{4}+3\a _{5}+2\a _{6}+\a
_{7}+2\a _{8}\,, \\
\theta _{E_{7}} &=&\a _{2}+2\a _{3}+3\a _{4}+4\a
_{5}+3\a _{6}+2\a _{7}+2\a _{8}\,, \\
\theta _{E_{8}} &=&2\a _{1}+3\a _{2}+4\a _{3}+5\a
_{4}+6\a _{5}+4\a _{6}+2\a _{7}+3\a _{8}\,.
\end{eqnarray*}%
\newline
ii) \underline{Fundamental weights of $\eten$:}
The expression, on the set of simple roots, of the fundamental weights of $\eten$ defined by $(\Lambda^{i}|\a_{j})=\delta _{j}^{i}$ for $i,j=-1,0,1,..,8$ is obtained by inverting 
$\Lambda ^{i}=(A(\e_{10})^{-1})^{ij}\a _{j}$. In the physical basis, these weights have the particularly simple expression:
\begin{equation*}
\begin{array}{rcl|c|rcl|c}
&&& |\Lambda|^2  && & &|\Lambda|^2 \\ \hline
-\Lambda ^{-1}&=&(0,1,1,1,1,1,1,1,1,1)  &  0 &-\Lambda ^{4}  &=&(5,5,5,5,5,5,6,6,6,6)& -30\\[2pt]
-\Lambda ^{0} &=&(1,1,2,2,2,2,2,2,2,2)  &  -2  &-\Lambda ^{5}  &=&(6,6,6,6,6,6,6,7,7,7)& -42\\[2pt]
-\Lambda ^{1}  &=&(2,2,2,3,3,3,3,3,3,3) &  -6  & -\Lambda ^{6} &=&(4,4,4,4,4,4,4,4,5,5)&-18 \\[2pt]
-\Lambda ^{2}  &=&(3,3,3,3,4,4,4,4,4,4) &   -12  &-\Lambda ^{7} &=&(2,2,2,2,2,2,2,2,2,3)& -4\\[2pt]
-\Lambda ^{3}  &=&(4,4,4,4,4,5,5,5,5,5) &  -20 & -\Lambda ^{8}  &=&(3,3,3,3,3,3,3,3,3,3)& -10
\end{array}
\end{equation*}
For their expression in the root basis, see, for instance, \cite{KacMooWaki}. It can be recast
in the following recursion relations:
\[
\begin{array}{rclcrcl}
\Lambda^{-1}&=&-\delta \,, & &\Lambda ^{4}&=&2\Lambda ^{3}-\Lambda ^{2}-\a _{3}\,,
\\[2pt]
\Lambda^{0}&=&-(\a _{-1}+2\delta )\,, & &\Lambda ^{5}&=&2\Lambda ^{4}-\Lambda
^{3}-\a _{4} \,,\\[2pt] 
\Lambda^{1}&=&2\Lambda ^{0}+\theta _{E_{8}}\,, & &\Lambda ^{6}&=&\Lambda
^{3}+\theta_{D_{5}}\,, \\[2pt] 
\Lambda^{2}&=&2\Lambda ^{1}-\Lambda ^{0}-\a _{1}\,, & &\Lambda ^{7}&=&2\Lambda
^{6}-\Lambda^{5}-\a _{6} \,,\\[2pt]
\Lambda^{3}&=&2\Lambda ^{2}-\Lambda ^{1}-\a _{2} \,,& &\Lambda ^{8}&=&\Lambda
^{2}+\theta _{E_{6}} \,.
\end{array}%
\]

\section{The U-duality group for 11D supergravity}
\label{AppendixB}

This appendix is meant as a complement to Section \ref{sec:sing}, and
reviews the U-duality transformations for finite U-duality groups as
presented in \cite{OberPiol1}, which trivially extends to the hyperbolic case in $D=1$.

It has been shown in \cite{Elitz,Ob1,Ob2}, that the 
$E_{11-D|11-D}(\Z)$ chain of U-duality group relevant for M-theory incorporates a
generalized T-duality symmetry, which exchanges not only the radii of
10-dimensional IIA theory among themselves, but also the 11-dimensional
radius $R_{10}$ with any of them, leading to the transformation:

\begin{equation}
\mathcal{T}_{ijk}:R_{i}\rightarrow \frac{1}{M_{P}^{3}R_{j}R_{k}}\,,\text{ \
\ }R_{j}\rightarrow \frac{1}{M_{P}^{3}R_{i}R_{k}}\,,\text{ \ \ }
R_{k}\rightarrow \frac{1}{M_{P}^{3}R_{i}R_{j}}\,,\text{ \ \ }
M_{P}^{3}\rightarrow M_{P}^{6}R_{i}R_{j}R_{k}\,.  \label{Transf}
\end{equation}
for $i,j,k\in \{D,..,10\}$. To get the whole Weyl group of $E_{11-D}$, one
must supplement the transformation (\ref{Transf}) with the permutation of
all radii (belonging to the $SL(11-D,\Z)$ modular group of the
torus)%
\begin{equation*}
\mathcal{S}_{ij}:R_{i}\leftrightarrow R_{j}\,,
\end{equation*}
which is part of the permutation group $\mathcal{S}_{11-D}$ generated by the 
neighbour to neighbour permutations $\{S_{i,i+1}\}_{i=D,..,9}$. 
Then, taking the closure of the latter with the generator 
$\mathcal{T}_{89\,10}$ leads to the Weyl group:
\begin{equation}
W(E_{11-D})=\Z_{2}\overline{\times }\mathcal{S}_{11-D}
\label{WE10}
\end{equation}
with $\Z_{2}=\{\Id,\mathcal{T}_{89\,10}\}$. This gives the
whole set of Weyl generators in terms of their action on the M-theory radii.

If we compactify to IIA string theory by setting $M_{P}R_{10}\rightarrow 0$,
then the generators
\begin{equation*}
\mathcal{T}_{\hat{\imath}\hat{\jmath}10}:R_{\hat{\imath}}\rightarrow 
\frac{1}{M_{s}^{2}R_{\hat{\jmath}}}\,,\text{ \ \ }R_{\hat{\jmath}}\rightarrow 
\frac{1}{M_{s}^{2}R_{\hat{\imath}}}\,,\text{ \ \ }g_{A}\rightarrow
\frac{g_{A}}{M_{s}^{2}R_{\hat{\imath}}R_{\hat{\jmath}}}
\end{equation*}
for $\hat{\imath},\hat{\jmath}\in \{D,..,9\}$, represent a double T-duality
symmetry mapping IIA string theory to itself. Likewise, the group of
permutations is reduced to $\mathcal{S}_{10-D}$, generated by 
$\{S_{\hat{\imath},\hat{\imath}+1}\}_{\hat{\imath}=D,..,8}$, 
which belong to the $SL(10-D,\Z)$ modular group of the IIA torus.

In $D=1$, this setup naturally extends to the dilaton vector $H_{R}\in \h(E_{10})$. 
The permutation group $\mathcal{S}_{10}$ acts as $H_{R}^{i}\rightarrow H_{R}^{j}$, 
for $i,j=1,..,10$, which corresponds to the dual Weyl transformation: 
$r_{\a}^{\vee}(H_{R})=H_{R}-\langle H_{R},\a \rangle \a^{\vee}$ for 
$\a=\a_{i-2}+\ldots+\a_{j-3} \in \Pi (A_{9})$.

The $\Z_{2}$ factor in expression (\ref{WE10}) on the other hand,
corresponds to a Weyl reflection with respect to the electric coroot:%
\begin{eqnarray*}
r_{8}^{\vee}(H_{R}) &=&H_{R}-\langle H_{R},\a_{8}\rangle \a_{8}^{\vee} \\
&=&\left(H_{R}^{1}+\frac{1}{3}\Delta H,H_{R}^{2}+\frac{1}{3}\Delta
H,..,H_{R}^{7}+\frac{1}{3}\Delta H,H_{R}^{8}-\frac{2}{3}\Delta H,H_{R}^{9}-
\frac{2}{3}\Delta H,H_{R}^{10}-\frac{2}{3}\Delta H\right)\,,
\end{eqnarray*}
with $\Delta H=H_{R}^{8}+H_{R}^{9}+H_{R}^{10}$.

On the generators of $\eten$, the Weyl group will act as 
$\sigma_{\a}=\exp \left[ \frac{\pi i}{2}(E_{\a}+F_{\a})\right]$ 
or alternatively as $\tilde{\sigma}_{\a}=\exp \left[ \frac{\pi}{2}
(E_{\a}-F_{\a})\right] $, $\forall \a \in \Delta_{+}(E_{10})$, 
depending on the choice of real basis. In particular, a $\Z_{4}$ orbifold of M-theory
can be represented in our language by a Weyl reflection, and is thus 
naturally incorporated in the U-duality group.

As mentioned in Section \ref{sec:sing}, from the point of view of its moduli
space, the effect of acting with the subgroup $W(E_{11-D})$ of the U-duality
group on the objects of M-theory on $T^{10}$ will typically be to exchange
instantons which shift fluxes, with instantons that induce topological
changes. On the cosmological billiard, a Weyl transformation will then exchange
the corresponding walls among themselves.

The rest of the U-duality group is given by the Borel generators. These act on
the expectation values $\mathcal{C}_{\a}$, $\a\in \Delta_+(E_{11-D})$, appearing, in particular, as
fluxes in expression (\ref{actInst}). Picking, in a given basis, a root $\b\in \Delta_+(E_{11-D})$, 
its corresponding Borel generator $B_{\b}$ will act on the (infinite) set $\{\mathcal{C}_{\a}\}_{\a\in \Delta_+(E_{11-D})}$ typically as \cite{OberPiol1,Gan}:
\begin{equation}
\label{spectral}
B_{\b}\,:\quad \mathcal{C}_{\b}\rightarrow \mathcal{C}_{\b}+1\,\qquad 
\mathcal{C}_{\c}\rightarrow \mathcal{C}_{\c}+\mathcal{C}_{\c-\b}\,, \text{ if } \c-\b \in \Delta_+(E_{11-D})\,.
\end{equation}
If $\c-\b \notin \Delta_+(E_{11-D})$, then $B_{\b}\,:\; \mathcal{C}_{\c}\rightarrow \mathcal{C}_{\c}$.
The first transformation in eqn.(\ref{spectral}) is the M-theory spectral flow \cite{OberPiol1}, generated
by part of the Borel subalgebra of the arithmetic group $E_{11-D}(\Z)$. Invariance 
under such a unity shift reflects the periodicity of the expectation values of the
fields $\mathcal{A}^i_{\phantom{i}1}$, $C_3$, $\widetilde{C}_6$ and $\widetilde{\mathcal{A}}^i_{\phantom{i}7}$.

\section{Conventions and involutive automorphisms for the real form $\widehat{\so}(8,6)$}\label{AppendixC}

i) \underline{{\it Conventions} for $\dn_7$:} we recall the conventions used in Section 
\ref{AffineT4} to label the basis of simple roots of the finite $\dn_{7}\subset \hat{\dn}_{7}\subset \ginv$ 
Lie algebra for the $T^{5}\times T^{4}/\Z_{n>2}$ orbifold of M-theory: 
\begin{equation}
\b_{1}\equiv \a_{-}\,,\quad \b_{2}\equiv \tilde{\a}\,,\quad
\b_{3}\equiv \a_{+}\,,\quad \b_{4}\equiv \a_{3}\,,\quad
\b_{5}\equiv \a_{2}\,,\quad \b_{6}\equiv \a_{1}\,,\quad
\b_{7}\equiv \c \,.  \label{convD7}
\end{equation}
The affine $\widehat{\dn}_{r}$ will be described by the following Dynkin
diagram:
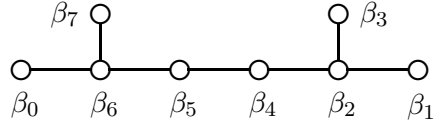
\begin{figure}[!h]
\hspace{2.5cm}
\begin{picture}(200,30)
\thicklines
\put(104,3.5){\line(0,1){14}}
\put(104,21){\circle{6.5}}
\multiput(74,0)(30,0){6}{\circle{6.5}}
\multiput(77,0)(30,0){5}{\line(1,0){23.5}}
\put(194,3.5){\line(0,1){14}}
\put(194,21){\circle{6.5}}
\put(70,-16){$\b_0$} 
\put(85,18){$\b_7$}  
\put(100,-16){$\b_6$}      
\put(130,-16){$\b_5$}       
\put(160,-16){$\b_4$}      
\put(190,-16){$\b_2$}
\put(202,18){$\b_3$}    
\put(220,-17){$\b_1$}    
\end{picture}
\vspace{0.7cm}
\caption{Dynkin diagram of $\hat{\dn}_7$ in the $\b$-basis}
\label{betbas}
\end{figure}
The lexicographic order used in convention (\ref{convD7}) is meant to
naturally extend the $\an_{3}\subset \ginv$ subalgebra appearing for the 
$T^4/\Z_{n>2}$ orbifold in $D=5$. In particular, we define 
$E_{\underline{i\cdots 4123}}\doteq [E_{\underline{i}},\ldots 
[E_{\underline{4}},E_{\a_{-}+\tilde{\a}+\a_{+}}]]\ldots ]$ for $i\geqslant 4$.

For non-simple roots of level 2 in $\b_{2}$, the corresponding ladder
operator is defined by commuting two successive layers of simple root ladder
operators, as, for instance, in: 
\begin{equation*}
E_{\underline{765^{2}4^{2}12^{2}3}}\doteq [E_{\underline{5}},[E_{\underline{4}},
[E_{\underline{2}},E_{\underline{7654123}}]]]\,.
\end{equation*}
This implies in particular the useful relation
\begin{equation*}
\mathcal{N}_{\b_{i},\b_{i-1\cdots j^{2}(j-1)^{2}\cdots
4^{2}12^{2}3}}=\mathcal{N}_{\b_{j},\b_{i\cdots j(j-1)^{2}\cdots
4^{2}12^{2}3}}\,,
\end{equation*}
which, combined with $\mathcal{N}_{-\a ,-\b }=-\mathcal{N}_{\a
,\b }$ and $\mathcal{N}_{\a ,\c -\a }=-\mathcal{N}_{\a
,-\c }$, induces 
\begin{eqnarray*}
\left[ F_{\underline{i}},E_{\underline{i\cdots (j+1)j^{2}\cdots 4^{2}12^{2}3}}\right] &=&
E_{\underline{i-1\cdots (j+1)j^{2}\cdots 4^{2}12^{2}3}}\,, \\
\left[ F_{\underline{j}},E_{\underline{i\cdots (j+1),j^{2}\cdots 4^{2}12^{2}3}}\right] &=&
E_{\underline{i\cdots j+1,j(j-1)^{2}\cdots 4^{2}12^{2}3}}\,.
\end{eqnarray*}
\newline
ii) \underline{{\it The representation $\Gamma $}:} the inner involutive automorphism
written in the form (\ref{U}) acts on elements of the algebra $\dn_{7}$ in the 
representation $\Gamma \{1,0,\ldots ,0\}$ (see~\cite{GTP2}) defined as follows.
For general $r$, let the basis of simple roots $\dn_{r}$
characterized by the Dynkin diagram of Figure~\ref{betbas} be recast in terms of the
orthogonal basis $\varepsilon_{i}$, $i=1,..,r$
\begin{equation}\label{beta1}
\begin{array}{c}
\b_{1} =\varepsilon_{r-1}-\varepsilon_{r}\,,\text{ \ \ \ \ }
\b_{2}=\varepsilon_{r-2}-\varepsilon_{r-1}\,,\text{ \ \ \ \ }
\b_{3}=\varepsilon_{r-1}+\varepsilon_{r}\,,   \\[2pt]
\b_{i} =\varepsilon_{r+1-i}-\varepsilon_{r+2-i}\,,\text{ \ \ }
\forall i=4,\ldots ,r\,. 
\end{array}
\end{equation}
The remaining non-simple roots can be reexpressed as follows: 
for $1\leqslant i<j\leqslant r-3$, we have
\begin{equation}\label{beta2}
\begin{array}{rclcrcl}
\multicolumn{7}{c}{\b_{r+1-i}+\ldots +\b_{r+1-j} =\varepsilon_{i}-\varepsilon_{j+1}\,,} \\[2pt]
\multicolumn{7}{c}{\b_{r+1-i}+\ldots +\b_{4}+\b_{2} = \varepsilon_{i}-\varepsilon_{r-1}\,,} \\[2pt]
\b_{r+1-i}+\ldots +\b_{4}+\b_{2}+\b_{1} &=&\varepsilon_{i}-\varepsilon_{r}\,,
& &\b_{2}+\b_{1}&=&\varepsilon_{r-2}-\varepsilon_{r}\,,   \\[2pt]
\b_{r+1-i}+\ldots +\b_{4}+\b_{3}+\b_{2} &=&\varepsilon_{i}+\varepsilon_{r}\,,
&&\b_{3}+\b_{2}&=&\varepsilon_{r-2}+\varepsilon_{r}\,,  \\[2pt]
\b_{r+1-i}+\ldots +\b_{4}+\b_{3}+\b_{2}+\b_{1}
&=&\varepsilon_{i}+\varepsilon_{r-1}\,, &&
\b_{3}+\b_{2}+\b_{2}&=&\varepsilon_{r-2}+\varepsilon_{r-1}\,,  
\end{array}
\end{equation}
while roots of level 2 in $\b_{2}$ decompose as 
\begin{eqnarray*}
\b_{r+1-i}+\ldots +\b_{r+1-j}+2(\b_{r-j}+\ldots +\b_{4}+\b_{2})+\b_{3}+\b_{1} &=&
\varepsilon_{i}+\varepsilon_{j+1}\,,\text{ \ \ \ }1\leqslant i<j\leqslant r-4\,, \\
\b_{r+1-i}+\ldots +\b_{4}+2\b_{2}+\b_{3}+\b_{1}
&=&\varepsilon_{i}+\varepsilon_{r-2}\,,\text{ \ \ \ }1\leqslant i\leqslant r-4\,.
\end{eqnarray*}
Introducing the elementary matrices $\mathcal{E}_{i,j}$, with components $(\mathcal{E}_{i,j})_{kl}=\d_{ik}\d_{jl}$,
the Cartan subalgebra of $\dn_{r}$ may be cast in the form
\begin{eqnarray*}
\Gamma (H_{\underline{1}}) &=&\frac{1}{\sqrt{r(r-1)}}
\left( \mathcal{E}_{r-1,r-1}-\mathcal{E}_{r,r}+\mathcal{E}_{r+1,r+1}-
\mathcal{E}_{r+2,r+2}\right) \,, \\
\Gamma (H_{\underline{2}}) &=&\frac{1}{\sqrt{r(r-1)}}\left( 
\mathcal{E}_{r-2,r-2}-\mathcal{E}_{r-1,r-1}+\mathcal{E}_{r+2,r+2}-
\mathcal{E}_{r+3,r+3}\right) \,, \\
\Gamma (H_{\underline{3}}) &=&\frac{1}{\sqrt{r(r-1)}}\left( 
\mathcal{E}_{r-1,r-1}+\mathcal{E}_{r,r}-\mathcal{E}_{r+1,r+1}-
\mathcal{E}_{r+2,r+2}\right) \,, \\
\Gamma (H_{\underline{i}}) &=&\frac{1}{\sqrt{r(r-1)}}\left(\mathcal{E}_{r+1-i,r+1-i}-
\mathcal{E}_{r+2-i,r+2-i}+\mathcal{E}_{r-1+i,r-1+i}-\mathcal{E}_{r+i,r+i}\right) 
\,,\text{ \ \ }\forall i=4,\ldots ,r\,.
\end{eqnarray*}
The matrices representing the ladder operators of $\dn_{r}$, and
solving in particular 
$[\Gamma (H_{\underline{i}}),\Gamma (E_{\underline{j}})]=
A_{\underline{ji}}\Gamma (E_{\underline{j}})$, can be determined to be (see \cite{ClarkCorn2})
\begin{equation}\label{Ed7}
\begin{array}{l}
{\ds \Gamma (E_{\varepsilon_{i}-\varepsilon_{j}}) =\frac{1}{\sqrt{r(r-1)}}
\left( \mathcal{E}_{i,j}+(-1)^{i+j+1}\mathcal{E}_{2r+1-j,2r+1-i}\right) \,,}
 \\[3pt]
{\ds \Gamma (E_{\varepsilon_{i}+\varepsilon_{j}}) =\frac{1}{\sqrt{r(r-1)}}
\left(\mathcal{E}_{i,2r+1-j}+(-1)^{i+j+1}\mathcal{E}_{j,2r+1-i}\right) \,. }
\end{array}
\end{equation}
Raising and lowering operators in the basis $\{\b_{i}\}_{i=1,..,r}$ then
readily follow from relations (\ref{beta1}) and (\ref{beta2}) and
expressions (\ref{Ed7}).

Finally, this representation of $\dn_{r}$ preserves the metric $%
G_{D_{r}}=\left( 
\begin{array}{cc}
\text{$\mathbbm{O}$} & g_{D_{r}}^{\top } \\ 
g_{D_{r}} & \text{$\mathbbm{O}$}%
\end{array}%
\right) $, where the off-diagonal blocs are given by 
$g_{D_{r}}=$offdiag$\{1,-1,1,-1,\ldots ,(-1)^{r-1}\}$. It can be checked that 
indeed: $\Gamma(X)^{\top }G_{D_{r}}+G_{D_{r}}\Gamma (X)=0$, for $X\in \dn_{r}$.
\newline
\newline
iii) \underline{Four involutive automorphisms for the real form $\so(8,6)$:} the set 
${\Delta }_{(+1)}$ of roots generating the
maximal compact subalgebra of the real form $\so(8,6)$ appearing
in Section \ref{AffineT4} is determined for the four involutive
automorphisms (\ref{tablAuto}). Since $\dim {\Delta}_{+}(\dn_{7})=42$, 
and since all four cases have $\dim {\Delta}_{(+1)}=18$, the
corresponding involutive automorphisms all have signature $\sigma=5$, and thus
determine isomorphic real forms, equivalent to $\mathfrak{so}(8,6)$. This
construction lifts to the affine extension $\hat{\dn}_{7}$ through
the automorphism (\ref{U}) building the Cartan decomposition (\ref{kD7+})
and (\ref{pD7+}).

Herebelow, we give the set of roots $\Delta_{(+1)}$ for the four cases
(\ref{tablAuto}) explicitly. We remind the reader that these four involutive
automorphisms all have $e^{\b'_{2}(\overline{H})}=+1$ and 
$e^{\b'_{i\neq 2,4,6}(\overline{H})}=-1$. Moreover, the set 
$\Delta_{(+1)}$ generating the non-compact vector space $\p$ 
(\ref{pD7+}) can be deduced from $\Delta_{(-1)}=\Delta_{+}\backslash \Delta_{(+1)}$, 
where $\Delta_{+}$ is obtained from the system (\ref{beta1}) and (\ref{beta2})
by setting $r=7$. In this case obviously $\dim {\Delta}_{(-1)}=24$.

The first involutive automorphism defined by $e^{\b'_{4}(\overline{H})}=
e^{\b'_{6}(\overline{H})}=+1$ has
\begin{eqnarray}
\Delta_{(+1)}=\big\{\b'_{6},\b'_{4},\b'_{2},\b'_{42},\b'_{765},\b'_{123},
\b'_{7564},\b'_{5412},\b'_{5423},\b'_{4123},\b'_{76542},\b'_{65412}, && \notag \\
\b'_{65423},\b'_{412^{2}3},\b'_{7654123},\b'_{765412^{2}3},
\b'_{7654^{2}12^{2}3},\b'_{65^{2}4^{2}12^{2}3}\big\}\,.
&& \label{Delta1}
\end{eqnarray}
The second, defined by $e^{\b_{4}^{\prime }(\overline{H})}=-e^{\b'_{6}(\overline{H})}=+1$ has
\begin{eqnarray}\notag
\Delta_{(+1)}=\big\{\b'_{4},\b'_{2},\b'_{76},\b'_{65},\b'_{42},\b'_{654},\b'_{123},
\b'_{6542},\b'_{5412},\b'_{5423},\b'_{4123},\b'_{412^{2}3}, && \\
\b'_{765412},\b'_{765423},b'_{654123},\b'_{65412^{2}3},\b'_{654^{2}12^{2}3},
\b'_{765^{2}4^{2}12^{2}3}\big\}\,. && \label{Delta2}
\end{eqnarray}
The third, defined by $e^{\b'_{4}(\overline{H})}=-e^{\b'_{6}(\overline{H})}=-1$ has
\begin{eqnarray}\notag
\Delta_{(+1)}=\big\{\b'_{6},\b'_{2},\b'_{54},\b'_{765},\b'_{654},
\b'_{542},\b'_{412},\b'_{423},\b'_{123},\b'_{6542},\b'_{54123},\b'_{765412}, && \\
\b'_{765423},\b'_{654123},\b'_{5412^{2}3},\b'_{65412^{2}3},\b'_{7654^{2}12^{2}3},
\b'_{65^{2}4^{2}12^{2}3}\big\}\,.
&& \label{Delta3}
\end{eqnarray}
The fourth, defined by $e^{\b'_{4}(\overline{H})}=e^{\b'_{6}(\overline{H})}=-1$ has
\begin{eqnarray}\notag
{\Delta }_{(+1)}=\big\{\b'_{2},\b'_{76},\b'_{65},\b'_{54},\b'_{542},\b'_{412},\b'_{423},
\b'_{123},\b'_{7654},\b'_{76542},\b'_{65412},\b'_{65423}, && \\
\b'_{54123},\b'_{5412^{2}3},\b'_{7654123},\b'_{765412^{2}3},\b'_{654^{2}12^{2}3},
\b'_{765^{2}4^{2}12^{2}3}\big\}\,. 
&&\label{Delta4}
\end{eqnarray}
The four of them lead as expected to $\dim {\Delta}_{(+1)}=18$.

\bibliographystyle{maison}
\bibliography{orbibib}

\end{document}